\documentclass[11pt]{article}
\pdfoutput=1
\usepackage[margin=1.0in]{geometry}
\usepackage{amsmath,amssymb,graphicx}
\usepackage{hyperref}
\usepackage{slashed}

\usepackage[T1]{fontenc}

\usepackage[utf8]{inputenc}

\usepackage[greek,english]{babel}

\usepackage{framed}

\usepackage[font=small,labelfont=bf]{caption}
\usepackage{cite}

\newcommand{\iu}{{\mathrm i}}
\newcommand{\E}{{\mathrm e}}
\newcommand{\cpi}{\text{\greektext p}}

\newcommand{\dif}[1]{\ensuremath{\operatorname{d}\!{#1}}}

\newcommand{\Uone}{{\mathrm{U(1)}}}
\newcommand{\SU}{{\mathrm{SU}}}
\newcommand{\SUN}{{\mathrm{SU}(N)}}
\newcommand{\BminusL}{{\textsc{B} - \textsc{L}}}
\newcommand{\BplusL}{{\textsc{B} + \textsc{L}}}

\DeclareMathOperator{\sgn}{sgn}

\newcommand{\bQ}{{\overline{Q}}}

\newcommand{\be}{\begin{equation}}
\newcommand{\ee}{\end{equation}}

\usepackage{xcolor}
\usepackage{bm}

\title{\bf \huge  TASI Lectures:\\ (No) Global Symmetries to Axion Physics}
\author{Matthew Reece\\
{\small \color{gray} \texttt{mreece~(@g.harvard.edu)}}\\
{\small Department of Physics, Harvard University, Cambridge, MA, 02138}}

\begin{document}
\maketitle

\begin{abstract}
These notes are an expanded version of lectures given at the 2022 TASI summer school in Boulder, Colorado. One goal of these lecture notes is to (partially) bridge the gap between what one learns in typical introductory quantum field theory classes and what one needs to understand to follow modern developments in particle theory beyond the Standard Model. Topics covered include global and gauge symmetries, charge quantization, instantons, chiral anomalies, the Strong CP problem, axion models from 4d and from higher dimensions, the expected absence of global symmetries in quantum gravity, and some phenomenological implications thereof. If these topics seem to be at best loosely related, I hope that reading the notes will convince you otherwise. Recent developments in quantum field theory have shown that ordinary gauge theories exhibit a much wider range of (generalized) global symmetries than were previously understood, while recent work in quantum gravity has provided sharper arguments that global symmetries should not exist. This sets up an interesting tension, the resolution of which can have implications for particle physics in the real world. Axion physics is one setting in which these ideas can guide phenomenology. The TASI audience comprised particle phenomenology students whom I hoped to convince of the importance of learning more about quantum field theory and quantum gravity. These notes may also be of interest to formal theory readers seeking closer connections to real-world particle physics.
\end{abstract}

\tableofcontents

\newpage

\section{Introduction}

These notes are extended version of lectures I gave at the TASI 2022 summer school. They cover aspects of symmetries in quantum field theory and quantum gravity, and applications to particle physics. The central phenomenological topic of these lectures is the physics of axions and their role in solving the Strong CP problem. I emphasize some topics that are not commonly emphasized in phenomenological treatments, related to charge quantization, the axion-gauge field coupling as a special case of a Chern-Simons term, and the viewpoint that an axion is a kind of gauge field (and indeed originates, in many theories, as a mode of a higher-dimensional gauge field). 

There are a number of review articles on the Strong CP problem and axion physics that get right to the details, without the lengthy review of QFT and symmetries I have here. There are reviews of generalized symmetries and the absence of such symmetries in quantum gravity, with little to say about axions. Why am I combining these topics?

The TASI audience in 2022 comprised mostly phenomenologically-oriented students. I think that it is important that particle theorists, even those who are most oriented toward experiments, have a thorough understanding of conceptual aspects of quantum field theory. Part of my aim with these lectures, then, is to try to equip hep-ph physicists with more of a common language to understand recent developments in hep-th. I hope that the lecture notes will also be of interest to hep-th readers who want to learn more about phenomenological applications.

The landscape of high-energy theory looks rather different today than it did when I was a student at TASI, in 2006. At that time, my sense is that there was a higher degree of uniformity of interests in the field. The LHC was due to turn on in the near future, and there was widespread interest in how electroweak symmetry was broken, whether the Higgs boson really existed, and whether superpartners might be found near the TeV scale. Even a number of string theorists whose careers had mostly been focused on formal topics were briefly entranced by collider physics. Since the LHC confirmed that a Standard Model-like Higgs boson exists, but has (so far) found no evidence of physics beyond the Standard Model, the field has fragmented. Astroparticle physics and cosmology have become more central, and a wider range of smaller-scale precision or high-intensity experiments are being pursued. This is healthy, as we have no way of knowing where the first discovery of new physics might arise. On the other hand, this wider range of phenomenological activity sometimes means that today's hep-ph students learn less quantum field theory than their predecessors did. Sometimes, this leads to misguided theoretical claims, and sometimes these even lead to misguided experimental proposals. It is important that we continue to ground particle theory in a solid understanding of quantum field theory.

Axion physics is ideally situated at the intersection of particle phenomenology, fundamental theory, and experiment. Axions have a longstanding bottom-up phenomenological motivation as a solution of the Strong CP problem, and naturally provide a compelling dark matter candidate. Axion dark matter will be a major target of experimental investigation in the coming decade. Furthermore, the theory of axions is very rich. They are intimately connected to the chiral anomaly. Their couplings to gauge fields are a type of Chern-Simons term. Axions and axion-like particles appear ubiquitously in string theory constructions of particle physics models, and there are reasons to think this isn't a ``lamppost effect,'' but that their existence is required by deep principles of quantum gravity. These are all positive reasons to invest some time exploring axion physics and how it connects to various important threads running through modern quantum field theory and quantum gravity. From a different viewpoint, we should situate axion physics within this larger field theoretic context because it helps us get the physics right. In some cases, axion physics can be quite subtle, and it is important that as theorists we do not mislead experimentalists into searching for effects that are simply theoretically inconsistent.

Most importantly, I think that this is an area where the best is yet to come. I am optimistic that the future will bring experimental discoveries and deeper theoretical understanding that will allow us to infer aspects of physics at very high energy scales from such experiments. I hope that these lecture notes will set some readers on a path to participate in such exciting future developments.

\subsection{Outline}

These lecture notes are divided into five parts. There were only four in-person lectures at TASI. Parts four and five of these notes contain a significantly expanded treatment of the material from the fourth lecture. Throughout the notes, there are more details and added topics that could not be included in the lectures due to time constraints. 

Below is a brief outline of the topics of the five parts, together with suggested reading for those wanting an alternative introduction or more depth on the topic.  Many more references are given throughout the text of these notes, although this is not a comprehensive review article and does not aim to cite or survey the entirety of the literature on these topics.

\begin{itemize}
\item \hyperref[sec:lectureone]{Part One}. 
In the first part, we will discuss global and gauge symmetries, focusing especially on the group $\Uone$ and the quantization of charge and flux. Before diving into the physics, we provide a brief review of the notation of differential forms. Although symmetries are discussed in standard quantum field theory textbooks, many of the aspects treated here, including symmetry operators and charge quantization, are usually not. (We return to the topic of generalized symmetries and symmetry operators in Part Four.)

Further reading: Our viewpoint on symmetries is based on ``Generalized Global Symmetries'' by Gaiotto, Kapustin, Seiberg, and Willett, especially sections 2 through 4.2~\cite{Gaiotto:2014kfa}. Although many of the detailed examples discussed in that paper draw on formal QFT topics that may be unfamiliar to a particle phenomenology readership, some of the exposition of the main ideas should be readable. A very recent pedagogical review of higher-form symmetries is~\cite{Gomes:2023ahz} by Pedro R.~S.~Gomes. Another recent review of generalized global symmetries is~\cite{McGreevy:2022oyu} by John McGreevy. Although it focuses on applications to condensed matter physics, it uses the language of quantum field theory and should be accessible to particle theorists. A recent Snowmass white paper~\cite{Cordova:2022ruw} gives a (very) brief introduction to this topic and extensive pointers to recent related literature.

\item \hyperref[sec:lecturetwo]{Part Two}. 
The second part begins with an exposition of the quantization of instanton number, i.e., of the integral of $\mathrm{tr}(F \wedge F)$. This includes a discussion of the classical BPST instanton solution to the Yang-Mills equations. We then turn to the chiral anomaly, explain the conceptual difference between ABJ and 't Hooft anomalies, and discuss how the path integral measure changes when fermion fields are rephased. We explain a conclusion that is important for axion physics, namely that the couplings of a periodic scalar field to $F \wedge F$ are quantized (an example of a Chern-Simons term). This section concludes with a rapid overview of Chern-Simons terms in general.

Further reading: Instantons (and other semiclassical physics) are discussed extensively in textbooks by Shifman~\cite{Shifman:2022shi} and by Erick Weinberg~\cite{Weinberg:2012pjx}. For the chiral anomaly, see sections 19.1 to 19.4 of the textbook by Peskin and Schroeder~\cite{Peskin:1995ev}, chapter 20 of the recent textbook by Fradkin~\cite{fradkin2021quantum}, and TASI lectures on anomalies by Jeff Harvey~\cite{Harvey:2005it}.

\item \hyperref[sec:lecturethree]{Part Three}. The third part is an introduction to the Strong CP problem and models of axions. We will see how the $\theta(x) \mathrm{tr}(G \wedge G)$ coupling solves the Strong CP problem, and how it can arise in different types of models, including classic 4d axion models (KSVZ and DFSZ) and in qualitatively different models in which the axion arises as a mode of a higher-dimensional gauge field. In my opinion, the latter class of models is strongly favored both by bottom-up considerations (as a solution to the axion quality problem) and top-down considerations (both from examples in string theory, and from quantum gravity arguments covered in later parts).

Further reading: There are earlier TASI lectures on the Strong CP problem by Michael Dine~\cite{Dine:2000cj} and on the Strong CP problem and axions by Anson Hook~\cite{Hook:2018dlk}. Outside the TASI context, a very good review article is~\cite{Kim:2008hd}.   To fully understand the derivation of the axion potential, you need to know about the chiral Lagrangian for pion physics, which is reviewed in (for example)~\cite{Georgi:weak, Leutwyler:1994fi, Manohar:1996cq}. Ben Safdi's lectures from this TASI school also discuss axions, with an emphasis on axion dark matter and its consequences for astrophysics and cosmology~\cite{Safdi:2023wsk}. I do not explore in much detail the cosmology of axions (see~\cite{Marsh:2015xka} for more) or experimental searches for them (see~\cite{Graham:2015ouw,Jaeckel:2022kwg}).

\item \hyperref[sec:lecturefour]{Part Four}. The fourth part introduces the concept that there are no fundamental global symmetries. This is a longstanding idea about quantum gravity, which has also served as an aesthetic guide to model-building in quantum field theory: it is better to find an accidental global symmetry (one which simply cannot be violated by low-dimension operators) than to postulate a nearly unbroken global symmetry without further explanation. In recent years, arguments against global symmetries in quantum gravity have been put on a stronger foundation, and in some cases we can quantify the extent to which we expect a symmetry to be broken. The Weak Gravity Conjecture (WGC) is one statement along these lines, which has itself been sharpened into more useful conjectures. Recent years have also seen significant generalizations of the concept of global symmetry in quantum field theory. Combined with the expectation that such symmetries are absent in quantum gravity, this sets up a tension: how does quantum gravity manage to break (or gauge) all the candidate symmetries?

Many of the topics discussed in this part fall under the rubric of the ``Swampland program,'' which searches for universal features of quantum gravity theories and obstructions to embedding effective field theories in quantum gravity~\cite{Vafa:2005ui}. A global symmetry is one such obstruction, an idea that was widely explored for decades before the term ``Swampland'' was coined. In these lectures I will stick to what I view as the most well-established parts of the Swampland program.

Further reading: The Harlow-Ooguri holographic argument against global symmetries is a long paper, but along the way reviews many important aspects of physics related to these lectures in great detail~\cite{Harlow:2018tng}. (You might at least read the short version,~\cite{Harlow:2018jwu}, though it omits much of the useful pedagogical content.) For the Weak Gravity Conjecture, you should read the original paper~\cite{Arkani-Hamed:2006emk}. More recent work on the WGC is thoroughly reviewed in a recent article that I wrote with Daniel Harlow, Ben Heidenreich, and Tom Rudelius~\cite{Harlow:2022gzl}. 

Readers wishing to learn about other aspects of the Swampland program beyond those discussed here may be interested in the reviews~\cite{Palti:2019pca, Agmon:2022thq}.

\item \hyperref[sec:lecturefive]{Part Five}. The fifth part explores the consequences of no global symmetries for particle physics. The quantum gravity perspective can shed light on various questions, such as whether the photon has a mass, and potential origins of neutrino masses. In my opinion, the Strong CP problem is currently an especially exciting place to look for a confrontation between quantum gravity principles and phenomenology. I discuss an emerging picture of why axion-like fields have a crucial role to play in quantum gravity, which could inform the search for axions and axion-like particles in the real world.

Further reading: There is a recent Snowmass white paper on phenomenological implications of quantum gravity that I wrote with Patrick Draper and Isabel Garcia Garcia~\cite{Draper:2022pvk}.

\end{itemize}

David Tong's lecture notes on gauge theory~\cite{Tong:GT} are excellent and have substantial overlap with a number of topics discussed here, though in a different enough order that it's hard to point to specific places to dip into them. I highly recommend finding the time to read them all.

\subsection*{Acknowledgments} 

First, I thank the scientific organizers of the 2022 TASI summer school (JiJi Fan, Stefania Gori, and Lian-Tao Wang) for inviting me to give these lectures and providing the impetus for writing up these notes, and the TASI local organizers (Tom DeGrand, Oliver DeWolfe, and Ethan Neil) for providing an excellent venue and ensuring that the school ran smoothly. I also thank the TASI students, too many to name individually, who asked many insightful questions and stopped me whenever I wasn't being clear or making sense. This has been a great help in figuring out what to say, and how to say it, in this written form of the lectures. Some of the material in these notes has similarly been tested on colloquium audiences at the University of Chicago, Caltech, and the IFT in Madrid, who have provided useful feedback. I thank Daniel Aloni, Liam Fitzpatrick, Ben Heidenreich, Jake McNamara, and Tom Rudelius for feedback on the draft of these lecture notes.

I've learned much of the material I present here over the years from many collaborators with whom I've worked on axion physics, the Weak Gravity Conjecture, and the absence of global symmetries in quantum gravity. In alphabetical order: Prateek Agrawal, Manuel Buen-Abad, JiJi Fan, Katie Fraser, Daniel Harlow, Ben Heidenreich, Jake McNamara, Miguel Montero, Tom Rudelius, John Stout, Chen Sun, Irene Valenzuela, and Lian-Tao Wang. Others who have taught me about some of the material discussed here include Nima Arkani-Hamed, Tom Banks, Cliff Cheung, Kiwoon Choi, Clay C\'ordova, Patrick Draper, Thomas Dumitrescu, Isabel Garcia Garcia, Grant Remmen, Shu-Heng Shao, Cumrun Vafa, and Sasha Zhiboedov. I'm sure that I'm unintentionally omitting others who deserve thanks. These lecture notes are my particular way of deconstructing and re-assembling a large body of collective knowledge, and I hope that others will mine them for their own pedagogical writing in the future.

My work is partially supported by the DOE grant DE-SC0013607, the Alfred P.~Sloan Foundation Grant No.~G-2019-12504, and the NASA Grant 80NSSC20K0506.

\newpage

\section*{\Large Part One: Global and Gauge Symmetries, Charge Quantization}
\label{sec:lectureone}
\addcontentsline{toc}{section}{\nameref{sec:lectureone}}

\section{Differential forms: notation}
\label{sec:differentialforms}

{\noindent \em Readers who are comfortable with the concept and notation of differential forms should be able to freely skip this section (but may want to glance at the actions in~\S\ref{subsec:differentialformexamples} to be sure they are familiar).} 

\medskip

In studying gauge theory, it is useful to use the notation of  {\em differential forms}, which I will briefly review in this section. Differential forms are objects that can be integrated over $p$-dimensional manifolds. Essentially, a differential $p$-form is a $p$-index antisymmetric tensor. However, many formulas are simpler and clearer to think about when written in the language of differential forms instead of tensors. For example, this notation can make it easier to think about whether terms in an action are topological, meaning that they do not depend on the metric tensor that we use to measure distances. Topological terms are essential to axion physics, which is one reason why we will make extensive use of differential form notation in these lectures. 

\subsection{What is a differential form?}

Let's start with one example to illustrate the general point. A gauge field $A_\mu$ can be integrated along the worldline $\gamma$ of a charged particle (i.e., its path through spacetime). This integral might be written as $\int_\gamma A_\mu \mathrm{d}x^\mu$. What does this mean? One approach is to parametrize the curve $\gamma$ in terms of a function from a parameter $\lambda$ to spacetime, $x^\mu(\lambda)$. Then we could compute  $\int_\gamma A_\mu \mathrm{d}x^\mu$ as $\int_{-\infty}^\infty \mathrm{d}\lambda\,\frac{\mathrm{d}x^\mu}{\mathrm{d}\lambda} A_\mu$. However, because the answer is independent of the specific choice of parametrization (we could replace $\lambda$ by a monotonic function $\mu(\lambda)$, and get  the same result from an integral over $\mu$), we write the integral simply as $\int_\gamma A_\mu \mathrm{d}x^\mu$. Differential form notation takes this a step further, writing the integral simply as $\int_\gamma A$, where the (dimensionless) object $A$ is a {\em differential 1-form}, defined as $A_\mu \mathrm{d}x^\mu$. It packages the field to be integrated together with the differential $\mathrm{d}x^\mu$ that tells us about the integration measure.

A second example is a gauge field strength $F_{\mu \nu}$, an antisymmetric rank-2 tensor that we can integrate over a {\em surface} to calculate a flux. In this case, we define the differential form $F$ (again, a  dimensionless object) as
\begin{equation} \label{eq:Fform}
F = \frac{1}{2} F_{\mu \nu} \dif x^\mu \wedge \dif x^\nu.
\end{equation}
The {\em wedge product} $\wedge$ is an antisymmetrized tensor product,
\begin{equation}
\dif x^\mu \wedge \dif x^\nu = \dif x^\mu \otimes \dif x^\nu - \dif x^\nu \otimes \dif x^\mu.
\end{equation}
The fact that a given term, like $\dif x^1 \otimes \dif x^2$, appears twice here (in different orderings) accounts for the normalization factor of $1/2$ in the definition of $F$ relative to $F_{\mu \nu}$. Why should a differential form be {\em antisymmetric}? It's because the area element on a surface involves each of the {\em independent} coordinates, e.g., $\dif x \dif y$ in the plane; integrating an object $\dif x \dif x$ is meaningless. The antisymmetrization ensures that a $p$-dimensional integrand really involves $p$ {\em different} directions. Rephrasing our field strength example may help: think about calculating a magnetic flux through a surface. You may have encountered this in introductory physics in the form of an integral of a vector $\vec B$ dotted into a unit normal vector to a surface, $\int {\vec B} \cdot {\hat n} \dif S$, with $\dif S$ an area element on the surface. However, the antisymmetrization was lurking here in the fact that to obtain a vector like $\vec B$ from the field strength $F$, we use an antisymmetric symbol, $B^k = \frac{1}{2} \epsilon^{ijk} F_{ij}$. The combination $\epsilon^{ijk} {\hat n}_k \dif S$ is then the area element $\dif x^i \wedge \dif x^j$.

A differential form of degree $p$, or {\em a differential $p$-form} (or even just {\em $p$-form}, for short) is a simple generalization of this idea, with more indices and correspondingly more differentials:
\begin{equation}  \label{eq:pform}
\omega_p = \frac{1}{p!} \omega_{\mu_1 \cdots \mu_p} \dif x^{\mu_1} \wedge \dif x^{\mu_2} \wedge \cdots \wedge \dif x^{\mu_p},
\end{equation}
where $\omega_{\mu_1 \cdots \mu_p}$ are components of a $p$-index antisymmetric tensor. The subscript $p$ in the name $\omega_p$ is just a reminder that this is a $p$-form; it is not an index. Here the wedge product of the forms is a signed sum over permutations,
\begin{equation} \label{eq:wedgeproduct}
\dif x^{\mu_1} \wedge \dif x^{\mu_2} \wedge \cdots \wedge \dif x^{\mu_p} = \sum_{\sigma \in S_p} \epsilon(\sigma) \dif x^{\sigma(\mu_1)} \otimes \dif x^{\sigma(\mu_2)} \otimes \cdots \otimes \dif x^{\sigma(\mu_p)},
\end{equation}
where $\sigma$ denotes a permutation of the $p$ integers $\mu_1, \ldots \mu_p$ and $\epsilon(\sigma)$ is its {\em sign}, i.e., $1$ if it can be accomplished by an even number of swaps of two integers and $-1$ if it involves an odd number of swaps.  If this abstract notation about permutations is unfamiliar, one example might help to illustrate the idea:
\begin{align}
\dif x \wedge \dif y \wedge \dif z &= \dif x \otimes \dif y \otimes \dif z - \dif x \otimes \dif z \otimes \dif y + \dif z \otimes \dif x \otimes \dif y \\
& - \dif z \otimes \dif y \otimes \dif x + \dif y \otimes \dif z \otimes \dif x - \dif y \otimes \dif x \otimes \dif z.
\end{align}
The $1/p!$ in~\eqref{eq:pform} is a normalization factor correcting for the counting of the different permutations in which the same differential elements can appear. It's worth explicitly mentioning one special case, for clarity: a differential $0$-form is the degenerate case without any differentials $\dif x$; in other words, it's just a function $f(x)$.

The wedge product of differential forms, in general, is determined by applying linearity together with the definition~\eqref{eq:wedgeproduct} of the wedge product of the differential elements. Notice that changing the order of a wedge product depends on the degree of the forms. For instance, two 1-forms $A_1$ and $B_1$ have $A_1 \wedge B_1 = -B_1 \wedge A_1$, but a 1-form $A_1$ and a 2-form $F_2$ have $A_1 \wedge F_2 = F_2 \wedge A_1$. The reason is that, in the latter case, we have differentials appearing in the form $\dif x^\mu \wedge (\dif x^\nu \wedge \dif x^\rho) = -\dif x^\nu \wedge \dif x^\mu \wedge \dif x^\rho = +(\dif x^\nu \wedge \dif x^\rho) \wedge \dif x^\mu$. In general, any $p$-form with {\em even} $p$ will commute with other forms under the wedge product, because we move each differential past an even number of others and do not acquire a net sign. This can be summarized with the rule: given a $p$-form $\omega_p$ and a $q$-form $\eta_q$, we have
\begin{equation}
\omega_p \wedge \eta_q = (-1)^{p q}\, \eta_q \wedge \omega_p.
\end{equation}
This holds because $(-1)^{pq}$ is equal to $-1$ if and only if {\em both} $p$ and $q$ are odd.

\subsection{The exterior derivative}
\label{subsec:exteriorderiv}

The exterior derivative $\mathrm{d}$ is an operation that maps $p$-forms to $(p+1)$-forms. Starting with a 0-form, the components of the exterior derivative are those of the {\em gradient}:
\begin{equation}
\dif f = \frac{\partial f}{\partial x^\mu} \dif x^\mu,
\end{equation}
with the sum over $\mu$ implicit as usual. The exterior derivative of forms of higher degree follows a similar pattern: given a $p$-form $\omega$ as in~\eqref{eq:pform}, its exterior derivative is the $(p+1)$-form
\begin{equation}
\dif \omega = \frac{\partial \omega_{\mu_1 \cdots \mu_p}}{\partial x^\mu}\, \dif x^\mu \wedge \dif x^{\mu_1} \wedge \dif x^{\mu_2} \wedge \cdots \wedge \dif x^{\mu_p}.
\end{equation}
In words, $\dif \omega$ is an antisymmetrized derivative whose components consist of all of the derivatives of the components of $\omega$.

In particular, if $A = A_\mu \dif x^\mu$, the exterior derivative is
\begin{equation}
\dif A = \frac{\partial A_\nu}{\partial x^\mu} \dif x^\mu \wedge \dif x^\nu = F,
\end{equation}
where $F$, as in~\eqref{eq:Fform}, has components $F_{\mu \nu} = \partial_\mu A_\nu - \partial_\nu A_\mu$. This is just the usual field strength of a vector field. 

\medskip
\noindent\centerline{\rule{\textwidth}{0.5pt}}
\noindent {\em Exercise:} In 3-dimensional space, we can use the antisymmetric symbol $\epsilon^{ijk}$ to convert a 2-form like $F$ to a vector. Check that the resulting object can be identified with the {\em curl} ${\vec \nabla} \times {\vec A}$.\\
\noindent\centerline{\rule[6.0pt]{\textwidth}{0.5pt}}

The exterior derivative obeys a version of the usual product rule for derivatives, but potentially with an extra minus sign due to its antisymmetric property:
\begin{equation}
\dif(\omega_p \wedge \eta_q) = (\dif \omega_p) \wedge \eta_q + (-1)^p \omega_p \wedge (\dif \eta_q).
\end{equation}
In particular, you should be careful about this sign when integrating by parts to derive equations of motion!

The antisymmetry property further implies that $\mathrm{d}^2 = 0$, i.e., 
\begin{equation} \label{eq:dsquared}
\dif (\dif \omega_p) = 0, 
\end{equation}
for {\em any} $p$-form $\omega_p$. Whenever $\dif \omega_p = 0$, we say that $\omega_p$ is a {\em closed} $p$-form. We will see below that such forms should be thought of as {\em conserved currents}. When there exists a $(p-1)$-form $\eta_{p-1}$ such that $\omega_p = \dif \eta_{p-1}$, we say that $\omega_p$ is an {\em exact} $p$-form. Because of~\eqref{eq:dsquared}, all exact forms are closed, but not all closed forms are exact. The space of closed forms modulo exact forms is known as {\em de Rham cohomology}, and is a useful tool for analyzing the topology of the spaces on which differential forms are defined~\cite{BottTu}.

We urge readers to be somewhat cautious in interpreting~\eqref{eq:dsquared} in physics: sometimes the ``forms'' that we encounter are not true forms when they are gauge-dependent quantities, and as a result this equation fails to hold. For example, given a vector potential  $A$, we have a field strength $F = \dif A$; the above equation leads us to expect that $\dif F = 0$. This result, known as the {\em Bianchi identity}, can be violated in physics when the gauge field becomes ill-defined at a singular point, like the core of a magnetic monopole. Furthermore, despite the equation $F = \dif A$, we should not think of $F$ as an exact form in general. The reason is that there is no globally-defined 1-form $A$; we must patch together definitions of $A$ in different coordinate charts that are related by nontrivial gauge transformations on overlaps. As a result, $A$ is not really a 1-form (formally, it is a connection on a $\Uone$ principal bundle); $F$, however, is an honest 2-form that is globally defined (because electromagnetic fields, unlike gauge potentials, are physical observables). A classic example of this, the Dirac magnetic monopole, is discussed below in \S\ref{subsec:monopole}.

\subsection{Integration and Stokes's theorem}

Given a differential form $\omega_p$, one can integrate it over a closed $p$-dimensional surface. In fact, there is a small subtlety here: we have to pick an {\em oriented} surface. For example, if we want to integrate the 1-form $x \dif x$ over the manifold $(0,1) \subset \mathbb{R}$, we can choose the conventional orientation and calculate (as in introductory calculus!) that $\int_0^1 x \dif x = \frac{1}{2}$, or we can choose the opposite orientation and calculate $\int_1^0 x \dif x = - \frac{1}{2}$. The results of integration with different orientations differ by a sign.

\medskip

Let me make a brief aside here, which will not be relevant for the remainder of these lectures but which actually does have physical importance. The need for an orientation may bother you, and it should. For instance, the surface area of a M\"obius strip is a perfectly well-defined quantity, even though the space is non-orientable. Why should we need an orientation to compute an integral? In fact, we don't. We can integrate an object known variously as a differential {\em pseudo-form}~\cite{Frankel} or a {\em twisted} differential form~\cite{BottTu} (more verbosely, a form twisted by the orientation line bundle). You have encountered such objects in physics before, even if the terminology is new to you. An example is a pseudoscalar. It is defined only up to a sign; when swapping  the orientation of our coordinates (as in a parity transformation), we also change the sign of the field. Pseudo-forms can be integrated over non-orientable manifolds by picking a set of coordinate charts. On any given coordinate chart, we choose an orientation. We can choose whichever one we like, because if we change our choice, the pseudo-form we are integrating changes sign so that the integral does not. Then we patch together the integrals on different charts using a partition of unity (i.e., a set of functions defined on overlapping charts such that they add up to one everywhere when summed over all the charts). From this, we see that the volume element of a $p$-dimensional space is most naturally a {\em pseudo}-$p$-form, rather than an ordinary $p$-form.

A quantum field theory that can be defined without reference to an orientation is said to have {\em parity} as a global symmetry. Parity can also be a gauge symmetry, in the context of quantum gravity, which means that the path integral includes a sum over both orientable and non-orientable spacetimes. For more on this topic, see~\cite{McNamara:2022lrw}.

\medskip

Differential forms allow a simple statement of the generalized Stokes's theorem. Given a $p$-dimensional manifold $M$ with boundary $\partial M$ (the notation $\partial$ is commonly used in mathematics for boundaries, not just for derivatives!), and a $(p-1)$-form $\omega_{p-1}$, 
\begin{equation}
\int_{\partial M} \omega_{p-1} = \int_M \dif \omega_{p-1}.
\end{equation}
(A little fine print: $M$ has an orientation and this {\em induces} an orientation on $\partial M$, so that the two sides of the equation are computed with compatible orientations.) This one statement encodes many familiar statements in vector calculus: Green's theorem, Stokes's theorem, the divergence theorem, and even Cauchy's integral formula can all be understood as special cases.

\medskip
\noindent\centerline{\rule{\textwidth}{0.5pt}}
{\noindent \em Exercise: }work out how each of the aforementioned classic theorems can be understood in terms of the generalized Stokes's theorem.\\
\noindent\centerline{\rule[6.0pt]{\textwidth}{0.5pt}}

Notice that our discussion of integration has not referred to a metric on the manifold that we integrate over. Given a differential form, you can integrate it without needing to be told anything about a metric. This is one reason why differential forms play a prominent role in the study of topological effects in field theory. Of course, in order to find the components of a differential form in the first place, you might need to use the metric. For example, a $d$-dimensional manifold equipped with a metric tensor $g_{\mu \nu} \dif x^\mu \otimes \dif x^\nu$ (which is a {\em symmetric} tensor and hence {\em not} a differential form!) has a {\em volume form},
\begin{equation} \label{eq:volform}
\mathrm{vol}_d = \begin{cases} \sqrt{+\det g} \dif x^1 \wedge \cdots \wedge \dif x^d, & \text{Euclidean}; \\
\sqrt{-\det g} \dif x^0 \wedge \cdots \wedge \dif x^{d-1}, & \text{Minkowski}.
\end{cases}
\end{equation}
We will also write simply $|g|$ for $|{\det g}|$. In either case, we will also the shorthand $\mathrm{vol}_d = \sqrt{|g|} \dif{^{d}x}$. This is the familiar measure we integrate against when defining actions in $d$-dimensional spacetime. The $d$-form $\mathrm{vol}_d$ is  sometimes referred to as the Levi-Civita tensor. (In general, a $d$-form on a $d$-dimensional manifold is sometimes called a ``top form.'')

Note that in the Minkowski case we always write the time coordinate $x^0$ first. This is a choice of orientation, with the convention that $\int \sqrt{|g|} \dif{^{d}x}$ is always positive.

\subsection{The Hodge star}

The {\em Hodge star} is an operation that, on $d$-dimensional spaces (or spacetimes), takes a $p$-form to a $(d-p)$-form. It is defined only for manifolds equipped with a metric tensor, so it is geometric, rather than purely topological. In terms of the antisymmetric tensor components, you can think of this operation as essentially contracting a tensor with the Levi-Civita tensor with all $d$ indices raised. In terms of the differentials appearing in a given form, you can think of the Hodge star as replacing all those that appear in a given term with all the {\em others}; e.g., in flat 3-dimensional space, $\dif x$ is replaced by $\dif y \wedge \dif z$.

Specifically, the Hodge star is defined by:
\begin{equation}
\star(\dif x^{\mu_1} \wedge \cdots \wedge \dif x^{\mu_p}) = \frac{1}{(d-p)!} \sqrt{|g|} \epsilon_{\nu_1 \cdots \nu_d} g^{\mu_1 \nu_1} \cdots g^{\mu_p \nu_p} \dif x^{\nu_{p+1}} \cdots \dif x^{\nu_d},
\end{equation}
where $\epsilon$ is a fully antisymmetric symbol with
\begin{align} \label{eq:epsilonconvention}
\epsilon_{12 \cdots d} &\equiv +1 \quad \text{(Euclidean),} \nonumber \\ 
\epsilon_{01 \cdots (d-1)} &\equiv +1 \quad \text{(Minkowski).} 
\end{align}
With this sign convention, the Hodge star operation gives a very compact expression for the volume form~\eqref{eq:volform}:
\begin{equation}
\mathrm{vol}_d = \star 1.
\end{equation}
As is clear from the definition, the Hodge star operation depends explicitly on the metric.

The differential form $\star \omega_p$ is sometimes referred to as the ``Hodge dual'' of $\omega_p$. From the action of the Hodge star on differential elements, it is clear that acting twice with the Hodge star approximately returns us to what we started with. Working this out more carefully, one finds that if $\omega_p$ is a $p$-form in a $d$-dimensional space with $m$ minus signs in the metric signature, we have
\begin{equation}
\star(\star \omega_p) = (-1)^{p(d-p) + m} \omega_p.
\end{equation}

\medskip
\noindent\centerline{\rule{\textwidth}{0.5pt}}
\noindent {\em Exercise:} In 3-dimensional space, given a 1-form $V$, explain how to combine exterior derivatives and Hodge stars to  write a differential form expression equivalent to the {\em divergence} ${\vec \nabla} \cdot {\vec V}$.
\noindent\centerline{\rule[6.0pt]{\textwidth}{0.5pt}}

\subsection{Examples of actions in differential form notation}
\label{subsec:differentialformexamples}

Let's compare some familiar physics quantities in tensor index notation and in differential form notation. The action for a real scalar field (in mostly-minus metric signature) is given by
\begin{equation} \label{eq:scalaractionforms}
\int \mathrm{d}^d x\sqrt{|g|}\,\frac{1}{2} g^{\mu \nu} \partial_\mu \phi \partial_\nu \phi = \frac{1}{2} \int \dif \phi \wedge \star \dif \phi.
\end{equation}
The first thing to note is that the familiar measure $\mathrm{d}^d x\sqrt{|g|}$ is {\em not} written on the right-hand side, because it's built into the definitions of the differential forms that appear there! If you're used to writing every integral sign with a $\mathrm{d}^d x$ following it, when you work with differential forms you have to get used to the fact that it doesn't show up. The next thing to notice is that the expression on the left has a metric tensor $g^{\mu \nu}$ used to raise an index so that the two scalar gradients can be dotted into each other. In the expression on the right, we don't see this. Instead, what has happened is that $\star \dif \phi$ is now a {\em 3-index} tensor constructed by contracting $\partial_\nu \phi$ with an {\em antisymmetric} $\epsilon$-symbol. This is then combined with $\dif \phi$ using the {\em antisymmetric} wedge product. These two antisymmetrizations effectively cancel each other out, and lead to the contraction we're used to.

\medskip
\noindent\centerline{\rule{\textwidth}{0.5pt}}
\noindent {\em Exercise:} Verify the equivalence of the two sides in~\eqref{eq:scalaractionforms}, and the other two examples~\eqref{eq:gaugeactionforms} and~\eqref{eq:thetaactionforms} below.\\
\noindent\centerline{\rule[6.0pt]{\textwidth}{0.5pt}}

Our next example is a gauge field kinetic term. We will write this in a commonly used, but non-canonical, normalization where the gauge coupling $e$ appears in front of the kinetic term in a factor of $1/e^2$. This normalization of the gauge field is especially convenient for discussing charge quantization, as we will do below. With this normalization, $F_{\mu \nu}$ always has mass dimension 2 (whereas $F$, the differential form, is dimensionless) and $e^2$ has mass dimension $4-d$. We have:
\begin{equation} \label{eq:gaugeactionforms}
\int \mathrm{d}^d x\,\sqrt{|g|}\left(-\frac{1}{4e^2} F_{\mu \nu} F^{\mu \nu}\right) = \int \left(-\frac{1}{2e^2} F \wedge \star F\right).
\end{equation}
Similar to the scalar action, we again accomplish the dotting together of indices through a combination of a wedge product and a Hodge star. Notice that the left-hand side has a factor of 4 where the right-hand side has a factor of 2. In general, for a $p$-index antisymmetric tensor field strength, the tensor index notation would have an extra factor of $1/p!$ relative to the differential form notation, which always has simply a factor of $1/2$.

The final example will be one of the major players in the lectures below: the theta angle. In four dimensions, the Hodge dual of a 2-form field strength is again a 2-form. We can write it in components, which are often denoted $\widetilde{F}_{\mu \nu}$:
\begin{equation}
\star F = \frac{1}{2} \widetilde{F}_{\mu \nu} \dif x^\mu \wedge \dif x^\nu, \quad \text{where} \quad \widetilde{F}_{\mu \nu} = \frac{1}{2} \sqrt{|g|} \epsilon_{\mu \nu \rho \sigma} F^{\rho \sigma}.
\end{equation}
When $F$ is the electromagnetic field strength, this operation is electric-magnetic duality. 
We should be careful when raising indices: the $\epsilon$ symbol by itself is {\em not} a tensor; the Levi-Civita tensor is the combination $\sqrt{|g|} \epsilon_{\mu \nu \rho \sigma}$. The proper statement about raising indices is that
\begin{equation}
\sqrt{|g|} \epsilon_{\mu \nu \rho \sigma} g^{\mu \alpha} g^{\nu \beta} g^{\rho \gamma} g^{\sigma \delta} = \frac{(-1)^m}{\sqrt{|g|}} \epsilon^{\alpha \beta \gamma \delta},
\end{equation}
where $m$ is the number of minus signs in the metric signature and $\epsilon$ is as in~\eqref{eq:epsilonconvention}. Hence, in Minkowski space we have
\begin{equation}
\widetilde{F}^{\mu \nu} = -\frac{1}{2\sqrt{|g|}} \epsilon^{\mu \nu \rho \sigma} F_{\rho \sigma}.
\end{equation}
The ``$\theta$ term'' that is written in tensor index notation by dotting $F$ into $\widetilde{F}$ can thus be seen, in differential forms notation, to involve simply the wedge product of $F$ with itself:
\begin{equation} \label{eq:thetaactionforms}
\int \mathrm{d}^4 x\, \sqrt{|g|} \frac{\theta}{16\cpi^2} F_{\mu \nu}\widetilde{F}^{\mu \nu} =  - \int \frac{\theta}{8\cpi^2} F \wedge F.
\end{equation}
(We have chosen the sign for Minkowski space here; in Euclidean signature, there will also be an important  factor of $\iu$ from Wick rotation.) Here $\theta$ may be a constant or a pseudoscalar field; both cases will appear in the lectures below. In this case, the differential form notation reveals that the somewhat complicated-looking term appearing on the left is, in fact, {\em topological}: the right-hand side depends on $\theta$ and the 2-form $F$ but has no dependence whatsoever on the metric tensor! (Among other things, this means that the term does not contribute to the stress-energy tensor, which we compute by varying the action with respect to the metric.) The relative simplicity of this term in differential form notation is one of the reasons we will make extensive use of this formalism below.

\section{Ordinary global symmetries; currents and charges}
\label{sec:globalsymmetries}

\subsection{The symmetry group $\Uone$}

A $\Uone$ global symmetry is one associated with {\em counting} some kind of stuff, e.g., the number of particles of some type. A familiar example is a theory of a complex scalar field with a potential that depends only on the magnitude of the scalar:
\begin{equation} \label{eq:complexscalarU1}
\frac{1}{\sqrt{|g|}} {\cal L} = \partial_\mu \phi^\dagger \partial^\mu \phi - m^2 \phi^\dagger \phi - \frac{\lambda}{4} (\phi^\dagger \phi)^2.
\end{equation}
This theory is invariant under the symmetry transformation
\begin{equation}
\phi(x) \mapsto \E^{\iu \alpha} \phi(x),
\end{equation}
with constant $\alpha$. Notice that this symmetry operation only depends on $\alpha~\textrm{mod}~2\cpi$. This is the defining feature of $\Uone$.

The group $\Uone$ is simply the group of complex numbers of unit magnitude, i.e., numbers of the form $\E^{\iu \alpha}$ for $\alpha \in \mathbb{R}$. The group operation is multiplication of these numbers:
\begin{equation}
\E^{\iu \alpha} \cdot \E^{\iu \beta} = \E^{\iu (\alpha + \beta)}.
\end{equation}
This is a {\em group} because it has an identity element (the number 1), every element has an inverse ($\E^{\iu \alpha} \cdot \E^{-\iu \alpha} = 1$), and the multiplication law is associative.

There is a {\em different}, but closely related, group: $\mathbb{R}$, the real numbers viewed as a group under addition. Locally these groups look the same. They have the same Lie algebra, $\mathfrak{u}(1)$, which is also the group of real numbers under addition. (Lie algebras are equipped with another operation, the commutator, but because these groups are abelian the commutators in $\mathfrak{u}(1)$ are all zero.) To make the relationship precise, we can map an element $\alpha \in \mathbb{R}$ to the element $\E^{\iu \alpha} \in \Uone$. This is a {\em group homorphism}, i.e., it commutes with the group operation. However, it is a many-to-one map: $\Uone$ consists of the real numbers modulo addition of integer multiples of $2\cpi$. Notice that we could, via this homomorphism, say that the Lagrangian~\eqref{eq:complexscalarU1} admits an $\mathbb{R}$ global symmetry. However, in physics it is more natural to study theories with $\Uone$ symmetries; we will discuss some reasons for this below.

\subsection{$\Uone$ charges}

Consider a field $\Phi$ charged under a $\Uone$ symmetry, with some charge $q$. This means that a group element $g = \E^{\iu \alpha} \in \Uone$ acts on $\Phi$ according to the rule
\begin{equation}
\Phi(x) \mapsto \E^{\iu q \alpha} \Phi(x).
\end{equation}
Now, we have to remember that the group element $\E^{\iu \alpha}$ is {\em the same} as the group element $\E^{\iu (\alpha + 2 n \cpi)}$, for integer $n$. The right hand side of the above rule for how $\Phi$ transforms has to be well-defined, which means that we need
\begin{equation}
\E^{\iu q \alpha} \Phi(x) = \E^{\iu q (\alpha + 2 n \cpi)} \Phi(x), \quad \forall n \in \mathbb{Z}.
\end{equation}
This means that we need {\em quantized} charges!
\begin{equation}
q \in \mathbb{Z}.
\end{equation}
In mathematical terminology, the {\em representations} of the group $\Uone$ are labeled by integers.

This is a very basic mathematical fact, but you can find many books or papers by particle physicists that deny its validity, so let me put it in a box to emphasize it:

\begin{framed}
\noindent
The allowed charges under a $\Uone$ symmetry are quantized. They must be integers.
\end{framed}

If someone tells you otherwise, it means that they are really thinking of an $\mathbb{R}$ symmetry. One sometimes encounters the opinion that in physics we should only label symmetry groups by their Lie algebra. But as we will see, the global structure of the group has important implications, so it is useful to be precise about our language and label a theory by the group that acts on the fields.

\subsection{$\Uone$ conservation laws and currents}

\begin{figure}[!h]
\centering
\includegraphics [width = 0.3\textwidth]{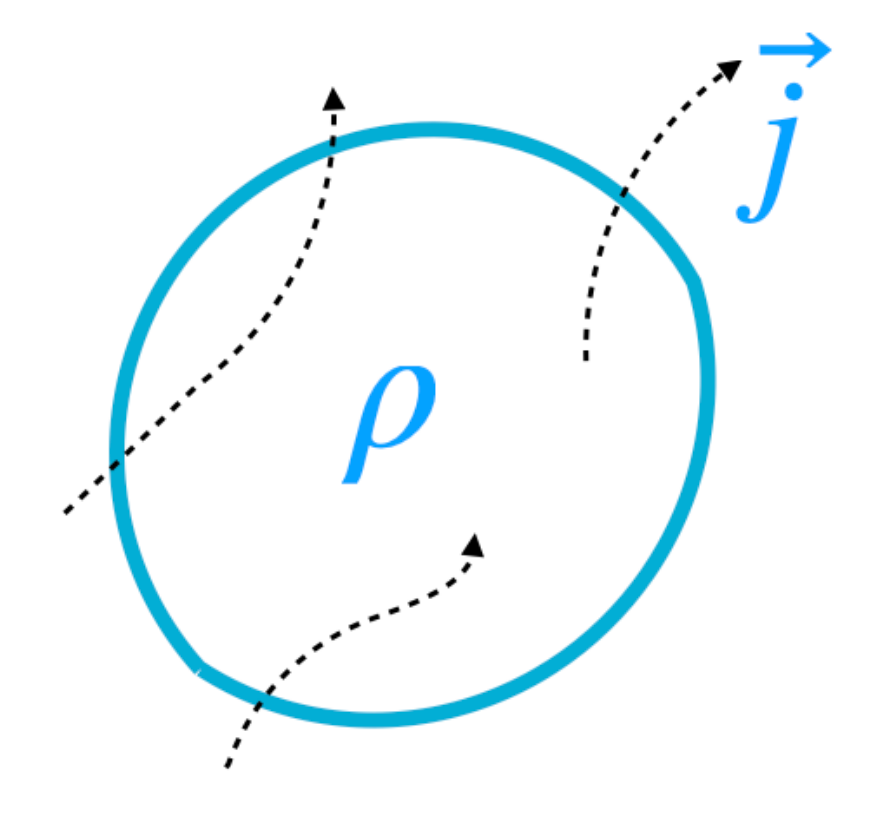}
\caption{A $\Uone$ conservation law relates a density $\rho$ of some quantity inside a region to the flux through the boundary of the region, as determined by the current ${\vec j}$.
} \label{fig:conservedstuff}
\end{figure}

Let's first recall what the conservation of any locally conserved ``stuff'' (e.g., mass of a fluid, or charge) looks like. In introductory physics classes you encountered many of these equations, which say that if a quantity is conserved then the rate of change of the amount of stuff inside some volume is given by the integrated flux of the stuff through the surface bounding the volume:
\begin{equation}
\frac{\dif Q}{\dif t} = - \oint {\vec \Phi} \cdot \dif{\vec S}.
\end{equation}
Such an equation has a {\em differential} formulation in terms of a density $\rho$ of stuff and a current ${\vec j}$ describing the motion of the stuff, generally known as a {\em continuity equation}:
\begin{equation}
\frac{\partial \rho(t, {\vec x})}{\partial t} = - {\vec \nabla} \cdot {\vec j}(t, {\vec x})
\end{equation}
See Fig.~\ref{fig:conservedstuff}. In the relativistic context, $\rho$ can be viewed as the time component of a 4-vector whose spatial components are $\vec j$, and these conservation equations take on the particularly simple form $\partial_\mu j^\mu = 0$. 

A Lagrangian field theory with a $\Uone$ global symmetry will have such a conserved current $j^\mu$. This is a special case of {\em Noether's theorem}, which tells us that any continuous global symmetry in a Lagrangian theory gives rise to a conserved current. (This applies to internal symmetries; the story for spacetime symmetries is a little more subtle, and we won't have a need to go into it here.) Noether's theorem has a straightforward derivation. Consider a field variation that {\em would} be a symmetry if $\epsilon$ is constant:
\begin{equation}
\phi(x) \mapsto \phi(x) + \epsilon(x) \xi(x).
\end{equation}
By definition of a symmetry, the action doesn't change if $\epsilon$ is constant, which means that the change in the action due to our field variation must come from {\em derivatives} of $\epsilon(x)$, i.e.,
\begin{equation} \label{eq:actionvarsym}
S[\phi] \mapsto S[\phi] + \int \dif{^d x}\, \sqrt{|g|} j^\mu(x) \partial_\mu \epsilon(x),
\end{equation}
for {\em some} quantity $j^\mu(x)$. In other words, we vary the action and simply {\em define} $j^\mu(x)$ to be whatever appears multiplying the derivative of $\epsilon$ (with a factor of $\sqrt{|g|}$ pulled out). Now, we recall that for a Lagrangian theory, a field configuration that solves the equations of motion of the theory is one for which $\delta S = 0$ for {\em any} small variation of the fields. The only way that we can have $\delta S = 0$ for the particular variation of the action~\eqref{eq:actionvarsym} is to have $\partial_\mu (\sqrt{|g|} j^\mu) = 0$, which is therefore an equation of motion obeyed by the fields. The reason that we don't include the factor of $\sqrt{|g|}$ in the definition of $j^\mu$ is that our chosen definition gives us the {\em covariant} form of the current conservation law for $j^\mu$ in curved spacetime:\footnote{If you haven't seen this before, or if your general relativity is rusty, you can verify the identity for Christoffel symbols with two indices contracted, $\Gamma^\mu_{\mu \nu} = \frac{1}{\sqrt{|g|}} \partial_\nu\sqrt{|g|}$, from which this form of $\nabla_\mu j^\mu$ follows.}
\begin{equation}
\nabla_\mu j^\mu = \frac{1}{\sqrt{|g|}} \partial_\mu(\sqrt{|g|} j^\mu) = 0.
\end{equation} 
We emphasize that this is {\em not} an identity that follows from the form of $j^\mu$ for {\em any} field configuration; it holds only for those specific field configurations that solve the classical equations of motion. In particular, in the quantum theory, this will not be true of every field configuration we sum over in the path integral, only the saddle-point configurations.

\medskip
\noindent\centerline{\rule{\textwidth}{0.5pt}}
\noindent {\em Exercise:} following the logic above, verify that the Noether current associated with the $\Uone$ symmetry of~\eqref{eq:complexscalarU1} is
\begin{equation}
j^\mu = \iu  \left[(\partial^\mu \phi^\dagger) \phi - \phi^\dagger (\partial^\mu \phi)\right].
\end{equation}
\noindent\centerline{\rule[6.0pt]{\textwidth}{0.5pt}}

\begin{figure}
\centering
\includegraphics [width = 0.4\textwidth]{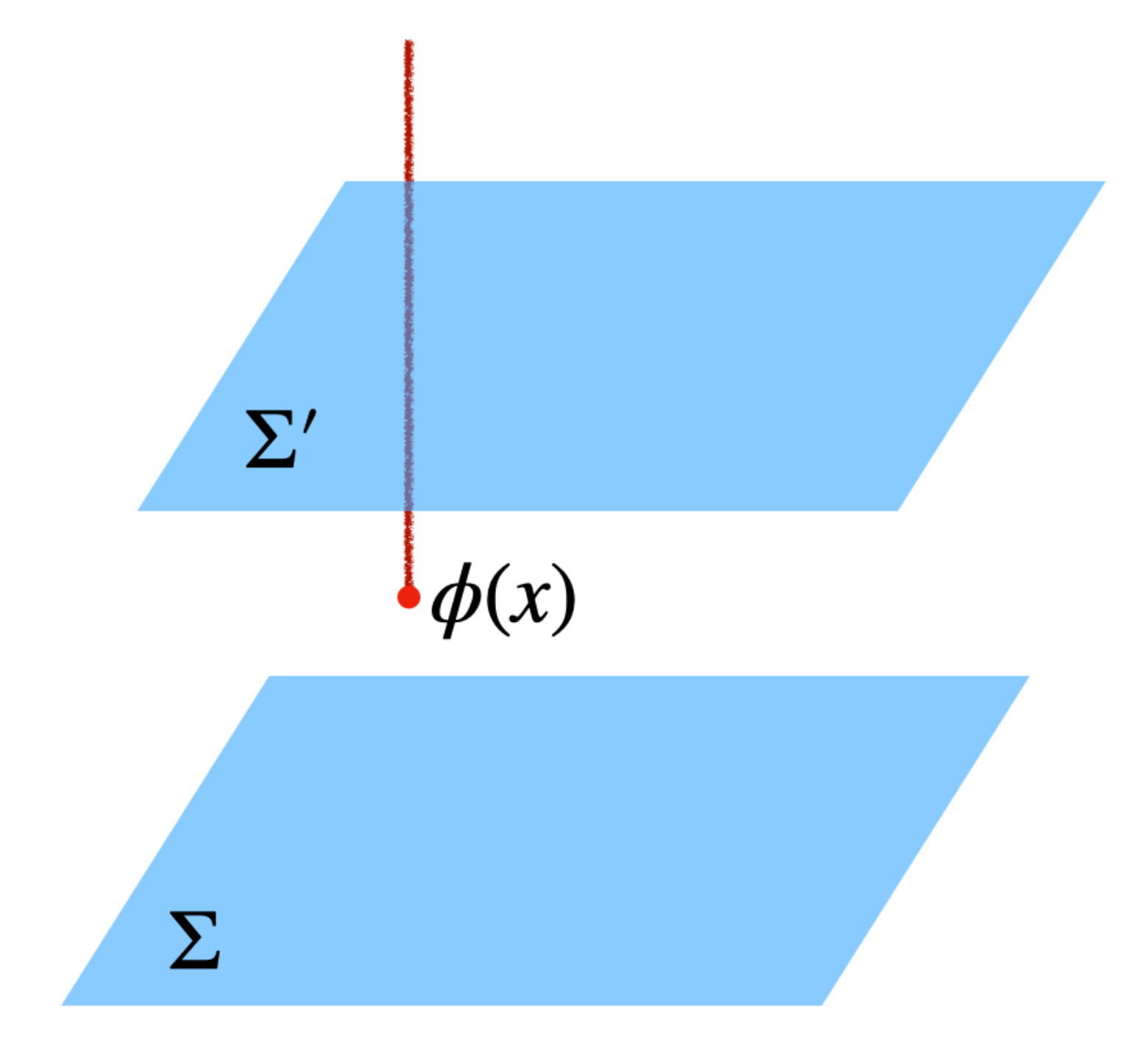}
\caption{We can measure the global $\Uone$ charge $Q$ along two different spatial slices, $\Sigma$ and $\Sigma'$. An insertion of an operator carrying global charge $q$, like $\phi(x)$, at an intermediate time will change the result: $Q(\Sigma') - Q(\Sigma) = q$. One way to think about this is that the local operator $\phi(x)$ can create a particle of charge $q$, which exists on the surface $\Sigma'$ (its worldline, depicted as a fuzzy red line emanating from the operator insertion, passes through the surface).
} \label{fig:chargemeasurement}
\end{figure}

Given a conserved current $j^\mu(x)$, we can find the total charge in the space by integrating the charge density $\rho = j^0$ over all of space. In $d$ spacetime dimensions, we have
\begin{equation}
Q = \int \mathrm{d}^{d-1}x \sqrt{|g|} j^0(x).
\end{equation}
The charge is a conserved quantity according to the equations of motion, but if we evaluate a correlation function involving insertions of charged operators $\phi(t, {\vec x})$ at some time, these can change the charge, because they effectively insert a worldline of a charged particle (see Fig.~\ref{fig:chargemeasurement}).

The fact that we integrate the charge density over a $(d-1)$-dimensional slice of spacetime is a hint that thinking of currents as vectors is sometimes not optimal. The object that can naturally be integrated over a $(d-1)$-dimensional manifold is a $(d-1)$-form. How is this related to the current we have discussed? We can view the current as defining a 1-form, $j = j_\mu \dif x^\mu$. The Hodge dual of this one form,
\begin{equation}
J = \star j,
\end{equation}
is a $(d-1)$-form. The charge, then, is simply
\begin{equation}
Q = \int_\Sigma J.
\end{equation}
This is the most mathematically convenient way to formulate a conserved current. The conservation law is simply the statement that this $(d-1)$-form current is {\em closed}:
\begin{equation}
\dif J = 0.
\end{equation}
This is such a useful perspective to keep in mind that I frame it in another box:

\begin{framed}
\noindent
Conserved currents are often best thought of as closed differential forms, $\dif J = 0$.
\end{framed}

\medskip
\noindent\centerline{\rule{\textwidth}{0.5pt}}
\noindent {\em Exercise:} convince yourself that the equations $\nabla_\mu j^\mu = 0$ and $\dif J = 0$ are equivalent.\\
\noindent\centerline{\rule[6.0pt]{\textwidth}{0.5pt}}

\subsection{Symmetry operators}

An important part of the modern perspective on global symmetries is that global symmetries {\em act locally}. This is already evident in the fact that local operators transform under these symmetries, though we usually consider the action of the symmetry on all local operators everywhere in spacetime at once. However, there is in fact a family of {\em surface operators} that implement the action of global symmetries within a limited region of spacetime~\cite{Gaiotto:2014kfa}. These are called {\em symmetry operators} or sometimes {\em charge operators} (we prefer the former, since the latter {\em sounds} almost like ``charged operators,'' but means something different). We can construct these operators in the Euclidean theory, since we will treat space and time directions on an equal footing in their construction instead of only integrating currents over fixed-time spatial slices. Essentially, the idea is to take the two spatial slices shown in Fig.~\ref{fig:chargemeasurement} and bend them around to meet each other, surrounding the operator insertion $\phi(x)$. 

\begin{figure}[!h]
\centering
\includegraphics [width = 0.7\textwidth]{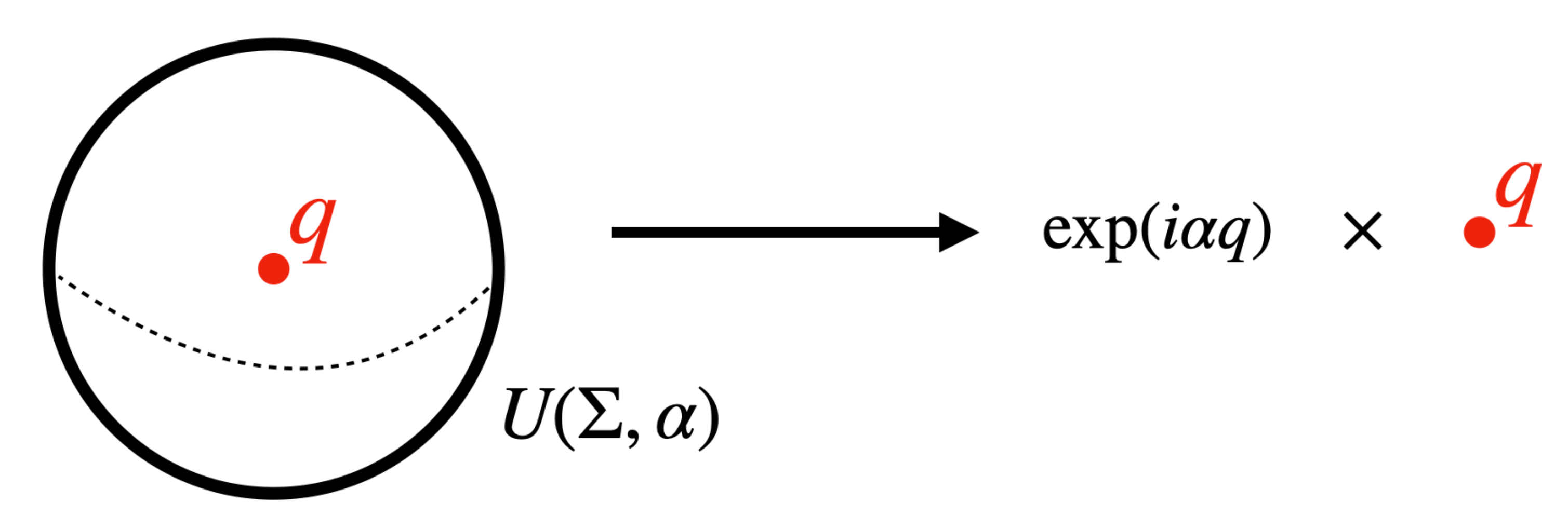}
\caption{For a global $\Uone$ symmetry, a {\em charge operator} or {\em symmetry operator} supported on a 3-dimensional slice $\Sigma$ of spacetime measures the charge $q$ of a pointlike operator, via this Ward identity.
} \label{fig:chargeoperator}
\end{figure}

The symmetry operators for a $\Uone$ global symmetry are labeled by a closed $(d-1)$-dimensional manifold $\Sigma$ and an element $\E^{\iu \alpha} \in \Uone$ (or simply by the phase $\alpha$, understood to be defined mod $2\cpi$). These operators are defined by exponentiating the integrated current over $\Sigma$, with coefficient $\alpha$:
\begin{equation}
U(\Sigma, \alpha) \equiv \exp\left(\iu \alpha \int_\Sigma J\right).
\end{equation}
See Fig.~\ref{fig:chargeoperator}. If the surface $\Sigma$ surrounds an operator insertion of a charged local operator $\phi(x)$ of charge $q$, and no other charged operators, then the symmetry operator inserted in a correlation function is equivalent to rephasing the charged operator. In other words, there is an OPE, $U(\Sigma, \alpha) \phi(x) = \E^{\iu \alpha q} \phi(x)$. This is illustrated in Fig.~\ref{fig:chargeoperator}.

\begin{figure}[!h]
\centering
\includegraphics [width = 0.5\textwidth]{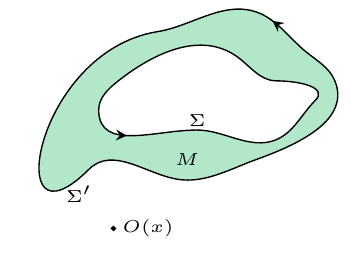}
\caption{Illustration of a topological operator (in 2d, for convenience). If we insert a symmetry operator $U(\Sigma, g)$ or a symmetry operator $U(\Sigma', g)$, their correlation functions will be identical provided that all charged operator insertions like $O(x)$ are outside the (shaded green) region $M$ swept out when deforming $\Sigma$ into $\Sigma'$.
} \label{fig:topological}
\end{figure}

The symmetry operators are {\em topological operators}, meaning that when inserted in correlation functions, the surface $\Sigma$ can be deformed arbitrarily without changing the answer, {\em provided} that no charged operators cross through $\Sigma$ as it is deformed. Why are the operators topological? Suppose that we deform the surface $\Sigma$ to a different surface $\Sigma'$. Then the operators are related by
\begin{equation}
U(\Sigma', \alpha) = \exp\left(\iu \alpha \int_{\Sigma - \Sigma'} J\right) U(\Sigma, \alpha).
\end{equation}
Fill in the interior of the region swept out by deforming $\Sigma$ to $\Sigma'$ and call it $M$ (see Fig.~\ref{fig:topological}). If no operators were crossed by the surface as we deformed it, we have $\partial M = \Sigma - \Sigma'$. Then by Stokes's theorem, $\int_{\Sigma - \Sigma'} J = \int_M \dif J$. The latter is zero by current conservation, which tells us that $U(\Sigma', \alpha)$ and $U(\Sigma, \alpha)$ are equivalent. However, a local operator insertion is a singularity: if we insert a charged operator $\phi(x)$ at a point that lies inside $M$, then  $\partial M = (\Sigma - \Sigma') \cup S^{d-1}_x$, where $S^{d-1}_x$ is a little sphere around the singular point with the operator insertion. This means that correlation functions of $U(\Sigma, \alpha)$ and $U(\Sigma', \alpha)$ will differ precisely due to localized contributions from the operator insertions in the region $M$ in between. This is just a description of the usual Ward identity in terms of symmetry operators in the path integral. Physically, the picture is that $\phi(x)$ {\em creates} a charged state, as in Fig.~\ref{fig:chargemeasurement}, whose propagation crosses $\Sigma'$ but not $\Sigma$. 

There is a nice discussion in section 2 of David Simmons-Duffin's TASI lectures on the conformal bootstrap~\cite{Simmons-Duffin:2016gjk} on how topological operators implement symmetries, albeit focused on the case of spacetime symmetries.

The formalism of symmetry operators is very powerful. For example, a discrete symmetry group like $\mathbb{Z}_N$ has no associated conserved current. It {\em does}, however, have a family of symmetry operators, $U(\Sigma, \alpha)$ where now the phase $\alpha$ takes on only the discrete values $\{0, 2\cpi/N, 4\cpi/N, \ldots 2\cpi(N-1)/N\}$. More generally, {\em any} global symmetry group $G$ comes with a family of surface operators $U(\Sigma, g)$ associated with $(d-1)$-dimensional surfaces $\Sigma$ and group elements $g \in G$. Inserted in correlation functions, these act on local $G$-charged operators surrounded by $\Sigma$ by transforming these operators according to the appropriate {\em representation} of $G$. (In the general case, this will not be simply a rephasing of the operator, but will replace an operator with a linear combination of other operators).

A variant of this formalism allows us to describe generalized global symmetries that act on extended operators. For instance, a 1-form global symmetry acts on 1-dimensional operators like Wilson loops; the charged {\em objects} would be strings, not point particles. There are a family of symmetry operators now labeled by $(d-2)$-dimensional surfaces, which can link with 1-dimensional loops. In general, $p$-form generalized global symmetries act on $p$-dimensional charged operators and are implemented by $(d-p-1)$-dimensional surface operators. We will discuss such generalized symmetries, known as higher-form symmetries, in \S\ref{sec:generalizedsymmetry}. This is part of a modern understanding of symmetries that dramatically enlarges the range of possible symmetries in quantum field theory and condensed matter theory, which I believe will play a large role in all areas of theoretical physics, including particle phenomenology, in the future. These include not only higher-form symmetries, but more complicated cases known as higher-group symmetries and non-invertible symmetries. It is well worth your effort to invest time in learning more. 

\section{U(1) gauge theory}

{\em ``U(1) is very deep.'' -- Lian-Tao Wang}

\subsection{Basic definitions; charge quantization}

Gauge symmetries have a very different interpretation than global symmetries. A global symmetry maps one physical state to a different physical state. A gauge symmetry, by contrast, is not really a symmetry at all: it is a {\em redundancy} in our description of the theory. It allows us to describe precisely the same physical state in different ways. This conceptual distinction has many important physical implications. For instance, when a continuous global symmetry is spontaneously broken, we can excite a long wavelength mode that transitions from one vacuum state to another. This is a Nambu-Goldstone boson. The same is not true in gauge theory, because there are no different states to transition between! This is why the Nambu-Goldstone boson disappears from the spectrum in the Higgs mechanism. As we will see later in these lectures, gauge symmetries are perfectly consistent with quantum gravity, whereas global symmetries are not.

As we discussed above, the group $\Uone$ consists of phases $\exp(\iu \alpha)$ under multiplication, and as such is distinguished from the group $\mathbb{R}$ of real numbers under addition by the $2\cpi$ periodicity of $\alpha$. These groups are {\em locally} the same, and both have the Lie algebra $\mathfrak{u}(1) \cong \mathbb{R}$. However, they are {\em globally} different, and have different physics. In particular, $\Uone$ gauge theory has quantized charge, and admits magnetic monopoles. On the other hand, $\mathbb{R}$ gauge theory does not have quantized charge, and forbids magnetic monopoles. We will derive these statements shortly, at least at a somewhat heuristic level. Readers seeking a more explicit mathematical treatment, still aimed a broad physics audience, can refer to~\cite{Alvarez:1984es} or~\cite{Nakahara:2003nw}.

When we carry out a gauge transformation, we choose a gauge group element $g(x)$ for every point $x$ in spacetime, and specify an action of $g(x)$ on our field. For $\Uone$ gauge theory, a field $\psi(x)$ of charge $q$ transforms under the gauge transformation $g(x) = \exp(\iu \alpha(x))$ according to
\begin{equation}
\psi(x) \mapsto \psi^g(x) \equiv \exp(\iu q \alpha(x)) \psi(x).
\end{equation} 
Because $\alpha(x)$ is only defined modulo $2\cpi$, this expression only makes sense  if $\exp(2\cpi \iu q) = 1$, i.e., if $q \in \mathbb{Z}$ is an integer. Thus, {\em by definition}, $\Uone$ gauge charge is quantized, as we already emphasized for $\Uone$ global charge.

By contrast, for the gauge group $\mathbb{R}$, we can consider a choice of gauge transformation $g(x) = \alpha(x)$ with $\alpha(x) \in \mathbb{R}$, together with a gauge transformation rule of precisely the same form, $\psi(x) \mapsto \exp(\iu q \alpha(x)) \psi(x)$. This time, $\alpha$ is simply a real number, and this expression is always well-defined. Hence, charges in $\mathbb{R}$ gauge theory need not be quantized: any $q \in \mathbb{R}$ is allowed. 

\subsection{Gauge fields}

In either $\Uone$ or $\mathbb{R}$ gauge theory, we introduce a gauge field $A_\mu(x)$ that transforms under gauge transformations via
\begin{equation} \label{eq:U1gaugefieldtransform}
A_\mu \mapsto A_\mu -\iu \mathrm{e}^{-\iu \alpha(x)} \partial_\mu \mathrm{e}^{\iu \alpha(x)} = A_\mu + \partial_\mu \alpha(x).
\end{equation}
In the $\Uone$ case, the gauge transformation is defined by the $\Uone$ element $\E^{\iu \alpha(x)}$; that is, $\alpha(x)$ is only defined mod $2\cpi$. This allows for winding of $\E^{\iu \alpha(x)}$ around a circle, which will play an important role below.

The gauge-invariant field strength $F_{\mu \nu}(x)$ is defined by 
\begin{equation}
F_{\mu \nu}(x) = \partial_\mu A_\nu(x) - \partial_\nu A_\mu(x).
\end{equation}
In the language of differential forms that we reviewed in \S\ref{sec:differentialforms}, $A(x) = A_\mu(x)  \dif x^\mu$ is a 1-form and $F(x) = \frac{1}{2} F_{\mu \nu}(x) \dif x^\mu \wedge \dif x^\nu$ is a 2-form. The gauge transformation is $A \mapsto A + \dif \alpha$, and the gauge invariance of $F$ is an automatic consequence of the general mathematical fact that $\mathrm{d}^2 = 0$ (as a consequence of antisymmetry).

In fact, we have been a bit too hasty in calling $A$ a 1-form. In general, the gauge field $A(x)$ need not be well-defined over all of spacetime. Spacetime can be covered with multiple coordinate patches, and we define fields separately on each patch. We'll give an explicit example below in \S\ref{subsec:monopole} when we discuss the Dirac monopole, which should make this idea clear, but let me state the general abstract formalism once for completeness. If we have two overlapping patches $U$ and $V$ with associated gauge fields $A_U(x)$ and $A_V(x)$, the fields only have to agree with each other up to a gauge transformation. That is, there should be a gauge transformation $g_{U \to V}(x) = \exp(\iu \alpha_{U \to V}(x))$ on the overlap $U \cap V$ under which $A_U$ maps to $A_V$ as in~\eqref{eq:U1gaugefieldtransform}. Furthermore, if we have {\em three} overlapping regions $U$, $V$, and $W$, we need a compatibility condition on the triple overlap $U \cap V \cap W$: $g_{V \to W}(x) \cdot  g_{U \to V}(x) = g_{U \to W}(x)$. (The mathematical jargon for this is a ``cocyle condition.'') It turns out---and here I will just point you to the more mathematical literature cited above, rather than trying to give an argument---that we can stop at triple overlaps; we don't have to worry further about quadruple overlaps and so on.

The collection of coordinate charts $U, V, \ldots$ together with choices of gauge field $A_U, A_V, \ldots$ on the charts and gauge group elements $g_{U \to V}, \ldots$ on the pairwise overlaps determines what is known as a {\em U(1) gauge bundle with connection} on spacetime (also known as a principal $\Uone$ bundle with connection). The gauge field $A$ is known as the {\em connection}: like the metric connection in general relativity (represented by the Christoffel symbols), it tells us how to parallel transport particles around loops, this time for charged particles. The field strength $F$ is also called the  {\em curvature} of the connection $A$, in much the same way that the metric curvature (measured by the Riemann tensor) comes from derivatives of the Christoffel symbols.

The definition of gauge bundles that we have given applies to any gauge group $G$, not just $\Uone$. The only difference is that the gauge transformations like $g_{U \to V}$ take values in $G$, and the gauge fields like $A_U$ take values in the Lie algebra of $G$. Even a {\em discrete} group $G$, like $\mathbb{Z}_N$, can be used to define a gauge theory in this way. In that case, the connection is trivial, so we only need the overlap transformations $g_{U \to V}$, which are locally constant. Discrete gauge theories tend to get little attention in introductions to quantum field theory for particle physicists, especially as theories in their own right rather than remnants of a higgsed continuous gauge theory. For a little more detailed introduction to discrete gauge theory, I refer you to \S2 of~\cite{McNamara:2022lrw}, where Jake McNamara and I recently tried to give a clear pedagogical summary.

\subsection{Wilson loops, quantized magnetic flux}
\label{subsec:wilsonloops}

In abelian gauge theory, we can define a family of gauge-invariant {\em Wilson loops} associated with closed loops $\gamma$ in spacetime and charges $q$,
\begin{equation}
W_q(\gamma) \equiv \exp\left(\iu q \oint_\gamma A_\mu \dif x^\mu\right) = \exp\left(\iu q \oint_\gamma A\right).
\end{equation}
As written, this is a classical expression, which can be inserted in a path integral to compute a correlation function. If the curve $\gamma$ extends in the time direction, we can think of $W_q(\gamma)$ as inserting a very heavy particle of charge $q$ with worldline $\gamma$, too heavy to move, which acts as a static probe of the theory. When inserted along a spatial slice in a path integral, the Wilson loop can also be thought of as an operator acting on Hilbert space. In the general case, the definition of $W_q(\gamma)$ requires a {\em path-ordering} of operators along the curve $\gamma$, sometimes written with a $P$ in front of the exponential. This is just like the time-ordered exponential you have encountered in quantum mechanics.

For gauge transformations by a well-defined $0$-form $\alpha$, the Wilson loop $W_q(\gamma)$ is clearly invariant under $A \mapsto A + \dif \alpha$, because $\oint_\gamma \dif \alpha = 0$, using Stokes' theorem and the fact that $\gamma$ is a closed loop without boundary. For $\mathbb{R}$ gauge theory, this establishes the gauge invariance of the Wilson loop, for {\em any} charge $q$. For $\Uone$ gauge theory, we must be a bit more careful (recall the note of caution at the end of~\S\ref{subsec:exteriorderiv}) about the case where $\alpha$ itself is not well-defined but $\E^{\iu \alpha}$ and $\dif \alpha$ are. In other words, we must also consider shifts of $A$ by a general flat connection or, equivalently, cases in which $\alpha(x)$ has a nontrivial winding number around the circle, $\alpha(2\cpi) = \alpha(0) + 2\cpi n$.
 In this case, we have
\begin{equation}
W_q(\gamma) \mapsto W_q(\gamma) \exp\left[\iu q \left(\alpha(2\cpi) - \alpha(0)\right)\right] = W_q(\gamma) \exp\left(\iu q 2\cpi n\right) = W_q(\gamma).
\end{equation}
We see that only $q \in \mathbb{Z}$ is a consistent charge assignment for Wilson loops in $\Uone$ gauge theory, consistent with our earlier remark that charge is quantized in $\Uone$ gauge theory.

\begin{figure}[!h]
\centering
\includegraphics [width = 0.6\textwidth]{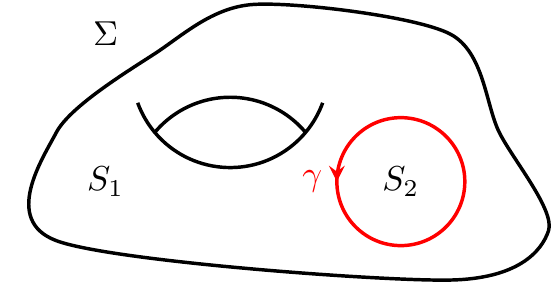}
\caption{Illustration of the magnetic flux quantization argument discussed in the text: we divide a two-dimensional surface $\Sigma$ along a curve $\gamma$ into two oppositely-oriented regions $S_1$ and $S_2$. The Wilson loop over $\gamma$ is well-defined, which implies that the surface operators $\exp\left(\iu q \int F\right)$ over $S_1$ and $-S_2$ agree, which implies the flux of $F$ through all of $\Sigma$ is quantized.
} \label{fig:fluxquantization}
\end{figure}

Consider a Wilson loop in either $\Uone$ or $\mathbb{R}$ gauge theory. If the closed loop $\gamma$ is the boundary of a two-dimensional surface $S$, a relationship denoted $\partial S = \gamma$, then we can use Stokes's theorem to write the Wilson loop observable in terms of the gauge-invariant field strength $F$ integrated over $S$:
\begin{equation}
W_q(\gamma) = \exp\left(\iu q \oint_{\partial S} A\right) = \exp\left(\iu q \int_S F\right).
\end{equation}
We have to be a bit careful about this: Stokes's theorem works when $A$ is a well-defined 1-form in the region of interest, but in general, gauge fields need not be single-valued. However, we can carry out a gauge transformation so that the ``Dirac string'' where $A$ is poorly defined (see \S\ref{subsec:monopole}) is outside the region $S$.
Suppose that we consider two {\em different} surfaces, $S_1$ and $S_2$, both bounded by $\gamma$. Because the integral of $F$ over both surfaces corresponds to the same Wilson loop, we can combine the two to form a trivial operator. That is, we make a closed surface $\Sigma$, without boundary, by combining $S_1$ with the orientation-reversed surface $-S_2$: $\Sigma = S_1 \cup (-S_2)$, as depicted in Fig.~\ref{fig:fluxquantization}. Thus, we have
\begin{equation}
\exp\left(\iu q \int_\Sigma F\right) = \exp\left(\iu q \int_{S_1} F\right)\exp\left(-\iu q \int_{S_2} F\right) = \exp\left(\iu q \oint_{\partial S} A\right)\exp\left(-\iu q \oint_{\partial S} A\right) = 1.
\end{equation}
Conversely, given any closed surface $\Sigma$, we can draw a closed loop $\gamma \subset \Sigma$ dividing $\Sigma$ up into two regions and use this argument to conclude that $\exp\left(\iu q \int_\Sigma F\right) = 1$ for any allowed charge $q$. (When there is nonzero flux, we have to be a bit careful, defining each integral in a gauge where the Dirac string does not go through the surface of interest.) From this we learn that:
\begin{itemize}
\item In $\mathbb{R}$ gauge theory, where $q$ can take on any real value, we must have $\int_\Sigma F = 0$ for any closed two-dimensional surface $\Sigma$. In other words, {\em magnetic flux vanishes} in $\mathbb{R}$ gauge theory. The theory forbids magnetic monopoles.
\item In $\Uone$ gauge theory, we require $q \int_\Sigma F \in 2\cpi \mathbb{Z}$ for any integer $q$. This is only possible if {\em magnetic flux is quantized}, i.e., 
\begin{equation}
\frac{1}{2\cpi} \int_\Sigma F  \in \mathbb{Z}.    \label{eq:fluxquant}
\end{equation}
\end{itemize}
These facts together form the statement of ``Dirac quantization.'' It is commonly said that the existence of magnetic monopoles requires electric charge to be quantized. Here we have approached this from the other direction: beginning by specifying our gauge group as $\Uone$, we learn that electric charge is quantized and magnetic monopoles are allowed. On the other hand, non-quantized electric charge requires a gauge group of $\mathbb{R}$, which we have found to be incompatible with magnetic monopoles.

Importantly, the quantization condition~\eqref{eq:fluxquant} is a property of any $\Uone$ bundle, meaning that it is obeyed by {\em every field configuration} that is summed over in the path integral. Electric flux quantization holds in a quite different way, as we will see below.

\subsection{Brief aside on topology}

Before I give an example of nonzero magnetic flux by using a space with a nontrivial topology, a brief remark is in order. The modern viewpoint on quantum field theory is that a QFT should have the ability to be defined not just in Minkowski space, but on arbitrary spacetime manifolds (or at least, those within some very general class). Formal quantum field theorists tend to take this as a given, and it has also become a commonplace in condensed matter theory, where many interesting phases of matter can be classified by the ground states that they have on spaces of nontrivial topology. However, the assumption that we can analyze QFTs by studying them on nontrivial spaces seems to be viewed with some skepticism (and occasional outright hostility) by some particle physicists. I think that the simplest justification that I can give you is that we live in a world with gravity, and we believe that a quantum theory of gravity is described (at least in a semiclassical limit) by a path integral that sums over different spacetime manifolds (including those with different topologies). Thus, any QFT that we eventually hope to couple with gravity should be compatible with spacetimes of nontrivial topology. Another justification is that conclusions that we can draw by studying QFT on nontrivial spacetimes often coincide with conclusions that we can draw from QFTs in the presence of interesting dynamical objects like magnetic monopoles or cosmic strings. Which objects are allowed (even as static probes of the theory, like Wilson lines) depends on the global structure of the gauge group, e.g., SU(2) versus SO(3) (and the classification can even depend on additional discrete data~\cite{Aharony:2013hda}). As we will discuss later, it is expected that in quantum gravity, all charged objects allowed by a gauge group actually exist, so these global choices have real physical implications. From this viewpoint, the claim (sometimes found in textbooks) that only projective representations matter in quantum field theory is too glib; it discards a useful mathematical tool for distinguishing between different theories.

Without further apology, I will proceed to analyze the structure of $\Uone$ gauge theory by placing it on topologically nontrivial spacetimes.

\subsection{Configurations with magnetic flux}

\subsubsection{Example 1: flux on a torus} \label{ex:torusflux}

As a first example of $\Uone$ magnetic flux, we consider a case where there is no magnetically charged object, but a magnetic flux arises because our theory is defined on a spacetime with nontrivial topology. Suppose our spacetime contains two periodic directions forming a torus $T$, parametrized by $(x_1, x_2)$ where $x_i \cong x_i + 2\cpi r_i$ ($i = 1,2$). Then a field configuration with nonzero flux over the torus is given by a constant field strength $F_{12}$,
\begin{equation}
\frac{1}{2\cpi} \int_{T} F = \frac{1}{2\cpi}  \int_0^{2\cpi r_1} \dif x_1 \int_0^{2\cpi r_2} \dif x_2 \,F_{12} = 2\cpi r_1 r_2 F_{12}.
\end{equation}
Based on our general reasoning above, we know that this field configuration is allowed only if $F_{12}$ is an integer multiple of $\frac{1}{2\cpi r_1 r_2}$. To understand why only these values of $F_{12}$ are allowed, let us try to construct a gauge field $A$ whose field strength is $F$. Because $F_{12} = \partial_1 A_2 - \partial_2 A_1$, we can consider a case where $A_2$ increases linearly with $x_1$, i.e., $A = F_{12} x_1 \dif x_2$. This is not a well-defined {\em function} because $x_1$ is only defined modulo $2\cpi r_1$. However, it does define a {\em connection} if the difference between the original $A$ and a transformed $A$ with $x_1 \mapsto x_1 + 2\cpi r_1$ is {\em gauge equivalent} to zero, i.e., if we can write
\begin{equation}
2\cpi r_1 F_{12} \dif x_2 = -\iu \mathrm{e}^{-\iu \alpha(x)} \dif{\mathrm{e}^{\iu \alpha(x)}}  \label{eq:fluxconditiontocheck}
\end{equation}
for some $\Uone$-valued function $\mathrm{e}^{\iu \alpha(x)}$. To accomplish this, we consider gauge transformations that wind around the $x_2$ direction, i.e.,
\begin{equation}
\exp(\iu \alpha(x)) = \exp\left(\iu n \frac{x_2}{r_2}\right), \quad n \in \mathbb{Z}.    \label{eq:windingtransf}
\end{equation}
Comparing this to~\eqref{eq:fluxconditiontocheck}, we see that we have a sensible connection $A$ in the case that
\begin{equation}
2\cpi r_1 F_{12} \dif x_2 = \frac{n}{r_2} \dif x_2 \quad \Rightarrow \quad F_{12} = \frac{n}{2\cpi r_1 r_2}.
\end{equation}
This establishes that our field configuration obeys~\eqref{eq:fluxquant}.

For $\mathbb{R}$ gauge theory, $\alpha(x)$ lives on the real line, not on a circle. As a result, winding configurations like~\eqref{eq:windingtransf} do not exist, because no well-defined choice of $\alpha(x)$ can satisfy the equation. As a result, only $F_{12} = 0$ is permitted.

\subsubsection{Example 2: the Dirac monopole}
\label{subsec:monopole}

A magnetic monopole is an object that carries magnetic charge. If we integrate the magnetic flux around a monopole of magnetic charge $q_m \in \mathbb{Z}$, we should obtain $\int_{S^2} F = 2\cpi q_m$. If we locate our magnetic monopole at the origin, one choice of $F$ that achieves the correct magnetic flux is proportional to the volume form on $S^2$, namely
\begin{equation}
F = \frac{1}{2} q_m \mathrm{vol}_{S^2} = \frac{1}{2} q_m \sin\theta \dif \theta \wedge \dif \phi.
\end{equation}
We could try to integrate this to find a gauge field $A$ such that $F = \dif A$, but it turns out that no global solution works. For instance, you might try to write $A = -\frac{1}{2} q_m \phi \sin \theta \dif \theta$ or $A = -\frac{1}{2} q_m \cos \theta \dif \phi$. However, neither of these are actually well-defined 1-forms on the whole space. The problem is that $\phi$ is not globally defined, and not single-valued where it is defined. In particular, $\phi$ degenerates along the $z$-axis where $\theta = 0, \cpi$. This doesn't cause any problem for $F$, because the prefactor $\sin \theta$ is $0$ precisely where the $\dif \phi$ factor stops making sense. However, it is a problem for $A$.

What we can do is define two {\em different} choices of $A$, each of which is valid in part of our space, and which are gauge-equivalent in the region where both are valid. Specifically, we define:
\begin{align} \label{eq:diracmonopole}
A_N &= \frac{1}{2} q_m (+1 - \cos \theta) \dif \phi, && \theta \neq \cpi. \nonumber \\
A_S &= \frac{1}{2} q_m (-1 - \cos\theta) \dif \phi, && \theta \neq 0. \nonumber \\
A_N - A_S &= q_m \dif \phi = -\iu \E^{-\iu q_m \phi} \dif(\E^{\iu q_m \phi}), && \theta \neq \{0,\cpi\}.
\end{align}
This gauge field configuration is known as the Dirac monopole. We have $\dif A_{N, S} = F$ wherever they are defined. When we pick a particular gauge, the locus where the gauge field is not valid in that gauge is known as the ``Dirac string.'' It is not a physical object, just an artifact of a choice of gauge. The ``northern'' gauge field $A_N$ is valid everywhere except a Dirac string at the ``south pole'' at $\theta = \cpi$; the ``southern'' gauge field $A_S$ is valid everywhere except a Dirac string at the ``north pole'' at $\theta = 0$.  In particular, $A_N \to 0$ at the north pole and $A_S \to 0$ at the south pole, so they are well-defined at these points even though $\dif \phi$ isn't. Everywhere that both are valid, they are related by a gauge transformation as in~\eqref{eq:U1gaugefieldtransform}. As in the torus example from \S\ref{ex:torusflux}, the gauge field transformation has nontrivial {\em winding}, this time in the azimuthal angle $\phi$. Unlike the case of the torus, the spacetime itself does not have any 1-cycles for the gauge transformation to wind around. However, the region on which the gauge fields overlap does: it is a slice of spacetime with the $z$-axis removed, allowing winding around the axis. The integer winding number $q_m$ is the same as the number of units of magnetic charge carried by the monopole.

The Dirac monopole is a singular field configuration, in the sense that the energy density stored in the magnetic field diverges at the location of the monopole. This is nothing special about monopoles; it is really the same problem as the classical self-energy puzzle for the electron. However, as we will discuss in \S\ref{sec:WGC}, the expected resolution of the puzzle is different. For the electron, the self-energy puzzle is resolved by quantum mechanics. For monopoles, we expect that it is resolved by the monopole having new physical structure inside a core region. A famous example is the {\em 't Hooft--Polyakov monopole}, which is a classical solution in the theory of an SU(2) gauge field higgsed to U(1) by the VEV of an adjoint scalar. In this case, the monopole core radius is of order $m_W^{-1}$; the solution behaves like the Dirac monopole~\eqref{eq:diracmonopole} at larger radii, but involves the full set of SU(2) fields inside the core. The classical solution has finite energy. However, it is important to realize that this is just one example of how a magnetic monopole can arise. Not every U(1) gauge theory has a non-abelian UV completion. There are other known examples where magnetic monopoles are really fundamental objects, like D-branes, and EFT breaks down completely inside the radius of their core.

\subsubsection{Flux in integral cohomology}

When we do not consider magnetic monopoles, we have $\dif F = 0$, i.e., $F$ is a {\em closed} differential form. This means that $F$ has an image in the de Rham cohomology of our spacetime manifold $M$, $[F] \in H^2_{\mathrm{dR}}(M, \mathbb{R})$, which consists of closed forms modulo {\em exact} forms (those with $F = \dif A$ with $A$ a well-defined 1-form). In fact, the quantization condition~\eqref{eq:fluxquant} implies that $\frac{1}{2\cpi} F$ is a representative of a class in the {\em integral} cohomology, 
\begin{equation}   \label{eq:FinH2Z}
\left[\frac{F}{2\cpi}\right] \in H^2(M, \mathbb{Z}).
\end{equation} 
This will be important below when we discuss quantization of axion couplings. When we {\em do} have magnetic monopoles in the theory, we could consider $M$ to be our spacetime manifold with monopole worldlines excised, and these statements will still apply. (However, there is also more to say in cases with degrees of freedom localized on monopole worldlines.)

\subsection{Electric flux quantization; the Witten effect}
\label{subsec:electricfluxquantization}

We have seen that electric charge is quantized in $\Uone$ gauge theory. This also leads to a quantization of electric flux. To discuss electric flux quantization, we have to include the kinetic term of the gauge field and its coupling to an electric current $j_\mathrm{el}$. We can write this action in two equivalent forms, one with tensor index notation and one with differential form notation. As we discussed in the case of global symmetries, it is actually most natural to define a current as a $(d-1)$-form, hence the 3-form $J_\mathrm{el}$ in 4d $\Uone$ gauge theory. Thus we have:\footnote{In general, we may have to amend our definition of the current to generate ``seagull terms'' like the $A_\mu A^\mu |\phi|^2$ term that is familiar in scalar QED.}
\begin{align}  \label{eq:SQED}
S &= \int \mathrm{d}^4x \sqrt{-g}\, \left(-\frac{1}{4 e^2} F_{\mu \nu} F^{\mu \nu} - A_\mu j_\mathrm{el}^\mu\right) \nonumber \\
&= \int \left(-\frac{1}{2 e^2} F \wedge \star F - A \wedge J_\mathrm{el}\right).
\end{align}
Note that the gauge field here is not canonically normalized, due to the $1/e^2$ in front of the kinetic term. We have been using this non-canonical normalization all along, because it makes formulas like the Wilson line simple. Charge is manifestly quantized in integer units in this normalization. One can easily translate between these normalizations; the canonical gauge field ${\hat A} = A/e$.

From the action~\eqref{eq:SQED}, we can derive the equation of motion for $A$:
\begin{equation}  \label{eq:maxwelleq}
\frac{1}{e^2} \dif{\star F} = J_\mathrm{el}.
\end{equation}
This is just the familiar Maxwell equation (in curved spacetime) $\frac{1}{e^2} \nabla^\mu F_{\mu \nu} = j_\nu$, written in the language of differential forms. An immediate consequence of this equation is that, in QED, given a 3-manifold $\Omega$ in spacetime we have
\begin{equation} \label{eq:QOmega}
Q(\Omega) = \int_\Omega J_\mathrm{el} = \frac{1}{e^2} \int_\Omega \dif{\star F} = \frac{1}{e^2} \int_{\partial \Omega} \star F.
\end{equation}
In particular, if $\Omega$ is a closed manifold (i.e., compact and without boundary), then $Q(\Omega) = 0$. This is just Gauss's law: on a compact space without boundary, there can be no net electric charge, because all field lines must end somewhere. Notice that Gauss's law follows from the fact that $J_\mathrm{el}$ is not just a closed form (i.e., a conserved current), but an {\em exact} one: it is $\mathrm{d}$ of another form. Generic closed forms define global symmetries, but exact forms define {\em gauged} symmetries. This is worth emphasizing.

\begin{framed}
\noindent
A conserved current $J$ that is not just closed but {\em exact}, $J = \dif K$, is the hallmark of a symmetry that has been {\em gauged}. There is a corresponding Gauss's law constraint.
\end{framed}

Another consequence is the quantization of electric flux. Suppose that $\Sigma$ is a closed surface in spacetime that does not intersect any charged objects or nonzero currents, which bounds a 3-manifold $\Omega$.\footnote{Unlike for magnetic flux quantization, we will not discuss the case of homologically nontrivial $\Omega$ here. The general story will appear later, in \S\ref{sec:generalizedsymmetry}, when we discuss the 1-form electric symmetry.} Then~\eqref{eq:QOmega} implies
\begin{equation} \label{eq:elecfluxquant}
\frac{1}{e^2} \int_\Sigma \star F \in \mathbb{Z}.
\end{equation}
Notice that, unlike {\em magnetic} flux quantization~\eqref{eq:fluxquant}, electric flux quantization depends on the spacetime metric (via the Hodge star) and on the prefactor $1/e^2$ for the gauge kinetic term. A related, important remark is that electric flux quantization,~\eqref{eq:elecfluxquant}, is a statement that holds {\em for field configurations that obey the equations of motion of the theory}, as evidenced by the prominent role played by Maxwell's equation~\eqref{eq:maxwelleq} in our discussion above. It is {\em not} a valid statement about arbitrary configurations that are summed over in the path integral. This is important, because otherwise one would have been able to view the gauge field kinetic term as a topological invariant, formed from the wedge product of the electric and magnetic flux densities. Instead, quantization of $\int F$ is topological but quantization of $\int \star F$ is dynamical: it holds only for those special field configurations which are saddle points of the path integral. In theories with electric-magnetic duality, like free Maxwell theory, these roles can be interchanged. It is important to realize, however, that to define the path integral in any duality frame, one must specify what $\Uone$ gauge bundles are being summed over, which will always impose a topological constraint~\eqref{eq:fluxquant} on that $\Uone$ field strength's fluxes in every field configuration in the path integral.

In fact, electric flux quantization in general depends on the detailed form of the Lagrangian. Given the coupling $A \wedge J$, we can insert a quantized charge $J = q\, \delta^{(3)}(\vec x)$ and then work out which flux it sources. For example, if we added terms proportional to $(F_{\mu \nu}F^{\mu \nu})^2$ or ${\bar \Psi} \sigma^{\mu \nu} \Psi F_{\mu \nu}$ to our Lagrangian, then the quantized quantity on the left-hand side of~\eqref{eq:elecfluxquant} would be corrected. You might recall that the electric field in free Maxwell theory is the canonical conjugate of the gauge field. This generalizes to the statement that the ``electric flux'' whose integral is quantized is $\delta {\cal L} / \delta F$. A useful example is provided in the case of two $\Uone$ gauge fields with kinetic mixing, as discussed in \S\ref{sec:kineticmixing} below. Another important example arises when we add a $\theta$ term~\eqref{eq:thetaactionforms} to the theory. In this case, the variation of ${\cal L}$ with respect to $F$ acquires a new term, and so the electric flux quantization condition takes the form
\begin{equation} \label{eq:witteneffect1}
\int_\Sigma \left(\frac{1}{e^2} \star F + \frac{\theta}{4\cpi^2} F\right) \in \mathbb{Z}.
\end{equation}
This fact is known as the Witten effect~\cite{Witten:1979ey}. It implies that if we have a magnetic monopole of magnetic charge $p \in \mathbb{Z}$, which sources a magnetic flux $\frac{1}{2\cpi} \int_\Sigma F = p$, then the electric charge of the monopole, defined as $Q = \int_\Sigma \frac{1}{e^2} \star F$, is necessarily nonzero for generic $\theta$:
\begin{equation} \label{eq:witteneffect2}
Q = n - p \frac{\theta}{2\cpi}, \quad n \in \mathbb{Z}.
\end{equation}
The minimal electric charge is, in general, fractional. This does not violate charge quantization, in the sense that there is still a discrete set of possible charge assignments labeled by a lattice of integers $(n, p)\in \mathbb{Z}^2$. It does mean that the quantity we usually refer to as electric charge---what you would infer if you placed an electron near the object and measured the Coulomb force---no longer takes only integer values, because it is an irrational combination~\eqref{eq:witteneffect2} of the integers $n$ and $p$. 

Another simple derivation of the Witten effect is given by considering the theory in a spatially varying background $\theta(x)$ that gradually turns on at some radius away from a magnetic monopole, and solving the modified Maxwell's equations, then taking a limit where $\theta$ becomes constant. You can find this in~\cite{Wilczek:1987mv} or the lecture notes~\cite{Tong:GT}.

\subsection{Canonically normalized gauge fields and the Ward identity}

In the normalization in which we are working, $J_\mathrm{el}$ and $A$ have normalizations that are fixed by topology: $J_\mathrm{el}$ is normalized so that the charges obtained by integrating it are integers, and $A$ is normalized so that it transforms as~\eqref{eq:U1gaugefieldtransform} under $\Uone$ gauge transformations for which $\alpha(x) \cong \alpha(x) + 2\cpi$. Thus, quantum corrections can only change the coefficient $\frac{1}{e^2}$ in front of the kinetic term, and indeed they do. The familiar QED beta function shows up in this kind of shift. At one loop,
\begin{equation}
S \mapsto \int \left[-\frac{1}{2} \left(\frac{1}{e^2(\Lambda)} + \frac{b}{8\cpi^2} \log\frac{\Lambda}{\mu}\right) F \wedge \star F - A \wedge J_\mathrm{el}\right],
\end{equation}
where we identify the quantity in parentheses as the running coupling $1/e^2(\mu)$:
\begin{equation} \label{eq:runningU1coupling}
\frac{1}{e^2(\mu)} = \frac{1}{e^2(\Lambda)} + \frac{b}{8\cpi^2} \log\frac{\Lambda}{\mu}.
\end{equation}
Here $b$ is a beta function coefficient, equal to $\frac{2}{3} \sum_i q_i^2$ in a theory with a collection of Dirac fermions of charge $q_i$. 

This way of understanding the running of the electromagnetic coupling is one advantage of working in a non-canonical normalization where charge quantization is manifest. Compare the textbook approach: one introduces separate rescaling factors for the kinetic term $F_{\mu \nu}^2$, the kinetic term of charged fields, and the coupling term $A_\mu J_\mathrm{el}^\mu$. It seems not at all obvious that the renormalization of the three-point electron-positron-photon vertex should have anything to do with the vacuum polarization diagram that renormalizes the photon two-point function! However, eventually, one finds that the Ward identity guarantees that renormalization of the gauge coupling $e$ is completely determined by the renormalization of the photon kinetic term. This is often argued by complicated diagrammatic analysis. When we work in the normalization with manifest $\Uone$ charge quantization, it is just obvious: there is no coupling in front of the $A \wedge J_\mathrm{el}$ term, only in front of the $F \wedge \star F$ term, so the latter is the only thing that can run! After working this out, we are then always free to go back to canonical normalization.

\section{Kinetic mixing of U(1)s}
\label{sec:kineticmixing}

In recent years, ``millicharged particles'' have been frequently considered in the study of dark matter or physics beyond the Standard Model more generally. In this context, the ``milli-'' prefix just means ``very small,'' not specifically $10^{-3}$. I have told you that $\Uone$ charge is quantized, so how could we have a millicharge? It arises in theories with multiple gauge fields that kinetically mix with each other~\cite{Galison:1983pa, Holdom:1985ag}, and is perfectly compatible with charge quantization.

Consider a theory with gauge group $\Uone_A \times \Uone_B$, where the group elements for the gauge transformations are parametrized by $\E^{\iu \alpha(x)}$ and $\E^{\iu \beta(x)}$ respectively. We label the gauge fields $A$ and $B$. A field $\psi$ can transform in the $(q_A, q_B)$ representation, with
\begin{equation}
\psi(x) \mapsto \E^{\iu \left[q_A \alpha(x) + q_B \beta(x)\right]} \psi(x).
\end{equation}
From this it is clear that {\em both} $q_A, q_B \in \mathbb{Z}$ are quantized, since $\alpha$ and $\beta$ are only defined modulo $2\cpi$. 

The Lagrangian for the $A$ and $B$ gauge fields can have a kinetic mixing parameter $\kappa$, and currents $J_A, J_B$ coupled to the gauge fields:
\begin{equation}
-\frac{1}{2e_A^2} F_A \wedge \star F_A - \frac{\kappa}{e_A e_B} F_A \wedge \star F_B -\frac{1}{2e_B^2} F_B \wedge \star F_B - A \wedge J_A - B \wedge J_B
\end{equation}
In this basis charge conservation is completely manifest, $\int_{\Sigma_3} J_{A,B} \in \mathbb{Z}$, and each gauge field couples to a quantized charge. In particular, we have the usual magnetic flux quantization conditions:
\begin{equation}
\frac{1}{2\cpi} \int_{\Sigma_2} F_A \in \mathbb{Z}, \qquad \frac{1}{2\cpi} \int_{\Sigma_2} F_B \in \mathbb{Z}.
\end{equation}
Notice that these hold independently for each $\Uone$ gauge group.

Nonetheless, a particle charged under $A$ and a particle charged under $B$ can scatter through the kinetic mixing. There is a propagator that connects an $A$ vertex on one end to a $B$ vertex on the other. In this sense, the scattering amplitude of an electron $e^-$ charged under $A$ and a dark fermion $f$ charged under $B$ will {\em appear} as if $f$ carries a small charge under $A$. One manifestation of this is that the {\em electric} flux quantization conditions are nonstandard. In particular, a particle charged under $A$ will source a $B$ field, despite having no direct coupling. The equations of motion tell us that
\begin{equation}
\int \left(\frac{1}{e_A^2} \star F_A + \frac{\kappa}{e_A e_B} \star F_B\right) \in \mathbb{Z}, \qquad \int \left(\frac{1}{e_B^2} \star F_B + \frac{\kappa}{e_A e_B} \star F_A\right) \in \mathbb{Z}. 
\end{equation}
This makes clear that if we insert a particle of charge $q$ under $B$, sourcing a flux $q = \int \frac{1}{e_B^2} \star F_B$, we will inevitably also source a small flux of $A$, namely (to order $\kappa$)
\begin{equation}
\int \frac{1}{e_A^2} \star F_A = - \int \frac{\kappa}{e_A e_B} \star F_B = -\frac{\kappa e_B}{e_A} q.
\end{equation}
This is the ``millicharge'' that $B$-charged particles carry under the photon $A$. (Notice the close similarity to the discussion of the Witten effect~\eqref{eq:witteneffect1}, although that case was a mixing of magnetic and electric charge under a single $\Uone$, rather than of electric charges under two different $\Uone$s.)

Another way to see that particles carrying different charges interact through the kinetic mixing is to do a field redefinition. We generally choose to maintain the definition of $A$ as the field to which the electron couples. However, we can redefine $B \mapsto B - \frac{\kappa e_B}{e_A} A$, which cancels the kinetic mixing term. This choice gives familiar-looking Feynman rules where each particle has an independent propagator. However, it also means that the gauge fields in the new basis no longer couple to quantized charges: we now have a coupling $A \wedge J_B$ with an irrational coefficient (the millicharge). Another consequence is that the {\em magnetic} flux quantization conditions now take on an unusual form: instead of $\frac{1}{2\cpi} \int F_B \in \mathbb{Z}$, we now have $\frac{1}{2\cpi} \int \left(F_B - \frac{\kappa e_B}{e_A} F_A\right) \in \mathbb{Z}$. There is nothing wrong with such a basis. It is well-suited for perturbative calculations, but it does make questions of charge quantization more obscure.

\medskip
\noindent\centerline{\rule{\textwidth}{0.5pt}}
\noindent {\em Exercise:} Consider a case where the second photon, $B$, is massive. This could happen through the Higgs mechanism, but for the current problem you can simply add an explicit $B_\mu B^\mu$ term (which you can write in Stueckelberg form to make it manifestly gauge invariant, if you like). Explain how a field redefinition can simultaneously diagonalize the kinetic and mass terms. Investigate the charge and flux quantization conditions in the new frame. This is the well-studied case of a ``dark photon'' particle, often studied as a potential mediator between the Standard Model and dark matter.\\
\noindent\centerline{\rule[6.0pt]{\textwidth}{0.5pt}}

Notice that, although the preferred choice of basis might shift, the physics of the case of massless and massive $B$ is completely continuous as the mass is taken to zero. (At least one popular review article about dark photons is highly misleading on this point.)

To summarize, in a theory with multiple $\Uone$ gauge groups, the charges (both magnetic and electric) are always quantized in a lattice. However, particles carrying different charge can scatter with each other through Coulomb interactions mediated by the off-diagonal kinetic terms. Depending on the physical question that one wants to ask, one basis or another might be better suited, and in a given basis it might appear that there can be a non-quantized millicharge. None of this changes the underlying group theoretic fact of charge quantization.

\newpage

\section*{\Large Part Two: Instantons and Chiral Anomalies}
\label{sec:lecturetwo}
\addcontentsline{toc}{section}{\nameref{sec:lecturetwo}}

\section{U(1) fields with nonzero $\int F \wedge F$}

We saw above that U(1) gauge theory obeys a flux quantization condition: $\frac{1}{2\cpi}\int F \in \mathbb{Z}$. Much of our following discussion will be about axion fields that couple to $F \wedge F$, where a generalized ``squared-flux quantization'' condition will play an important role, both for U(1) gauge fields and for $\SUN$ gauge fields. A full mathematical treatment of these conditions is beyond the scope of these lectures, but can be found in textbooks that discuss characteristic classes, e.g.,~\cite{milnor1974characteristic, Nakahara:2003nw}. Rather than giving a general proof, let's look at an example that happens to give us the right answer.

In \S\ref{ex:torusflux}, we showed that in a theory where two spatial dimensions form a torus, configurations exist with flux $n$, given by $F = \frac{n}{2\cpi r_1 r_2} \dif x_1 \wedge \dif x_2$. Then it is clear that we can define a configuration on a four-dimensional torus that has a nonzero $\int F \wedge F$ simply by taking
\begin{equation}
F = \frac{n}{2\cpi r_1 r_2} \dif x_1 \wedge \dif x_2 + \frac{m}{2\cpi r_3 r_4} \dif x_3 \wedge \dif x_4, \quad n, m \in \mathbb{Z}.
\end{equation}
Indeed, direct computation shows that 
\begin{equation}
F \wedge F = \frac{n m}{2\cpi^2 r_1 r_2 r_3 r_4} \dif x_1 \wedge \dif x_2 \wedge \dif x_3 \wedge \dif x_4,
\end{equation}
and hence
\begin{equation}
\int F \wedge F = (2\cpi)^4 r_1 r_2 r_3 r_4 \frac{n m}{2\cpi^2 r_1 r_2 r_3 r_4} = 8 \cpi^2 n m.
\end{equation}
Thus we see that, for this class of field configurations, the smallest possible value of $\int F \wedge F$ is $8\cpi^2$, and the attainable values are all integer multiples of this.

It turns out that, even though we focused on a particular example, this conclusion is the correct one for our purposes. The full mathematical story is somewhat more subtle. The quantization of magnetic flux~\eqref{eq:FinH2Z} implies that
\begin{equation} \label{eq:FcupF}
\left[\frac{F}{2\cpi}\right] \smile \left[\frac{F}{2\cpi}\right] \in H^4(M, \mathbb{Z}),
\end{equation}
and hence that
\begin{equation}
\frac{1}{4\cpi^2} \int_M F \wedge F \in \mathbb{Z}, \quad \text{for any closed }M.
\end{equation}
Thus, the base unit of $\int F \wedge F$ is, in complete generality, $4\cpi^2$ rather than $8\cpi^2$. Any example, as in our torus context, that constructs $M$ as a product of two two-manifolds with flux will lead to $\frac{1}{4\cpi^2} \int F \wedge F$ being an {\em even} integer. However, it is possible to define a gauge field configuration on the complex projective space $\mathbb{CP}^2$ (which does not have such a product form) for which $\frac{1}{4\cpi^2} \int F \wedge F = 1$. On the other hand, $\mathbb{CP}^2$ does not admit a spin structure, meaning that we cannot consistently define a quantum field theory with spinor fields on this manifold.

It is a mathematical fact that, for any {\em spin} 4-manifold, the integral of the square of the magnetic flux is an even integer, or equivalently 
\begin{equation} \label{eq:FwedgeFquantization}
\frac{1}{8\cpi^2} \int_M F \wedge F \in \mathbb{Z}, \quad \text{for any closed, spin }M.
\end{equation}
Since we will exclusively be interested in theories with fermions, as in the real world, this is the quantization condition that will be relevant for us.

There are no U(1) gauge field configurations on $S^4$ for which $\int F \wedge F \neq 0$; the structure~\eqref{eq:FcupF} implies that we only find such configurations on spaces with 2-cycles. As we will see in section \S\ref{sec:instantons}, the story is different for the gauge group $\SUN$.

\section{Non-abelian gauge fields}
\label{sec:nonabelian}

We will not give a detailed review of non-abelian gauge theory in these notes, as it is covered adequately in many quantum field theory textbooks. It is useful, however, for us to give a very quick summary of how some of the standard formulas look when written in differential form notation, and in particular we should see how differential form notation meshes with familiar non-abelian matrix notation. A (continuous) nonabelian gauge group has a set of generators $T^a$ whose commutation relations define the structure constants $f^{abc}$:
\begin{equation}
[T^a, T^b] = \iu f^{abc} T^c.
\end{equation}
The vector space of real linear combinations of these generators, together with the commutation relations, defines a Lie algebra. We define an appropriate field strength $F$ that transforms linearly under a gauge transformation, $F \mapsto g F g^{-1}$, with components
\begin{equation}
F^a_{\mu \nu} = \partial_\mu A^a_\nu - \partial_\nu A^a_\mu + f^{abc} A^b_\mu A^c_\nu.
\end{equation}
Our gauge field $A$ is (locally) a 1-form taking values in the Lie algebra, 
\begin{equation}
A \equiv A_\mu \dif x^\mu \equiv A^a_\mu T^a \dif x^\mu.
\end{equation}
The field strength $F$ is then a Lie algebra-valued 2-form,
\begin{equation}
F = \frac{1}{2} F_{\mu \nu} \dif x^\mu \wedge \dif x^\nu = \frac{1}{2}\left(\partial_\mu A_\nu - \partial_\nu A_\mu - \iu [A_\mu, A_\nu]\right) \dif x^\mu \wedge \dif x^\nu = \dif A - \iu A \wedge A.
\end{equation}
The final step gives the most convenient differential form notation for the field strength, which implicitly encodes the matrix algebra because
\begin{equation}
A \wedge A = A^a_\mu T^a A^b_\nu T^b \dif x^\mu \wedge \dif x^\nu = \frac{1}{2} A^a_\mu A^b_\nu \iu f^{abc} T^c \dif x^\mu \wedge \dif x^\nu.
\end{equation}
The standard kinetic term for a non-abelian gauge theory is
\begin{equation} \label{eq:SUNgaugetheory}
\int \left(-\frac{1}{g^2} \mathrm{tr}(F \wedge \star F)\right),
\end{equation}
where for $\SUN$ gauge theory we conventionally write the generators $T^a$ in the fundamental representation normalized such that $\mathrm{tr}(T^a T^b) = \frac{1}{2} \delta^{ab}$. (It is because of this $1/2$ that we simply have $-1/g^2$ in~\eqref{eq:SUNgaugetheory}, rather than the $-1/(2e^2)$ factor in the $\Uone$ case~\eqref{eq:gaugeactionforms}.) 

In working with non-abelian gauge theories, we frequently encounter traces. When one has differential forms appearing inside a trace, one must be careful with minus signs when applying standard identities like the cyclic property of the trace. For example, if $\omega$ and $\eta$ are respectively Lie algebra-valued $p$- and $q$-forms, then we have
\begin{equation}
\mathrm{tr}(\omega \wedge \eta) = (-1)^{pq} \mathrm{tr}(\eta \wedge \omega).
\end{equation}
This combines the familiar graded commutativity of the differential forms with the familiar cyclic property of the trace. Simple applications are:
\begin{align}
\mathrm{tr}(A \wedge F) &= \mathrm{tr}(F \wedge A), \nonumber \\
\mathrm{tr}(A \wedge A) &= - \mathrm{tr}(A \wedge A) = 0, \nonumber \\
\mathrm{tr}(A \wedge A \wedge A \wedge A) &= 0.
\end{align}
This last identity will make an appearance below, in \S\ref{sec:CSgeneral}.

\section{Instantons}
\label{sec:instantons}

{\em ``You can't eat an instanton.'' -- John Stout}

\subsection{The BPST instanton solution and its properties}

We have just constructed field configurations with nonzero $F \wedge F$ in a $\Uone$ gauge theory. This was relatively straightforward, once we allow ourselves to work on a space of nontrivial topology, because we can just exploit the winding of $\Uone$ gauge transformations around a circle. For $\SUN$ gauge theory, there is a similar story, but the topology that we exploit is a bit different. In this case, it is possible to construct a classical solution to the Euclidean Yang-Mills equations of motion on $S^4$, which is {\em localized} in Euclidean spacetime and has a nonzero $\int \mathrm{tr}(F \wedge F)$. This is known as the BPST (Belavin, Polyakov, Schwarz, Tyupkin) instanton solution~\cite{Belavin:1975fg}. For the gauge group $\SU(2)$, an explicit solution for a family of such solutions is given that depends on five real parameters: a four-vector of positions $x_0^\mu$ and a ``size modulus'' $\rho$. (When a classical solution depends on parameters in this way, they are variously referred to as ``zero modes,'' ``moduli,'' or ``collective coordinates.'') Namely,
\begin{equation} \label{eq:BPST}
A_\mu^a(x) = \frac{2 \eta^a_{\mu \nu} (x - x_0)^\nu}{(x-x_0)^2 + \rho^2}.
\end{equation}
Here $\mu$ is a (Euclidean) spatial index and $a$ is an $\SU(2)$ adjoint gauge index ($a \in \{1,2,3\}$). Here $\eta^a_{\mu \nu}$ is known as the ``'t Hooft symbol''~\cite{tHooft:1976snw} and is defined as
\begin{equation} \label{eq:tHooftsymbol}
\eta^a_{\mu \nu} = \begin{cases} \epsilon^{a\mu \nu}, & \mu, \nu \in \{1,2,3\} \\ -\delta^{a\nu}, & \mu = 0 \\ +\delta^{a\mu}, & \nu = 0 \\ 0, & \mu = \nu = 0.\end{cases}
\end{equation}
The first thing you might notice is that we have written this as a solution in flat Euclidean space $\mathbb{R}^4$. However, it can be thought of as a solution on the compact space $S^4$, with a point at infinity added to $\mathbb{R}^4$. The reason is that this solution decays away quickly at $|x| \to \infty$, as $F \sim |x|^{-4}$, so the field strength goes to zero at the extra point at infinity (where $A_\mu^a$ asymptotes to a pure gauge field configuration).

This solution has the special property that it is {\em self-dual}, i.e., 
\begin{equation}
F^a_{\mu \nu} = \widetilde{F}^a_{\mu \nu}.
\end{equation}
There is a different solution which is {\em anti-self-dual}, $F^a_{\mu \nu} = -\widetilde{F}^a_{\mu \nu}$. It looks exactly the same except that the 't Hooft symbol is replaced by a different symbol ${\overline \eta}^a_{\mu \nu}$, which exchanges the role of $\mu$ and $\nu$ in~\eqref{eq:tHooftsymbol}. The 't Hooft symbol may seem mysterious. Some rough intuition for where it comes from is the following: the Euclidean theory has an SO(4) rotational symmetry. There is a double cover of SO(4) by the group $\SU(2) \times \SU(2)$. One can search for a classical solution that breaks the product of {\em one} spatial $\SU(2)$ and the gauge $\SU(2)$ to the diagonal. This is loosely what is going on with the 't Hooft symbol, which mixes up spatial and internal $\SU(2)$ indices.\footnote{A reader might wonder why I am discussing this in terms of $\SU(2)$ rather than SO(3); the reason is that $\SUN/\mathbb{Z}_N$ has a different quantization of instanton number than $\SUN$, allowing ``fractional instantons.'' This is related to all sorts of interesting physics that I will omit to keep these lecture notes to a vaguely manageable length.}

The self-dual instanton solution has two key properties that will be at the core of the physics of axions that we will discuss soon. First, it has a nontrivial integral $\int_{S^4} \mathrm{tr}(F \wedge F)$:
\begin{equation} \label{eq:instantontopcharge}
\left. \int_{S^4} \mathrm{tr}(F \wedge F) \right|_{\mathrm{inst.}} = 8 \cpi^2.
\end{equation}
The anti-self-dual instanton solution has $-8\cpi^2$ for this integral. In fact, one can show that {\em any} $\SUN$ gauge field configuration on any closed 4-manifold has a quantized integral of $\mathrm{tr}(F \wedge F)$ which is an integer multiple of this base unit:
\begin{equation} \label{eq:instantonnumberquantization}
\frac{1}{8\cpi^2} \int_\Sigma \mathrm{tr}(F \wedge F) \in \mathbb{Z}.
\end{equation}
Like the quantization of magnetic flux, this is a property of {\em any field configuration} that we sum over in the path integral, not just of solutions to the equations of motion. We call this integer the {\em instanton number} of the field configuration. (It is also sometimes called the ``topological charge,'' though this is a bit more vague because it could have other meanings in different contexts.)

The second key property is that the instanton solution has a large Euclidean action, diverging in the limit $g \to 0$:
\begin{equation} \label{eq:instantonaction}
\left. \int_{S^4} \left(\frac{1}{g^2} \mathrm{tr}(F \wedge \star F)\right) \right|_{\mathrm{inst.}}= \frac{8\cpi^2}{g^2}.
\end{equation}
This shows that the instanton is a non-perturbative effect: the {\em canonically} normalized gauge field solution scales as $1/g$. (Notice that the Euclidean expression for the kinetic term has the opposite sign from the Minkowski action in~\eqref{eq:SUNgaugetheory}.)

We can understand the special role of self-dual field configurations in this story using a Bogomol'nyi trick. This is a rewriting of the action as a sum of a perfect square and a topological invariant, which is useful in studying a number of interesting topological objects in quantum field theory. In this context, the Euclidean action can be rewritten as
\begin{equation}  \label{eq:Bogomolnyi}
\int \frac{1}{g^2} \mathrm{tr}(F \wedge \star F) = \int \frac{1}{g^2} \left\{\frac{1}{2} \mathrm{tr}\left[(F \mp \star F) \wedge \star(F \mp \star F)\right] \pm \mathrm{tr}(F \wedge F)\right\}.  
\end{equation}
The Bogomol'nyi trick makes it obvious that a self-dual field is always a solution to the full Yang-Mills equations of motion. This is because $\int \mathrm{tr}(F \wedge F)$ is a topological invariant, so a small variation away from a self-dual field cannot change the second term in~\eqref{eq:Bogomolnyi}. On the other hand, the first term (with the upper sign choice) vanishes for a self-dual field, and is positive otherwise, so it can only increase when the field is varied. As a result, a self-dual field is necessarily a local minimum of the Euclidean action. The same holds for an anti-self-dual field, with the lower sign choice. Because self-dual and anti-self-dual field configurations have positive and negative instanton number $n$ respectively, we see that they all have $S = 8\cpi^2|n|/g^2$.

\subsection{Instantons in the path integral}

The path integral for non-abelian gauge theory sums over all field configurations in spacetime. Because the instanton number $n$ in~\eqref{eq:instantonnumberquantization} is a topological invariant, one can decompose the path integral into a discrete sum over $n \in \mathbb{Z}$ together with a continuous integral over topologically trivial differences in field configurations for every $n$. The sum over nontrivial topologies is necessary: restricting to the $n = 0$ sector is actually inconsistent with locality and unitarity. For example, a spacetime that pinches off into two separate spacetimes can have a total instanton number zero but have nonzero instanton numbers in its constituent parts. Thus, compatibility with basic axioms of QFT requires a sum over topological sectors. However, it was appreciated in recent years that it is possible to restrict the sum to only values of $n$ that are multiples of a fixed nonzero integer $p \in \mathbb{Z}$~\cite{Seiberg:2010qd}. Such theories have identical local physics, but differ in the correlation functions of nonlocal operators; in particular, they have a $\mathbb{Z}_p$ 3-form symmetry~\cite{Tanizaki:2019rbk} (see \S\ref{sec:generalizedsymmetry} for this terminology, although not this application).

When instantons were first discovered, there were hopes that they might help to analytically understand the strongly coupled physics of confinement, because they contribute calculable nonperturbative effects in the path integral~\cite{Callan:1976je}. One can treat them with semiclassical methods: the instanton solutions are saddle points in the path integral, and we can integrate over their collective coordinates or moduli, like the instanton position and size. Unfortunately, these methods do not take us very far. The semiclassical method works well when the instanton action $8\cpi^2/g^2$ is large, i.e., when $g$ is small, and so we should expect that in the infrared, where a non-abelian gauge theory confines because $g$ becomes large, semiclassical methods will also fail. In fact, when one computes the measure for integrating over the instanton size parameter $\rho$, one finds that it modifies the instanton measure in the path integral for $\SUN$ Yang-Mills theory with a prefactor~\cite{Schafer:1996wv, Weinberg:2012pjx, Shifman:2022shi}
\begin{equation}
\left(\frac{8\cpi^2}{g^2}\right)^{2 N} \frac{1}{\rho^5} \left(M \rho\right)^{11 N/3} \exp\left(-\frac{8\cpi^2}{g^2}\right).
\end{equation}
Here $M$ is a UV regulator mass scale, and we should understand the coupling $g$ to be the running coupling evaluated at the scale $M$. The exponent $11 N/3$ should look familiar: it is the one-loop beta function coefficient in Yang-Mills theory. This is no accident, as it allows the $M$-dependent prefactor to be combined with the exponential of the instanton action to form
\begin{equation}
\left(M \rho\right)^{11N/3} \exp\left(-\frac{8\cpi^2}{g(M)^2}\right) = \exp\left(-\frac{8\cpi^2}{g(M)^2} + \frac{11N }{3} \log(M\rho)\right) = \exp\left(-\frac{8\cpi^2}{g(\rho^{-1})^2}\right).
\end{equation}
In other words, the measure for integrating over the size modulus effectively replaces the coupling $g$ evaluated at the UV scale $M$ in the instanton action with the coupling evaluated at the scale $\rho^{-1}$. For large values of $\rho$, this running coupling becomes large, and the calculation breaks down. In particular, the integral over $\rho$ with this measure diverges. We say that a ``dilute instanton gas approximation'' is valid for small $\rho$, but not for large $\rho$. This limits the phenomenological utility of semiclassical instanton calculations in QCD. We will see a consequence when we discuss the axion solution to the Strong CP problem in \S\ref{subsec:axioncoreidea}: there is a loose sense in which the axion potential is generated ``by instantons'' (as is often said colloquially), which is that axions couple to the instanton number density $\mathrm{tr}(F \wedge F)$. However, the axion potential cannot be calculated using semiclassical instanton methods, and is instead obtained from a conceptually different approach using the chiral Lagrangian. 

\subsection{Comments on instantons}
\label{subsec:commentsoninstantons}

The sage words of my collaborator John Stout at the opening of this section, ``you can't eat an instanton,'' refer to the fact that an instanton is not an object that exists in real time. It is localized in Euclidean time as well as in space. It has no worldline. It lives at one {\em instant}---hence the name, due to 't Hooft. (BPST called it a ``pseudoparticle,'' but 't Hooft's term became the standard one.) However, in more than four dimensions, BPST instantons {\em do} become ordinary dynamical objects. For example, in a 5d theory, we can consider the solution~\eqref{eq:BPST} as a function of four spatial coordinates, and just tack on extra time coordinate that the solution does not depend on. This now describes a kind of {\em particle}, of size $\rho$, localized at $x_0$ and staying put for all time. Such a particle is an honest, dynamical, solitonic object in the theory. If you lived in 5d, you {\em could} eat an instanton. In 6d a BPST instanton is a string; in 7d it is a membrane with two spatial dimensions (a ``2-brane''); and so on. (In 10d it is a 5-brane, which might ring a bell for any of you who have studied string theory; in fact, NS5 branes are intimately related to BPST instantons in heterotic string theory.) Here we should mention an annoying feature of language: in {\em any} number of spacetime dimensions, the word ``instanton'' can be used to refer to a localized solution of the Euclidean equations of motion: a zero-dimensional object, localized in spacetime. On the other hand, the term ``BPST instanton'' or ``Yang-Mills instanton'' (or even just ``instanton'') can refer to the specific solution~\eqref{eq:BPST}, which is a $(d-4)$-dimensional object. This is one of those ambiguities in language that you just have to learn to resolve from contextual clues (or directly asking someone which they mean).

\section{The chiral ABJ and 't Hooft anomalies}
\label{sec:chiralanomalies}

I suggest that even readers who are thoroughly familiar with the chiral anomaly read~\ref{sec:chiralanomalytakeaway}, which summarizes the key way in which we will be making use of the anomaly to discuss axion physics in the subsequent sections.

\subsection{Introductory remarks}

In this section we will briefly review the {\em chiral anomaly}, in two forms: the Adler--Bell--Jackiw or ABJ anomaly (explicit symmetry breaking)~\cite{Adler:1969gk, Bell:1969ts} and the 't Hooft anomaly (obstruction to gauging)~\cite{tHooft:1979rat}. There are several different QFT calculations that reveal these anomalies, which require defining a regulator and using it carefully to obtain a regulator-independent physical result in the end. These calculations are somewhat subtle and are discussed in great detail in standard textbooks, so I will not reproduce them in depth here. In particular, the first four sections of chapter 19 of the textbook by Peskin and Schroeder~\cite{Peskin:1995ev} contain several calculations of the chiral anomaly from different viewpoints. I strongly encourage you to work through these calculations carefully. Here, I will simply summarize these standard arguments and highlight some conceptual aspects of these anomalies. I will also outline a derivation by Nielsen and Ninomiya, which is less of a standard textbook treatment, and which gives a picture of the anomaly in terms of a physical process of particle production~\cite{Nielsen:1983rb}.

We will study these anomalies in two different, related QFTs. The first is simply a theory of two free, massless Weyl fermions,
\begin{equation} \label{eq:Lag2F}
\frac{1}{\sqrt{|g|}} {\cal L} =  \psi^\dagger \iu \slashed{\partial} \psi + \overline{\psi}^\dagger \iu \slashed{\partial} \overline{\psi},
\end{equation}
with $\slashed{\partial} \equiv \overline{\sigma}^\mu \partial_\mu$. Notice that here $\psi$ and $\overline{\psi}$ are just {\em names} of two {\em different} left-handed Weyl fermions; I use the notation $\psi^\dagger$ for the hermitian conjugate of $\psi$, so don't interpret the bar over $\overline{\psi}$ as a conjugate.
This theory has a $\Uone_L \times \Uone_R$ global symmetry, where the two $\Uone$ factors rotate the two fermions {\em independently}: $\psi \mapsto \E^{\iu \alpha} \psi$ and $\overline{\psi} \mapsto \E^{-\iu \beta} \overline{\psi}$. In other words, the charges $(q_L, q_R)$ of $\psi$ and $\overline{\psi}$ under this group are $(1,0)$ and $(0,-1)$, respectively. There is a $\Uone$ {\em diagonal} subgroup of this symmetry, denoted $\Uone_V$, given by the elements where $\beta = \alpha$. Under $\Uone_V$, $\psi$ has charge $1$ and $\overline{\psi}$ has charge $-1$. The $V$ subscript here stands for ``vector-like''; if we package $\psi$ and $\overline{\psi}^\dagger$ into  a single Dirac fermion $\Psi$, this would simply  be the rotation $\Psi \mapsto \E^{\iu \alpha} \Psi$ that would be preserved by a Dirac mass term $m \overline{\Psi} \Psi$ (where here $\overline{\Psi}$ {\em does} mean the Dirac conjugate of $\Psi$). (The idea of packaging the two fields into a single Dirac fermion is also the origin of our labels ``L'' and ``R,'' as well as the sign convention we put on $\beta$, but from the point of the theory~\eqref{eq:Lag2F} taken on its own merits, which after all describes two completely unrelated and non-interacting fields, these conventions seem like idiosyncratic historical artifacts.) These symmetries have corresponding Noether currents
\begin{equation}
j_L^\mu = \psi^\dagger \overline{\sigma}^\mu \psi, \quad 
j_R^\mu = -\overline{\psi}^\dagger \overline{\sigma}^\mu \overline{\psi}, \quad j_V^\mu = j_L^\mu + j_R^\mu.
\end{equation}
We will see that this theory has an {\em 't Hooft anomaly}: although $\Uone_L \times \Uone_R$ is a perfectly good global symmetry of the quantum field theory, the full group {\em cannot be gauged}.

In the second theory, we also have a $\Uone$ gauge field, which has gauged only part of our original symmetry group, namely $\Uone_V$:
\begin{equation} \label{eq:Lag2Fgauged}
\frac{1}{\sqrt{|g|}} {\cal L} =  -\frac{1}{4e^2} F_{\mu \nu}F^{\mu \nu} + \psi^\dagger \iu \slashed{D} \psi + \overline{\psi}^\dagger \iu \slashed{D} \overline{\psi},
\end{equation}
where the covariant derivatives are
\begin{equation}
\slashed{D} \psi \equiv \overline{\sigma}^\mu (\partial_\mu - \iu A_\mu) \psi, \quad 
\slashed{D}{\overline \psi} \equiv \overline{\sigma}^\mu (\partial_\mu + \iu A_\mu) {\overline \psi}.
\end{equation}
Taken as a classical Lagrangian, this theory has a global symmetry that acts as $\Uone_L$, i.e., it acts as $\psi \mapsto \E^{\iu \alpha} \psi$ but does not touch $\overline{\psi}$. In the quantum field theory, this is not a symmetry at all! Its associated Noether current is not actually conserved. This is an example of an {\em ABJ anomaly}, when a symmetry of the classical action is not a symmetry of the quantum field theory. (We could say the same thing about $\Uone_R$.)

These two different types of anomalies are clearly related to each other. The ABJ anomaly, in a sense, triggers the 't Hooft anomaly. If we tried to gauge the full $\Uone_L \times \Uone_R$ symmetry of~\eqref{eq:Lag2F}, we could do so by first gauging the subgroup $\Uone_V$. We would not encounter an obstruction. But if we then tried to gauge the rest of the group, we would be faced with the ABJ anomaly of~\eqref{eq:Lag2Fgauged}: the current is not conserved, and so we can't couple a gauge field to it. The 't Hooft anomaly of~\eqref{eq:Lag2F} can alternatively be thought of as a {\em precursor} to the ABJ anomaly of~\eqref{eq:Lag2Fgauged}.

\subsection{The particle production calculation}

The first argument I want to present for the chiral anomaly is arguably the most physical. It computes how particles are produced in the presence of a time-dependent gauge field background, and how the resulting particle production violates chiral symmetry. It was first published (as far as I am aware) by Nielsen and Ninomiya~\cite{Nielsen:1983rb}, and is discussed in a recent textbook by Fradkin~\cite{fradkin2021quantum}. A nice treatment of it can also be found in lectures by David B.~Kaplan~\cite{Kaplan:2009yg}.

\subsubsection{The $(1+1)$d anomaly}

For this calculation it is easiest to explain the $(1+1)$d case first; this is an interesting result in its own right, as well as a key input to the $(3+1)$d calculation. Our goal is to study the behavior of chiral fermions in the presence of a background electric field. In $(1+1)$d, space is a line, and the analogue of chirality is whether the fermion is left-moving or right-moving. A left-moving Weyl fermion $\psi_-$ and right-moving Weyl fermion $\psi_+$ coupled to a background gauge field $A_\mu$ have an action
\begin{equation}
\int \mathrm{d}^2x\,\sqrt{|g|} \left(\iu \psi_+^\dagger D_- \psi_+ + \iu \psi_-^\dagger D_+ \psi_-\right),
\end{equation}
where
\begin{equation}
D_\pm = (\partial_t \pm \partial_x) + \iu (A_t \pm A_x).
\end{equation}
In particular, if we turn off the gauge field $A$, we have equations of motion $\partial_+ \psi_- = 0$ and $\partial_- \psi_+ = 0$, which are solved by allowing $\psi_-$ to be an arbitrary function of $t + x$ and $\psi_+$ to be an arbitrary function of $t - x$ (hence, left- and right-moving respectively). There is an important difference between $(1+1)$d chirality and $(3+1)$d chirality: in $(3+1)$d, the conjugate $\psi^\dagger$ of a left-handed Weyl fermion is a right-handed Weyl fermion, whereas in $(1+1)$d, the conjugate $\psi_-^\dagger$ of a left-mover is still a left-mover.

The classically conserved 1-form currents in this theory are
\begin{align}
J_+ &= \sqrt{|g|} \left(+\psi_+^\dagger \psi_+ \dif{t} + \psi_+^\dagger \psi_+ \dif{x}\right), \nonumber \\
J_- &= \sqrt{|g|} \left(-\psi_-^\dagger \psi_- \dif{t} + \psi_-^\dagger \psi_- \dif{x}\right).
\end{align}
The gauge field $A$ couples as $A \wedge (J_+ + J_-)$.

Using the ansatz $\psi \propto \exp(-\iu \omega t + \iu p x)$, the equations of motion $\partial_t \psi_+ = \partial_x \psi_+$ and $\partial_t \psi_- = - \partial_x \psi_-$ indicate that the right- and left-moving fermions have dispersion relation $\omega(p) = -p$ and $\omega(p) = +p$, respectively. It is useful to think in terms of the ``Dirac sea'' picture, so that initially all negative-frequency states, corresponding to $p \geq 0$ for $\psi_+$ and $p \leq 0$ for $\psi_-$, are filled. In general, I tend to think of the Dirac sea as a historical artifact that confuses more than it explains, but in this particular context it seems to be the simplest way to discuss the physics. If you are uncomfortable with this, you should be able to restate all of the following in terms of Bogoliubov coefficients and particle creation.

\begin{figure}[!h]
\centering
\includegraphics [width = 0.8\textwidth]{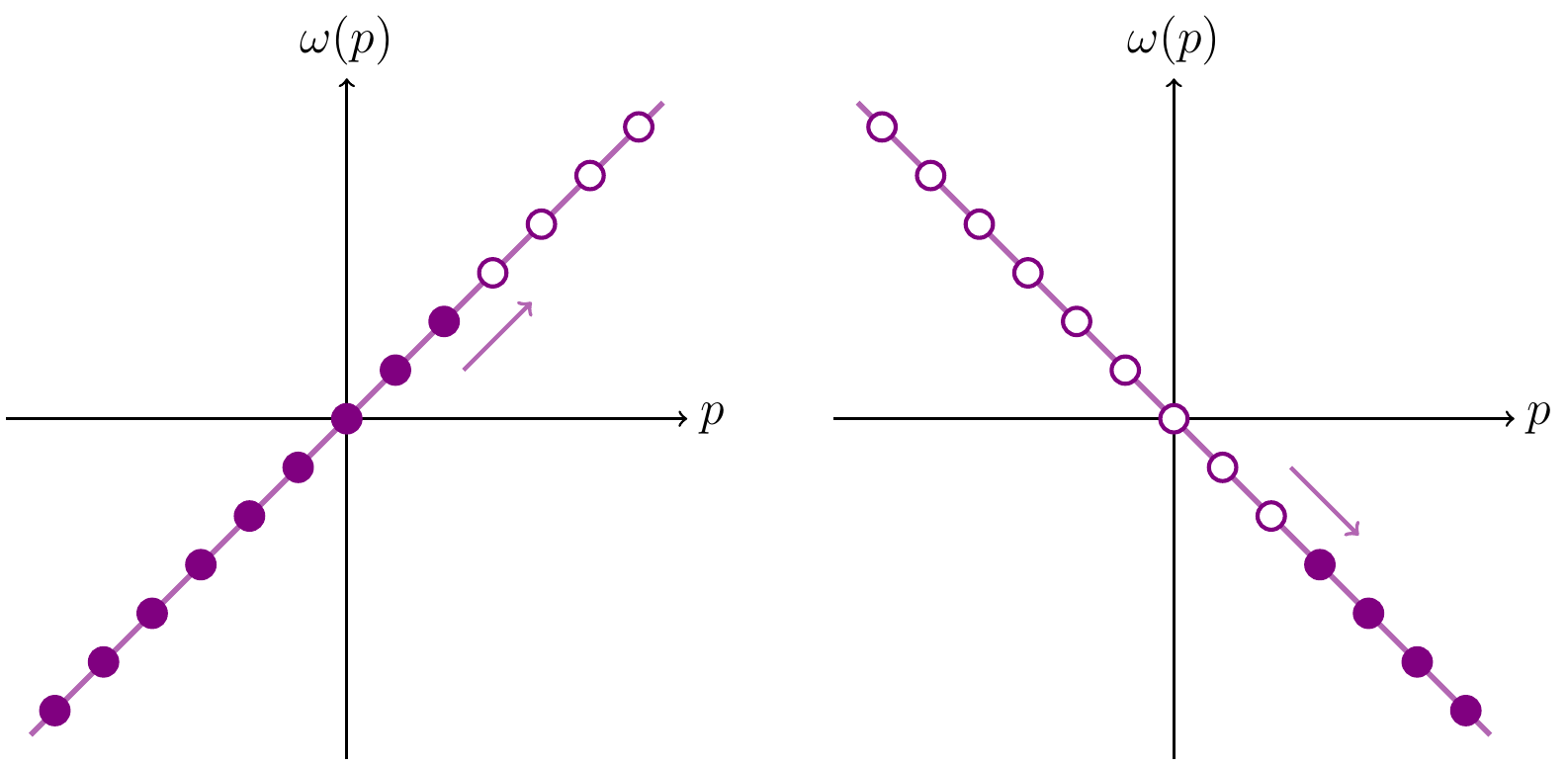}
\caption{Fermi surface for left- and right-handed fermion modes in the $(1+1)$d theory~\cite{Nielsen:1983rb}. Shaded circles are populated levels and white circles are unpopulated levels. An applied electric field increases the population of left-moving states and decreases the population of right-moving states, as indicated by the arrows.
} \label{fig:fermionlevels1d}
\end{figure}

Now, consider the theory in a background electric field $F = E(t) \dif t \wedge \dif x$, where $E(t)$ is adiabatically turned on at some time, and subsequently adiabatically turned off. In the presence of the electric field, the charged particles accelerate: their momentum changes according to $\dif p/\dif t = E(t)$. For concreteness, let's suppose that $p_E \equiv \int E(t) \dif t \geq 0$. For $\psi_-$, we initially had populated all states with $p \leq 0$. Each of these states increases in momentum, so at the time the electric field is switched off, we have populated a {\em larger} range of states, up to $p_E$. On the other hand, for $\psi_+$, we initially had populated all states with $p \geq 0$. Each of these states increases in $p$, so at the time the electric field is switched off, we have populated a {\em smaller} range of states, $p \geq p_E$. The result is a net depletion of right-moving states and a net increase in the population of left-moving states, i.e., a creation of a chiral asymmetry. This is depicted in Fig.~\ref{fig:fermionlevels1d}.

This story sounds a bit odd, because I've phrased it as a population of particle states that individually shift in momentum, which doesn't {\em sound} like it should be creating or destroying any particles. What makes it work, as Kaplan explains~\cite{Kaplan:2009yg}, is a sort of ``Hilbert hotel'': the states in the Dirac sea extend all the way out to negative infinity in $\omega$, so for every $\psi_-$ state we have populated by moving a formerly negative-energy state to positive energy, there is another to take its place.

How many particles have we actually created? To answer this question it is useful to temporarily introduce an infrared cutoff. We imagine that our particles live on a circle of length $L$, so that the available momentum modes are discretized: $p_k = 2\cpi k/L$, for $k \in \mathbb{Z}$. Then the number of new left-moving particles we have introduced is given by
\begin{equation} \label{eq:particlecreation}
n_- = \frac{p_E}{p_1} = \frac{L}{2\cpi} \int E(t) \dif t = \frac{1}{2\cpi} \int F,
\end{equation}
where the final integral is over all of spacetime. We see that this result is actually independent of our infrared cutoff, which we can now dispense with. Eq.~\eqref{eq:particlecreation} corresponds to the local violation of current conservation (and the analogous equation for the left-movers),
\begin{equation} \label{eq:2danomaly}
\dif J_\pm = \mp \frac{1}{2\cpi} F.
\end{equation}
This is the anomaly equation in $(1+1)$d. 

In the above discussion, we have characterized the gauge field $A$ as a {\em background} gauge field, an external classical source that drives quantum particle creation. In this interpretation, equation~\eqref{eq:2danomaly} corresponds to an 't Hooft anomaly. This is an obstruction to gauging: we cannot gauge the full $\Uone \times \Uone$ symmetry generated by the currents $J_+$ and $J_-$. If we do choose to gauge the diagonal combination $J_+ + J_-$ with a {\em dynamical} gauge field $A$, then~\eqref{eq:2danomaly} is an operator equation that tells us that the individual currents $J_+$ and $J_-$ are simply not conserved in the gauge theory. In that case, we characterize this is as an ABJ anomaly.

Note that the integral of the right-hand side of~\eqref{eq:2danomaly} over a closed Euclidean spacetime is always (for every field configuration in the path integral) an integer, by~\eqref{eq:fluxquant}. The anomaly equation implies that, when the equation of motion is satisfied, the flux is actually {\em zero} because it is the integral of a total derivative. Structurally, this is similar to what we said about the electric charge in light of Maxwell's equation~\eqref{eq:maxwelleq}: one might say that in the $(1+1)$d theory with chiral fermions, there is now a ``Gauss law constraint'' imposing that the net electromagnetic flux through a closed Euclidean spacetime is zero. We will have more to say along similar lines when discussing axions and instantons in the next lecture.

\subsubsection{Landau levels and the $(3+1)$d anomaly}

Now, continuing to follow Nielsen and Ninomiya, we consider an analogous calculation in $(3+1)d$. The anomaly involves $F \wedge F \propto {\vec E} \cdot {\vec B}$, so we want to turn on parallel electric and magnetic fields to see the effect. We begin by considering fermions in a uniform magnetic field along the $z$-direction, which we achieve by taking
\begin{equation} \label{eq:magfieldA}
A = B x \dif y, \quad B > 0.
\end{equation}
We can solve for the states of a massless charged Weyl fermion in such a background. This is a familiar problem of {\em Landau levels}. 

Specifically, we aim to solve the Dirac equation for a left-handed Weyl fermion $\psi$ of charge $q$, $\overline{\sigma}^\mu D_\mu \psi = 0$. If we can solve the auxiliary equation 
\begin{equation} \label{eq:auxiliary}
\overline{\sigma}^\mu D_\mu(\sigma^\nu D_\nu \phi) = 0,
\end{equation}
then we can obtain a solution as $\psi = \sigma^\nu D_\nu \phi$. The auxiliary equation takes a useful form assuming the ansatz $\phi(t,{\vec x}) = \E^{-\iu \omega t} \phi({\vec x})$ and a time-independent magnetic field, namely,
\begin{equation} \label{eq:Landauleveleq}
\omega^2 \phi({\vec x}) = \left[\left({\vec p} - q {\vec A}\right)^2 - 2 q {\vec B} \cdot {\vec S}\right] \phi({\vec x}),
\end{equation}
where ${\vec p} = -\iu {\vec \nabla}$ and ${\vec S} = \frac{1}{2} {\vec \sigma}$ is the spin operator. This has precisely the same form as the non-relativistic Schr\"odinger equation for a charged particle in a magnetic field, with the exception that the eigenvalue is $\omega^2$ rather than simply $\omega$. For the choice~\eqref{eq:magfieldA} of gauge field, the operators $p_y$, $p_z$, and $S_z$ all commute with the Hamiltonian. The solution proceeds in the familiar way (see, e.g.,~\cite{weinberg2015lectures}): we can replace $p_y$ by its eigenvalue and then complete the square to obtain harmonic-oscillator solutions. 

The $\omega^2$ eigenvalues are then determined by a continuous momentum eigenvalue $p_z$, an integer $n$, and an $S_z$ eigenvalue $s_z = \pm 1/2$, with corresponding dispersion relation
\begin{equation} \label{eq:Landaulevels}
\omega_n(p_z, s_z)^2 = p_z^2 + (2n+1) |q| B - 2 q B s_z.
\end{equation}
The corresponding solutions depend on $p_y$, but the eigenvalues do not (i.e., there is an infinite degeneracy for each one). For $n > 0$, or $n = 0$ and $s_z = -\frac{1}{2} \sgn(q)$, the solutions to this equation form a hyperbola in the $(p_z, \omega)$ plane with two branches, one where $\omega > 0$ and one where $\omega < 0$. In the Dirac sea, the $\omega < 0$ branch is completely filled. Turning on an electric field will shift states within such a band, but will not create any new particles. The $\omega > 0$ branch is completely empty, and will remain so when the electric field is turned on. Thus, none of these solutions is relevant for the chiral anomaly.

\begin{figure}[!h]
\centering
\includegraphics [width = 0.4\textwidth]{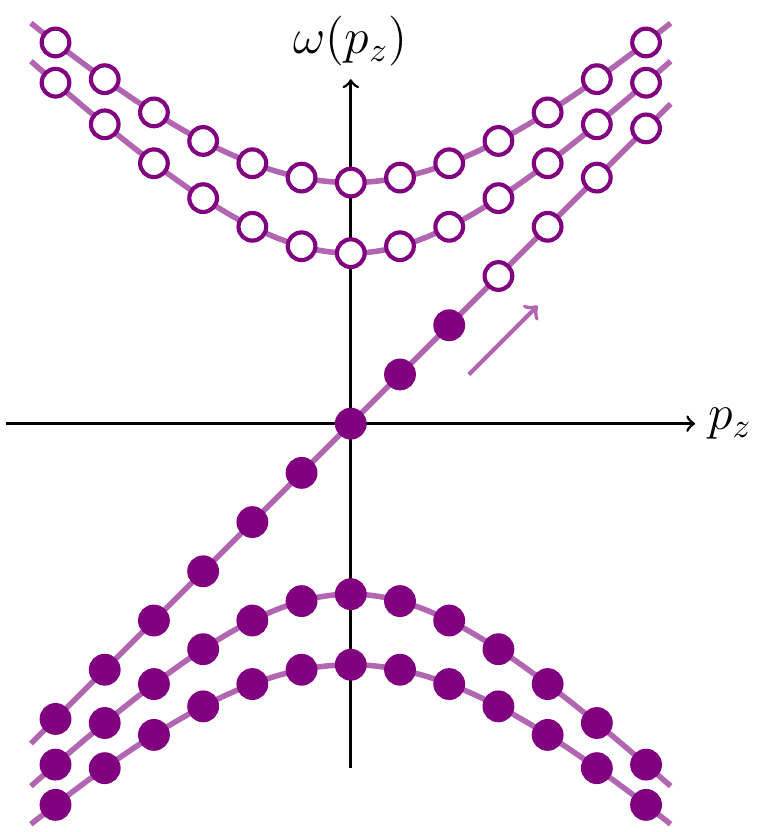}
\caption{Fermi surface for a $q < 0$ fermion in the $(3+1)$d theory~\cite{Nielsen:1983rb}. The hyperbolic branches of the Landau levels~\eqref{eq:Landaulevels} are either fully populated or empty, and the states within these branches simply rearrange under an applied electric field. Only the linear $n = 0$ branch~\eqref{eq:Landauzeromodes} changes in the presence of an applied field, effectively reducing to the $(1+1)$d problem.
} \label{fig:fermionlevels3d}
\end{figure}

The special case that remains is $n = 0, s_z = +\frac{1}{2} \sgn(q)$, for which the dispersion relation becomes simply $\omega^2 = p_z^2$. In this case, the hyperbola degenerates to the two intersecting lines, $\omega(p_z) = \pm p_z$. However, we should be a bit more careful: these are the solutions to~\eqref{eq:Landauleveleq} for $\phi({\vec x})$. We must act on them with $\sigma^\nu D_\nu$ to get $\psi$. We find that if $q > 0$ and $\omega = p_z$, or if $q < 0$ and $\omega = -p_z$, we obtain $\psi = 0$. Thus, this solution to~\eqref{eq:Landauleveleq} doesn't correspond to an actual state; instead of two intersecting lines, we have only a single line, as depicted in Fig.~\ref{fig:fermionlevels3d}. Summarizing, the nontrivial solutions of interest are:
\begin{align} \label{eq:Landauzeromodes}
q > 0: & \quad n = 0, \quad s_z = +\frac{1}{2}, \quad \omega = -p_z; \nonumber \\
q < 0: & \quad n = 0,  \quad s_z = -\frac{1}{2}, \quad \omega = +p_z.
\end{align}
Now this looks just like the $(1+1)$d case! Our $(3+1)$d fermions $\psi$, $\overline{\psi}$ in~\eqref{eq:Lag2F} with opposite charges $+1$ and $-1$ behave like $(1+1)$d right- and left-moving fermions, respectively, albeit with a huge degeneracy associated with $p_y$. If we adiabatically turn on an electric field in the $z$-direction, it will create particles according to~\eqref{eq:particlecreation}.

\medskip
\noindent\centerline{\rule{\textwidth}{0.5pt}}
\noindent {\em Exercise.} Fill in the details of the derivation of~\eqref{eq:Landaulevels} and~\eqref{eq:Landauzeromodes}. Carefully check what happens in the degenerate case~\eqref{eq:Landauzeromodes}: substitute the solutions back into the $(3+1)$d action and show that the modes behave in the $t-z$ space the way we expect $(1+1)$d fermions to behave.\\
\noindent\centerline{\rule[6.0pt]{\textwidth}{0.5pt}}

The remaining detail to understand is the density of states: how do we relate the $(1+1)$d particle creation rate to a $(3+1)$d particle creation rate? For a given set of labels $(p_z, n, s_z)$ for eigenvalues in~\eqref{eq:Landaulevels}, there is a large degeneracy corresponding to the other, continuous, quantum number $p_y$. Precisely as in the familiar non-relativistic context, the density of states corresponds to $\frac{1}{2\cpi} B$ times the transverse area. As a result, we have
\begin{equation}
\frac{\dif n_R}{\dif t} = \frac{1}{4\cpi^2} \int \dif{^3x}\, EB
\end{equation}
which corresponds to the $(3+1)$d anomaly equation
\begin{equation}\label{eq:4danomaly}
\dif J_R = \frac{1}{8\cpi^2} F \wedge F, \quad \text{and similarly} \quad \dif J_L = -\frac{1}{8\cpi^2} F \wedge F.
\end{equation}
This is the chiral anomaly, from a physical perspective: in the presence of parallel electric and magnetic fields, charged fermions of one chirality are created and those of the opposite chirality are destroyed. When we consider $A$ as a background gauge field,~\eqref{eq:4danomaly} shows that there is an 't Hooft anomaly, and we cannot gauge the full $\Uone \times \Uone$ symmetry generated by $J_L$ and $J_R$. When $A$ is a dynamical gauge field coupling to the diagonal $\Uone$ symmetry, this becomes an ABJ anomaly, and $J_L$ and $J_R$ are simply not conserved.

Now that we have derived the chiral anomaly in a physically transparent manner, let's quickly summarize the two more common textbook derivations in the next two subsections.

\subsection{The triangle anomaly calculation}

The three-point function of $\Uone$ currents, computed via a loop of fermions, is the classic calculation that revealed the existence of the chiral anomaly. Because it can be found in so many places, I will not give a detailed treatment here. That shouldn't be taken as an indication that the details are unimportant. This is a calculation that every particle physicist should work through, carefully, at least once. I especially recommend the treatment in terms of Weyl fermions found in~\cite{Dreiner:2008tw}.

\begin{figure}[!h]
\centering
\includegraphics [width = 0.9\textwidth]{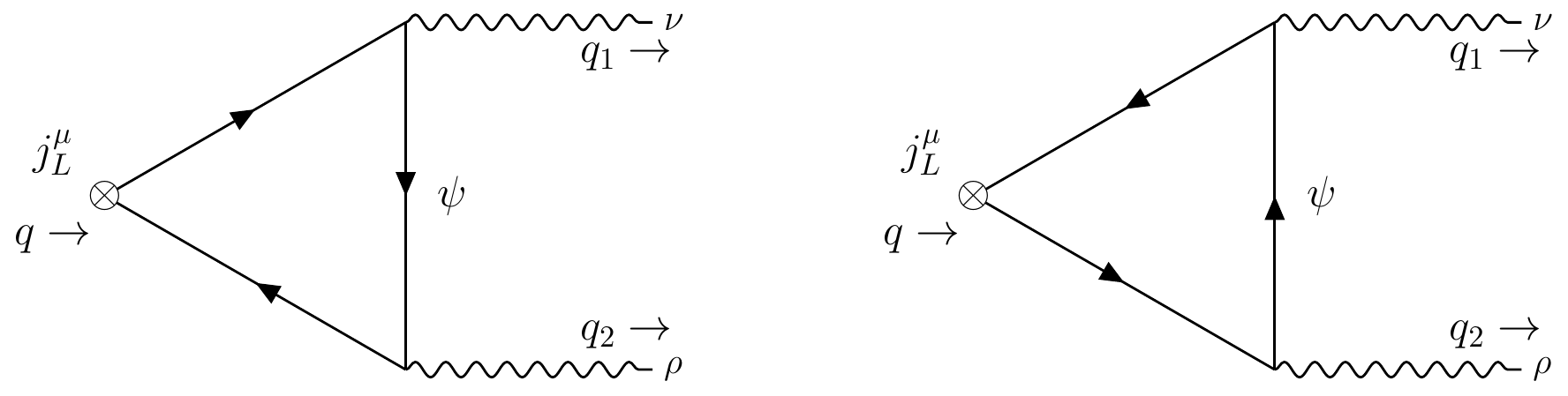}
\caption{Triangle diagrams for the chiral anomaly: here, the ABJ anomaly between the chiral current $j_L^\mu = \psi^\dagger \overline{\sigma}^\mu \psi$ and electromagnetism, induced by a loop of the chiral fermion $\psi$. The two diagrams differ by the orientation of the fermion loop in the triangle (or, equivalently, by crossing the two exterior photons).
} \label{fig:chiralanomalydiagrams}
\end{figure}

The same triangle diagram calculation sheds light on multiple anomalies. We can compute a three-point function of currents, each of which is bilinear in fermions, from a triangle loop with a current insertion at each vertex. We can also replace some of these external currents with gauge bosons. In Fig.~\ref{fig:chiralanomalydiagrams} we show such diagrams. We can interpret this as a calculation in which two of the external currents have been gauged, so there are now external gauge boson particles. This calculation reveals the ABJ anomaly, an explicit breaking of the remaining current $j^\mu_L$ in the gauge theory. On the other hand, if we think of the external gauge bosons as merely {\em background} gauge fields, the calculation reveals a non-conservation of the remaining current in the presence of certain classical backgrounds. This is an 't~Hooft anomaly: it is an obstruction to gauging, but the current remains conserved in general spacetime backgrounds as long as we don't turn on nontrivial background gauge fields as well. (You might wonder: what if we consider a case with two external currents, and a single gauge field? The answer is more subtle, involving what is known as a 2-group symmetry~\cite{Cordova:2018cvg}.)

In evaluating the triangle diagrams in Fig.~\ref{fig:chiralanomalydiagrams}, naively we have
\begin{align} 
\iu \mathcal{M}^{\mu \nu \rho} = & - e^2 \int \frac{\dif{^4k}}{(2\cpi)^4} \mathrm{tr}\left[(-\iu \overline{\sigma}^\mu)\frac{\iu}{\slashed{k}-\slashed{q_2}}(-\iu \overline{\sigma}^\rho)\frac{\iu}{\slashed{k}}(-\iu \overline{\sigma}^\nu)\frac{\iu}{\slashed{k}+\slashed{q}_1}\right] \nonumber \\
&- e^2 \int \frac{\dif{^4k}}{(2\cpi)^4} \mathrm{tr}\left[(-\iu \overline{\sigma}^\mu)\frac{\iu}{\slashed{k} - \slashed{q}_1}(-\iu \overline{\sigma}^\nu)\frac{\iu}{\slashed{k}}(-\iu \overline{\sigma}^\rho)\frac{\iu}{\slashed{k} + \slashed{q}_2}\right].
\label{eq:trianglediagram}
\end{align}
We are using notation $1/{\slashed{k}} \equiv (k_\mu \sigma^\mu)/k^2$. If current conservation holds, one should be able to contract $\mathcal{M}^{\mu \nu \rho}$ with $q_1^\nu$, $q_2^\rho$, or $q^\mu = (q_1 + q_2)^\mu$ and find zero, according to the Ward identity. However, one must be careful with this procedure. The integrals in~\eqref{eq:trianglediagram} are linearly divergent. Such integrals are notoriously ill-defined. Under a shift $k_\mu \mapsto k_\mu + k^{(0)}_\mu$ in the integration variable, the integral acquires a term proportional to the constant shift vector $k^{(0)}_\mu$. We can carry out such shifts separately in the two terms in~\eqref{eq:trianglediagram}. These constant shifts, then, must be chosen according to some physical principle to give a well-defined value to $\cal{M}^{\mu \nu \rho}$. Making this choice correctly is the sort of fiddly apparent regulator-dependence that usually shows up, in one way or another, in calculations of the chiral anomaly.

Because we have taken the momenta $q_1$ and $q_2$ to correspond to external photons, consistency of QED requires that the Ward identity hold for these. One can then show that there is a unique consistent choice in the shifts of the loop momentum variables, which leads to the equation
\begin{equation}
(q_1 + q_2)_\mu \mathcal{M}^{\mu \nu \rho} = -\frac{e^2}{4\cpi^2} \epsilon^{\nu \rho \alpha \beta} (q_1)_\alpha (q_2)_\beta.
\end{equation}
This reveals a non-conservation of the chiral current $j^\mu_L$ that, when translated into an operator equation, precisely agrees with~\eqref{eq:4danomaly}.

The triangle anomaly calculation has the virtue of being easily adapted to general symmetry currents, not just U(1). If we insert symmetry generators $T^a_{\bm{R}}$, $T^b_{\bm{R}}$, and $T^c_{\bm{R}}$ for a fermion in a group representation $\bm{R}$ at the three vertices of the triangle, we find that the corresponding matrix element is proportional to a group-theoretic invariant that is symmetric in $a$, $b$, and $c$:
\begin{equation}
D^{abc}(\bm{R}) \equiv \frac{1}{2} \mathrm{tr}\left(\left\{T^a_{\bm{R}}, T^b_{\bm{R}}\right\} T^c_{\bm{R}}\right).
\end{equation}
The only simple Lie groups for which there are representations $\bm{R}$ with nonzero $D^{abc}(\bm{R})$ are the $\SUN$ groups with $N > 2$. Furthermore, it turns out that independent of $a$, $b$, and $c$, there is a proportionality $D^{abc}(\bm{R}) = A({\bm{R}}) D^{abc}(\Box)$, where $\Box$ denotes the fundamental representation of $\SUN$. The relative factor $A({\bm{R}})$ is known as the {\em anomaly coefficient} of the representation ${\bm{R}}$. Consistency of gauge theory then requires that all $\SUN^3$ anomalies, $\Uone^3$ anomalies, and $\Uone \SUN^2$ anomalies vanish. In fact, there is a further related criterion arising from a possible {\em gravitational} anomaly of a $\Uone$ current, in which $\dif J$ can be proportional to $\mathrm{tr}(R \wedge R)$ where $R$ is the Riemann curvature 2-form.

\subsection{Fujikawa's path integral calculation}

Another way to find the anomaly is to consider the path integral measure. A path integral for the theory~\eqref{eq:Lag2Fgauged} has the form
\begin{equation}
\int {\cal D}A\,{\cal D}{\overline{\psi}}\,{\cal D}\psi\,\E^{\iu S[A, \psi, \overline{\psi}]}.
\end{equation}
A classical symmetry operation on the fields, by definition, leaves the last $\exp(\iu S)$ factor invariant. Thus, the only way that a symmetry of the classical theory can fail to be a symmetry of the quantum theory is if it does not leave the {\em measure} invariant. This is what Fujikawa calculated, by carefully regulating the path integral measure~\cite{Fujikawa:1979ay, Fujikawa:1980eg}.

We will only very briefly sketch the idea. We expand the field $\psi(x)$ in a set of orthonormal modes, $\psi(x) = \sum_m a_m \phi_m(x)$, where the $\phi_m(x)$ which are eigenstates of the Dirac operator, $i \slashed{D} \phi_m = \lambda_m \phi_m$. Then ${\cal D}\psi$ is $\prod_m \dif a_m$. The hard work is entirely in understanding how to properly regulate this product, and I encourage you to seek out either the original literature or textbook accounts to understand the details. It turns out (carefully regulating) that $\prod_m \dif a_m$ has a nontrivial Jacobian under the field redefinition $\psi(x) \mapsto \E^{\iu \alpha(x)} \psi(x)$. The end result is that under this field redefinition, the path integral measure changes according to
\begin{equation}
{\cal D}\psi \mapsto {\cal D}\psi\, \E^{\iu \int \frac{1}{8\cpi^2} \alpha(x) F(x) \wedge F(x)}.
\end{equation}
Effectively, this has added a new term to the action, where $\alpha(x)$ couples to $F \wedge F$. 

I have stated this argument for the case where we have a dynamical gauge field in the theory, where it reproduces the ABJ anomaly. But we could also do the calculation without integrating ${\cal D}A$, but keeping $A$ in the action as a fixed classical background field sourcing the fermion currents. The final conclusion is the same: the path integral measure changes in a way that is sensitive to the background field. This is the hallmark of an 't Hooft anomaly.

\subsection{The takeaway message} \label{sec:chiralanomalytakeaway}

The chiral anomaly will play a {\em central} role in our explanation of axion physics, so let me emphasize the one major result that we will need for both $\Uone$ and $\SUN$ gauge groups.

A field redefinition on a Weyl fermion $\psi$,
\begin{equation}
\psi(x) \mapsto \E^{\iu \alpha(x)} \psi(x),
\end{equation}
not only produces the obvious changes from plugging this into the path integral (e.g., rephasing mass or Yukawa terms, generating a derivative coupling $\psi^\dagger \overline{\sigma}^\mu \psi \partial_\mu \alpha$ from the kinetic term), but {\em also}, via the chiral anomaly, adds a term to the action of the form
\begin{equation} \label{eq:SUNanomalyterm}
\int 2 I({\bm{R}}_\psi)\frac{1}{8\cpi^2} \alpha(x) \mathrm{tr}(F \wedge F)
\end{equation}
for every $\SUN$ gauge group $\psi$ is charged under, where ${\bm{R}}_\psi$ is the representation of $\psi$ under the group and $I({\bm{R}}_\psi)$ is the Dynkin index of that representation (e.g., $1/2$ for the fundamental representation of $\SUN$), as well as a term
\begin{equation} \label{eq:U1anomalyterm}
\int q^2 \frac{1}{8\cpi^2} \alpha(x) F \wedge F
\end{equation}
for every $\Uone$ group $\psi$ is charged under, with $q$ the charge of $\psi$ under that group.

\subsection{One-loop exactness}
\label{sec:oneloopexactness}

An important fact about the anomaly is that it is {\em one-loop exact}: the operator equation~\eqref{eq:4danomaly} for current non-conservation receives no corrections beyond one loop. One way to see this is to consider the case $\alpha = 2\cpi$ in~\eqref{eq:SUNanomalyterm} or~\eqref{eq:U1anomalyterm}. Clearly this is a trivial rephasing, doing nothing to the fermion, but it seems that it adds a nontrivial contribution to the action. How is this possible? It makes sense only if $\E^{\iu S}$ is unchanged by this addition to $S$. This follows from the quantization conditions~\eqref{eq:instantonnumberquantization} and~\eqref{eq:FwedgeFquantization} that we have discussed previously. If we allowed for a more general function of the gauge coupling modifying the relationship between the left- and right-hand sides of~\eqref{eq:4danomaly}, it would no longer be true that a $2\cpi$ rephasing of the fermion would leave $\E^{\iu S}$ invariant. This tells us that there are no perturbative corrections to~\eqref{eq:4danomaly} involving a series in the gauge couplings.

A related fact is that instantons contribute to amplitudes that violate chiral symmetries, converting some fermion modes into other fermion modes. In particular, fermions have zero modes in an instanton background, which lead to an effective interaction known as the 't~Hooft vertex, which involves one mode for every chiral fermion. This is a direct consequence of~\eqref{eq:4danomaly}: an instanton has a nonzero integral of $\frac{1}{8\cpi^2} F \wedge F$, which in turn corresponds to nonzero integrals of $\dif J_R$ and $\dif J_L$, i.e., to a net change in the number of right- and left-handed fermions. The 't~Hooft vertex is simply the effective operator that encapsulates the associated multi-fermion interaction. Again, this makes it clear that the instanton effect is one-loop exact: if we had a generic coefficient between the left and right sides of~\eqref{eq:4danomaly}, an instanton would correspond to a {\em fractional} change in the number of fermion modes, and no sensible interpretation in terms of a chirality-violating process could be associated with it.

\subsection{Further remarks on anomalies and QCD}

The one-loop exactness of chiral anomalies is a consequence of their topological nature. This makes anomalies an extremely powerful tool for analyzing nonperturbative properties of quantum field theory. An example is 't~Hooft anomaly matching: given a set of symmetry currents in a theory that have an 't~Hooft anomaly, this anomaly will be preserved under RG flows that do not explicitly break any of the symmetries~\cite{tHooft:1979rat}. In QCD (with massless quarks), the nonabelian flavor symmetry currents for $\SU(N_f)_\textsc{L} \times \SU(N_f)_\textsc{R}$ have 't~Hooft anomalies. This implies that confinement cannot fully gap the spectrum: there should be massless composite states in the infrared that match these anomalies. These could potentially involve new composite chiral fermions, or they can involve massless Nambu-Goldstone bosons (the pions) with Wess-Zumino-Witten interactions that match the 't~Hooft anomalies from the UV theory. With some additional assumptions, one can argue that only the latter possibility is consistent, so the theory must spontaneously break $\SU(N_f)_\textsc{L} \times \SU(N_f)_\textsc{R} \to \SU(N_f)_\textsc{V}$~\cite{Frishman:1980dq, Coleman:1980mx}. This is consistent with experiment: the pions, kaons, and $\eta$ meson appear to have the properties of pseudo-Nambu-Goldstone bosons parametrizing the coset $[\SU(3)_\textsc{L} \times \SU(3)_\textsc{R}]/\SU(3)_\textsc{V}$.

Another important application of the chiral anomaly in QCD is the resolution of the $\Uone$ puzzle.  Naively, one would expect an additional light pseudo-Nambu-Goldstone boson for the broken $\Uone_\textsc{A}$ symmetry, but this is not what we see: the $\eta'$ is the candidate, but it is much heavier than one would expect if it were a pseudo-Nambu-Goldstone boson. This is because the $\Uone_\textsc{A}$ symmetry has an ABJ anomaly with the strong interactions themselves, so strong dynamics badly breaks the would-be symmetry and produces a large mass for the $\eta'$.

\section{Axion couplings to gauge fields}
\label{sec:axiongaugefieldcouplings}

The next part of these lecture notes will be all about the phenomenology of fields known as {\em axions}. One important piece of axion physics follows from our discussions of the quantization laws~\eqref{eq:FwedgeFquantization} and~\eqref{eq:instantonnumberquantization}, so we explain it here. For now, the term ``axion'' refers to a compact scalar field $\theta(x)$. By ``compact'' we mean that the field values live on a circle, which is to say that the value of $\theta(x)$ is only defined modulo $2\cpi$. This means that $\theta(x) \mapsto \theta(x) + 2\cpi$ is a {\em gauge redundancy} of our theory.

This gauge redundancy places strong limitations on the Lagrangian for our theory. We can write arbitrary interactions involving $\partial_\mu \theta$, because this is gauge-invariant. We can also write interactions involving $2\cpi$-periodic functions like $\cos(k \theta)$ (with $k \in \mathbb{Z}$), which can give rise to a mass for the axion. (As we will see in the next part, axion model-building primarily revolves around finding ways to explain why such terms can naturally have tiny coefficients.) What we cannot do is write terms that involve $\theta$ without a derivative acting on it and without packing it in a periodic function. However, there is one remarkable exception, which is the key to all of axion physics. This is a term of the form $\theta F \wedge F$ (or $\theta\, \mathrm{tr}(F \wedge F)$, for nonabelian gauge fields). 

Let $F$ be a $\Uone$ gauge field strength, and consider the Euclidean action
\begin{equation}
S_E = \iu \,c \int \theta\, F \wedge F.
\end{equation}
Here $c$ is a real coeffiicient; the factor of $\iu$ in front arises from Wick rotation from Minkowski signature to Euclidean signature, because $F \wedge F$ has a single time-index in it. (Such factors of $\iu$ generally accompany terms that, in tensor index notation, are written with an odd number of Levi-Civita tensors.) This Euclidean action is manifestly not gauge-invariant. Under the operation $\theta \mapsto \theta + 2\cpi$, we have 
\begin{equation} \label{eq:SEthetashift}
S_E \mapsto S_E' = S_E + 2\cpi\iu \,c \int F \wedge F.
\end{equation}
In fact, this is not a fatal problem for the theory. In order for calculations of physical quantities to be gauge invariant, it is not necessary for $S_E$ to be gauge invariant. The requirement is that the path integral measure be gauge invariant, and hence that $\exp\left[-S_E\right] = \exp\left[-S_E'\right]$. In other words, we require
\begin{equation}
\exp\left[-2\cpi \iu\, c \int F \wedge F\right] = 1.    \label{eq:FFcondition}
\end{equation}
Equivalently, for any $F$, we must have $c \int F \wedge F \in \mathbb{Z}$. In $\mathbb{R}$ gauge theory, we have $\int F \wedge F = 0$ and this is automatically satisfied for any value of $c$. In $\Uone$ gauge theory, however, $\int F \wedge F$ can take on nonzero values, as in~\eqref{eq:FwedgeFquantization}. This immediately implies that $c$ is quantized. Considering the smallest possible value of $F \wedge F$, the condition~\eqref{eq:FFcondition} becomes
\begin{equation}
\exp\left[2\cpi \iu \cdot c \cdot 8 \cpi^2\right] = 1 \quad \Rightarrow \quad c = \frac{k}{8\cpi^2}, \quad k \in \mathbb{Z}.
\end{equation}
In other words, we have learned that consistently defining the theory in the presence of general $\Uone$ gauge bundles requires that the axion coupling to $\Uone$ gauge fields be an integer multiple of $\frac{1}{8\cpi^2}$. 

The story is precisely the same for $\SUN$ gauge fields, due to the quantization of instanton number~\eqref{eq:instantonnumberquantization}. Reverting to Minkowski signature, we can summarize by saying that the coupling of an axion $\theta$ to a $\Uone$ gauge field strength $F$ and an $\SUN$ gauge field strength $G$ is
\begin{equation} \label{eq:axioncouplingquantization}
S = \frac{k_F}{8\cpi^2} \int \theta\,F \wedge F + \frac{k_G}{8\cpi^2} \int \theta\,\mathrm{tr}(G \wedge G), \qquad k_F, k_G \in \mathbb{Z}.
\end{equation}
Such a coupling of an axion to $\SU(3)$ gauge fields is the crucial element in the axion solution to the Strong CP problem, discussed in detail in the next part of these lecture notes.

This argument relies on one assumption that has gone unstated so far: in~\eqref{eq:SEthetashift} we took the gauge field strength $F$ to be invariant under the gauge transformation $\theta \mapsto \theta + 2\cpi$. This is not a completely innocuous assumption. For example, in free $\Uone$ gauge theory, there is an $\mathrm{SL}(2,\mathbb{Z})$ duality group whose action mixes up $F$ and $\widetilde{F}$. As a consequence of the Witten effect~\eqref{eq:witteneffect1}, the magnetic gauge field strength shifts when $\theta \mapsto \theta + 2\cpi$. One could imagine a theory in which the shift of $\theta$ is accompanied by such a duality operation that changes the {\em electric} field strength, so that $F \wedge F$ itself would not be invariant, and our argument above would not apply. It turns out that such a loophole does not apply to real-world electromagnetism, because it would require that the electron obtain a magnetic monopole charge in the presence of an axion background, which would be a phenomenological disaster.\footnote{I thank Ben Heidenreich and Jake McNamara for clarifying discussions on this topic, more details of which will be published elsewhere.}

One last consequence of the Witten effect that I would like to mention here is that loops of magnetic monopoles can generate a mass for the axion. In order for a magnetic monopole to acquire an electric charge in an axion background, as in~\eqref{eq:witteneffect1}, the monopole must have a dyon collective coordinate. (Such a mode was originally found for the 't~Hooft-Polaykov monopole~\cite{Jackiw:1975ep}, but it must exist for any monopole in a theory with an axion.) This is a quantum mechanical degree of freedom living on the monopole worldline that takes the form of a particle on a circle, which has a quantized momentum. Giving momentum to this degree of freedom is equivalent to giving the magnetic monopole an electric charge. This compact degree of freedom can wind around a loop of monopole worldline. This can be viewed as a $\Uone$ instanton, closely analogous to the $\SUN$ instantons we have discussed. One can integrate out monopole worldlines in the path integral to see that such a $\Uone$ instanton generates a periodic potential for the axion~\cite{Fan:2021ntg}. (Related physics is also reviewed in~\cite{Heidenreich:2020pkc,Choi:2022fgx}.)

It is also possible to have additional axion couplings to gauge fields via higher-dimension operators that are explicitly gauge invariant, starting with $(\Box \theta)\,F \wedge F$. However, via equations of motion, this is suppressed by the mass of the axion squared, and in most cases will be extremely subdominant. An important exception is the QCD axion's coupling to photons via mixing with the pion, which is large because the axion and pion mass arise through the same strong dynamics. We will discuss this further below, in \S\ref{subsec:axionpionmixing}.

Let us sum this up, now quoting the result in canonical normalization. To canonically normalize, we assume the axion kinetic term has the form $\frac{1}{2} f^2 \dif \theta \wedge \star \dif \theta$, where the numerical factor $f$ is known as the ``axion decay constant.'' The gauge field is canonically normalized with a factor of the gauge coupling. This implies:

\begin{framed}
\noindent
The leading coupling of a light axion to gauge fields is {\em quantized} in integer multiples of a loop factor divided by the axion decay constant, $g^2/(8\cpi^2 f)$, in canonical normalization. Corrections are necessarily suppressed by the square of the axion mass.
\end{framed}

\section{Chern-Simons terms}
\label{sec:CSgeneral}

The couplings of a compact scalar to gauge fields that we have just described are an example of a larger class of interactions known as {\em Chern-Simons terms}. The characteristic feature of a Chern-Simons term in an action is that it is not gauge-invariant, but that $\exp[\iu S]$ is gauge-invariant, and so the path integral is well-defined. Chern-Simons terms are topological and have a quantized coefficient. Here we discuss some other examples. This section lies somewhat outside the main line of reasoning of these lectures and can be skipped by readers eager to get to axion physics, though we will refer back to some parts of it in later sections.

\subsection{Chern-Simons terms in 3d gauge theory}
\label{subsec:CSterm3d}

Perhaps the most familiar and well-studied examples of Chern-Simons terms are those appearing in 3d gauge theories. A Chern-Simons term for a $\Uone$ gauge theory takes the form 
\begin{equation} \label{eq:CS3dU1}
S^{\Uone}_\textsc{CS} = \frac{k}{4\cpi} \int_M A \wedge \dif A,
\end{equation}
and for a non-abelian gauge theory takes the form
\begin{equation} \label{eq:CS3dNA}
S^{\SUN}_\textsc{CS} = \frac{k}{4\cpi} \int_M \mathrm{tr}\left(A \wedge \dif A - \iu \frac{2}{3} A \wedge A \wedge A\right).
\end{equation}
If we view the integrands of these 3d Chern-Simons terms from the 4d viewpoint, their derivatives are the familiar instanton number densities that we have discussed above. In the $\Uone$ case we obviously have $\dif(A \wedge \dif A) = \dif A \wedge \dif A = F \wedge F$. In the nonabelian case, we have
\begin{align}
\mathrm{d}\left[\mathrm{tr}\left(A \wedge \dif A - \frac{2}{3} \iu A \wedge A \wedge A\right)\right] &= \mathrm{tr}\left(\dif A \wedge \dif A -2 \iu \dif A \wedge A \wedge A\right) \nonumber \\
&= \mathrm{tr}\left(\dif A \wedge \dif A -2 \iu \dif A \wedge A \wedge A - A \wedge A \wedge A \wedge A\right) \nonumber \\
&= \mathrm{tr}(F \wedge F),
\end{align}
where we have made repeated use of the graded cyclic property of the trace reviewed in \S\ref{sec:nonabelian} and the last step uses the formula $F = \dif A - \iu A \wedge A$.

The Chern-Simons term has a quantized coefficient, which is a consequence of its lack of gauge invariance. At first glance, you might think that under a gauge transformation (in the $\Uone$ case)
\begin{equation} \label{eq:AdAgaugetransform}
A \wedge \dif A \mapsto (A + \dif \alpha) \wedge \dif A = A \wedge \dif A + \dif(\alpha \dif A),
\end{equation} 
so the integrand in the action shifts by a total derivative and the action is invariant. As usual, the subtleties are all about topology: what if the gauge transformation $\alpha$ winds around a circle, or if $A$ is not single-valued? Giving a proper treatment of this subject is beyond the scope of these lecture notes, but we can give a partial argument based on the connection to $F \wedge F$ terms in one higher dimension. Specifically, suppose that we want to make sense of~\eqref{eq:CS3dU1} on a 3d manifold $M$, despite its lack of gauge invariance. What we really need, to define a path integral, is not the action itself but $\exp(\iu S)$. Suppose that $M$ is the boundary of a 4d manifold $X$ and that we can extend our gauge field configuration over $X$.\footnote{This is not an innocuous assumption; a $d$-dimensional manifold that can be written as the boundary of a $(d+1)$-dimensional manifold is said to be ``null-bordant.'' It is a nontrivial fact that all closed 3-manifolds are null-bordant, but this is special to 3 dimensions. Even when the spacetime is null-bordant, it may not be possible to extend the gauge bundle over the higher-dimensional manifold. Discussing this properly would require a significant enlargement of these lectures. See~\cite{Dijkgraaf:1989pz}.} Then we can replace our apparently ill-defined formula~\eqref{eq:CS3dU1} with a manifestly well-defined, gauge-invariant formula in 4d:
\begin{equation} \label{eq:CSvia4d}
\exp\left(\iu S^\Uone_\textsc{CS}\right) = \exp\left(\iu \frac{k}{4\cpi} \int_X F \wedge F\right).
\end{equation}
This is motivated by Stokes's theorem, which suggests $\int_X F \wedge F = \int_X \dif(A \wedge F) = \int_M A \wedge F$, even though this isn't as simple as it looks since $A$ may not be single-valued. In any case, if we try to define the path integral directly via~\eqref{eq:CSvia4d} instead of~\eqref{eq:CS3dU1}, we can ask whether we would get the same answer for a different choice of 4-manifold $X'$ that {\em also} has boundary $M$. Now, because $X$ and $X'$ have the same boundary, we can glue $X$ to the orientation-reversal $-X'$ to make a {\em closed} 4-manifold. (If this argument is sounding familiar, it's essentially the flux-quantization argument that we gave in~\S\ref{subsec:wilsonloops}, run backwards. In fact, a Wilson loop can be thought of as a 1d Chern-Simons term.) We know that the integral of $F \wedge F$ over any closed 4-manifold is $8 \cpi^2 n$ for $n \in \mathbb{Z}$. Thus we have
\begin{equation}
\exp\left(\iu \frac{k}{4\cpi} \int_{X \cup (-X')} F \wedge F\right) = \exp\left(2 \cpi \iu k n\right),
\end{equation}
which is always $1$ if 
\begin{equation}
k \in \mathbb{Z}.
\end{equation}
Precisely the same argument goes through for the case of $\SUN$ and the non-abelian Chern-Simons term~\eqref{eq:CS3dNA}. Thus, we conclude (up to all the mathematical subtleties I've swept under the rug along the way!) that the coefficient of 3d Chern-Simons terms is quantized in units of $\frac{1}{4\cpi}$ in order for us to obtain a well-defined path integral. You can consider winding gauge transformations on simple spacetimes like $S^1 \times S^2$ to gain more confidence in this conclusion. 

This argument follows very similar logic to the argument that led us to conclude that axion couplings to gauge fields are quantized: in both cases, we have an action that is not gauge invariant, but $\exp(\iu S)$ is gauge invariant for specific choices of coupling.

Chern-Simons terms in 3d have important physical implications. For example, they lead to an effective photon mass, as you can derive in the following exercise.

\medskip
\noindent\centerline{\rule{\textwidth}{0.5pt}}
\noindent {\em Exercise.} Consider a 3d $\Uone$ gauge theory with both a Maxwell kinetic term and a Chern-Simons term, 
\begin{equation}
S = \int \left(-\frac{1}{2e^2} F \wedge \star F + \frac{k}{4\cpi} A \wedge \dif A\right).
\end{equation}
Derive the equation of motion for the gauge field, and show that it implies that the field strength obeys a massive wave equation,
\begin{equation} \label{eq:massivephoton3d}
\left(\Box + m_A^2\right) F^{\mu \nu} = 0,
\end{equation}
for some constant $m_A$. What is $m_A$ in terms of the parameters $e$ and $k$? How many independent, propagating polarization states does a massless gauge field (with $k = 0$) have in $2+1$ dimensions? What about the gauge field with a Chern-Simons mass, $k \neq 0$? Explain your answers, and comment on how this differs from the familiar $(3+1)$-dimensional case.
\\
\noindent\centerline{\rule[6.0pt]{\textwidth}{0.5pt}}

Chern-Simons terms can also be related to $\theta$ terms in one lower dimension, as you can see by example in the following exercise.

\medskip
\noindent\centerline{\rule{\textwidth}{0.5pt}}
\noindent {\em Exercise.} Now suppose that we compactify the theory from the previous exercise on a circle $x^3 \cong x^3 + 2\cpi L$, and consider an ansatz where we turn on a constant $A_3 \neq 0$ so that
\begin{equation}
\varphi \equiv \int_0^{2\cpi L} \dif x^3 A_3 \neq 0.
\end{equation}
Show that in the dimensionally reduced $(1+1)$-dimensional theory, the gauge field has a 2d theta term,
\begin{equation}
\frac{\theta}{2\cpi} \int F.
\end{equation}
How is $\theta$ related to $\varphi$? Explain how the periodicity of the 2d coefficient $\theta$ is compatible with the 3d origin of the term.
\\
\noindent\centerline{\rule[6.0pt]{\textwidth}{0.5pt}}

Our discussion of how the Chern-Simons term can be well-defined crucially assumed that the 3-manifold $M$ had no boundary. On a 3-manifold with boundary, even topologically trivial gauge transformations as in~\eqref{eq:AdAgaugetransform} do not give rise to a gauge-invariant action, as we acquire a term $\int_{\partial M} \frac{k}{4\cpi} \alpha \dif A$ when carrying out the gauge transformation $A \mapsto A + \dif \alpha$. However, this can be compensated if there is an {\em anomalous} 2d gauge theory living on the boundary $\partial M$. This is a common property of Chern-Simons theories, sometimes referred to as {\em anomaly inflow}~\cite{Callan:1984sa}: consistency of the theory on spaces containing boundaries or other defects requires the existence of localized, charged degrees of freedom on the defect. A familiar example arises in condensed matter physics, where quantum Hall systems are described by $(2+1)$d Chern-Simons effective theories in the bulk and admit charged {\em edge modes} that provide an anomalous boundary theory that cancels the non-gauge-invariant terms from the bulk.

\subsection{Chern-Simons masses and Stueckelberg masses}
\label{sec:CSStueckelberg}

In 4d gauge theory, we can't add a photon mass with a term of the form $A \wedge \dif A$ as in 3d. However, there is a very similar type of photon mass term, sometimes called a ``BF term.'' In this case, we add a new 2-form $\Uone$ gauge field $B$ to the theory. That is, there is an antisymmetric tensor field $B_{\mu \nu}$, with $B = \frac{1}{2} B_{\mu \nu} \dif x^\mu \wedge \dif x^\nu$ having a gauge invariance $B \mapsto B + \dif \lambda$, where $\lambda$ is a 1-form. This is a $\Uone$ gauge symmetry in the sense that it obeys quantization laws similar to those for ordinary $\Uone$ gauge fields; in particular, the field strength $H = \dif B$ has quantized flux when integrated over any closed 3-manifold $\Omega$,
\begin{equation}
\frac{1}{2\cpi} \int_\Omega H \in \mathbb{Z},
\end{equation}
directly parallel to the usual magnetic flux quantization~\eqref{eq:fluxquant}. Similarly, $B$ is invariant under large or ``winding'' gauge transformations, in the sense that if $[\omega] \in H^2(M, \mathbb{Z})$ is a class in integral cohomology, there is an extended gauge invariance under $B \mapsto B + \omega$ even though $\omega$ can only {\em locally} be written as $\dif \lambda$. Such higher-form gauge fields are ubiquitous in string theory and quantum field theories in more than four spacetime dimensions, but they can also be useful in 4d as well. In particular, the following action describes a massive photon field:
\begin{equation} \label{eq:BFtheory}
S = \int \left( -\frac{1}{2e^2} F \wedge \star F - \frac{1}{2 g^2} H \wedge \star H + \frac{k}{2\cpi} B \wedge F\right).
\end{equation}
Here $g$ has dimensions of mass and can be thought of as the coupling constant of the $B$ field, and the last term is a Chern-Simons term that is only well-defined when $k \in \mathbb{Z}$. You can see that the photon is massive in essentially the same way that you derived~\eqref{eq:massivephoton3d} in the exercise above; in this case, its mass is proportional to $k$, $e$, and $g$.

There is a different way to formulate the theory of a photon with a BF term mass as a theory with a Stueckelberg mass. This reflects a more general way of recasting Chern-Simons terms as Stueckelberg terms. This is very useful physics to familiarize yourself with, although it will play only a small role in the remainder of these lectures. We begin with the idea of a {\em Hodge dual} to a gauge field, which is a sort of generalization of electric-magnetic duality. A free $p$-form $\Uone$ gauge field $C_p$ with action
\begin{equation}
\int \left(- \frac{1}{2 e_p^2} \dif C_p \wedge \star \dif C_p\right)
\end{equation}
can be dualized to a $(d-p-2)$-form gauge field $\widetilde{C}_{d-p-2}$, via the map
\begin{equation}
\frac{1}{2\cpi} \dif {\widetilde{C}}_{d-p-2} = \frac{1}{e_p^2}\star \dif C_p.
\end{equation}
The dual theory has action
\begin{equation}
\int \left(- \frac{1}{2 \widetilde{e}_p^2} \dif{\widetilde{C}}_{d-p-2} \wedge \star \dif{\widetilde{C}}_{d-p-2}\right)
\end{equation}
with the gauge coupling ${\widetilde{e}}_{d-p-2} = 2\cpi/e_p$, as in the usual Dirac quantization formula relating electric and magnetic couplings. One can also derive that if $C_p$ is a $\Uone$ gauge field (with the associated flux quantization condition on its field strength), then so is $\widetilde{C}_{d-p-2}$, with the dualilty trading a magnetic flux quantization condition like~\eqref{eq:fluxquant} for an electric flux quantization condition like~\eqref{eq:elecfluxquant}. This can be derived explicitly at the level of the path integral by introducing auxiliary Lagrange multiplier fields and then integrating out the original gauge field; see, e.g.,~\cite{Witten:1995gf, Deligne:1999qp}. As an example, in 4d, the Hodge dual of a 2-form gauge field $B$ is a 0-form field $\theta$ with quantized fluxes $\int_C \dif \theta = 2\cpi n$ ($n \in \mathbb{Z}$) around closed curves. This is just a periodic scalar field, $\theta \cong \theta + 2\cpi$.

In the presence of a Chern-Simons term as in~\eqref{eq:BFtheory}, the Hodge dualization procedure is more subtle. One would like to write $\frac{1}{2\cpi} \dif \theta = \frac{1}{g^2} \star H$. However,~\eqref{eq:BFtheory} implies the equation of motion
\begin{equation} \label{eq:modifiedBianchi}
\dif{\left(\frac{1}{g^2} \star H\right)} = \frac{k}{2\cpi} F.
\end{equation}
Because $\star H$ isn't closed, we can't locally write it as $\mathrm{d}$ of some quantity $\theta$. Instead, what we can do is rewrite the above equation (locally) as
\begin{equation}
\dif{\left(\frac{1}{g^2} \star H - \frac{k}{2\cpi} A\right)} = 0.
\end{equation}
We can thus identify the quantity in parentheses with $\frac{1}{2\cpi} \dif \theta$, or in other words, we have
\begin{equation}
\frac{1}{2\cpi} \left(\dif \theta + k A\right) = \frac{1}{g^2} \star H,
\end{equation}
with a compact scalar $\theta \cong \theta + 2\cpi$. The twist is that only the combination $\dif \theta + k A$ is gauge invariant. This means that $\theta$ must be a Stueckelberg field, which {\em shifts} under a gauge transformation of $A$:
\begin{equation}
A \mapsto A + \dif \alpha, \quad \theta \mapsto \theta - k \alpha.
\end{equation}
The compact scalar field $\theta$, dual to $B$, is eaten to provide a mass for $A$.  The action dual to~\eqref{eq:BFtheory} is the standard Stueckelberg action:
\begin{equation} \label{eq:Stueckelberg}
S = \int \left( -\frac{1}{2e^2} F \wedge \star F - \frac{1}{2} f^2 (\dif \theta + k A) \wedge \star(\dif \theta + k A)\right),
\end{equation}
with $f = g/(2\cpi)$. This formulation of a massive gauge field is likely more familiar to you than the BF term, but they are completely equivalent. Notice that it no longer has a Chern-Simons term! The ability to trade a Chern-Simons term in one formulation of a theory for a Stueckelberg term in a dual formulation, in which one gauge field shifts under the gauge transformation of a {\em different} gauge field, is quite general. In fact, there is another such action lurking in this example as well. We could have started with~\eqref{eq:BFtheory} and dualized $A$ to a magnetic 1-form gauge field $\widetilde{A}$, obtaining a theory in which the $B$ field has a standard kinetic term but $\widetilde{A} \mapsto \widetilde{A} + k \lambda$ when $B \mapsto B + \dif \lambda$, where the mass term takes the form $|\dif{\widetilde{A}} - k B|^2$, which we could think of as the $B$ field eating the magnetic photon. (For a more extended review of this theory including its global symmetry properties and the possible addition of objects carrying various electric and magnetic charges, see~\cite{Heidenreich:2020pkc}.)

\subsubsection{A magnetic photon mass}
\label{subsubsec:magneticmass}

The possibility of a mass for the photon is often studied in the real world, and is highly constrained experimentally. We will discuss the implications of quantum gravity for this possibility below, in \S\ref{subsec:photonmass}. Recently attention was drawn to the possibility that the photon could have a {\em magnetic} mass~\cite{Hook:2022pcf}. One way to formulate this theory is in terms of a magnetic dual photon $\widetilde{A}$, with a Stueckelberg mass term of the form $|\dif \theta + k \widetilde{A}|^2$. Attempts to treat both standard electric photon couplings to $A$ and terms involving the magnetic dual $\widetilde{A}$ within the same action are cumbersome.

The discussion above suggests an equivalent formulation, which avoids the need to refer to the magnetic gauge field $\widetilde{A}$.\footnote{I thank Ben Heidenreich for a discussion on this topic.} Instead, we introduce a 2-form gauge field $B$ with field strength $H = \dif B$, with an action of the form
\begin{equation}
S = \int \left(- \frac{1}{2g^2} H \wedge \star H - \frac{1}{2e^2} \left(\dif A - k B\right) \wedge \star \left(\dif A - k B\right)  \right).
\end{equation}
This is a dual description of a photon with magnetic mass $kg/e$: the usual gauge field $A$ is eaten by $B$. It has the crucial feature that there are two nontrivial gauge transformations acting on the gauge field $A$:
\begin{equation}
A \mapsto A + \dif \alpha, \qquad \textrm{and} \qquad 
B \mapsto B + \dif \lambda, \quad A \mapsto A + k \lambda.
\end{equation}
The novel gauge invariance, with $A$ shifting under a $B$ gauge transformation, spoils our ability to couple particles electrically to $A$ in the standard way. However, this is exactly what we expect a magnetic mass to do! A magnetic mass confines electrically charged particles, meaning that they come with strings attached: we can write a gauge invariant coupling of $A$ to a particle worldline $C$ only if $C$ is the boundary of a string worldsheet $\Sigma$, with
\begin{equation}
S = q \left(\int_C A - \int_\Sigma k B\right).
\end{equation}
These strings will generally have a tension, which we might expect to be of order $g^2$ but which can only be determined within a UV completion of the theory.

This formulation is useful for highlighting what the invariant physical challenge of modeling a magnetic photon mass is. The formulation in~\cite{Hook:2022pcf} makes it appear that the challenge is in simultaneously keeping track of electric and magnetic vector potentials describing the same underlying photon field. Instead, we see that the real challenge is to study a theory of dynamical strings coupled to all of the electrically charged particles we know. There's an important general principle here that is mostly orthogonal to the topics of these lectures, but let's highlight it anyway:

\begin{framed}
\noindent
Higgsing an electric gauge theory confines the dual magnetic charges, and vice versa.
\end{framed}

\subsection{Higher Chern-Simons terms}
\label{subsec:higherCS}

So far we have seen a few different examples of Chern-Simons terms: $A \wedge F$ in 3d, $B \wedge F$ in 4d, and $\theta F \wedge F$ in 4d. The general pattern is that these terms are not gauge invariant: they involve several gauge field strengths and a single additional gauge field with no derivative acting on it. The gauge fields are potentially higher degree $p$-forms (like the 2-form $B$ in 4d) or even a 0-form gauge field, i.e., a periodic scalar $\theta$. This pattern continues in higher dimensions. For example, in a 5d $\Uone$ gauge theory we might have a term $\frac{k}{4\cpi^2} \int A \wedge \dif A \wedge \dif A$, or if we have a $\Uone$ gauge field $A$ and an $\SUN$ gauge field with field strength $G$, we can have a 5d Chern-Simons term $\frac{k}{8\cpi^2} \int A \wedge \mathrm{tr}(G \wedge G)$. Such Chern-Simons terms always have a quantized coefficient $k \in \mathbb{Z}$ (when normalized appropriately). It is also true in general that if we lift a Chern-Simons term to one higher dimension and take $\mathrm{d}$ of it, we obtain a {\em theta} term in the higher-dimensional theory, which has a periodic coefficient $\theta$. We have seen this with terms of the form $F \wedge F$ or $\mathrm{tr}(F \wedge F)$, but other examples include $\frac{1}{2\cpi} \int F$ in 2d $\Uone$ gauge theory, $F \wedge F \wedge F$ terms in 6d, or even $\dif\sigma \wedge F$ terms in 3d where $\sigma \cong \sigma + 2\cpi$ is a compact scalar.

Along the lines that we saw in \S\ref{sec:CSStueckelberg}, general Chern-Simons terms involving a $\Uone$ $p$-form gauge field can have a dual formulation as Stueckelberg terms. This is due to non-conservation of electric flux in their presence, much as we saw in~\eqref{eq:modifiedBianchi} for the case of a BF term. To give another example, in a 4d theory with a $\frac{1}{8\cpi^2} \theta F \wedge F$ term, the compact scalar $\theta$ can be dualized to a 2-form gauge field $B$. However, our original theory has an equation of motion
\begin{equation}
f^2 \dif{\star \dif \theta} =-\frac{1}{8\cpi^2} F \wedge F,
\end{equation}
which in the dual picture is a ``modified Bianchi identity'' requiring that we introduce the dual 2-form via
\begin{equation} \label{eq:dualizeaxion}
\frac{1}{2\cpi} \dif B = f^2 \star\dif \theta + \frac{1}{8\cpi^2} A \wedge F.
\end{equation}
The kinetic term for $B$ in the dual formulation then takes the form
\begin{equation}
\int\left[-\frac{1}{2g^2} \left(\dif B - \frac{1}{4\cpi} A \wedge F\right) \wedge \star \left(\dif B - \frac{1}{4\cpi} A \wedge F\right)\right],
\end{equation}
which is gauge invariant because $B$ shifts as $B \mapsto B + \frac{1}{4\cpi} \alpha F$ under the gauge transformation $A \mapsto A + \dif \alpha$. In much the same way that an ordinary Chern-Simons term in 3d requires edge modes on boundaries, this structure in the 4d theory requires chiral charged modes to exist on ``axion strings,'' the objects that are charged under $B$~\cite{Callan:1984sa}. This is another example of the general concept of anomaly inflow, which we will discuss in more detail below.

\newpage

\section*{\Large Part Three: The Strong CP Problem and Axion Models}
\label{sec:lecturethree}
\addcontentsline{toc}{section}{\nameref{sec:lecturethree}}

\section{The Strong CP problem and the axion solution}
\label{sec:StrongCP}

\subsection{The problem and proposed solutions}

\begin{figure}[!h]
\centering
\includegraphics [width = 0.3\textwidth]{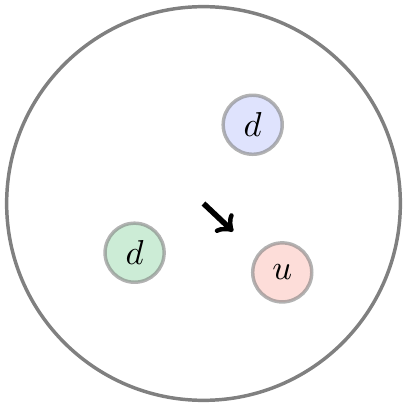}
\caption{A cartoon of a neutron, containing two down quarks and an up quark at generic positions in the interior. This configuration has a sizable electric dipole moment, as indicated by the arrow.
} \label{fig:neutronEDM}
\end{figure}

The Standard Model can have a $\theta$ term for the $\SU(3)_\textsc{C}$ gluon field, whose field strength I will denote $G$ to distinguish it from the generic field strengths $F$ that I have been referring to all along. This term takes the form
\begin{equation} \label{eq:thetaterm}
\frac{\theta}{8\cpi^2} \int \mathrm{tr}(G \wedge G).
\end{equation}
This is a CP-violating term. When quarks and gluons confine into hadrons at low energies, this term potentially has a variety of effects on hadron physics. Physical implications include a $\theta$-dependent vacuum energy, which is minimized at $\theta = 0$ (this is the Vafa--Witten theorem~\cite{Vafa:1984xg}); CP-violating pion-nucleon couplings, like $\pi^0 \overline{N} N$ or $(\overline{N}\tau^i N)\pi^i$ where $\tau^i$ is an SU(2) (global) isospin generator; and a CP-violating neutron electric dipole moment,
\begin{equation}
-\iu \frac{d_n}{2} \overline{N} \sigma_{\mu \nu} \gamma^5 N F^{\mu \nu}.
\end{equation}
This last effect has proven to be the easiest to constrain experimentally. The current bound is~\cite{Abel:2020pzs}
\begin{equation}
|d_n| \leq 1.8 \times 10^{-26}\,e\,\mathrm{cm} \qquad (90\%\,\text{C.L.}).
\end{equation}
We expect, just from dimensional analysis, a neutron EDM of order $\theta$ times the size of the neutron. At a cartoon level, this is simply because the neutron is made up of three valence quarks, two down quarks of charge $-1/3$ and one up quark of charge $2/3$, and a generic such configuration has an electric dipole moment, as sketched in Fig.~\ref{fig:neutronEDM}. To have no EDM at all, for this cartoon neutron, would require an unlikely configuration with the two down quarks precisely lined up on opposite sides of the up quark. (For a more extended discussion, including an interesting analogy to the CO$_2$ molecule, see~\cite{Hook:2018dlk}; another analogy involving a pool table, elaborate almost to the point of absurdity, can be found in~\cite{Sikivie:1995pz}.) A slightly more detailed estimate leads us to expect that
\begin{equation}
d_n \sim 10^{-16}\, \theta\,e\,\mathrm{cm}.
\end{equation}
Comparing expectations with data, we learn that
\begin{equation}
|\theta| \lesssim 10^{-10}.
\end{equation}
Why is this number so small? The only symmetry that it violates is CP, so one explanation could be that our universe is CP-symmetric. However, we know that to be false. The CKM matrix has an order-one CP-violating phase. So we would like to have a better explanation. The puzzle of small $\theta$ is known as the Strong CP Problem.

There are essentially three leading ideas that have been proposed to explain the Strong CP Problem. Namely,
\begin{itemize}

\item A massless up quark. If the up quark has no mass, we can rephase one chirality of the up quark, e.g., $u_L(x) \mapsto \E^{-\iu \theta} u_L(x)$, to remove the $\theta$ term from the theory (due to the chiral anomaly; see~\S\ref{sec:chiralanomalytakeaway}). The reason we focus on the {\em up} quark, rather than any of the others, is simply that empirically we know it is the lightest quark, so the closest to being massless. However, despite occasional attempts to resuscitate it, expert consensus is that this scenario has been ruled out by a combination of precision data and theory. (See, e.g.,~\cite{Kaplan:1986ru, Banks:1994yg, FlavourLatticeAveragingGroup:2019iem, Alexandrou:2020bkd}.)

\item CP is a fundamental symmetry of nature, which is spontaneously broken. The breaking must somehow generate a large CKM phase but {\em not} a large $\theta$. This requires some clever model-building. The paradigmatic example is the Nelson--Barr model~\cite{Nelson:1983zb,Barr:1984qx}, invented by Ann Nelson when she was a graduate student. A related class of models uses a generalized parity symmetry (which can exchange some fields with others, in addition to acting on spacetime) instead of CP (e.g.,~\cite{Babu:1989rb}). Unfortunately, a discussion of such models goes well beyond what I have time to discuss in these lectures.

\item {\bf \em Axions!} This solution, by far the most popular, is the subject of much of the remainder of these lectures. My current belief is that the existence of an axion field with at least approximately the properties needed to solve Strong CP is likely to be a {\em requirement} of a consistent theory of quantum gravity, for reasons we will come to later in these lectures. Thus, I find this to be one of the most promising scenarios for physics beyond the Standard Model.

\end{itemize}

\subsection{Axions for Strong CP: the core idea}
\label{subsec:axioncoreidea}

The core idea of using axions to solve the Strong CP problem emerged from a set of papers by Peccei, Quinn, Weinberg, and Wilczek~\cite{Peccei:1977ur, Peccei:1977hh, Weinberg:1977ma, Wilczek:1977pj}.\footnote{Wilczek named the particle the ``axion'' after a laundry detergent, presumably motivated by its relationship to axial symmetries, and because it ``cleans up'' the Strong CP problem. Weinberg, on the other hand, originally named the particle the ``higglet.''} The {\em only} necessary ingredients for this solution are a light, dynamical scalar field $\theta(x)$---the axion---which is periodic, $\theta \cong \theta + 2\cpi$, and has an approximate continuous shift symmetry $\theta  \mapsto \theta + \text{const.}$ (like a Nambu-Goldstone boson) that is (to very good approximation) {\em only} broken by a coupling to gluons:
\begin{equation} \label{eq:axionstrongCP}
S_\mathrm{axion} = \int \mathrm{d}^4 x\sqrt{|g|}\,\frac{1}{2} f^2 \partial_\mu \theta \partial^\mu \theta  + \frac{N}{8\cpi^2} \int \theta(x) \mathrm{tr}(G \wedge G).
\end{equation}
There can also be a similar $\theta F \wedge F$ coupling to photons, which is harmless for the Strong CP problem and potentially useful for experimental detection; in terms of the axion potential generated---via magnetic monopole loops---this is a very subleading effect to the QCD-generated potential we will shortly discuss. I will refer to the number $f$ as the axion decay constant, but be aware that some literature will instead refer to $f/N$ with that term. (We will shortly see why the combination $f/N$ is relevant.)\footnote{The term ``decay constant'' is an artifact of history. The analogous quantity $f_\pi$ in pion physics, which can be read off from the matrix element between an axial current and a one-pion state, has long been known as the pion decay constant. It should not be confused with the decay {\em width}. The charged pion decay width to leptons is proportional to $f_\pi^2$, for example. On the other hand, the axion decay width turns out to scale as $1/f^2$, for fixed axion mass.}

Let's (re)emphasize some important theoretical points, already discussed around~\eqref{eq:axioncouplingquantization} above. First, $\theta \mapsto \theta + 2\cpi$ is a {\em gauge} symmetry; these are two different labels for the same field configuration. Gauge invariant operators include $\partial_\mu \theta(x)$ or $\E^{\iu n \theta(x)}$ (with $n \in \mathbb{Z}$), but $\theta$ itself is {\em not} gauge-invariant. As a result, the action $S_\mathrm{axion}$ is not gauge invariant because the $\theta\, \mathrm{tr}(G \wedge G)$ term above changes when we shift $\theta$ by a multiple of $2\cpi$. However, what must be well-defined for the path integral to make sense is only $\exp(\iu S_\mathrm{axion})$, which transforms to $\exp[\iu S_\mathrm{axion} + 2\cpi \iu N \frac{1}{8\cpi^2} \int \mathrm{tr}(G \wedge G)]$. This is the same as $\exp(\iu S_\mathrm{axion})$ because of the quantization of instanton number,~\eqref{eq:instantonnumberquantization}, provided $N \in \mathbb{Z}$.

Why does this model solve the Strong CP problem? QCD dynamics generates a potential for $\theta$ which is minimized at the CP-preserving value $\theta = 0$. This follows from general principles~\cite{Vafa:1984xg}. However, we can actually go further. At high energies, $E \gg \Lambda_\textsc{QCD}$, small instantons (with size $\rho \sim E^{-1}$) generate calculable semiclassical contributions to the effective potential, proportional to $\E^{-8\cpi^2/g^2(E)} \left(\E^{\iu N \theta} + \mathrm{c.c.}\right)$. Recall that the QCD scale comes from RG running with a beta function coefficient $b = \frac{11}{3} N_c - \frac{2}{3} N_f$, so it is proportional to $\E^{-8\cpi^2/(b g^2)}$ with $b > 1$, and the small-instanton contributions are expected to be much smaller than contributions from the QCD scale.

At low energies, $E \lesssim \Lambda_\textsc{QCD}$, a description of the dynamics in terms of gluons and quarks is no longer valid; we should describe the axion's interactions with hadrons. Remarkably, we can obtain such an accurate description using the chiral Lagrangian! Here is a very quick review of the chiral Lagrangian, which exploits the fact that the only light composite states in QCD are the pseudo-Nambu-Goldstone bosons associated with spontaneous chiral symmetry breaking, which take values in the coset $[\SU(N_f)_\textsc{L} \times \SU(N_f)_\textsc{R}]/\SU(N_f)_\textsc{V}$. Thus, the low-energy EFT must be a sigma model living on this coset, which is accomplished by introducing the pseudo-Nambu-Goldstone boson fields $\pi^i$, packaged into a matrix
\begin{equation}
U(x) = \exp\left[2\iu \pi^a(x) T^a / f_\pi\right],
\end{equation}
with $T^a$ the hermitian generators of $\SU(N_f)$. We use the normalization
\begin{equation}
f_\pi \approx 92\,\mathrm{MeV}.
\end{equation}
The matrix-valued field $U(x)$ transforms under the flavor symmetry as $U(x) \mapsto L U(x) R^\dagger$ with $L \in \SU(N_f)_\textsc{L}, R \in \SU(N_f)_\textsc{R}$. One then builds up the chiral Lagrangian from flavor-symmetry invariants, like the kinetic term $\frac{1}{4} f_\pi^2 \mathrm{tr}[(D_\mu U)^\dagger D^\mu U]$. The quark masses explicitly break $\SU(N_f)_\textsc{L} \times \SU(N_f)_\textsc{R}$, and are incorporated as spurions in the form of a matrix $M$ that transforms as
\begin{equation}
M \mapsto L M R^\dagger,
\end{equation}
which at leading order can only appear in the chiral Lagrangian via a term
\begin{equation} \label{eq:masschiralLagrangian}
\mu^3 \,\mathrm{tr}\left[M U^\dagger + M^\dagger U \right],
\end{equation}
where $\mu$ is a dimensionful scale related to the scale of chiral symmetry breaking, $\mu^3 \sim \langle q {\bar q} \rangle$. Expanding out this term, we learn that pion masses scale like the square root of quark masses. If any of this is unfamiliar, I strongly encourage you to learn more~\cite{Georgi:weak, Leutwyler:1994fi, Manohar:1996cq}. The chiral Lagrangian is a key piece of Standard Model physics, and a beautiful example of how the logic of effective field theory and symmetry allows us to match aspects of UV physics to IR physics even across a strongly-coupled boundary region we don't understand. It is also vital to understanding the form of the axion potential, as we will now see.

At first glance, it might seem hopeless to match the gluonic operator $\mathrm{tr}(G \wedge G)$ onto a theory of effective interactions of pions and kaons, which after all are more closely related to quarks than to gluons. But in fact, there is a clever trick for doing so. We can remove the axion coupling to gluons using the by-now familiar trick of rephasing a fermion field and exploiting the chiral anomaly, i.e., we perform a field redefinition like $q_L \mapsto q_L \E^{\iu N \theta}$. This removes the axion--gluon coupling, produces a derivative coupling of the axion to the quark (not very important for our purposes at the moment), and changes the quark mass term:
\begin{equation}
m q_L \overline{q}_R + \mathrm{h.c.} \mapsto m \E^{\iu N \theta} q_L \overline{q}_R + \mathrm{h.c.}
\end{equation}
We now have a shift-symmetry violating coupling of the axion to quarks (at leading order in small field fluctuations, this is a Yukawa coupling). This is useful, because we know that quark mass terms appear in the chiral Lagrangian via the term~\eqref{eq:masschiralLagrangian}. Because of our field redefinition, the matrix $M$ is now not a constant but has $\theta(x)$ dependence inside it! We can expand this out and obtain an axion potential, which takes the form~\cite{DiVecchia:1980yfw, GrillidiCortona:2015jxo}
\begin{equation} \label{eq:axionpotential}
V_\textsc{QCD}(\theta) = - m_\pi^2 f_\pi^2 \sqrt{1 - \frac{4 m_u m_d}{(m_u + m_d)^2} \sin^2 \frac{N \theta}{2}}.
\end{equation}
Equation~\eqref{eq:axionpotential} is a good approximation to the axion potential generated by strong dynamics. There can be terms in the chiral Lagrangian involving higher powers of $M$, but because the quark masses are small, they should give small corrections. Expanding in small $\theta$, and recalling the factor of $f^2$ in the axion kinetic term to properly normalize the field, this translates into an axion mass
\begin{equation}
m_a = \frac{m_\pi f_\pi}{f/N} \sqrt\frac{m_u m_d}{2(m_u+m_d)^2} \approx 6\,\mu\mathrm{eV}\,\frac{10^{12}\,\mathrm{GeV}}{f/N}.
\end{equation}
Notice that this vanishes when a quark mass goes to zero, consistent with the massless up quark solution to Strong CP that we discussed earlier, when the value of $\theta$ has no physical effect. I've chosen a reference value of $10^{12}\,\mathrm{GeV}$ for $f/N$ because this turns out to be about the right value for axions to constitute all of the dark matter in the universe, in the simplest axion cosmology scenario. Thus, we expect the axion particle to be quite light, but also very weakly interacting, because all of its couplings are suppressed by the high energy scale $f$. Let's highlight the conclusion:

\begin{framed}
\noindent
The strong dynamics of QCD generates a potential for the axion, which relaxes it to the value that cancels the $\theta$ term, explaining why we do not see a nonzero neutron EDM. The axion mass is of order $m_\pi f_\pi / f$. The axion is very light and very weakly coupled when $f$ is a UV scale.
\end{framed}

\medskip
\noindent\centerline{\rule{\textwidth}{0.5pt}}
\noindent {\em Exercise.} Fill in the details, i.e., derive the potential~\eqref{eq:axionpotential} by carrying out an axion-dependent rephasing of a quark field and matching to the chiral Lagrangian.\\
\noindent\centerline{\rule[6.0pt]{\textwidth}{0.5pt}}

\subsection{Axions for Strong CP: further details}

I have streamlined the above discussion somewhat, in the interest of stating the core points about axion physics concisely. Here are a few further details and clarifications.

Above, I have phrased the whole discussion in terms of the $\theta$ term~\eqref{eq:thetaterm}. However, we know that due to the chiral anomaly such a term can be altered by a fermion field redefinition (see \S\ref{sec:chiralanomalytakeaway}). We used this to shift $\theta$ entirely into the quark mass matrix to compute the axion potential. However, in general, we might have {\em both} a $\theta$ term and a phase of the quark mass matrix $M$. The physically meaningful combination is
\begin{equation}
{\overline \theta} = \theta - \arg \det M.
\end{equation}
The chiral anomaly allows us to shift $\theta$ and $\arg \det M$ by the same constant, keeping $\overline \theta$ unchanged. Thus, the whole discussion surrounding~\eqref{eq:axionstrongCP} should be understood as one in which we have initially rephased the fermions to set $\arg \det M = 0$, and absorbed the corresponding constant shift in the $\theta$ term into the definition of our axion field $\theta(x)$.

We discussed the axion mass in terms of the chiral Lagrangian, which includes the pseudo-Nambu-Goldstone bosons of spontaneous chiral symmetry breaking. These are the pion fields, and (if we are interested in subleading effects) also the kaon and $\eta$ fields. The remaining pseudoscalar boson, the $\eta'$, obtains a significantly larger mass because of the $\Uone_\textsc{A}$ anomaly. Nonetheless, one can include it and its mixing with the pions and the axion by keeping the extra $\mathrm{U}(N_f)$ generator in the definition of the field $U(x)$ in the chiral Lagrangian (i.e., not requiring that $\det U = 1$, but allowing it to have a phase), and giving the extra mode a mass via a $(\log \det U)^2$ term in the chiral Lagrangian. In this case, $\log \det U$ shifts under a chiral rephasing of the quarks. You can read more in~\cite{Witten:1980sp, DiVecchia:1980yfw}.

\subsubsection{The axion-photon coupling from pion mixing}
\label{subsec:axionpionmixing}

The theory above the QCD scale (but below the electroweak scale) can contain a coupling 
\begin{equation} \label{eq:axionphotonUV}
\frac{E}{8\cpi^2} \int \theta\, F \wedge F,
\end{equation}
where $F$ is the field strength of electromagnetism, normalized so that the charge of the electron is $-1$ as usual. Here $E$ is quantized, as discussed around~\eqref{eq:axioncouplingquantization}. However, it is not necessarily quantized in integer units, because in our conventional normalization electric charge can be as small as $1/3$. This is a somewhat subtle point. Even if we assume that there are no particles in nature with hypercharge smaller than $1/6$, the Standard Model gauge group is still ambiguous: it takes the form $[\SU(3)_\textsc{C} \times \SU(2)_\textsc{L} \times \Uone_\textsc{Y}]/\Gamma$ where $\Gamma$ can be trivial, $\mathbb{Z}_2$, $\mathbb{Z}_3$, or $\mathbb{Z}_6$. This is because, given the Standard Model matter content alone, the centers of the nonabelian factors act on the fields in the same way as a hypercharge transformation.  An extensive discussion can be found in~\cite{Tong:2017oea}. If $\Gamma = \mathbb{Z}_3$ or $\mathbb{Z}_6$, the QED $\theta$ angle in its conventional normalization~\eqref{eq:thetaactionforms} has period $2\cpi$ and so we must have $E \in \mathbb{Z}$. On the other hand, if $\Gamma = \bm{1}$ or $\mathbb{Z}_2$ then the QED $\theta$ angle has period $2\cpi/9$. However, we've normalized the {\em axion} field in~\eqref{eq:axionphotonUV} to have period $2\cpi$, so $E$ is allowed to be any integer multiple of $1/9$. Currently we have no experimental evidence either way, so the most general statement we can make is (assuming the smallest hypercharge really is $1/6$) that $9E \in \mathbb{Z}$.

In any theory with the coupling~\eqref{eq:axionstrongCP}, the axion will couple to photons, {\em whether or not} it has a quantized coupling to electromagnetism of the form~\eqref{eq:axionphotonUV}. The reason is that, because of its coupling to gluons, the axion inevitably mixes with the $\pi^0$ meson~\cite{Kaplan:1985dv, Srednicki:1985xd, Georgi:1986df, Svrcek:2006yi}, and the $\pi^0$ couples to photons. 
Superficially, this coupling seems to violate the quantization condition that we derived on topological grounds in~\S\ref{sec:axiongaugefieldcouplings}. It does not, because this coupling is proportional to the axion mass, which is to say it is really a coupling of the form $(\Box \theta)F \wedge F$ or (using equations of motion) $\sin(\theta) F \wedge F$. Ordinarily, we expect that such contributions are highly subdominant, because they are proportional to $m_a^2$ and the axion mass is small. This case is an exception; since the axion mass and the pion mass are both generated at the QCD scale, we have $m_a^2 f^2 \sim m_\pi^2 f_\pi^2$, and it turns out that the coefficient of the axion coupling to photons induced by mixing with the pion is of order $(m_a^2 f^2)/(m_\pi^2 f^2)$, so it gives an $O(1)$ modification. For a more detailed discussion on this point, see Appendix A of~\cite{Agrawal:2017cmd}.

The end result is that, below the QCD scale, the axion has an effective coupling to photons of the form~\cite{Svrcek:2006yi, GrillidiCortona:2015jxo}
\begin{equation} \label{eq:axionphoton1}
\frac{1}{8\cpi^2} \left(E - \frac{2}{3} N \frac{4 m_d + m_u}{m_d + m_u}\right) \int \theta\, F \wedge F,
\end{equation}
up to subleading terms (e.g., those suppressed by the kaon mass). What one encounters most often in the phenomenological literature is a coupling $g_{a\gamma\gamma}$ written in terms of the canonically normalized axion field $\hat \theta = f \theta$ and the canonically normalized photon field strength $\hat F$, of the form 
\begin{align} \label{eq:gagammagamma}
{\cal L}_{a\gamma\gamma}&= -\sqrt{|g|}\frac{1}{4} g_{a\gamma\gamma} {\hat \theta} {\hat F}_{\mu \nu} \widetilde{{\hat F}}^{\mu \nu}, \nonumber \\
g_{a \gamma \gamma} &= \frac{\alpha}{2\cpi f/N} \left(\frac{E}{N} - 1.92(4)\right).
\end{align}
Here $\alpha$ is the fine structure constant. The factor $f/N$ is often simply called $f_a$, as it is the combination appearing in the coupling to gluons. The numerical value $-1.92(4)$ is a relatively recent estimate from~\cite{GrillidiCortona:2015jxo}, including subdominant terms beyond those in~\eqref{eq:axionphoton1}.

\subsection{Terminology: pseudo-Nambu-Goldstone bosons, QCD axions, and ALPs}
\label{subsec:terminology}

Before looking at detailed models of axions that solve the Strong CP problem, I want to make a brief aside about terminology. A pseudo-Nambu-Goldstone boson (PNGB) arises whenever an approximate continuous global symmetry is broken. If the global symmetry is {\em compact} (as it usually is), then the PNGB will parametrize a compact field space. For instance, pion fields in QCD-like theories take values in the compact coset manifold $[\SU(N_f)_\textsc{L} \times \SU(N_f)_\textsc{R}]/\SU(N_f)_\textsc{V}$. When an approximate $\Uone$ global symmetry is spontaneously broken, the field space is a circle, so the PNGB is a periodic scalar field, $\theta \cong \theta + 2\cpi$. It is common for such a periodic scalar field to be referred to as an axion, or axion-like particle, even outside of the context of axion solutions of the strong CP problem. Some people prefer to reserve the word ``axion'' for the original context of a periodic scalar coupled to $\mathrm{tr}(G \wedge G)$ and solving the Strong CP problem. Those who favor the strict use of the word ``axion'' favor the term ``axion-like particle'' or ``ALP'' for similar fields that do not interact with QCD and solve the Strong CP problem.\footnote{I have also seen ``ALF'' for ``axion-like field,'' though US readers of a certain age will associate this more strongly with the Alien Life Form from a late-1980s sitcom.} On the other hand, people who use the term ``axion'' more broadly will often say ``QCD axion'' to refer to the original case. I usually fall in the latter camp, although in this part of the lecture notes I will simply write ``axion'' since the context is the Strong CP problem.

Any periodic scalar field can be thought of as a PNGB, if only in the trivial sense that the low-energy theory contains the shift symmetry current $\partial_\mu \theta$ (or $\star \dif \theta$) that creates a single particle from the vacuum. In the original Peccei-Quinn scenario, as well as in other models we will review shortly, the QCD axion explicitly arose as a PNGB for a spontaneously broken approximate global $\Uone$ ``PQ'' symmetry. However, we will also discuss models in which the axion arises from a higher-dimensional gauge field, where there is no 4d PQ symmetry to break, so in such models axions are only PNGBs in the trivial sense.

In my usage, the term ``PNGB'' would generally be associated with pseudo-Nambu-Goldstone bosons for approximate symmetries broken by generic operators, whereas the term ``axion'' mostly refers to special cases where couplings of the form $\theta\, F \wedge F$ or $\theta\, \mathrm{tr}(F \wedge F)$ are the {\em dominant} sources of breaking of the scalar field's shift symmetry. In any case, you will quickly learn to infer from context what someone means by the term ``axion.'' 

\section{Classic 4d axion models; axion quality problem}

\subsection{The KSVZ model}
\label{subsec:KSVZ}

The KSVZ (Kim~\cite{Kim:1979if}; Shifman, Vainshtein, Zakharov~\cite{Shifman:1979if}) axion model is the simplest, most canonical model of a QCD axion. It consists of a complex scalar $\phi$ and two new fermion fields, $Q$ and ${\bQ}$, transforming in the $\bm{3}$ and $\bm{\overline{3}}$ of SU(3)$_\textsc{C}$ respectively. 
\begin{equation} \label{eq:LKSVZ}
\frac{\cal L_\textsc{KSVZ}}{\sqrt{-g}} = \frac{1}{2} \partial_\mu \phi^* \partial^\mu \phi + \iu Q^\dagger \overline{\sigma}^\mu D_\mu Q + \iu {\bQ}^\dagger \overline{\sigma}^\mu D_\mu {\bQ} + \left(y \phi Q {\bQ} + \mathrm{h.c.}\right) - V(\phi^* \phi).
\end{equation}
(As in \S\ref{sec:chiralanomalies}, the bar does not denote complex conjugation; $Q$ and $\bQ$ are {\em independent} Weyl fermion fields and the bar is just part of the name of the field $\bQ$.) In general, $Q$ and $\bQ$ can transform under $\SU(2)_\textsc{L} \times \Uone_\textsc{Y}$ as well, but the choice of representation may be model-dependent; this choice, of course, determines the form of the covariant derivatives $D_\mu$. The model in which $Q$ and $\bQ$ are neutral under electroweak interactions is the minimal model that you will often see  labeled simply ``KSVZ'' on plots of constraints on the axion parameter space.\footnote{If $Q$ and $\bQ$ have no hypercharge, they cannot decay to any Standard Model state due to the fractional hypercharge assignments of the ordinary quarks. This poses a potential cosmological problem, although the $Q$ or $\bQ$ particles would mostly annihilate away and any surviving asymmetric population must be bound into heavy hadrons. The stability of $Y = 0$ quarks is related to the question of the global structure of the Standard Model gauge group mentioned above in \S\ref{subsec:axionpionmixing}. Finding a color triplet particle with zero hypercharge would imply that the $\mathbb{Z}_6$ quotient is not the true global structure of the gauge group.} This Lagrangian, as written, has a classical global (0-form) symmetry $\Uone_\textsc{PQ} \times \Uone_\textsc{Q}$ acting as
\begin{equation}
\begin{aligned}
\Uone_\textsc{PQ}: & \quad  \phi \mapsto \E^{\iu \alpha} \phi, && Q \mapsto \E^{-\iu \alpha} Q; \nonumber \\
\Uone_\textsc{Q}: & \quad Q \mapsto \E^{\iu \beta} Q, && \bQ \mapsto \E^{-\iu \beta} \bQ.
\end{aligned}
\end{equation}
In a more complete model U(1)$_\textsc{Q}$ might be subsumed into a larger baryon number symmetry, or simply broken explicitly. It is U(1)$_\textsc{PQ}$ that will concern us here. We assume that $V(\phi^* \phi)$ has a symmetry-breaking form, so that $\langle \phi \rangle \neq 0$ and the U(1)$_\textsc{PQ}$ symmetry is spontaneously broken at the potential's minimum. Thus, at the classical level, there is a $\Uone$ Nambu-Goldstone boson $\theta$, which is the Peccei-Quinn axion mode in this theory, parametrizing a circular vacuum manifold. 

The expectation value of $\phi$ determines the axion decay constant. There is a corresponding massive radial mode of $\phi$, moving up the hill away from the minimum, which is sometimes called the ``saxion'' field or the {\em scalar partner} of the axion. (The terminology ``saxion'' has been mostly used in supersymmetric theories, where one adds an `s' at the beginning of the name of a fermion to refer to its scalar superpartner, e.g., squarks and sleptons; here, we are referring instead to a scalar partner of a {\em different} (pseudo)scalar, and the terminology need not be limited to supersymmetric theories.) Summing up,
\begin{equation}  \label{eq:complexfieldparam}
\phi(x) = \frac{1}{\sqrt{2}} \big(\underbrace{f}_\text{decay constant} + \underbrace{s(x)}_\text{saxion}\big) \exp(\iu \underbrace{\theta(x)}_\text{axion}).
\end{equation}
In this parametrization, it is manifest that $\theta(x)$ is a {\em periodic} scalar field, defined only modulo $2\cpi$ shifts. In more formal language, we can say that in the low-energy effective theory (below the mass scale of the saxion), there is an {\em emergent} $\mathbb{Z}$ gauge symmetry, $\theta \cong \theta + 2\cpi n$. This symmetry breaks down at the origin of field space, $\phi = 0$, where $\theta$ is no longer well-defined (not even mod $2\cpi$). However, that point is not accessible within the low-energy theory because $s(x)$ is massive.

Inspecting~\eqref{eq:LKSVZ}, we see that the fields $Q$ and $\bQ$ pair up into a massive Dirac fermion of mass
\begin{equation}
m_Q = \frac{1}{\sqrt{2}} y f.
\end{equation}
 while the saxion gets a mass from the potential. If the potential has a simple quartic form, $V(\phi^* \phi) = \lambda_\phi (\phi^* \phi - f^2/2)^2$, then the saxion mass is
\begin{equation}
m_s = \sqrt{2\lambda_\phi} f.
\end{equation}
At energies $\Lambda_\textsc{QCD} \ll E \ll f$, we can integrate out the heavy fields $Q$, $\bQ$, and $s$ to obtain an EFT of the axion $\theta(x)$ coupled to the Standard Model. However, there is a small complication. When we substitute the ansatz~\eqref{eq:complexfieldparam} into~\eqref{eq:LKSVZ}, we see that the mass term for the heavy quarks is $\theta$-dependent:
\begin{equation} \label{eq:thetadependentmQ}
m_Q \E^{\iu \theta(x)} Q\bQ(x) + m_Q \E^{-\iu \theta(x)} Q^\dagger \bQ^\dagger(x).
\end{equation}
We cannot simply set $Q = \bQ = 0$ in the Lagrangian due to this $\theta$-dependence. We have to integrate out the quarks more carefully. This can be done by computing loop diagrams, but a more efficient way is to use the chiral anomaly. We can eliminate $\theta(x)$ from~\eqref{eq:thetadependentmQ} by carrying out a field redefinition,
\begin{equation}
Q(x) \mapsto \E^{-\iu \theta(x)} Q(x).
\end{equation}
Substituting this field redefinition into~\eqref{eq:LKSVZ}, we find no non-derivative interaction between $\theta$ and the heavy fields, which we can then integrate out by simply setting them to zero.\footnote{We do find a derivative interaction of the axion with the heavy quarks, $Q^\dagger \overline{\sigma}^\mu Q \partial_\mu \theta$. This respects a continuous shift symmetry of $\theta$. When we integrate out the heavy quarks, this leads to derivative self-interactions of the axion, with effects that vanish at small momentum and have negligible impact on axion phenomenology.} However, due to the chiral anomaly, our field redefinition changes the path integral measure in a way that corresponds to adding a new term to the action, of the form~\eqref{eq:SUNanomalyterm}:
\begin{equation} \label{eq:thetacoupling}
- \frac{1}{8\cpi^2} \int \theta(x) \mathrm{tr}(G \wedge G),
\end{equation}
where we have used the Dynkin index $I_2(\bm{3}) = 1/2$. This coupling reflects the existence of a chiral  $\Uone_\textsc{PQ}\text{-}\SU(3)_\textsc{C}^2$ anomaly. (This is an ABJ anomaly: QCD explicitly breaks the would-be global symmetry $\Uone_\textsc{PQ}$, and in particular we cannot gauge it.) Thus we see that $\theta$ couples to gluons, and our effective theory has the form~\eqref{eq:axionstrongCP} that we have previously argued solves the strong CP problem (with $N = -1$). In the case that $Q$ and $\bQ$ carry electroweak charges, there would be additional interactions of $\theta$ with the electroweak gauge fields. 

The PQ symmetry is {\em not} an accidental symmetry of ${\cal L}_\textsc{KSVZ}$.\footnote{An accidental symmetry is one that can only be broken by irrelevant operators; see \S\ref{sec:protondecay}.}  Gauge invariance does not forbid relevant terms like  $M_Q Q \bQ + \mathrm{h.c.}$ or $b_\phi \phi^2 + \mathrm{h.c.}$ (where $M_Q$, $b_\phi$ are parameters of positive mass dimension), which would explicitly break the symmetry. This is an important point to which we will return later: we never expect global symmetries to be fundamental, so the theory must have more structure in order to explain why these terms are suppressed.

\medskip
\noindent\centerline{\rule{\textwidth}{0.5pt}}
\noindent {\em Exercise:} convince yourself that we can still talk about an EFT of an axion, even with (small) explicit breaking of Peccei-Quinn symmetry (beyond the ABJ anomaly). More explicitly: for a sufficiently small $b_\phi \phi^2$ term, show that the parametrization~\eqref{eq:complexfieldparam} is still sensible. We can integrate out the mode $s(x)$ and write an effective theory of $\theta(x)$. However, now $\theta(x)$ has a potential. What condition is required for this to be subdominant to the potential generated by QCD dynamics?\\ 
\noindent\centerline{\rule[6.0pt]{\textwidth}{0.5pt}}

In fact, the PQ symmetry is not even a symmetry of the {\em quantum} theory with Lagrangian ${\cal L}_\textsc{KSVZ}$, due to its ABJ anomaly with $\SU(3)_\textsc{C}$. It is an explicitly broken symmetry. This makes it even harder to see why the Lagrangian should have respected the symmetry in the first place.\footnote{Interestingly, it has recently been shown that in some cases, a type of generalized {\em non-invertible} symmetry remains even after an ABJ anomaly~\cite{Choi:2022jqy, Cordova:2022ieu}, though this is not the case for the mixed anomaly with QCD discussed here.} Because of the ABJ anomaly, the axion field $\theta$ is not a true Nambu-Goldstone boson but a {\em pseudo}-Nambu-Goldstone boson. As such, we expect it to acquire a mass; the coupling $\theta \,\mathrm{tr}(G \wedge G)$ indeed generates a mass, as discussed in \S\ref{subsec:axioncoreidea}.

\subsection{DFSZ model}

The DFSZ model (Zhitnitsky~\cite{Zhitnitsky:1980tq}; Dine, Fischler, Srednicki~\cite{Dine:1981rt}) is the second classic model of a weakly-coupled axion. In this model, the Standard Model is extended to a two Higgs doublet model (2HDM), and the PQ symmetry acts on the Higgs fields as well as on a heavy complex scalar $\phi$. Because the Higgs fields carry PQ charge and have Yukawa couplings to SM fermions, the fermion fields must also carry PQ charge. Thus, DFSZ models are more complicated than KSVZ models.

As an example, consider a Type II 2HDM, in which the two Higgs doublets $H_u$ and $H_d$ couple to Standard Model fermions via
\begin{equation} \label{eq:2hdmyukawas}
y_u H_u q {\bar u} + y_d H_d q {\bar d} + y_e H_d \ell {\bar e} + \mathrm{h.c.}
\end{equation}
This structure arises in the MSSM, though we can also consider it outside the context of supersymmetry. We take the fields $H_u$ and $H_d$ to have PQ charge $+1$, and the fields ${\bar u}$, ${\bar d}$, and ${\bar e}$ to have PQ charge $-1$. We also take the complex scalar $\phi$ to have PQ charge $+1$. This allows for a quartic coupling between the Higgs bosons and the $\phi$ field,
\begin{equation} \label{eq:DFSZquartic}
\lambda_{ud\phi} H_u H_d {\phi^\dagger}^2 + \mathrm{h.c.}
\end{equation}
Again, we assume that the field $\phi$ has a potential leading to a VEV as in~\eqref{eq:complexfieldparam}, with $f \gg v_u, v_d$. From this we immediately see that we either require $\lambda_{ud\phi} \lesssim v^2/f^2$, or the model must have some fine tuning to separate the weak scale from the scale of the axion decay constant. Of course, this is not unique to the DFSZ model; quite generally, extensions of the Standard Model involving heavy mass scales generate electroweak fine-tuning problems. In the KSVZ model, however, the corrections to the Higgs mass arise only at higher loops, whereas in the DFSZ case~\eqref{eq:DFSZquartic} is an integral part of the model and the problem is already apparent at tree level.

As in the KSVZ case, we would like to integrate out the radial mode of the heavy field $\phi(x)$ and treat its phase $\theta(x)$ as a pseudo-Nambu-Goldstone boson that survives in the low-energy theory. However, due to the coupling~\eqref{eq:DFSZquartic}, the phase $\theta(x)$ appears in an effective Higgs mass term below the scale $f$. We can perform a field redefinition to rephase the Higgs fields and eliminate this term, but then the phase will appear in the Yukawa couplings~\eqref{eq:2hdmyukawas}. Finally, we can rephase Standard Model fermions to eliminate the phase in the Yukawa couplings, but this generates a coupling of $\theta(x)$ to gluons via the chiral anomaly (as in~\S\ref{sec:chiralanomalytakeaway}) as well as derivative couplings of Standard Model fermions to the axion, of the form
\begin{equation} \label{eq:axionfermionDFSZ}
\frac{v_u^2}{v_u^2 + v_d^2} (\partial_\mu \theta) {\bar u}^\dagger \overline{\sigma}^\mu {\bar u} + \frac{v_d^2}{v_u^2 + v_d^2} (\partial_\mu \theta) {\bar d}^\dagger \overline{\sigma}^\mu {\bar d} + \cdots.
\end{equation}
From the experimental viewpoint, then, an important distinction between KSVZ models and DFSZ models is that the axion couples more strongly to Standard Model fermions in DFSZ models. Note that the derivative coupling to fermions cannot induce a non-derivative coupling to gauge fields like photons, because the former preserves a continuous shift symmetry, whereas the latter preserves only a discrete shift of $\theta$. As we saw in \S\ref{sec:axiongaugefieldcouplings}, the non-derivative couplings have quantized coefficients for topological reasons, so they cannot be generated by non-quantized couplings like~\eqref{eq:axionfermionDFSZ}.

A full discussion of the couplings in the DFSZ model is not very useful for the more conceptual points I want to focus on in this note, but I do want to make a few brief comments on how to analyze such models. Much of the literature on DFSZ models carries out field redefinitions on Standard Model fermions of the form $\psi(x) \mapsto \E^{\iu \alpha \theta(x)} \psi(x)$. If $\alpha \notin \mathbb{Z}$, this is not a mathematically sensible operation. Relatedly, many of these papers also refer to $\Uone$ charges that are irrational numbers. Despite such intermediate steps that (strictly speaking) make no sense, they tend to get the right answers. My collaborators and I tried to explain how to do these calculations carefully in~\cite{Buen-Abad:2021fwq}, a paper written in the context of a particular experimental anomaly but one that (I hope) may be a useful reference outside the context of the anomaly. In any case, it's a good exercise for you to work through all the details of the DFSZ model for yourself.

\medskip
\noindent\centerline{\rule{\textwidth}{0.5pt}}
\noindent {\em Exercise (somewhat open-ended):} Explore the EFT obtained by integrating out heavy fields in the DFSZ model. Notice that you have some choices along the way. For example, you might choose to eliminate the phase in~\eqref{eq:DFSZquartic} by carrying out a field redefinition only on $H_u$, or alternatively only on $H_d$. This would then lead to different field redefinitions on fermions to eliminate phases in~\eqref{eq:2hdmyukawas}. Understand why the results are physically equivalent despite such arbitrary choices.\\
\noindent\centerline{\rule[6.0pt]{\textwidth}{0.5pt}}

In the DFSZ model, the axion acquires a coupling to photons due to the chiral anomaly associated with the charged Standard Model fermions. This is unlike the (minimal) KSVZ model, where the PQ-charged quarks are electrically neutral. However, people tend to overstate the importance of this difference: one could consider a modified KSVZ model where $Q$ and $\bQ$ have hypercharge, so the size of the axion-photon coupling in the two models is not a real structural difference between them. In any case, it turns out that the contribution to the axion-photon coupling from the anomaly in the DFSZ model is $8/3$, but this must be combined with the contribution from the axion-pion mixing discussed in \S\ref{subsec:axionpionmixing}. The result is that in this model, we have, following~\eqref{eq:gagammagamma},
\begin{equation}
g_{a\gamma\gamma} \propto \frac{E}{N} - 1.92 = \frac{8}{3} - 1.92 \approx 0.75, 
\end{equation}
which is significantly smaller (in absolute value) than the pion mixing contribution alone. Thus, the (minimal) DFSZ model predicts a smaller axion-photon coupling than the (minimal) KSVZ model, and for this reason it is often taken as a target for ambitious experiments that wish to probe the full range of possible axion-photon couplings. (Of course, one could easily write down another model where $E/N = 2$, and then would have much more of a challenge!)

The value $E/N = 8/3$ appearing in the minimal DFSZ model is also characteristic of a large class of GUT models, in which the Standard Model gauge group embeds in an SU(5) subgroup of the GUT gauge group and an axion couples to $\mathrm{tr}(F \wedge F)$ for the full GUT group. You can find an extensive discussion of the phenomenology of axions in GUTs in the recent paper~\cite{Agrawal:2022lsp}. Note that a DFSZ model need not be a GUT model: the defining feature of DFSZ is Higgs fields that carry PQ charge, independent of the details of far-UV physics. Similarly, not every GUT model with an axion is a DFSZ model: models with no 4d Peccei-Quinn symmetry at all, along the lines we will discuss in \S\ref{sec:xdimaxion}, could be GUT models. 

For all of these reasons, I don't like the practice of labeling experimental plots with ``KSVZ'' and ``DFSZ'' lines. I would prefer to see labels like ``$E/N = 0$'' and ``$E/N = 8/3$'' that don't bias the interpretation toward specific UV completions, but there is enough inertia behind the current practice that it's unlikely to change.

\subsection{Axion quality problem}

Throughout the above discussion we've been assuming that we can impose the global PQ symmetry to restrict the terms that we write in the Lagrangian. As we already hinted in \S\ref{subsec:KSVZ}, there are two problems with this. The first is that $\Uone_\textsc{PQ}$ has an ABJ anomaly, which means that it is not a symmetry of the quantum theory at all. The second is that even without the anomaly, it is a global symmetry, and (as we will discuss extensively in \S\ref{sec:BHnoglobal} and \S\ref{sec:globalvsQG} below), we do not expect global symmetries to ever exist in theories of quantum gravity.

This is a severe problem, because explicit PQ-breaking terms can completely spoil the solution to the Strong CP problem! For example, we can consider the KSVZ scenario with an added Peccei-Quinn-violating (``PQV'') term in the Lagrangian of the form
\begin{equation}
\frac{1}{\sqrt{-g}} {\cal L}_{\textsc{PQV}} = \frac{c}{M_\mathrm{Pl}^{n-4}} \phi^n + \mathrm{h.c.},
\end{equation}
where the coefficient $c$ in general can have a complex phase $\varphi$,
\begin{equation}
c = |c| \E^{\iu \varphi}.
\end{equation}
Expanding around a $\phi$ VEV as in~\eqref{eq:complexfieldparam}, this becomes an effective axion potential,
\begin{equation} \label{eq:axionPQVpotential}
V_{\textsc{PQV}}(\theta) = \frac{|c|}{M_\mathrm{Pl}^{n-4}} \left(\frac{f}{\sqrt{2}}\right)^n \left[ \E^{\iu (\varphi + n \theta)} + \E^{-\iu (\varphi + n \theta)}\right] = 2 |c| M_\mathrm{Pl}^4 \left(\frac{f}{\sqrt{2} M_\mathrm{Pl}}\right)^n \cos\left(n \theta + \varphi\right).
\end{equation}
If CP is not a fundamental symmetry, there is no reason for $\varphi$ to be a small phase. As a result, such an effective potential term can shift the minimum of $V(\theta)$ away from $\theta = 0$. However, experimentally, we know that $|\theta| \lesssim 10^{-10}$, so either the magnitude of $V_{\textsc{PQV}}(\theta)$ or the phase $\varphi$ must be extremely small.

This problem is known as the {\em axion quality problem}, and it is extremely severe for simple models of KSVZ or DFSZ type. The QCD axion potential~\eqref{eq:axionpotential} is naturally exponentially small, because the overall size of the potential is set by the QCD scale, which arises from dimensional transmutation. By contrast, the PQV contribution~\eqref{eq:axionPQVpotential} is suppressed only by a {\em power} of $f/M_\mathrm{Pl}$. In order for a power-law suppressed potential to be many orders of magnitude smaller than an exponentially suppressed potential, we require the power to be large. For example, for $|c| = 1$ and $f = 10^{12}\,\mathrm{GeV}$ we compute 
\begin{equation}
V_\textsc{PQV}(\theta) \approx \begin{cases} (250\,\mathrm{TeV})^4\cos(8 \theta + \varphi), & n = 8, \\ (73\,\mathrm{MeV})^4\cos(12\theta + \varphi), & n = 12, \end{cases}
\end{equation}
showing that even moderately large values of $n$ would give contributions that overwhelm the QCD axion potential without further suppression in the coefficient or the phase. Indeed, for $O(1)$ values of $|c|$ and $\varphi$ we need $n \geq 14$ to avoid spoiling the solution of the Strong CP problem.

The axion quality problem, then, requires that we either forbid (using an exact gauge symmetry) or strongly (exponentially) suppress {\em many} dangerous operators so that the desired low-energy axion EFT~\eqref{eq:axionstrongCP} dominates the dynamics~\cite{Barr:1992qq, Kamionkowski:1992mf, Holman:1992us, Randall:1992ut}. Notice that the first priority is to forbid {\em relevant} operators, which are a major hazard in typical axion models without additional gauge symmetries. But because the problem is so severe, even highly irrelevant operators can be dangerous. 

Within 4d axion models, a solution to the axion quality problem necessarily requires a theory with extended gauge symmetry beyond the Standard Model. One of the simplest approaches is to invoke a discrete gauge symmetry, e.g., a $\mathbb{Z}_k$ subgroup of $\Uone_\textsc{PQ}$. For such a subgroup to be non-anomalous, given that $\Uone_\textsc{PQ}$ {\em is} anomalous, we require that the constant $N$ in~\eqref{eq:axionstrongCP} is a multiple of $k$. This is potentially a viable solution, though the values of $k$ that are required are awkwardly large. Another approach is to consider a model of a composite axion, replacing the complex Peccei-Quinn  field $\phi$ in the basic axion models with a composite operator of larger scaling dimension. If this operator dimension is high enough, the number of low-dimension operators we must forbid in the Lagrangian can be much smaller than in the simplest models. One can also combine discrete gauge symmetries and compositeness.

Over the years, a great deal of ingenuity has been applied in constructing 4d axion models that evade the quality problem. However, in my opinion, the best solution is to abandon models with a 4d Peccei-Quinn symmetry entirely, as discussed in the next section.

\section{Axions from higher-dimensional gauge fields}
\label{sec:xdimaxion}

A particularly elegant solution to the axion quality problem is to dispense with the whole idea of spontaneously breaking a 4d $\Uone_\textsc{PQ}$ symmetry, and instead to derive the basic low-energy axion action~\eqref{eq:axionstrongCP} from an entirely different UV starting point: a higher-dimensional gauge theory with a Chern-Simons term. Zero modes of a higher-dimensional gauge field can automatically have the key features of axions, with {\em exponentially good} control of the axion quality problem.

\subsection{Basic ingredients for extra-dimensional axions} 

This idea originated (shortly after the KSVZ and DFSZ papers) in string theory examples, where the gauge field in question was a higher $p$-form field~\cite{Witten:1984dg, Choi:1985je, Barr:1985hk}. To illustrate the core idea, we will present a simpler example arising from an ordinary 1-form gauge field in a 5d context (see~\cite{Cheng:2001ys,Arkani-Hamed:2003xts} and especially~\cite{Choi:2003wr} for similar phenomenological models). We consider a 5d theory of a $\Uone$ gauge field $A$ compactified to 4d on a circle $S^1$ with coordinate $x^5 \cong x^5 + 2\cpi R$. We identify the 4d axion $\theta$ as a Kaluza-Klein zero mode of $A$. That is:
\begin{equation}
\theta(x) = \int_{S^1} A(x) = \int_0^{2\cpi R} A_5(x,x^5) \dif{x^5}.
\end{equation}
This is a periodic variable, $\theta \cong \theta + 2\cpi$. To see this, consider a winding gauge transformation of $A$ in 5d, with the $\Uone$ element $g(x,x^5) = \exp(\iu x^5/R)$. Then $A \mapsto A + \dif\alpha$ where $\alpha(x^5) = \frac{x^5}{R}$ is not single-valued. Under such a gauge transformation, $\int_{S^1} A \mapsto \int_{S^1} A + 2\cpi$. This shows that $\theta$ is not a well-defined (gauge invariant) quantity, but it is defined modulo $2\cpi$, and in particular $\E^{\iu \theta}$ is well-defined. This discussion should be familiar: it's precisely how we described Wilson loops in \S\ref{subsec:wilsonloops}. Another way to explain the origin of the 4d axion field is that the gauge invariant Wilson loop around the 5d circle over the 4d point $x$, $W(S_x^1)$, is an element of $\Uone$. We identify this element as $\E^{\iu \theta(x)}$ to define the 4d periodic scalar $\theta(x)$.

Given that a 4d periodic scalar exists, our next task is to identify the origin of the two pieces of the action~\eqref{eq:axionstrongCP}: the kinetic term, from which we can read off the decay constant $f$, and the coupling to gluons that is crucial for solving the Strong CP problem. The kinetic term arises directly from the 5d kinetic term for a gauge field, where the gauge coupling squared $e_5^2$ has units of length,
\begin{equation}
\int_{M \times S^1} -\frac{1}{2e_5^2} F \wedge \star_{5\mathrm{d}} F,
\end{equation}
where $M$ is the 4d spacetime manifold. This becomes, with the ansatz $A = \theta \dif x^5/(2\cpi R)$, 
\begin{equation} \label{eq:5dto4df}
\int_M \frac{1}{2} f^2 \dif \theta \wedge \star \dif\theta, \qquad f^2 = \frac{1}{2\cpi R e_5^2}.
\end{equation}
Thus, we see that a small $f$ can be achieved when the compactification radius is large compared to the 5d length scale $e_5^2$. The second ingredient is a 5d Chern-Simons term, of the general type discussed in~\S\ref{sec:CSgeneral}. For this, we have to extend the Standard Model gauge fields over the extra dimension, so we suppose that there are $\SU(3)$ gluon fields $G$ also propagating in 5d, with coupling
\begin{equation} \label{eq:5dCS}
\frac{N}{8\cpi^2} \int_{M \times S^1} A \wedge \mathrm{tr}(G \wedge G).
\end{equation}
In 4d this directly becomes the usual axion $\theta\, \mathrm{tr}(G \wedge G)$ Chern-Simons term.

Because the gluon fields propagate in the extra dimensions, the 4d gluon kinetic term arises when the 5d gluon kinetic term
\begin{equation}
\int_{M \times S^1} -\frac{1}{2g_5^2} \mathrm{tr}(G \wedge \star_{5\mathrm{d}} G)
\end{equation}
becomes, with the simple ansatz that $G$ is independent of the 5th dimension,
\begin{equation}\label{eq:5dto4dG}
\int_M -\frac{1}{2 g^2} \mathrm{tr}(G \wedge \star G), \quad \text{where} \quad \frac{1}{g^2} = \frac{2\cpi R}{g_5^2}.
\end{equation}
Thus, we see that the smallness of the 4d $\SU(3)_\textsc{C}$ gauge coupling (at the compactification energy, where we match 5d to 4d) can potentially be explained by a relatively large volume of the internal dimension.

One major difference between such a higher-dimensional axion and the familiar 4d models is that we have not mentioned a $\Uone_\textsc{PQ}$ symmetry at all! In particular, the higher-dimensional gauge group is {\em not} the same thing as the Peccei-Quinn symmetry group. In these models there is no Peccei-Quinn symmetry breaking phase transition. Is the axion still a pseudo-Nambu-Goldstone boson? Yes, but only in the trivial sense mentioned in \S\ref{subsec:terminology}: there is an approximately conserved current that, when acting on the vacuum, produces a single-particle state of the axion. This current is simply $\partial_\mu \theta$ itself (or in differential form notation, $\star \dif \theta$), that is, the shift symmetry current. The sense in which the axion is a pseudo-Nambu-Goldstone boson in these theories is the same sense in which the photon is a pseudo-Nambu-Goldstone boson, which we will discuss later in \S\ref{subsubsec:photongoldstone}. The Peccei-Quinn symmetry is just the 4d remnant of the electric 1-form symmetry of the 5d theory.
 
The case of a 5d compactification on a circle is somewhat unappealing, because it also produces a massless $\Uone$ gauge boson (the Kaluza-Klein gauge field) in 4d. It is also unclear where 4d chiral fermions would originate in this construction. For a more realistic model, we could take two approaches. One is to compactify the 5d theory on an interval, rather than a circle. However, in this case we will not necessarily find a corresponding massless axion in 4d, depending on the boundary conditions. The other approach, which is realized in a large collection of string theory examples, is to consider a geometry with multiple extra dimensions and obtain the axion by reducing a $p$-form gauge field on a $p$-dimensional cycle within the extra dimensions. In this case, there is always a (perturbatively) massless mode for every such cycle. We will comment on this case further in \S\ref{subsec:evenhigherdim}. For now, let's continue to investigate the 5d toy model on a circle, as it provides the simplest setting in which to learn some important qualitative lessons.

\subsection{Quality problem for extra-dimensional axions}
\label{subsec:qualityxdim}

Because $A$ is a gauge field, it is exactly massless in the 5d theory. So there is no immediate mystery of why we don't have a large bare potential for $\theta$. (One could wonder about a Stueckelberg or $BF$-type mass for $A$, but these depend on an integer coefficient, so it is perfectly consistent to set any such coupling to zero and forget about it.) In general, 5d couplings that depend on the field strength $F = \dif A$ become 4d terms depending on $\dif \theta$, which do not generate an axion potential. Thus, the only terms that can possibly matter for the axion quality problem are 5d terms that depend {\em non-derivatively} on $A$. In gauge theories, such terms are very highly constrained. One such term is the 5d Chern-Simons term~\eqref{eq:5dCS}, which becomes the 4d axion coupling to gluons, and is crucial for generating the potential $V_\textsc{QCD}$. We could also potentially have 5d Chern-Simons couplings of $A$ to different gauge fields (e.g., hidden sector gluons). These would generate a 4d potential in a similar manner to QCD. In particular, such a potential would be exponentially small, and so it would not be surprising for such contributions to be small enough, relative to $V_\textsc{QCD}$, to not spoil the Strong CP solution.

\begin{figure}[!h]
\centering
\includegraphics [width = 0.7\textwidth]{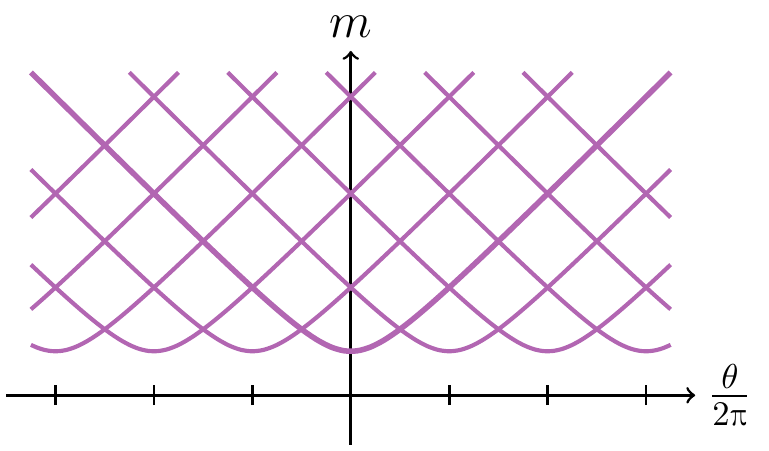}
\caption{Spectrum of KK mode masses~\eqref{eq:kkmasses} as a function of $\theta$, illustrating monodromy. The $n = 0$ mode is drawn with a thicker line; other curves are for different values of $n$. The spectrum of masses is $2\cpi$ periodic, but the mass of an individual mode is not.
} \label{fig:monodromy}
\end{figure}

The other 5d source of couplings depending non-derivatively on $A$ is the existence of 5d fields charged under the gauge field. In other words, the 5d covariant derivative for a field with $\Uone$ charge $q \in \mathbb{Z}$, which is $D = \dif{} + \iu q A$, leads to $\theta$-dependent terms in 4d. The axion field lives in $A_5$, so we specifically look at the 5th component:
\begin{equation}
D_5 = \underbrace{\partial_5}_{\iu \frac{n}{R},~\text{KK~number}} + \iu q \underbrace{A_5}_{\frac{\theta}{2\cpi R}}.
\end{equation}
A 5d field with mass $m_{5\mathrm{d}}$ becomes an infinite tower of 4d fields with $n \in \mathbb{Z}$ units of Kaluza-Klein  momentum around the circle, and an axion-dependent mass,
\begin{equation} \label{eq:kkmasses}
m_n^2 = m_{5\mathrm{d}}^2 + \frac{1}{R^2} \left(n + \frac{q \theta}{2\cpi}\right)^2.
\end{equation}
This is a manifestly non-derivative coupling of the axion. We also see that the mass of a given KK mode $m_n$ is {\em not} a periodic function of $\theta$, even though our theory is supposed to be gauge invariant under $\theta \mapsto \theta + 2\cpi$. It {\em is}, but in a nontrivial way: the mass of an individual mode is not invariant, but the mass of the mode with number $n - q$ shifts to match the mass that the mode with number $n$ originally had. The infinite tower rearranges itself so that the full spectrum is invariant under a $2\cpi$ shift of $\theta$. This phenomenon, illustrated in Fig.~\ref{fig:monodromy}, is generally referred to as {\em monodromy}. The simplest example, which you may have encountered before, is the quantum mechanical problem of a particle on a ring.

Once we have a set of modes with mass depending on the value of $\theta$, quantum effects will necessarily translate this into a $\theta$-dependent potential energy. This type of calculation is often referred to as a Coleman-Weinberg potential~\cite{Coleman:1973jx}: we think of $\theta(x)$ as a classical background in which the various KK modes are propagating, and simply sum up their vacuum bubbles in this background. When we integrate out the KK modes of a given 5d field, we have to sum over all $n$. This sum should give us a periodic function of $\theta$, because of the monodromy phenomenon. One mathematical trick for making this periodicity manifest is Poisson resummation. This is a frequently useful trick, so it is worth your time to become familiar with it by working through some exercises.

\medskip
\noindent\centerline{\rule{\textwidth}{0.5pt}}
\noindent {\em Exercise (Poisson summation formula):} This formula asserts that the sum of a (sufficiently well-behaved) function at integer values is the same as the sum of its Fourier transform at integer values, i.e.,
\begin{equation}
\sum_{n = -\infty}^{\infty} f(n) = \sum_{k = -\infty}^\infty {\hat f}(k), \quad {\rm where} \quad {\hat f}(k) \equiv \int_{-\infty}^{\infty} f(x) {\E}^{-2\cpi \iu k x}\,{\rm d}x.
\end{equation}
To see why this is true, notice that $F(x) \equiv \sum_{n = -\infty}^{\infty} f(x+n)$ is a periodic function, so it can be expanded in a Fourier series. The usual formula for the coefficients in a Fourier series leads almost directly to the desired identity; fill in the details.\\
\noindent\centerline{\rule[6.0pt]{\textwidth}{0.5pt}}

\medskip
\noindent\centerline{\rule{\textwidth}{0.5pt}}
\noindent {\em Exercise:} Use the Poisson summation formula and the general formula for a Gaussian integral to show that
\begin{equation}
\sum_{n = -\infty}^\infty \E^{-A n^2 - B n - C} = \sqrt{\frac{\cpi}{A}} \sum_{k = -\infty}^\infty \E^{-{\widetilde A} k^2 - {\widetilde B} k - {\widetilde C}},
\end{equation}
where the coefficients are related as
\begin{equation}
{\widetilde A} = \frac{\cpi^2}{A}, \quad 
{\widetilde B} = -\frac{\iu \cpi B}{A}, \quad
{\widetilde C} = C - \frac{B^2}{4A}.
\end{equation}
\noindent\centerline{\rule[6.0pt]{\textwidth}{0.5pt}}

In our context, the Poisson summation trick converts a sum over KK mode number $n$ to a sum over {\em winding} number $w \in \mathbb{Z}$. There is a particularly nice semiclassical understanding of this in the limit when the 5d mass of a charged field is large compared to the KK scale, $2\cpi m_{5\mathrm{d}} R \gg 1$. Instead of thinking of a path integral over the charged field, we can imagine the 5d particles as heavy objects with a 1-dimensional worldline $\gamma$ tracing their path through spacetime, and we can sum over these worldlines in the path integral. Attached to the worldline is an action for the particle, which for a Euclidean (spacelike) worldline is
\begin{equation}
S(\gamma) = \int_\gamma m_{5\mathrm{d}}\,\dif\tau + \iu q \int_\gamma A,
\end{equation}
where $\tau$ is the proper length along the worldline. The sum over worldlines includes a sum over topological sectors where $\gamma$ wraps $w$ times around the 5d circle. In particular, there are saddle-point contributions to the path integral for each such topological sector, and we can estimate that the sum of contributions from winding numbers $\pm w$ scales as
\begin{equation} \label{eq:worldlineheuristic}
\E^{-S(\gamma)} + \E^{-S(-\gamma)} \sim \E^{-2\cpi w m_{5\mathrm{d}} R} (\E^{-\iu q w \int_{S^1} A} + \mathrm{c.c.}) \sim \E^{-2\cpi w m_{5\mathrm{d}} R} \cos(w q \theta).
\end{equation} 
This heuristic expectation is correct, and in fact the sums can be done analytically (see, e.g.,~\cite{Hosotani:1983xw, Cheng:2002iz, Arkani-Hamed:2003xts, Arkani-Hamed:2007ryu}).

We have learned that a periodic potential for $\theta$ is generated when we integrate out KK modes of 5d charged particles, and this potential is {\em exponentially small} when the extra dimensions are large. The intuition behind this exponential is that any effect that can generate an axion potential must somehow know the correct periodicity of $\theta$; as we saw in~\eqref{eq:5dto4df}, the period of the canonically normalized 4d axion is sensitive to the size of the compactification, $f^2 \propto 1/R$. Thus, effects generating an axion potential are somehow {\em non-local} in the extra dimensions. This makes sense, as local effects would exist already in 5d, but we don't expect a 5d gauge field to become massive merely by coupling to charged particles. In the semiclassical worldline picture, this nonlocality is very explicit: the worldline of the charged particle must stretch all the way around the cycle we integrated over to get the axion. Propagation of massive particles over long distances always comes at a cost that is exponentially small in mass times distance, and this is precisely what we see. Such wrapped worldlines are sometimes referred to as ``Euclidean worldline instantons,'' and we can think of the exponent $2\cpi w m R$ as the action of such an instanton.

From~\eqref{eq:worldlineheuristic}, we see that the axion quality problem can be solved by an extra-dimensional axion provided that all the 5d charged particles are sufficiently heavy enough compared to the compactification radius. We are now comparing the exponentially small QCD scale to the exponentially small worldline instanton effect, rather than to a power-law suppression as in the KSVZ and DFSZ scenarios. We can summarize this:

\begin{framed}
\noindent
Extra-dimensional axion scenarios effectively take the log of the axion quality problem, making it much milder than in conventional 4d scenarios.
\end{framed}

\subsection{Brief comments on higher dimensions}
\label{subsec:evenhigherdim}

Although we have focused on a 5d example, the principles apply to a much larger class of theories. We consider a $p$-form gauge field $C_p$, which can be integrated over a $p$-dimensional closed manifold $\Sigma^{(p)}$ without boundary (a ``cycle'') in the extra dimensions:
\begin{equation}
\theta = \int_{\Sigma^{(p)}} C_p.
\end{equation}
More precisely, we take an ansatz $C_p(x, y) = \theta(x) \omega(y)$, where $\omega(y)$ is a harmonic $p$-form in the extra dimensions. That is, $\omega$ is a $p$-form living in the extra dimensional manifold $Y$ which is both closed ($\dif \omega = 0$) and co-closed ($\dif \star_Y \omega = 0$). This ansatz leads to a massless 4d field $\theta$. It is periodic because of generalized ``winding'' gauge transformations of $C_p$ around the cycle $\Sigma^{(p)}$. The axion coupling to gluons arises from a Chern-Simons coupling $\frac{N}{8\cpi^2} \int_{M \times \Sigma^{(p)}} C_p \wedge \mathrm{tr}(G \wedge G)$. Finally, we again obtain exponentially small contributions to the axion potential from objects charged under $C_p$. The difference is that this object now has a $p$-dimensional worldvolume. Such an object is conventionally called a $(p-1)$-brane, because it has $(p-1)$ spatial dimensions and one time dimension. However, the axion potential arises from a {\em Euclidean} brane where all $p$ worldvolume dimensions are spatial and are wrapped on the cycle $\Sigma^{(p)}$. In this case, the semiclassical sum over winding is much easier to understand than the sum over Kaluza-Klein modes. It tells us that the axion potential is proportional to a factor of $\exp(- {\cal T}\, \mathrm{Vol}(\Sigma^{(p)}))$, where ${\cal T}$ is the brane's tension. Again, the quality problem is potentially solved when the extra dimensions are large compared to the tension scale of the branes. As in~\eqref{eq:5dto4dG}, the smallness of the Standard Model gauge couplings in the UV can be explained if the cycle $\Sigma^{(p)}$ has large volume compared to fundamental scales, which makes it very plausible that the Euclidean brane instanton effects are small.

Importantly, we can have more than $p$ extra dimensions in total, as long as there is a $p$-dimensional cycle within the extra dimensions. For example, in Type IIB string theory, there are six extra dimensions but one might consider an axion field arising from $C_4$. The Standard Model gluons would then live not in the full ten dimensions, but on an 8-dimensional submanifold $M \times \Sigma^{(4)}$. More precisely, they would live on a localized object wrapping that submanifold: a stack of D7-branes. More generally, axion-like fields are ubiquitous in string theory constructions of 4d gauge theories that are at least vaguely Standard Model-like in the sense that they contain gauge fields and chiral matter. Such axions are widely studied; see, for instance,~\cite{Witten:1984dg, Barr:1985hk, Choi:1985je, Svrcek:2006yi, Conlon:2006tq, Arvanitaki:2009fg, Broeckel:2021dpz, Demirtas:2021gsq}.

\section{Perspective: axions as gauge fields}
\label{sec:axionasgauge}

Extra-dimensional axions are modes of gauge fields in the extra dimensions, but there is a more general sense in which {\em any} axion with couplings of the form~\eqref{eq:axionstrongCP} can be thought of as a type of gauge field. First, the axion field itself, like a gauge field $A$, is not single-valued; it has a gauge redundancy, $\theta \cong \theta + 2\cpi$, and ``Wilson point operators'' like $\exp(\iu\theta)$ are well-defined in much the same way that Wilson loop operators are for ordinary gauge fields. Furthermore, the axion has an equation of motion
\begin{equation} \label{eq:axionasgaugefield}
\dif(f^2 \star\dif\theta) = \frac{N}{8\cpi^2} \mathrm{tr}(G \wedge G).
\end{equation}
If we integrate both sides of this equation over all of (a closed) spacetime, we learn that
\begin{equation}
0 = N \cdot N_\mathrm{inst},
\end{equation}
where $N_\mathrm{inst}$ is the total instanton number. This is a type of ``Gauss's law constraint'': the axion equation of motion enforces that the net instanton number on spacetime is zero. The logic is completely parallel to the usual derivation of Gauss's law for electric charge on a closed {\em space}, derived from Maxwell's equation~\eqref{eq:maxwelleq}. In general, $\Uone$ gauge theories with $p$-form gauge fields lead to Gauss's law constraints on a charge that is evaluated over a $(d-p)$-dimensional slice of spacetime. From this perspective, it is reasonable to say that an axion $\theta$ is a 0-form gauge field, which serves to gauge the instanton number charge $\int \frac{1}{8\cpi^2} \mathrm{tr}(G \wedge G)$.

The axion quality problem is related to the possibility of additional terms on the right-hand side of~\eqref{eq:axionasgaugefield}. Such terms would tend to spoil the interpretation of the axion as a gauge field. On the other hand, in Maxwell's equations we have various currents on the right-hand side (the electron current, the muon current, and so on). These currents are not independently conserved, due to processes that convert one kind of charge into another (e.g., a muon decay transfers the electric charge to the electron). In models where the axion is a higher-dimensional gauge field, there may be additional terms on the right-hand side corresponding to other objects carrying gauge charge (e.g., Euclidean D-brane instantons), but the objects carrying such charges can be continuously deformed into gauge theory instantons~\cite{Witten:1995gx, Douglas:1995bn}, so there is ultimately only a single conserved charge. Thus, such models can maintain the interpretation of the axion as a gauge field for some generalization of instanton number, whereas a 4d model with generic Peccei-Quinn violating operators may not admit such an interpretation.

In higher dimensions, instanton number becomes a perfectly ordinary symmetry charge. For instance, for a 5d gauge theory, $\mathrm{tr}(G \wedge G)$ is still a 4-form, so one can integrate it over a spatial slice. Instantons are now charged particles: there is a BPST instanton solution in 5d which is just the 4d solution, taken to be independent of time, and interpreted as a kind of solitonic particle with a worldline extended through time, as we mentioned in \S\ref{subsec:commentsoninstantons}. Furthermore, in higher dimensions it becomes a well-defined question to ask if $\dif{\left[\mathrm{tr}(G \wedge G)\right]} = 0$, i.e., if instanton number is conserved. (In 4d, it is trivially true that $\mathrm{d}$ of {\em any} 4-form is zero, so this is not a very meaningful question.) In higher dimensions, the answer is yes:
\begin{equation} \label{eq:CWconserve}
\dif{\left[\mathrm{tr}(G \wedge G)\right]} = 2\,\mathrm{tr}(\dif{G \wedge G}) = 2\, \mathrm{tr}(\mathrm{D}G \wedge G) = 0,
\end{equation}
where $\mathrm{D}G$ is the {\em covariant} derivative and the final step follows from the nonabelian Bianchi identity $\mathrm{D}G = 0$. This computation plays a role in what is known as ``Chern-Weil theory,'' which is the mathematical framework that relates abstract topological invariants called characteristic classes with concrete formulas built out of differential forms, like $\int \frac{1}{8\cpi^2} \mathrm{tr}(G \wedge G)$. We learn from~\eqref{eq:CWconserve} that instanton number {\em is} a $\Uone$ global symmetry charge in 5d. My collaborators and I gave such symmetries, with conserved currents built out of gauge field strengths, the name {\em Chern-Weil global symmetries}~\cite{Heidenreich:2020pkc}.\footnote{Examples were previously discussed in various contexts, e.g.,~\cite{Lambert:2014jna,Tachikawa:2015mha,BenettiGenolini:2020doj,Apruzzi:2020zot,Bhardwaj:2020phs,Cordova:2020tij}.} This symmetry can be gauged by a Chern-Simons coupling of the form $C \wedge \mathrm{tr}(G \wedge G)$, where $C$ is an ordinary $\Uone$ gauge field. Such Chern-Simons couplings are ubiquitous in known quantum gravity theories, likely for fundamental reasons~\cite{Montero:2017yja,Heidenreich:2020pkc}. The 4d analogue of $C$ is an axion, and as we saw in~\S\ref{sec:xdimaxion}, one can obtain such 4d theories by dimensional reduction of the higher-dimensional theory. 

As we will discuss more extensively in subsequent parts of these lecture notes, quantum gravity theories do not admit global symmetries. There are reasons to believe that this applies to even generalized symmetries like instanton number in four dimensions. However, quantum gravity is perfectly consistent with gauge symmetries. This gives the axion a reason for being: it gauges instanton number, and thus eliminates a would-be generalized global symmetry that is incompatible with quantum gravity. I believe that this is the underlying explanation for why axion fields are so ubiquitous in string theory constructions of gauge theories like the Standard Model. (Though this is not a complete argument: one should ask when it is possible for the symmetry to simply be broken, rather than gauged.)

Summarizing, there are multiple reasons why I find extra-dimensional axions more compelling than 4d axion models:

\begin{framed}
\noindent
Extra-dimensional axion scenarios have a strong phenomenological motivation (solving the axion quality problem) and a strong motivation from deeper principles (gauging instanton number). They also appear ubiquitously in string theory compactifications.
\end{framed}

\section{Axions in cosmology}
\label{sec:axioncosmology}

Apart from the Strong CP problem and other theoretical motivations, one reason that axions have attracted a great deal of attention is that they are a very natural dark matter candidate. A light scalar field, in the early universe, can be frozen at a value away from the minimum of its potential (``misaligned''). When the Hubble expansion rate drops below its mass, it can begin to coherently oscillate. A coherently oscillating scalar field is essentially the same thing as a collection of massive particles at rest~\cite{Turner:1983he}. This is the misalignment mechanism for dark matter production, and it was appreciated at an early stage that it could give rise to QCD axion dark matter~\cite{Preskill:1982cy, Dine:1982ah, Abbott:1982af}.

Here, I want to very briefly mention a couple of important issues in axion cosmology. For a more complete discussion and references to the literature, see the review article~\cite{Marsh:2015xka}.

The cosmology of the axion is conventionally divided into the ``pre-inflationary scenario'' and the ``post-inflationary scenario,'' depending on whether the Peccei-Quinn phase transition occurs before (or during) inflation, or after inflation. One point that I would emphasize is that in my preferred axion models, those where the axion arises from a higher-dimensional gauge field as in \S\ref{sec:xdimaxion}, there is no Peccei-Quinn phase transition. The axion is not a 4d pseudo-Nambu-Goldstone boson, it is intrinsically a compact scalar, and there is no point in field space at which the symmetry is restored. As a result, these models are expected to have a cosmology of the ``pre-inflationary axion'' type, although the terminology is a bit misleading.

The pre-inflationary and post-inflationary axion scenarios each suffer from potential cosmological problems, but they are different problems. In the post-inflationary axion scenario, during the Peccei-Quinn phase transition the complex scalar $\phi$ rolls off its potential in different directions in different parts of the universe. Because the initial value of $\theta$ is randomized over the universe, one obtains a sharp prediction for the amount of axion dark matter for a given decay constant (and a given assumption about the thermal expansion history of the universe). The scrambling of initial values of $\theta$ at the PQ phase transition leads to the formation of topological defects called axion strings, that is, dynamical objects around which the field $\theta$ winds from $0$ to $2\cpi$. At later times, during the QCD phase transition, the axion acquires a periodic potential and domain walls can form separating different minima. In this context, the integer $N$ in~\eqref{eq:axionstrongCP} is very important. Across a domain wall, the value of $\theta$ changes by the range from one minimum of the potential to the next, which in this case is $2\cpi/|N|$. If $|N| = 1$, a domain wall can end on an axion string. The axion strings that formed at the PQ phase transition fill the universe, domain walls tend to form attached to strings, and the network of strings and walls destroys itself. However, if $|N| > 1$, then one has to put $|N|$ domain walls together to achieve a total winding of $2\cpi$. A network with multiple domain walls ending on a single string cannot tear itself apart. The domain walls overwhelm the energy density of the universe, and we have a cosmological disaster. As a result, in the post-inflationary axion scenario, a major challenge is to achieve a model where $|N| = 1$. 

In the pre-inflationary scenario, on the other hand, the axion is a dynamical mode during inflation and the radial mode is effectively heavy during inflation. As a result, at the end of inflation, it is a reasonable approximation to assume that $\theta$ has the same value everywhere in the universe. No domain walls form, because the field is homogeneous. In this case, there are two potential cosmological problems. The first is that, if the initial value of $\theta$ is too far from its minimum, then the theory could predict too much dark matter. However, unlike in the post-inflationary case, we can always decrease the predicted amount of dark matter simply by assuming that $\theta$ begins close to its minimum (i.e., at the cost of some tuning). A more difficult problem to avoid is the axion isocurvature problem: the axion field fluctuates, during inflation, by an amount $\delta \theta = \frac{1}{f} \delta {\hat \theta} \sim \frac{H}{2\cpi f}$. These fluctuations imply that the energy density stored in the axion field has small inhomogeneities across the universe, in a way that is decoupled from the adiabatic density perturbations sourced by the inflaton. Observations place powerful constraints on such isocurvature perturbations in the dark matter density. In the simplest cosmology, this implies that the Hubble scale during inflation must be very low relative to $f$. Non-standard cosmologies can ameliorate this problem. 

There is a large literature on axion cosmology, and saying more would take us outside the scope of these lectures. What I would like to emphasize, which is {\em not} commonly said in the literature, is that the extra-dimensional axion models that are most compelling from the theoretical viewpoint are necessarily of the pre-inflationary type. In this case, we have no axion domain wall problem and the value of $N$ is not a concern. However, the axion isocurvature problem can be rather severe in such models, and may point to a need for a non-minimal cosmology.

\newpage

\section*{\Large Part Four: No Global Symmetries in Quantum Gravity}
\label{sec:lecturefour}
\addcontentsline{toc}{section}{\nameref{sec:lecturefour}}

\section{Proton stability: symmetry, or not?}
\label{sec:protondecay}

Global symmetries are very useful tools for characterizing quantum field theories. However, the modern view is that we do not expect global symmetries to ever truly exist in theories that describe nature. We often find very good {\em approximate} global symmetries, but they are not exact. Below, I will review some arguments for why we expect that global symmetries do not exist in quantum theories that include gravity. First, though, it is worth recalling that particle physics as we know it does not require us to invoke any fundamental global symmetries. I will illustrate this point by discussing how a theorist seventy years ago might have thought about the stability of the proton, and contrast this with the modern perspective that emerged after the Standard Model was established.

Imagine that you were a physicist in the 1950s. A number of particles were known. The positron had been discovered by Carl Anderson in 1932. In 1930, Pauli postulated the existence of the neutrino in a desperate attempt to patch up energy conservation in beta decay of nuclei.\footnote{In fact, he called it the ``neutron''; it was only after the particle we now know as the neutron was later discovered that Fermi used the name ``neutrino'' for Pauli's lighter neutral particle, as a joke. Of course, the name stuck, and now the Italian diminutive ``-ino'' is ubiquitous in particle theory.} The neutron was discovered by Chadwick in 1932 and Fermi explained how its decay, $n \to p e^- {\bar \nu}$, could be accounted for with a four-fermion interaction Lagrangian, of the schematic form\footnote{``Schematic'' because I'm omitting the projection onto left-handed fermions, which historically came later.}
\begin{equation}
{\cal L}_\beta \sim G_F {\bar n}pe^- {\bar \nu} + \mathrm{h.c.},  
\label{eq:betadec}
\end{equation}
where the Fermi constant $G_F$ has mass dimension $-2$. The neutral pion was discovered in 1950 (a few years after the charged pion), having been predicted by Yukawa as a mediator of the strong force in 1935.

All of this sets the stage for a basic question: why is the proton stable, when the neutron is unstable? The neutron decays in a way that conserves electric charge and angular momentum. So could the proton, via the decay $p \to e^+ \pi^0$. We could even write down an interaction term in our Lagrangian that would allow it,
\begin{equation}
{\cal L}_\mathrm{dec} = y_p p e^- \pi^0 + \mathrm{h.c.}
\label{eq:pdec1}
\end{equation}
Observations tell us that the proton is extremely stable. There are enormous numbers of protons in our vicinity and we don't observe them decaying, so the proton lifetime $\tau_p$ (if indeed it decays at all) must be much longer than the age of the universe. In fact, experiments now tell us that $\tau_p \gtrsim 10^{34}\,\mathrm{yrs}$~\cite{Super-Kamiokande:2020wjk}. A back of the envelope estimate tells us that this requires $y_p \lesssim 10^{-32}$. This is an extraordinarily tiny number! A physicist in the 1950s would not have had such a stringent bound on the proton lifetime, but they would still have concluded that $y_p$ must be many orders of magnitude less than one. Why is it so small?

\begin{figure}[!h]
\centering
\includegraphics [width = 0.3\textwidth]{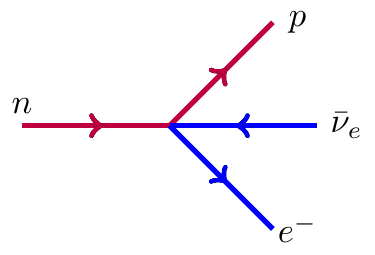}\quad\quad\includegraphics [width = 0.3\textwidth]{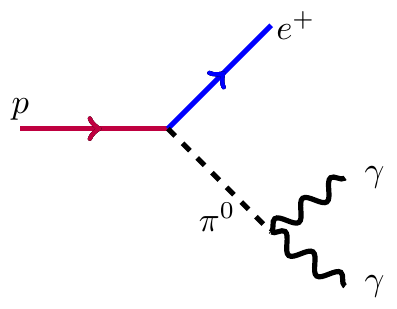}
\caption{How a 1950s physicist might have thought about (left) neutron decay, preserving baryon number symmetry (purple lines) and lepton number symmetry (blue lines); (right) proton decay, violating both baryon and lepton number symmetries, and hence forbidden.
} \label{fig:baryonconservationsymmetry}
\end{figure}

A clever theorist then might have observed\footnote{And presumably did; a full historical review is outside my scope.} that a possible answer would lie in {\em global symmetries}. Fermi's beta decay Lagrangian~\eqref{eq:betadec} is compatible with two global symmetries, namely ``baryon number'' and ``lepton number'':
\begin{equation}
\begin{aligned}
\Uone_\textsc{B}: & \quad  p \mapsto \E^{\iu \alpha} p, && n \mapsto \E^{\iu \alpha} n; \nonumber \\
\Uone_\textsc{L}: & \quad e^- \mapsto \E^{\iu \beta} e^-, && \nu \mapsto \E^{\iu \beta} \nu.
\end{aligned}
\end{equation}
On the other hand, the proton decay Lagrangian~\eqref{eq:pdec1} explicitly violates these symmetries (though it does preserve the subgroup with $\alpha = -\beta$, known as $\Uone_{\textsc{B}-\textsc{L}}$). Thus, one could explain the lack of observation of proton decay by postulating that baryon number is an exact global symmetry of the universe. In that case, $y_p = 0$, and there is no longer a puzzle. This understanding is depicted in Fig.~\ref{fig:baryonconservationsymmetry}.

\begin{figure}[!h]
\centering
\includegraphics [width = 0.3\textwidth]{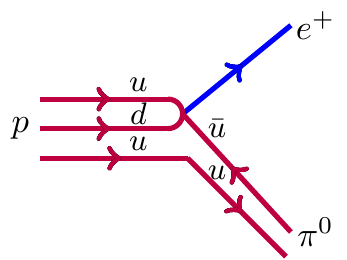}
\caption{Proton decay after the discovery of quarks: the process must involve a four-fermion interaction, potentially suppressed by a very large mass scale. There is no need to postulate an exact symmetry to explain why we do not observe the process.
} \label{fig:baryonconservationcompositeness}
\end{figure}

This was a completely reasonable expectation several decades ago, but it turned out to be the wrong way to think about proton decay. Instead, by the mid-1970s it was established that protons were in fact {\em composite objects}, built out of three quarks. In the fundamental theory, the operator that destroys a proton really looks like $uud$ (with appropriately antisymmetrized $\mathrm{SU(3)}_\textsc{C}$ indices). This turns out not to fundamentally change the structure of the beta decay Lagrangian, because beta decay involves a single down quark turning into an up quark, i.e., it has the schematic form $G_F {\bar d}ue^-{\bar \nu}$ (generated by integrating out a $W$ boson, and with only the left-handed chiral fermions in the interaction). Because the beta decay interaction is a dimension six operator in effective theories both above and below the QCD scale, there is no substantial change in how we understand the parametric prediction for the neutron lifetime. However, the compositeness of the proton {\em does} fundamentally change the way we think about proton decay. To obtain a description of proton decay in terms of quarks, the interaction~\eqref{eq:pdec1} should be replaced by something of the form: 
\begin{equation}
{\cal L}_\mathrm{dec} = \frac{1}{M_\mathrm{dec}^2} u u d e^- + \mathrm{h.c.}
\label{eq:pdec2}
\end{equation}
Now two quarks in the proton can scatter into a positron and an antiquark, with the outgoing antiquark and the remaining initial quark rearranging themselves into a pion. This process is depicted in Fig.~\ref{fig:baryonconservationcompositeness}. We can still use our old interaction Lagrangian~\eqref{eq:pdec1} as an effective theory for the proton decay process; however, by dimensional analysis, the matching between its parameter $y_p$ and the parameter $M_\mathrm{dec}$ in~\eqref{eq:pdec2} is schematically
\begin{equation}
y_p \sim \frac{\Lambda_\textsc{QCD}^2}{M_\mathrm{dec}^2} \sim 10^{-32} \left(\frac{10^{15}\,\mathrm{GeV}}{M_\mathrm{dec}}\right)^2.
\end{equation}
This suggests a completely different possible explanation for why the proton is so stable: there is no symmetry at all, but the symmetry-violating interactions are irrelevant at low energies. Symmetry violation could be mediated by very heavy particles, with mass larger than $10^{15}\,\mathrm{GeV}$~\cite{Georgi:1974sy}. This is relatively close to the Planck scale, so perhaps this is a fundamental process in a hypothetical theory incorporating gravity. At such high energies, the amplitude for baryon number violating processes could become large: the symmetry is {\em badly} broken. The insight that was missing for our hypothetical 1950s theorist was that the proton and pion were not elementary fields; compositeness can make the tiny coupling $y_p \sim 10^{-32}$ much less concerning. Or rather, to state the case more carefully, it translates the question of why $y_p$ is small into the question of why $\Lambda_\textsc{QCD}$ is small---a question that was answered by the discovery of asymptotic freedom and dimensional transmutation, which produces an exponentially small $\Lambda_\textsc{QCD}$ from a mildly small coupling $g_s$ at high energies.

What the discovery of the full structure of the Standard Model made clear was that baryon number could be an {\em accidental} symmetry. We say that a theory has an accidental symmetry when the lowest-dimension gauge-invariant operator that violates a symmetry is irrelevant. This is the case for baryon number in the Standard Model, which is first violated by dimension six operators like $uude$ and $qqq\ell$~\cite{Weinberg:1979sa}. It is also the case for lepton number in the Standard Model, though this is violated already at dimension five, by operators of the form $(h\ell)^2$, which correspond to Majorana neutrino masses at low energies. We do not know that neutrino masses take this form, but it is very plausible that they do; again, they are small parameters in the low-energy theory, but the perspective of accidental symmetries can render their smallness much less mysterious. The modern perspective on global symmetries is that we expect them to arise in the way that baryon and lepton number symmetries arise in the Standard Model: not as exact properties of the UV theory, but as emergent properties of the IR theory, enforced by gauge invariance. Indeed, we generally expect that they are not even approximate symmetries in the UV. Good approximate symmetries can emerge naturally from a complete lack of symmetry, simply by dimensional analysis, when the symmetry is accidental.

There is more to the story of baryon number violation, but the discussion above captures the key insight we will build on below. The additional important physics related to baryon number is that ABJ anomalies lead to nonperturbative violation of both baryon and lepton number in the Standard Model (but preserve $\Uone_{\textsc{B}-\textsc{L}}$), even {\em without} adding any irrelevant operators. Interestingly, the anomaly leads to processes that violate baryon number only by multiples of three units. Because of this, it is still a consistent possibility that our universe has a {\em discrete} gauge symmetry that stabilizes the proton, but  would allow three protons to collectively annihilate into leptons. Discrete gauge symmetry is an interesting topic that I do not have time to do justice to in these lectures.

In the real world, we sometimes encounter approximate symmetries that are {\em not} accidental. A good example is isospin, the symmetry of the strong interactions that rotates up and down quarks into each other. This is explicitly (but weakly) broken by electromagnetism, because the up and down quarks have different electric charge. It is also explicitly broken by the unequal up and down quark masses. This is said to be a {\em soft} breaking: a small breaking by relevant operators, which does not introduce divergent corrections to quantities in the IR. The fact that such a non-accidental, approximate global symmetry arises in the real world tells us that such symmetries might also arise in physics beyond the Standard Model. They can't be forbidden by any general principle. Still, we would like to understand them better. In the full Standard Model, the fact that the up and down quark masses are not arbitrarily big is partially explained by the fact that they arise from the Higgs vev that breaks electroweak symmetry (though why {\em that} is small is a great mystery). They are further suppressed by small Yukawa couplings. We expect the smallness of these couplings to be explained by some deeper principle: perhaps by the higgsing of some new gauge symmetry (like a discrete flavor symmetry) in the UV, or perhaps by localization of the left- and right-handed fermions at different places in extra dimensions. This example highlights that approximate but not accidental symmetries are allowed, but are also unsatisfying, and may be important clues to deeper structure in a theory.

Many model builders use {\em technical naturalness} as a guiding principle: parameters in a Lagrangian are allowed to be very small if a symmetry is restored when they are set to zero. Technical naturalness is important, as it tells us that quantum corrections are under control. However, it is far from sufficient. If the explanation of a small parameter invokes a very good approximate global symmetry, then the approximate global symmetry itself cries out for a deeper explanation. Technical naturalness can be a useful guidepost in searching for a theory, but it is far from the end of the story. Ultimately, we would like to have a satisfying explanation of small parameters in terms of $O(1)$ inputs or deeper principles.

\section{Black holes: mini-review}

The most convincing arguments about universal properties of quantum gravity rely on semiclassical properties of black holes, since these results can be derived simply by combining general relativity and quantum field theory, rather than relying on a specific theory of quantum gravity like string theory. A complete review of black holes in effective field theory would go far outside the scope of these lecture notes, so instead we will take a brief tour of key aspects of black hole physics. For a more thorough introduction to black hole physics, I recommend~\cite{Harlow:2014yka}.

\subsection{Classical black hole solutions}

The most prominent feature of a black hole is its {\em horizon}, the boundary through which signals from inside the black hole cannot escape. The horizon is more important than the {\em singularity}, which lies deep in the interior of the black hole. One might wonder if the singularity, where the curvature becomes infinite, is somehow resolved by unknown short-distance physics. On the other hand, for a large enough black hole, the horizon is a region of small curvature, within the domain where we expect effective field theory to be reliable. Thus, the interesting properties of black holes are to a large extent associated with the horizon. 

\begin{figure}[!h]
\centering
\includegraphics [width = 0.4\textwidth]{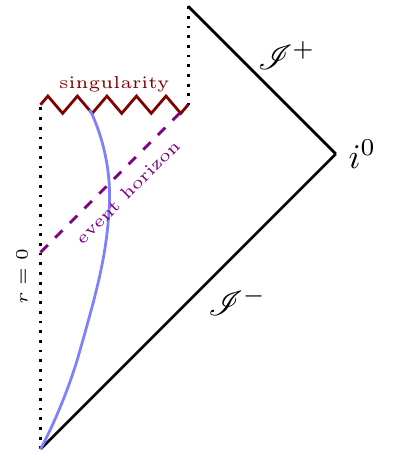}
\caption{Penrose diagram for a black hole formed by collapse of infalling matter (blue curve) and then evaporating. Time runs vertically and (radial) space horizontally, with an implicit angular two-sphere over every point in the diagram. The dashed purple line is the event horizon, where $f(r) = 0$, which is a lightlike boundary. The red jagged line is the spacelike singularity at $r = 0$. Dotted lines are coordinate singularities at $r = 0$, not physical singularities.
} \label{fig:penroseBH}
\end{figure}

The simplest black hole solution in general relativity is the neutral, spherically symmetric, static {\em Schwarzschild} black hole. The Schwarzschild metric takes the form
\begin{equation} \label{eq:sphericalBH}
\dif s^2 = - f(r) \dif t^2 + \frac{1}{f(r)} \dif r^2 + r^2 \dif \Omega^2,
\end{equation}
where for a four-dimensional black hole $\dif \Omega^2 \equiv \dif \theta^2 + \sin^2 \theta \dif \phi^2$ is the unit metric on the 2-sphere and 
\begin{equation} \label{eq:fSchwarzschild4d}
f(r) \equiv 1 - \frac{2 G M}{r}.
\end{equation}
Here $G$ is Newton's constant and $M$ is the black hole mass. In units where $\hbar = c = 1$, Newton's constant is the same as $(8 \cpi M_\mathrm{Pl}^2)^{-1}$ where $M_\mathrm{Pl} \approx 2.4 \times 10^{18}\,\mathrm{GeV}$ is the Planck mass. The coordinates $(\theta, \phi)$ parametrize the angular directions, forming an $S^2$ that shrinks to zero size at the singularity $r = 0$. This is not just a coordinate singularity. For example, the Kretschmann scalar, a curvature invariant defined by $R_{\mu \nu \rho \sigma}R^{\mu \nu \rho \sigma}$, blows up at $r \to 0$. The black hole {\em horizon} is located at $r = 2 G M$, where $f(r) = 0$. At this point, the signs in front of the $\dif t^2$ and $\dif r^2$ terms in the metric flip, so that time and space exchange roles at the horizon. This corresponds to the fact that light rays inside the horizon can never propagate to the exterior. The Schwarzschild metric is an approximate description of a realistic black hole which forms by collapse and eventually evaporates; Fig.~\ref{fig:penroseBH} shows a Penrose diagram depiction of the spacetime of such a black hole.

A black hole can carry electric or magnetic charge under a $\Uone$ gauge field. For the case of electric charge, this corresponds to the Reissner-Nordstr{\"o}m solution, for which the metric has the same form~\eqref{eq:sphericalBH} where
\begin{equation} \label{eq:fRN4d}
f(r) \equiv 1 - \frac{2 G M}{r} + \frac{G {\widetilde Q}^2}{r^2},
\end{equation}
and there is also a $\Uone$ gauge field
\begin{equation} \label{eq:A0RN4d}
A_0 = \frac{\widetilde Q}{r}.
\end{equation}
(Sometimes instead a gauge is chosen where a constant is subtracted in order to take $A_0 = 0$ at a special radius $r = r_+$, to be defined shortly.) The charge $\widetilde Q$ is normalized according to a commonly-used convention in GR, related to a $\Uone$ integer charge $Q$ with gauge coupling $e$ as:
\begin{equation}
{\widetilde Q}^2 = \frac{e^2}{4\cpi} Q^2.
\end{equation}
The Reissner-Nordstr{\"o}m black hole has {\em two} horizons where $f(r) = 0$, the outer horizon at $r_+$ and the inner horizon at $r_-$, where
\begin{equation}
r_\pm = G M \pm \sqrt{G^2 M^2 - G \widetilde{Q}^2}.
\end{equation}
This expression does not make sense for arbitrary $(M, \widetilde{Q})$. In particular, for a given charge there is a {\em minimum mass}:
\begin{equation} \label{eq:extremality}
M^2 \geq M_\mathrm{ext}(Q)^2 \equiv \frac{1}{G} \widetilde{Q}^2 = 2 e^2 Q^2 M_\mathrm{Pl}^2.
\end{equation}
Intuitively, this is because a charged black hole has a nonzero electric field outside the horizon, since $\partial_r A_0 \neq 0$. The electric field stores energy, which adds to the rest mass of the black hole, preventing it from being arbitrarily small. The bound~\eqref{eq:extremality} is known as the ``extremality bound.'' A charged black hole with a mass that saturates this bound is referred to as an extremal black hole. A black hole that satisfies the bound without saturating it is referred to as subextremal. An object that violates this bound is said to be superextremal.  Black hole solutions, at least to the classical theory with a standard two-derivative action, are never superextremal.

In the presence of higher derivative terms in the effective action, black hole solutions are modified, and the extremality bound generally changes. (The exception is in when extremal black holes are BPS, in theories with sufficient supersymmetry.) In general, the extremal mass-to-charge ratio obtains a series of corrections that can be written in inverse powers of the charge. If we define the extremality bound as the limiting case of asymptotically large black holes, these corrections can allow a finite-mass black hole to slightly violate the bound~\cite{Kats:2006xp}.

\medskip
\noindent\centerline{\rule{\textwidth}{0.5pt}}
\noindent {\em Exercise:} Find the $D$-dimensional generalization of the Schwarzschild and Reissner-Nordstr{\"o}m solutions, and of the extremality bound.\\
\noindent\centerline{\rule[6.0pt]{\textwidth}{0.5pt}}

\subsection{Black hole thermodynamics}

In the early 1970s, Bekenstein and Hawking began to study black holes semiclassically, that is, incorporating quantum effects. Important early insights into the quantum properties of black holes came from studying quantum field theory on a fixed black hole background spacetime (effectively, taking the strict $G \to 0$ limit) and by considering the Euclidean path integral for gravity at finite temperature. There is a rich and fascinating story to tell here, but in these notes I will only summarize a few key conclusions.

Famously, black holes were found to be thermodynamic objects, with an associated entropy and temperature~\cite{Bekenstein:1973ur, Bardeen:1973gs, Bekenstein:1974ax, Hawking:1974rv}. For a black hole of mass $M$ and (integer) charge $Q$, we define the {\em extremality ratio}
\begin{equation}
\xi \equiv \frac{2 e^2 Q^2 M_\mathrm{Pl}^2}{M^2}. 
\end{equation}
In this convention, $\xi = 1$ for an extremal black hole, and $0 \leq \xi \leq 1$ for any black hole. Bekenstein found that the {\em entropy} of a black hole is proportional to the area of its (outer) horizon, namely
\begin{align}
S = \frac{A_+}{4 G} &= 8 \cpi^2 r_+^2 M_\mathrm{Pl}^2 \nonumber \\
&= \frac{M^2}{4 M_\mathrm{Pl}^2} \left(1 + \sqrt{1 - \xi} - \frac{1}{2} \xi\right).
\end{align}
This entropy allows black holes to obey a Generalized Second Law, which is to say that the sum of this entropy and conventional thermodynamic entropy always increases. Hawking showed that black holes emit blackbody radiation with temperature~\cite{Hawking:1974sw}
\begin{align}
T &= \frac{1}{4\cpi} \left( \frac{1}{r_+} - \frac{G  \widetilde{Q}^2}{r_+^3}\right) \nonumber \\
&= \frac{4 M_\mathrm{Pl}^2}{M} \frac{\sqrt{1 - \xi}}{\left(1 + \sqrt{1 - \xi}\right)^2}.
\end{align}
Hawking radiation is thermal, but for a charged black hole the electric potential at the horizon serves as {\em chemical potential} favoring discharge,
\begin{equation}
\mu = \frac{\widetilde{Q}}{\sqrt{4\cpi} r_+} = \frac{\sqrt{2\xi}}{1 + \sqrt{1- \xi}} M_\mathrm{Pl} \sgn(Q).
\end{equation}
This is a chemical potential in the usual thermodynamic sense that it distorts the Boltzmann factor for emitted particles, so that the probability of emitting  a particle of energy $E$ and charge $q$ is proportional to $\exp[-(E - \mu e q)/T]$. In particular, the black hole is {\em more likely} to emit radiation of the same-sign charge as the  black hole, so that the magnitude of its charge decreases over time. In terms of an intuitive heuristic that Hawking radiation can be thought of as the production of particle/antiparticle pairs, one of which falls into the black hole horizon and one of which escapes, what is happening is that the electric field pushes the same-sign particle away from the black hole so that it escapes, while the opposite-sign particle is attracted and falls in, decreasing the charge of the black hole.

Let us emphasize a few qualitative properties of these equations. The black hole entropy is proportional to the horizon area, and remains nonzero for an extremal black hole. The black hole {\em temperature}, far from extremality, is of order the inverse radius of the horizon. In particular, large black holes are very cold and radiate slowly. However, at extremality, $T \to 0$. Thus, exactly extremal black holes do not emit Hawking radiation. However, if charged particles are present in the theory, the black hole can discharge by Schwinger pair production of particles in the electric field outside the horizon~\cite{Gibbons:1975kk}. For a clear recent discussion that treats Hawking radiation and Schwinger pair production together, see~\cite{Johnson:2019kda}.

\medskip
\noindent\centerline{\rule{\textwidth}{0.5pt}}
\noindent {\em Exercise:} Find the $D$-dimensional generalization of the formulas for black hole entropy and temperature.\\
\noindent\centerline{\rule[6.0pt]{\textwidth}{0.5pt}}

\medskip
\noindent\centerline{\rule{\textwidth}{0.5pt}}
\noindent {\em Exercise:} Calculate the lifetime of a four-dimensional Schwarzschild black hole of mass $M$ to evaporate via Hawking radiation. (Hint: neglecting ``greybody factors'' that describe how particles escape from the near-horizon region to the asymptotic region, you can assume that radiation is emitted from the horizon according to the Stefan-Boltzmann law.)\\
\noindent\centerline{\rule[6.0pt]{\textwidth}{0.5pt}}

\section{Black holes versus continuous global symmetries}
\label{sec:BHnoglobal}

From the time that Hawking radiation was discovered, it was used to argue that black hole physics could forbid global symmetry charges~\cite{Zeldovich:1976vq}. The general idea is that global symmetry charge, unlike gauge symmetry charge, is invisible to semiclassical black hole physics. If you throw an electron into a black hole, you can tell that it has acquired an electric charge because you can measure the electric field outside the black hole. As discussed above, the electric charge of a black hole also distorts the spectrum of its Hawking radiation, serving as a chemical potential. Global symmetry charge is entirely different. For example, $\textsc{B} - \textsc{L}$ is a non-anomalous global symmetry of the Standard Model with right-handed neutrinos, if we do not turn on Majorana neutrino masses or other explicit symmetry violation. Thus, it is a potential global symmetry of nature. Neutrons carry $\textsc{B} - \textsc{L}$ charge but not electric charge, so you could imagine throwing a large number of neutrons into a black hole to give it $\textsc{B} - \textsc{L}$ charge. However,  no measurement that you can do outside the black hole would tell you that it carries this charge. For a  global symmetry, there is no electric field to measure,  and  there is correspondingly no chemical  potential, so  the Hawking radiation will not care about the black hole's global charge. It will continue  to emit just as many neutrons as antineutrons, or  neutrinos as antineutrinos. This led many  people to strongly suspect that black holes violate global symmetry charges: they will simply eat such charges and then forget about them. This argument is not entirely convincing, however, because one could always imagine that the black hole does remember its global charge, but in a way that is invisible to semiclassical physics like Hawking radiation. Perhaps black holes with different $\textsc{B}-\textsc{L}$ charges are simply different states in the theory, which happen to appear the same to external observers.

\begin{figure}[!h]
\centering
\includegraphics [width = 0.65\textwidth]{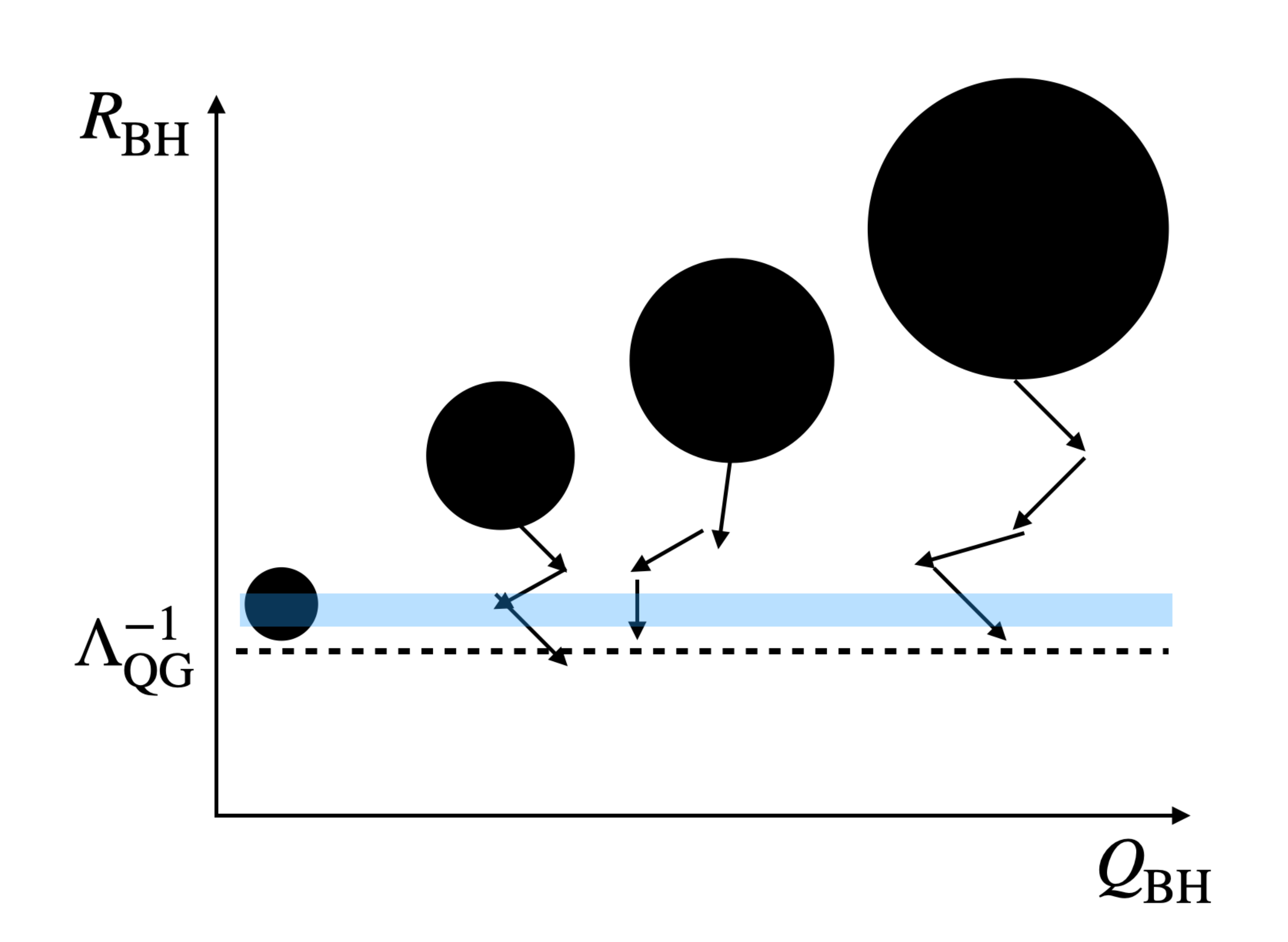}
\caption{The  Banks-Seiberg argument  against a global $\Uone$ symmetry. By creating large  black holes of  arbitrarily large global charge $Q_\mathrm{BH}$, then letting  them evaporate down to the radius $\Lambda_\textsc{QG}^{-1}$ where we stop trusting Hawking's semiclassical calculation, we can  construct infinitely many different states that all fit in a region of size slightly larger than $\Lambda_\textsc{QG}^{-1}$ (blue band),  in contradiction to any entropy bound.
} \label{fig:banksseiberg}
\end{figure}

A modern reformulation of the argument by Banks and Seiberg presents a sharp contradiction between this possibility and entropy bounds~\cite{Banks:2010zn}. It is believed that for any  bounded region of spacetime, there is a maximum possible entropy in quantum gravity, i.e., a finite number of  possible different microstates that can  exist in this region, corresponding to the entropy of a black hole that fills the region. This idea goes back to Bekenstein~\cite{Bekenstein:1980jp} and was given a precise covariant formulation by Bousso~\cite{Bousso:1999xy}. Banks and Seiberg pointed out that the existence of any continuous global symmetry---let's discuss $\Uone$, for simplicity, but the key point is that a continuous symmetry has infinitely many representations---would actually lead to an {\em infinite} entropy, and hence be in conflict with any such entropy bound. The argument is simple. We can construct black holes with arbitrarily large global symmetry charge $Q_\mathrm{BH}$, just by throwing many particles with  global charge (e.g., neutrons, in our  hypothetical  example  above) into a black hole. After constructing a black hole  of large global charge, simply wait. Hawking's calculation tells us that, as long as we trust semiclassical effective field theory, the black hole will shrink down to smaller size. Its global symmetry charge could go up or down, in a random walk, because  the black hole is just as likely to emit particles with either sign of this charge. Importantly, it is not preferentially driven to zero, because there is no chemical potential. Thus, if we start with a large black hole of large charge and wait, we will obtain a small black hole of large charge, with radius somewhat above $\Lambda_\textsc{QG}^{-1}$, the scale at which we stop trusting Hawking's calculation. We can carry out this process for states of charge as large as we like, and in this way we can obtain {\em infinitely} many different black hole states fitting in a region of fixed size---in blatant violation of the entropy bound. This is illustrated in Fig.~\ref{fig:banksseiberg}. This argument relies only on the assumption that we can trust Hawking's calculation of the evolution of the global symmetry charge as the black hole evaporates. One may doubt this; the black hole information problem famously tells us that effective field theory can get the answers to certain questions about black holes badly wrong. However, unlike in the black hole information context, we are not asking a subtle question here about entanglement  among many Hawking quanta. We are simply asking about the global charge of the black hole---a one-point function of a simple operator. If effective field theory fails to answer such a simple question, there is no reason to trust that it gets anything right, about the temperature or lifetime or any other basic property of the black hole. Rather than discard the elaborate, self-reinforcing structure of black hole thermodynamics as developed over the last fifty years, it seems more plausible that continuous global symmetries simply can't exist in theories with black holes. Let's highlight this key point:

\begin{framed}
\noindent
Black hole thermodynamics is incompatible with continuous global symmetries.
\end{framed}

\begin{figure}[!h]
\centering
\includegraphics [width = 0.55\textwidth]{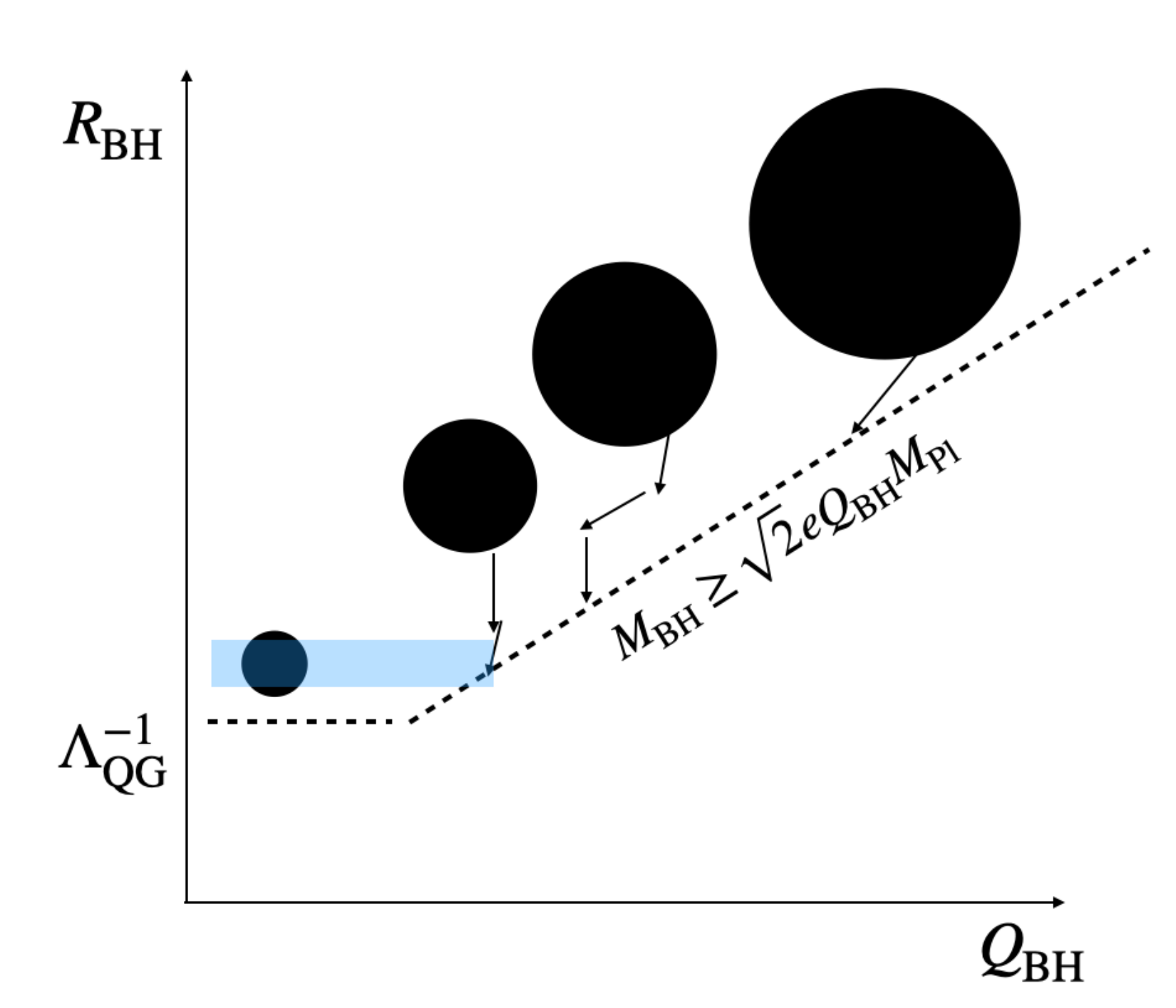}
\caption{The  Banks-Seiberg argument in  the case of gauge charge. The extremality bound now prohibits the existence of small black holes of large charge. As a result, in the $\Uone$ case, there is only a finite entropy in a region of a given size. However, for a noncompact gauge group like $\mathbb{R}$, charge becomes a continuous quantity  and there is still infinite entropy.
} \label{fig:banksseiberggauge}
\end{figure}

The Banks-Seiberg argument does not rule out theories with $\Uone$ {\em gauge} symmetry. In this case, because the gauge charge is associated with an electric field outside the black hole, an evaporating black hole preferentially sheds its charge as it evaporates. Eventually it either fully discharges or it  hits the extremality bound~\eqref{eq:extremality}, where it could get stuck if there are no sufficiently light particles for it to decay to by Schwinger pair production. In any case, the extremality bound imposes an upper limit on the charge of a black hole that can fit in a region of a given size, as indicated by the truncated blue band in Fig.~\ref{fig:banksseiberggauge}. The quantization of $\Uone$ charge implies that there are only finitely many states in this band, so there is  no  immediate contradiction with the existence of an entropy bound.\footnote{One might try to argue for a bound by comparing the finite number of states in this region with a {\em quantitative} entropy bound. The number of states in the region is  proportional to $Q_\mathrm{max} \sim 1/e$, shifting the entropy by $\log|e|$, thus requiring precise quantitative control over corrections to the leading black hole entropy calculation and the derivation of the bound.  Despite some efforts by various groups, to the best of my knowledge no convincing derivation of the Weak Gravity Conjecture (or similar result) has been given along these lines.} However, if the gauge group is $\mathbb{R}$, rather than $\Uone$, then there {\em are} infinitely many different possible charges---a continuum of them, in  fact, which we could densely fill by starting with any two particles with mutually irrational charges, e.g., $1$ and $\sqrt{2}$. In this way, the argument rules out not only continuous global symmetries in quantum gravity, but also noncompact, continuous {\em gauge} symmetries.

\begin{framed}
\noindent
Black hole thermodynamics is incompatible with noncompact, continuous gauge symmetries. In particular, abelian gauge groups in quantum gravity are always $\Uone$, never $\mathbb{R}$.
\end{framed}

\section{Generalized global symmetries}
\label{sec:generalizedsymmetry}

\subsection{$p$-form global symmetries}
\label{sec:pformglobalsym}

We discussed ordinary global symmetries in \S\ref{sec:globalsymmetries}. They acted on local operators living at a point (0-dimensional) or charged particles with 1-dimensional worldlines. They were associated with a charge $Q$ measured on a spatial slice. More generally, the charge can be measured on any codimension-1 slice $\Sigma$ through spacetime, e.g., a 3-manifold in 4d spacetime. (The ``codimension'' of a submanifold is just the number of dimensions {\em transverse} to the submanifold, i.e., the codimension of a $p$-manifold in $d$-dimensional spacetime is $d-p$.) An ordinary global symmetry comes with a family of symmetry operators $U(\Sigma, g)$ associated to a given symmetry group element $g$ and a closed 3-manifold $\Sigma$. For a $\Uone$ symmetry, there is a 3-form current $J$ which is conserved, $\dif J = 0$, and the charge is simply $\int_\Sigma J \in \mathbb{Z}$. Such a symmetry can be gauged by a 1-form gauge field $A$ with a coupling $A \wedge J$.

There are many familiar structures in quantum field theory that are almost completely parallel except that the dimensions are different. For example: in a $\Uone$ gauge theory without magnetic monopoles, we have $\dif F = 0$ (the Bianchi identity), and we also have a related integrated ``charge'' over a closed {\em 2-dimensional} manifold $\Sigma$, which is just the magnetic flux $\int_\Sigma \frac{1}{2\cpi} F \in \mathbb{Z}$. This is measured by a topological operator that depends on an angle $\alpha$ and a choice of 2-manifold $\Sigma$, $U(\Sigma, \alpha) \equiv \exp\left(\iu \alpha \int_\Sigma F\right)$. Notice that this manifold always has two dimensions, no matter what $d$ is, because $F$ is always a 2-form. We say that such an operator generates a $(d-3)$-form global symmetry. 

Similarly, in a $\Uone$ gauge theory without electrically charged particles, Maxwell's equations tell us that $\dif \star F = 0$. This is related to the conservation of electric flux, which we measure by integrating over a closed $(d-2)$-dimensional manifold, $\int \frac{1}{e^2} \star F \in \mathbb{Z}$. Again this is associated with a topological operator. It generates a 1-form global symmetry.  

Thus, we can have generalized conservation laws of the form $\dif J_p = 0$, and associated topological operators $U(\Sigma_p, \alpha) = \exp(\iu \alpha \int_{\Sigma_p} J_p)$. In general, we call this a $(d-p-1)$-form global symmetry. We could gauge such a symmetry by adding a $(d-p)$-form gauge field with a coupling $C_{d-p} \wedge J_p$, which respects a gauge symmetry of the form $C_{d-p} \mapsto C_{d-p} + \dif \lambda_{d-p-1}$. To complete the analogy, we should identify objects that are charged under this symmetry. In the gauge case, we should be able to integrate $C_{d-p}$ over the worldvolume of such an object, just as in the usual case we integrate $A$ over a particle's worldline. Thus, we expect that a current $J_p$ is related to an object with $(d-p-1)$ spatial dimensions and one time dimension, which can be created by an operator extended along $(d-p-1)$ spatial dimensions.

For example, in the case of $\Uone$ gauge theory with no electrically charged particles, we had a conserved current $\frac{1}{e^2} \star F$, which is a $(d-2)$-form. Thus we expect that the operators carrying the associated charge have dimension $d-(d-2)-1 = 1$. These must be line operators that have an associated electric flux. In fact, we already know about such operators: they are just the Wilson lines that we discussed in \S\ref{subsec:wilsonloops}. We refer to this symmetry as the ``electric 1-form symmetry'' of electromagnetism. Similarly, the symmetry with conserved current $\frac{1}{2\cpi} F$ is the ``magnetic $(d-3)$-form symmetry'' of electromagnetism, and the charged operators are 't Hooft operators, which are the magnetic analogue of Wilson lines. In the case $d = 4$, these are both 1-form symmetries, the Wilson and 't Hooft operators are both lines, and in fact there is a more general set of ``dyonic'' lines carrying both electric and magnetic flux and (in the theory of a free $\Uone$ gauge field) related to each other by $\mathrm{SL}(2,\mathbb{Z})$ duality transformations. 

Let's summarize: a $p$-form generalized global symmetry in a $d$-dimensional QFT is one that~\cite{Gaiotto:2014kfa}
\begin{itemize}
\item Acts on operators extended along $p$ dimensions, or equivalently dynamical objects with $(p+1)$ dimensional worldvolumes in spacetime. 
\item Is generated by symmetry operators $U(\Sigma, g)$ living on $(d-p-1)$-dimensional slices $\Sigma$ of spacetime. These are {\em topological} operators: we can deform the manifold $\Sigma$ without changing any correlation functions except when $\Sigma$ crosses the worldvolume of a charged operator.
\item In the special case that the symmetry group is $\Uone$, the symmetry is generated by a local conserved current operator which is a $(d-p-1)$-form obeying $\dif J_{d-p-1} = 0$. (Be aware that in some of the literature, what I call $J_{d-p-1}$ would be referred to as $\star j_{p+1}$.)
\item Can be gauged by a $(p+1)$-form gauge field. (Unless the symmetry has an 't Hooft anomaly, which prohibits gauging.)
\end{itemize}
The most familiar, ordinary symmetries are 0-form global symmetries.

\subsection{Breaking of $p$-form global symmetries}

\subsubsection{Explicit breaking}
\label{subsec:explicitbreakingpform}

Just for convenience, I will now specialize to $d = 4$. We have seen that a $\Uone$ gauge theory without any electrically or magnetically charged particles has a $\Uone_\mathrm{el} \times \Uone_\mathrm{mag}$ 1-form global symmetry, acting on Wilson and 't Hooft lines. However, these symmetries should be explicitly broken if charged objects exist, because the underlying conservation laws are no longer true:
\begin{equation}
\dif{\left(\frac{1}{e^2} \star F\right)} = J_\mathrm{el}, \quad \dif{\left(\frac{1}{2\cpi} F\right)} = J_\mathrm{mag}.
\end{equation}
These equations imply that the operators measuring electric and magnetic flux are no longer topological in a theory with charged objects.

\begin{figure}[!h]
\centering
\includegraphics [width = 0.8\textwidth]{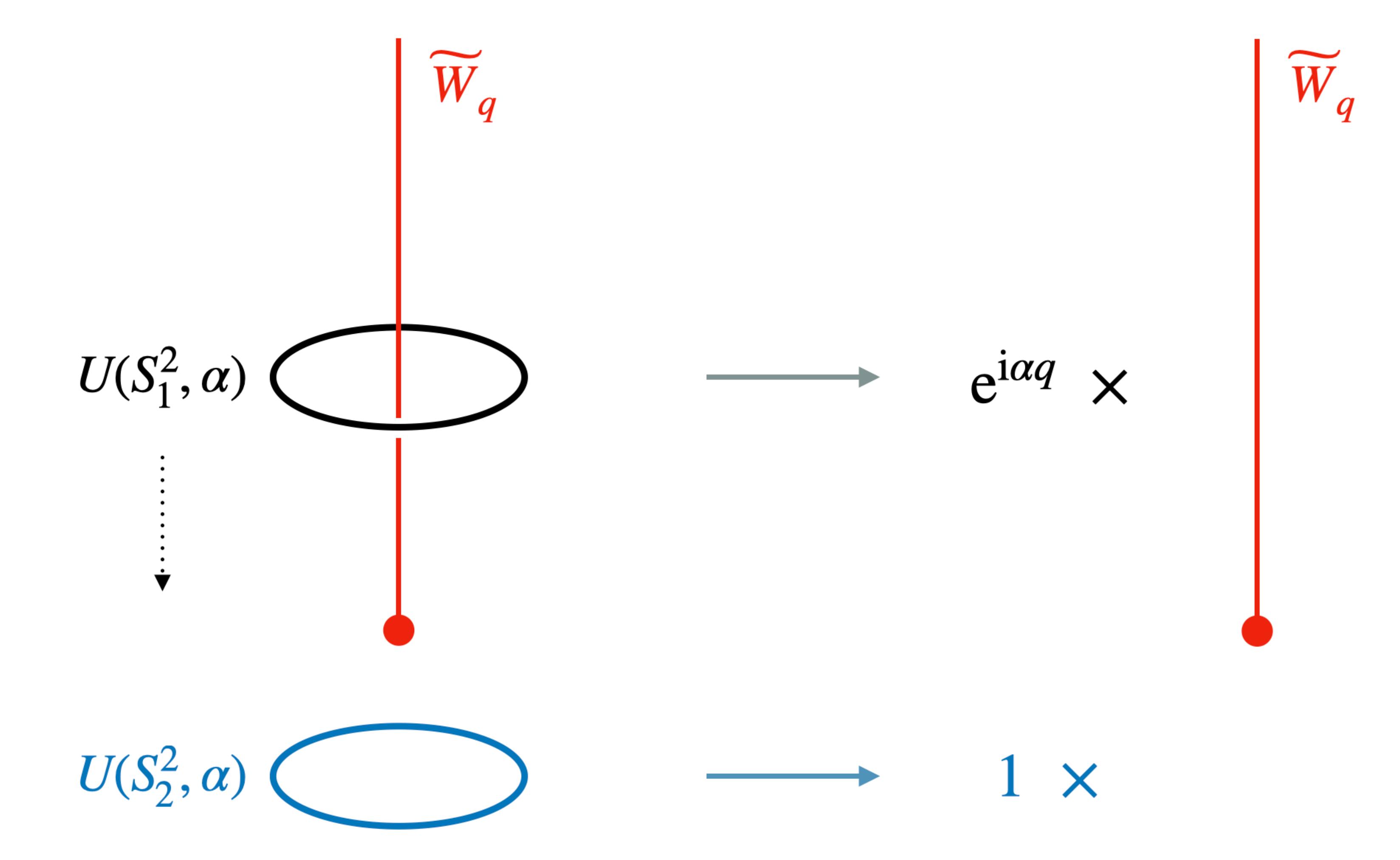}
\caption{Endability of a Wilson line causes the 1-form electric symmetry operator $U(\Sigma, \alpha)$ to no longer be topological. Left: symmetry operator inserted on two different spheres, $S^2_1$ and $S^2_2$. Right: corresponding OPE. The symmetry operator on $S^2_1$ sees an electric flux associated with charge $q$, and multiplies the Wilson line by a phase $\E^{\iu q \alpha}$. However, the symmetry operator on $S^2_2$ does not see a flux, and acts trivially. Hence, the electric 1-form symmetry is broken.
} \label{fig:endability}
\end{figure}

Let's see how this works more explicitly in the case of a Wilson line $W_q(\gamma)$ that carries charge under the 1-form electric symmetry, which is generated by an operator $U(\Sigma, \alpha)$. We will take the Wilson line to be supported on the timelike curve that sits at the origin of space, and the flux to be measured on a 2-sphere $\Sigma = S^2$ around the origin at some fixed time. Then we have an OPE
\begin{equation}
U(S^2, \alpha)W_q(\gamma) = \E^{\iu q \alpha} W_q(\gamma).
\end{equation}
This is completely parallel to the case of a 0-form symmetry depicted in Fig.~\ref{fig:chargeoperator}, and holds independent of what time $t$ we locate the 2-sphere at. However, if we add a charged operator $\phi$ of charge $q$ to the theory, then we can now consider a new operator which is the Wilson line ending at time $t = 0$ on a $\phi$ insertion, as depicted in Figure~\ref{fig:endability}. That is, we consider the Wilson operator with endpoint,
\begin{equation}
{\widetilde W}_q \equiv \exp\left[\iu q \int_{t = 0}^\infty A(t,0)\right] \phi^*(0,0).
\end{equation}
Now, if we insert the symmetry operator $U(S^2, \alpha)$ at $t > 0$ (as indicated by $S^2_1$ in the figure), it surrounds the Wilson line and acts with a nontrivial phase $\exp(\iu q \alpha)$, but if we insert the symmetry operator at $t < 0$ (as indicated by $S^2_2$ in the figure), it surrounds nothing at all and acts trivially. Thus, the {\em endability} of the Wilson line on a charged operator causes the symmetry operator to {\em no longer be topological}. Only topological operators can be thought of as obeying conservation laws. Said differently, acting on the vacuum with the local charged operator $\phi$ at some time creates an electric flux at subsequent times that was not present at earlier times, explicitly violating the conservation of electric flux.

In quantum gravity, we expect all global symmetries to be broken or gauged. To fully break the electric and magnetic 1-form global symmetries of $\Uone$ gauge theory, we need particles (or collections of particles) with all possible electric and magnetic charge to exist. We say that such a theory has a {\em complete spectrum}. It is a longstanding conjecture (the Completeness Hypothesis) that quantum gravity theories must have a complete spectrum~\cite{Polchinski:2003bq}. Here we see that, in the case of $\Uone$ gauge theory, the Completeness Hypothesis is implied by the absence of generalized global symmetries. It turns out that this does {\em not} generalize to all possible gauge groups: one can have an incomplete spectrum but no $p$-form generalized global symmetry~\cite{Harlow:2018tng}. However, if we broaden our notion of symmetry even more, we can revive the link between breaking of global symmetries and completeness. A gauge theory with an incomplete spectrum always contains topological operators, which can be said to generate a ``non-invertible global symmetry''~\cite{Rudelius:2020orz, Heidenreich:2021xpr}. Such topological operators obey a more complicated fusion algebra, rather than  group multiplication law as for standard symmetries. 

We saw that the breaking of a 1-form symmetry was associated with the endability of an operator linked by the symmetry operator. More generally, we expect that in quantum gravity, all extended operators should either be endable, or should themselves be the boundary of a different operator. The latter is the case where an operator is not itself gauge invariant (e.g., the local operator creating an electron is the end of a Wilson loop operator, without which it is not well-defined).

\subsubsection{Spontaneous breaking}
\label{subsubsec:photongoldstone}

A $p$-form global symmetry can also be spontaneously broken. For a 0-form symmetry, this happens when a charged local operator has a vacuum expectation value, $\langle O \rangle \neq 0$. For a 1-form global symmetry, the expected behaviors of the expectation value of a charged loop operator are that it scales with the exponential of minus the perimeter or the area of the loop. For area-law scaling, the expectation value of the loop operator goes rapidly to zero as the loop size increases. If it has the milder perimeter scaling, we say that the 1-form global symmetry is spontaneously broken. In this case, the operator can be redefined by adding a counterterm along the loop, so that it has a nonzero expectation value even for arbitrarily big loops~\cite{Gaiotto:2014kfa, Lake:2018dqm, Hofman:2018lfz}.

A concrete example is the expectation value of a Wilson loop or 't~Hooft loop in QED. When  we are in the phase where the photon is massless, the electric (or magnetic) 1-form symmetry is spontaneously broken. We can think of the photon as a Nambu-Goldstone boson for either spontaneously broken 1-form symmetry! For an ordinary Nambu-Goldstone boson, we expect that the spontaneously broken symmetry current can create a 1-particle Nambu-Goldstone boson state from the vacuum, $\langle \Omega | J_\mu(x) | p \rangle = p_\mu \E^{\iu p \cdot x} f$ for some constant $f$. An analogous statement holds for the photon, in terms of the electric or magnetic 1-form symmetry current, e.g., 
\begin{equation}
\langle \Omega | F_{\mu \nu} | p, \epsilon \rangle \propto \epsilon_{[\mu} p_{\nu]} \E^{\iu p \cdot x}.
\end{equation}
It may seem somewhat mysterious to think of the photon as a Nambu-Goldstone boson of the 1-form symmetry, because this symmetry is explicitly broken! If there are both electrically and magnetically charged particles in the universe, {\em both} 1-form symmetries are explicitly broken. When we explicitly break a 0-form symmetry, we are used to the fact that the pseudo-Nambu-Goldstone boson acquires a mass. But the photon is (as far as we know) exactly massless! This is something special about the explicit breaking of higher $p$-form symmetries, where $p > 0$: a pseudo-Nambu-Goldstone boson remains massless. This point is discussed further in~\cite{McGreevy:2022oyu, Iqbal:2021rkn}. (The enhanced robustness of higher $p$-form symmetries is closely related to the exponential improvement in the axion quality problem when the axion arises from a higher-dimensional gauge field, as discussed in \S\ref{subsec:qualityxdim}.)

\section{Global symmetries versus quantum gravity}
\label{sec:globalvsQG}

In \S\ref{sec:BHnoglobal} we saw that black hole thermodynamics provides arguments against continuous global symmetries. A much stronger statement is believed to be true: quantum gravity theories do not have {\em any} global symmetries. This includes ordinary (0-form) discrete global symmetries, but also $p$-form global symmetries (continuous or discrete) and even less familiar examples like non-invertible (or categorical) symmetries. In all of these cases, there are symmetry operators that implement the symmetry, which are {\em topological}. A very general expectation is that quantum gravity does not admit topological operators. Heuristically, this is because the gravitational path integral sums over all spacetimes, including those of nontrivial topology, and admits topology-changing transitions. Topology is not expected to be an invariant property in quantum gravity, and so it should not be possible to construct well-defined operators that correspond to topological invariants.

The state of the art in {\em proving}, from a well-defined starting point, that quantum gravity does not admit global symmetries is a holographic argument due to Harlow and Ooguri~\cite{Harlow:2018jwu, Harlow:2018tng}. This argument only applies in asymptotically AdS spacetimes, and exploits the fact that quantum gravity in such a spacetime is equivalent to a conformal field theory defined on the conformal boundary of spacetime. Reviewing the argument in detail is beyond the scope of these lectures, but I will give a very brief summary of the main idea. It relies on two key facts. The first is entanglement wedge reconstruction: an operator acting in a boundary region $R$ of the CFT can only influence a limited part of the bulk of AdS, the corresponding entanglement wedge $W_R$. The second is splittability: a symmetry operator on the boundary can be factored into a product of symmetry operators associated with a collection of subregions $R_i$. By breaking apart a global symmetry operator $U(R, g)$ on the boundary into a product of $U(R_i, g)$ over sufficiently many small regions $R_i$, we can make their domain of influence in the bulk, the union of the $W_{R_i}$, as small as we like. This shows that operators deep in the bulk cannot carry charge under the global symmetry---a contradiction, since a global symmetry of the type we discussed in \S\ref{sec:globalsymmetries} acts locally via topological operators that we should be able to pull out to the boundary. This argument rules out $p$-form global symmetries (continuous or discrete) in $d$-dimensional asymptotically AdS quantum gravity, for $0 \leq p \leq d-2$ (the case $p = d-2$ is slightly subtle, and less rigorous than the rest).  

The Harlow and Ooguri argument also establishes that the gauge group in asymptotically AdS quantum gravity must be compact, and that objects (or collections of objects) should exist in the bulk theory transforming in all representations of the gauge group (i.e., the Completeness Hypothesis).

Although this argument relied on AdS/CFT, there is a widespread expectation that the conclusions apply much more broadly to any realistic theory of quantum gravity.  The caveat ``realistic'' is important here, as a number of theories of quantum gravity in low dimensions lack many features of higher-dimensional quantum gravity; they may have global symmetries, but they also do not have unitary black hole evaporation, for example~\cite{Harlow:2020bee}. 

These modern arguments supplement a number of older arguments. For example, in perturbative string theory, Banks and Dixon showed that any putative continuous global symmetry actually is a gauge symmetry~\cite{Banks:1988yz}: given a global symmetry current in spacetime, one can construct a worldsheet vertex operator that creates a gauge field in spacetime. In AdS/CFT, a global symmetry of the boundary CFT is a gauge symmetry of the bulk~\cite{Witten:1998qj}. Other arguments, related to wormholes, have both early incarnations~\cite{Giddings:1988cx, Abbott:1989jw, Coleman:1989zu, Kallosh:1995hi} and very recent (and sharper) ones~\cite{Chen:2020ojn, Hsin:2020mfa, Belin:2020jxr, Bah:2022uyz}.

Gauge symmetries are common in quantum gravity, unlike global symmetries. However, we should remember that a global symmetry maps one state of the theory to another, but a gauge symmetry is a redundancy of our description, not really a symmetry at all. Let's sum this up with two crucial slogans to remember.

\begin{framed}
\noindent
There are no symmetries in quantum gravity except for gauge symmetries.

\noindent
Gauge symmetries are not symmetries.
\end{framed}

Any would-be global symmetry in quantum gravity must be either explicitly broken or gauged (a spontaneously broken symmetry is still a symmetry). Saying that global symmetries do not exist in quantum gravity is all well and good, but it's not very useful for doing real-world physics. As we discussed in \S\ref{sec:protondecay}, for particle physics we often care about {\em approximate} symmetries, like baryon or lepton number, or the approximate shift symmetry of an axion field or of an inflaton in the early universe. The arguments that we have discussed so far tell us that these symmetries cannot be exact, but to have a really {\em useful} statement about quantum gravity we need to be able to quantify how much a symmetry should be broken. This has been the focus of significant attention in recent years, which we will now review.

\section{Weak Gravity Conjecture}
\label{sec:WGC}

One way to obtain an approximate global symmetry is simply to have a very weakly coupled gauge theory. If we could start with QED and dial the electric coupling constant $e$ to be extremely small, then photons would almost decouple and the U(1) phase invariance of the electron would behave very much like a global symmetry. The principle that quantum gravity does not admit global symmetries then suggests that something should prevent us from taking the strict $e \to 0$ limit of a gauge coupling. The Weak Gravity Conjecture (WGC) is a quantitative statement of what happens in this limit. More precisely, there is in fact a family of related conjectures, some of which are not very constraining and some of which can have powerful implications for phenomenology. In this section we will discuss the original Electric and Magnetic WGCs, before turning to more powerful related conjectures in the next section.

\begin{figure}[!h]
\centering
\includegraphics [width = 0.8\textwidth]{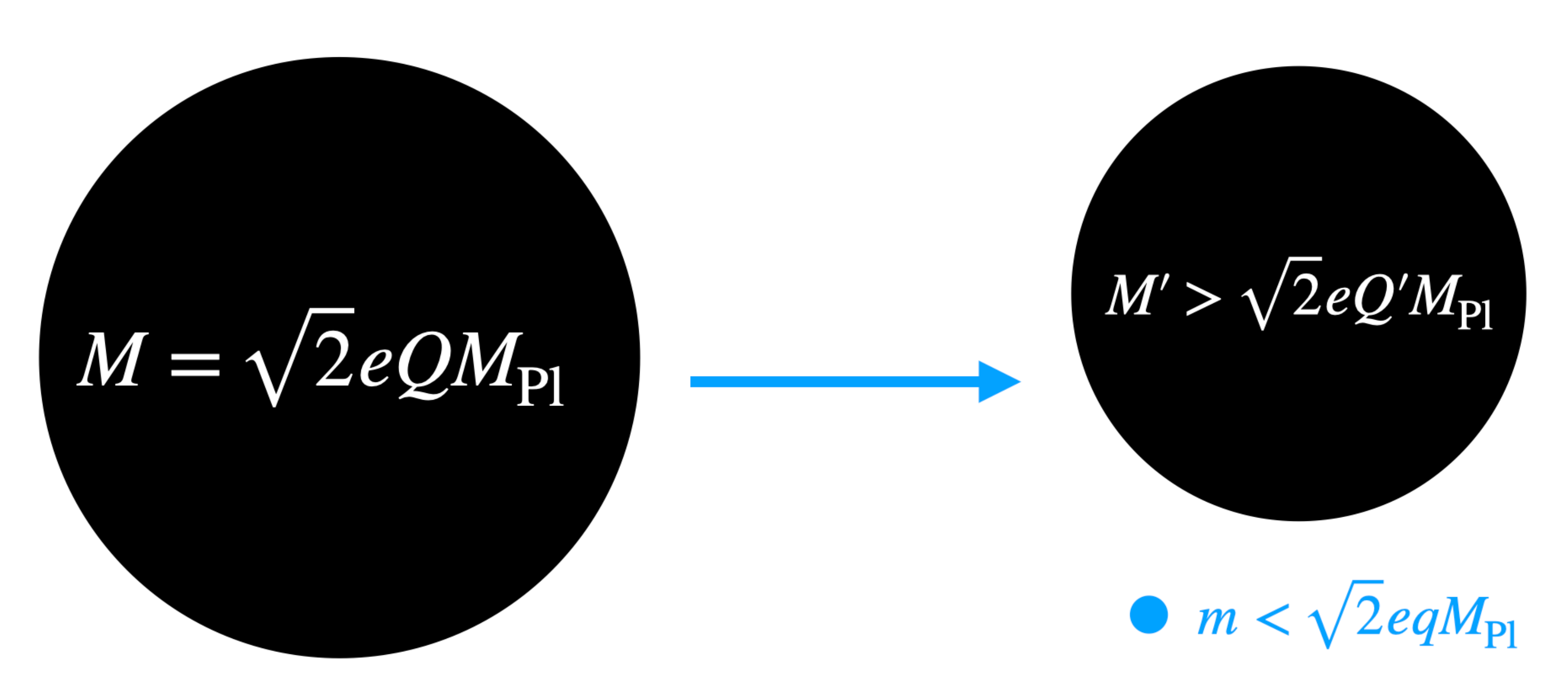}
\caption{An extremal black hole can decay to a WGC-obeying, or superextremal, particle (in blue, with $m < \sqrt{2} e q M_\mathrm{Pl}$) and a subextremal black hole.
} \label{fig:BHdecayWGC}
\end{figure}

The WGC story begins with the {\em Electric} WGC, proposed by Arkani-Hamed, Motl, Nicolis, and Vafa (henceforth ``AMNV'') in 2006~\cite{Arkani-Hamed:2006emk}. This is simply the statement that a charged particle exists that is not too heavy: given a U(1) gauge field in a theory of quantum gravity, there exists a particle of nonzero electric charge $q$ with a mass $m$ obeying the inequality
\begin{equation} \label{eq:electricWGC}
m < \sqrt{2} e |q| M_\mathrm{Pl}.
\end{equation}
Here $e$ is the U(1) gauge coupling, and we recognize that this inequality is the {\em opposite} of the black hole extremality bound~\eqref{eq:extremality}. One way to understand this condition is that it is a necessary kinematic requirement for extremal black holes to be able to shed their charge. As depicted in Fig.~\ref{fig:BHdecayWGC}, an extremal black hole with $M = \sqrt{2} e Q M_\mathrm{Pl}$ can decay (by Schwinger pair production) to a WGC-satisfying particle and a subextremal black hole with $M' > \sqrt{2} e Q' M_\mathrm{Pl}$. Thus, one way to understand the WGC is as a statement that black holes never get ``stuck'' with large charge that they can't get rid of. However, note that if they {\em did}, nothing would be obviously wrong with this, as we explained in Fig.~\ref{fig:banksseiberggauge} and surrounding discussion. This is why the WGC remains a conjecture, although versions of it have been proved in various contexts and with various additional assumptions.

Notice that for the one U(1) gauge theory that we know of in nature, the WGC is satisfied with many orders of magnitude to spare: for the electron, we have
\begin{equation}
\frac{m_e}{\sqrt{2} e M_\mathrm{Pl}} \approx 5 \times 10^{-22}. 
\end{equation}
It would have been more exciting if this was a close call, but at least the conjecture passes this test.

The Electric WGC by itself doesn't tell us that anything particularly bad or dramatic happens as we try to create a global symmetry by sending $e \to 0$. There must be some charged particle in the theory with a mass that tends to zero in this limit, but that seems innocuous. In particular, the WGC is a ``there exists'' statement, not a ``for all'' statement; only one particle needs to become light, for the minimal version of the conjecture.

A more useful constraint on weakly-coupled gauge theory arises from the {\em Magnetic} WGC, which was also explained in the original AMNV paper. The Magnetic WGC is just the WGC applied to magnetic charge. Due to Dirac quantization, when an electric field couples with strength $e$, the dual magnetic field couples with strength $2\cpi/e$. Thus, we can just repeat~\eqref{eq:electricWGC} but now conclude that there must exist a {\em magnetic monopole} with (integer) magnetic charge $q_\textsc{M}$, coupling strength $2\cpi/e$, and mass $m_\textsc{M}$: 
\begin{equation} \label{eq:magneticWGC}
m_\textsc{M} < \sqrt{2} \frac{2\cpi}{e} |q_\textsc{M}| M_\mathrm{Pl}.
\end{equation}
So far this is essentially just a relabeling of the Electric WGC, but the crucial physics insight was that electrically charged objects (with weak coupling $e \ll 1$) and magnetically charged objects (with strong coupling $2\cpi/e \gg 1$) are qualitatively different. 

The key difference is in the classical versus quantum (Compton) radii of the objects. Recall from classical electrodynamics that the {\em classical radius} of an electron is the radius at which the classical self-energy stored in the electric field is equal to the electron's mass, i.e., it is a radius $R_C$ at which
\begin{equation}
\int^\infty_{R_C} r^2 \dif r\, \dif\Omega\,\left(\frac{eq}{4\cpi r^2}\right)^2 = m_e.
\end{equation}
The integral is divergent at short distances, proportional to $1/R_C$. Solving for $R_C$, we see that 
\begin{equation}
R_C = \frac{\alpha q^2}{m_e},
\end{equation}
where $\alpha = \frac{e^2}{4\cpi}$ is the fine structure constant. By comparison, the Compton radius of the object, which we will denote by $R_Q$ with $Q$ for ``quantum,'' is
\begin{equation}
R_Q = \frac{1}{m_e}.
\end{equation}
Because the electric interaction is weakly coupled ($\alpha \ll 1$), we see that $R_C \ll R_Q$ (for $q \sim O(1)$). Thus, quantum mechanics resolved the classical self-energy puzzle of electromagnetism. Classically, one could argue that at the radius $R_C$, new physics should arise that explains why the electron mass is not extraordinarily large due to the energy stored in the electric field. In quantum mechanics, however, we see that the electron already behaves as a ``fuzzy'' object at the {\em much larger} distance scale $R_Q = \frac{1}{\alpha} R_C$, associated with virtual electron-positron pairs that screen the effective charge in QED. (This is perhaps the most dramatic example among the many places in physics where, in modern language, one could say that a ``naturalness argument'' successfully predicted new physics.)

\begin{figure}[!h]
\centering
\includegraphics [width = 0.9\textwidth]{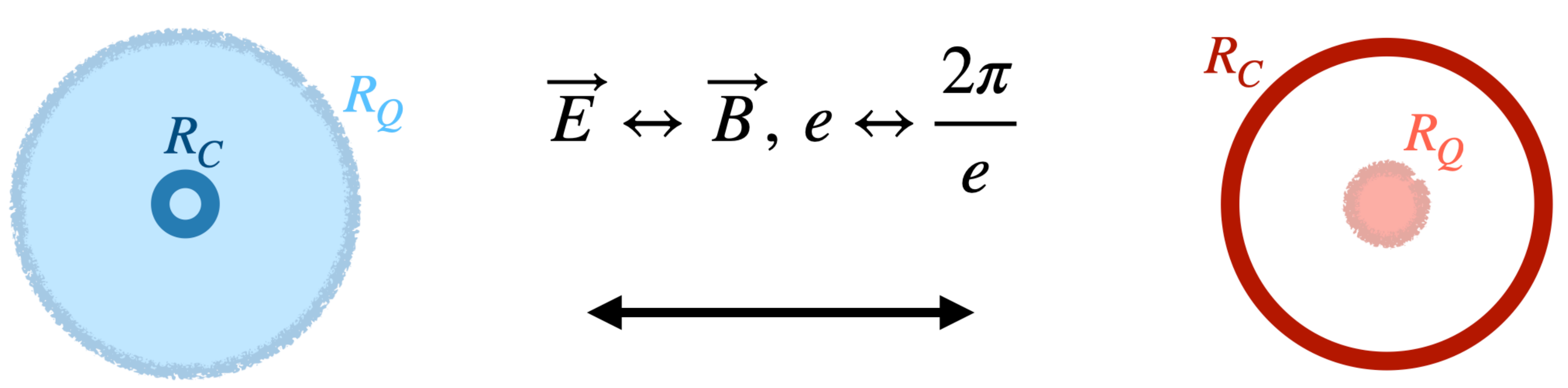}
\caption{Illustration of the classical radius $R_C$ and quantum (Compton) radius $R_Q$ of electrically (left, blue) and magnetically (right, red) charged objects.
} \label{fig:cqradii}
\end{figure}

Now consider the case of magnetically charged objects. Again, we define the Compton radius simply as the inverse mass, $R_Q = 1/m_\textsc{M}$. However, the classical radius now comes from integrating the energy stored in the {\em magnetic} field. This amounts to sending $e \to 2\cpi/e$ in the electric formula, so we have
\begin{equation}
R_C = \frac{\cpi q_\textsc{M}^2}{e^2 m_\textsc{M}} = \frac{q_\textsc{M}^2}{4\alpha} R_Q.
\end{equation}
In other words, for magnetically charged objects, the classical radius is {\em much larger} than the quantum radius! (At least for the integer charge $q_\textsc{M} \sim O(1)$.) The behavior of $R_Q$ and $R_C$ in the electric and magnetic cases is illustrated in Fig.~\ref{fig:cqradii}. In the magnetic case, wse can no longer argue that quantum fuzziness will save us from a naturalness problem. Instead, we expect that new physics will come in at distances of order $R_C$, corresponding to an {\em ultraviolet cutoff}
\begin{equation} \label{eq:LambdaRC}
\Lambda_\Uone \lesssim R_C^{-1} = \frac{e^2}{\cpi q_\textsc{M}^2} m_\textsc{M}.
\end{equation}
The details of this new physics will be model-dependent. In the classic example of the 't Hooft--Polyakov monopole, for instance, this is the scale at which we see that the gauge group is $\SU(2)$. In the case of the Kaluza-Klein monopole, this is the scale of the extra dimensions. In any case, we expect that at the energy scale $\Lambda_\Uone$, the physics will no longer be described by a 4d $\Uone$ gauge theory.

The conclusion~\eqref{eq:LambdaRC} relating the UV cutoff of the theory to the mass of a magnetic monopole is purely a statement about effective field theory, unrelated to gravity. However, if we combine it with the Magnetic WGC~\eqref{eq:magneticWGC}, we obtain a constraint relating the UV cutoff to the gauge coupling and the Planck scale:
\begin{equation} \label{eq:magWGCcutoff}
\Lambda_\Uone \lesssim e M_\mathrm{Pl}.
\end{equation}
This is now a much more useful statement than the Electric WGC! It doesn't just tell us that a charged particle becomes light when we send $e \to 0$, it says that the EFT description in terms of a 4d $\Uone$ gauge theory breaks down in this limit.

\begin{framed}
\noindent
Magnetic Weak Gravity Conjecture Cutoff: a weakly-coupled $\Uone$ gauge theory with small coupling $e$ breaks down at or below the energy $e M_\mathrm{Pl}$.
\end{framed}

\medskip
\noindent\centerline{\rule{\textwidth}{0.5pt}}
\noindent {\em Exercise:} Find the condition for the classical radius $R_C$ of a magnetically charged object to be larger than its Schwarzschild radius. Why is it reasonable to expect that some magnetically charged object exists that obeys this condition? (See discussions in~\cite{Arkani-Hamed:2006emk, delaFuente:2014aca}.)\\
\noindent\centerline{\rule[6.0pt]{\textwidth}{0.5pt}}

\section{Extensions and refinements of the Weak Gravity Conjectures}

The minimal electric WGC has little direct relevance for phenomenology. The magnetic WGC is a more interesting statement, because it tells us about when an EFT can break down at weak coupling. The last several years have led to a much sharper understanding of how and why such a breakdown of EFT can occur, due to {\em towers} of charged particles. Our main goal in this section is to summarize these new developments, which have enhanced the potential relevance of the WGC to phenomenological models. However, we will first turn to a relatively technical comment on the extension of the WGC to theories with multiple $\Uone$ gauge fields, because an analysis of such a case is one of the arguments supporting the new qualitative picture of the WGC and towers. 

\subsection{Multiple gauge fields}

In a theory with multiple $\Uone$ gauge groups, an extremal black hole can carry a combination of all the $\Uone$ charges. In particular, given a collection of $n$ gauge fields each normalized with integer magnetic flux $\frac{1}{2\cpi} \int F^i \in \mathbb{Z}$, with a kinetic term allowing for general mixing (recall \S\ref{sec:kineticmixing})
\begin{equation}
\int \left( - \frac{1}{2} K_{ij} F^i \wedge \star F^j \right),
\end{equation}
the black hole extremality condition for a 4d black hole carrying integer charges $Q_i$ under the gauge fields $A^i$ becomes
\begin{equation} \label{eq:extremalmulti}
M^2 \geq M_\mathrm{ext}(Q)^2  \equiv 2K^{ij} Q_i Q_j M_\mathrm{Pl}^2,
\end{equation}
where $K^{ij}$ is the inverse matrix of $K_{ij}$ (the analogue of $e^2$ in~\eqref{eq:extremality}). One way to extend the minimal electric WGC~\eqref{eq:electricWGC} is to generalize the kinematic criterion it captures: what is a necessary condition for all extremal black holes, with any combination of charges, to be able to shed their charge by emitting light charged particles? The answer is often referred to as the convex hull criterion~\cite{Cheung:2014vva}. A particle with mass $m$ and charges $q_i \in \mathbb{Z}$ is represented by the charge-to-mass vector $z_i = \sqrt{2} M_\mathrm{Pl} q_i / m$. The extremality bound~\eqref{eq:extremalmulti} tells us that there is a region at small $z$, $K^{ij} z_i z_j \leq 1$, that can be occupied by black holes. The convex hull condition says that there must exist a collection of particles in the theory with $z$ vectors whose convex hull, as measured by the metric $K^{ij}$, contains the black hole region. 

One could imagine the convex hull condition being satisfied by a finite number of particles, but it could also be satisfied by an infinite number of states of different charges that hug the exterior of the black hole region. A general statement, valid in this limit where infinitely many different particles are relevant, is: for every direction ${\hat q}$ in the space of possible $\Uone^n$ charges allowed by Dirac quantization, there is a superextremal multiparticle state with $z \propto {\hat q}$. By a superextremal multiparticle state, we simply mean a collection of particles whose total-charge to total-mass ratio vector $z_i$ lies outside the black hole region. This formulation is also valid in the case when massless scalar fields affect the form of the black hole solution, which changes the quantitative form of~\eqref{eq:extremalmulti} but not the qualitative existence of a black hole region in the space of $z$ vectors.

\subsection{Cutoffs take the form of towers of particles}
\label{sec:cutofftowers}

The magnetic WGC~\eqref{eq:magWGCcutoff} suggests that, in quantum gravity, weak coupling for a gauge theory comes at the cost of a low UV cutoff, below the Planck scale. But what happens at this UV cutoff? If, for example, it only requires us to embed a $\Uone$ gauge theory in a weakly-coupled non-abelian gauge theory, this would be a very mild form of cutoff and would not necessarily impose an interesting constraint for phenomenology.

A large body of work in the last several years has built up a compelling picture for a much more substantial sort of cutoff, involving a {\em tower} of charged particles~\cite{Heidenreich:2015nta, Heidenreich:2016aqi, Montero:2016tif, Andriolo:2018lvp}. Specifically:

\begin{framed}
\noindent
For a gauge theory with a weak coupling $e$ in quantum gravity, there is an {\em infinite} tower of charged particles of different charge, {\em each} obeying the electric WGC bound $m < \sqrt{2} e q M_\mathrm{Pl}$. In the non-abelian case, we have an infinite tower of particles in different representations of the gauge group obeying the WGC with respect to the Cartan $\Uone$ subgroups.
\end{framed}

One of the sharpest such statements consistent with known evidence, the Sublattice WGC, holds that there is a sublattice of the charge lattice of the theory on which a WGC-obeying particle exists at every site~\cite{Heidenreich:2016aqi, Montero:2016tif}. That is, given any charge vector $\vec q$ in the charge lattice, there is a small integer $n$ (the {\em coarseness} of the sublattice) such that there is a superextremal particle of charge $n {\vec q}$. In known examples, $n$ does not exceed 3.

Towers of particles are associated with UV cutoffs in two different ways. There is an obvious weak sense that an EFT breaks down if it fails to include massive particles that exist. But there is a more dramatic sense that an infinite tower of particles eventually leads to a complete failure of EFT altogether: all of these particles couple to gauge fields and to gravity, and run in loops, and eventually drive the whole theory to strong coupling. In particular, a gravitational theory with $N$ weakly coupled light particle species in 4d is expect to break down by an energy scale $\Lambda_\textsc{QG} \lesssim M_\mathrm{Pl}/\sqrt{N}$, sometimes referred to as the ``species bound'' or ``species scale''~\cite{Veneziano:2001ah, Dvali:2007hz, Dvali:2007wp}. One familiar example arises from extra dimensions that are large compared to the 4d Planck length; in this case, the higher-dimensional Planck mass $M_D$ is parametrically below the 4d Planck mass, precisely by a factor of the square root of the volume in higher-dimensional Planck units, i.e., the square root of the number of Kaluza-Klein modes below the cutoff. You can see this just by dimensional reduction of the higher-dimensional Einstein-Hilbert term:
\begin{equation}
\int_{M_4 \times X} \mathrm{d}^Dx\,\sqrt{|g|} M_D^{D-2} {\cal R} \mapsto \int_{M_4} \mathrm{d}^4x\,\sqrt{|g|} M_\mathrm{Pl}^2 {\cal R}\qquad\text{where}\qquad M_\mathrm{Pl}^2 = M_D^{D-2} \mathrm{Vol}(X).
\end{equation}
One way to define the quantum gravity cutoff $\Lambda_\textsc{QG}$ is to say that $\Lambda_\textsc{QG}^{-1}$ is the smallest radius that a black hole admitting a good semiclassical description can have. One argument for the species bound is that a black hole with radius $\Lambda_\textsc{QG}^{-1}$ can emit Hawking radiation quanta of all $N$ light species. Unless $N \lesssim M_\mathrm{Pl}^2/\Lambda_\textsc{QG}^2$, the lifetime of the black hole would be smaller than its radius, a clear breakdown of the semiclassical description~\cite{Dvali:2007wp}.

\begin{figure}[!h]
\centering
\includegraphics [width = 0.6\textwidth]{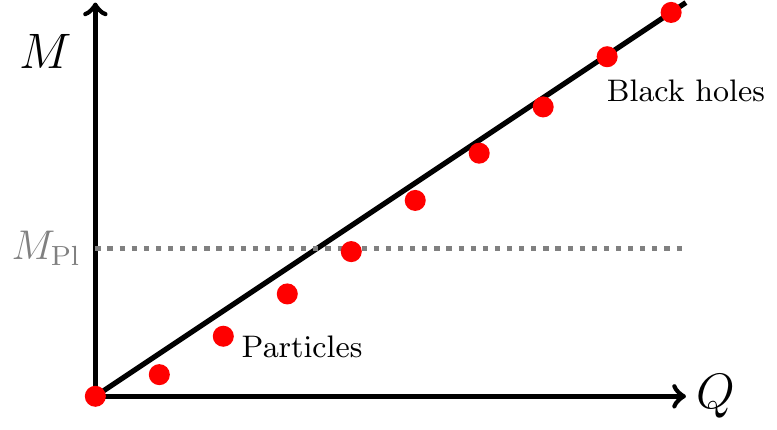}
\caption{Mass and charge spectrum of a tower of particles (red dots). The diagonal line indicates the black hole extremality bound, which is derived from the two-derivative action.  Any state lying below this line obeys the Weak Gravity Conjecture. At large $(M, Q)$, there are states lying along the line, but they can approach it from below if higher-derivative corrections have appropriate signs. At small $M$, below the Planck scale, the states are interpreted as light particles.
} \label{fig:towerinterpolation}
\end{figure}

Formulations of the WGC in which a tower of charged particles exists beginning at the scale $e M_\mathrm{Pl}$ thus imply a powerful statement about how local quantum field theory breaks down at high energies, eventually requiring a full theory of quantum gravity to understand physics at the scale $\Lambda_\textsc{QG}$. Let's briefly summarize several of the independent arguments pointing in this direction:

\medskip

\noindent
{\bf Interpolation from particles to black holes.} Extremal black holes are {\em already} an infinite tower of states, and even the smallest semiclassical black hole has a very large entropy if $\Lambda_\textsc{QG} \ll M_\mathrm{Pl}$. This suggests that there should be many more charged states, at lower masses, that continuously interpolate into black hole states, as depicted in Fig.~\ref{fig:towerinterpolation}. For small black holes, higher derivative corrections modify the extremality bound, and there are reasons to think they do so in the direction of smaller mass (see, e.g.,~\cite{Kats:2006xp, Cheung:2018cwt, Hamada:2018dde, Charles:2019qqt, Arkani-Hamed:2021ajd}; this is now a large body of research).

\medskip

\noindent
{\bf Consistency under dimensional reduction.} If we reduce a theory with gravity from $D$ dimensions to $d = D-1$ dimensions by compactifying on a circle, $x^D \cong x^D + 2\cpi R$, then we can see by matching the Einstein-Hilbert term that the lower-dimensional Planck scale $M_d$ is related to the higher-dimensional Planck scale $M_D$ via
\begin{equation}
M_d^{d-2} = 2\cpi R M_D^{D-2}.
\end{equation}
Similarly, if a $\Uone$ gauge theory in $D$ dimensions has coupling $e_D$ and reduces to a $\Uone$ gauge theory in $d$ dimensions with coupling $e_d$, we match the gauge kinetic term to find
\begin{equation}
\frac{1}{e_d^2} = \frac{2\cpi R}{e_D^2}.
\end{equation}
From these relations, we see that {\em parametrically} a WGC bound of the form $m^2 \leq \gamma_D e_D^2 q^2 M_D^{D-2}$ is maintained under dimensional reduction. The prefactor $\gamma_D$, read off from the black hole extremality condition, turns out to depend not only on $D$ but on how scalars couple to the gauge field. Under dimensional reduction there is always a scalar radion that affects the answer in precisely the right way that the extremality bound in $D$ dimensions reduces to the extremality bound in $d$ dimensions. This can be thought of as a consistency check of the WGC: we might hope that a good conjecture about quantum gravity would hold independently of the number of spacetime dimensions.

However, the full story is more subtle than this. When we dimensionally reduce the theory on a circle, we also obtain a new $\Uone$ gauge boson, the Kaluza-Klein gauge boson arising from the $g_{\mu D}$ components of the metric tensor. The graviton has a tower of charged Kaluza-Klein excitations (i.e., modes with different amounts of momentum in the $x^D$ direction), which saturate the WGC for the Kaluza-Klein $\Uone$. If we have a charged particle in $D$ dimensions, it gives rise to a tower of particles in $d$ dimensions that all carry the original charge as well as any number of units of additional Kaluza-Klein charge. If the $D$-dimensional particle obeys the WGC, all of these particles lie outside the black hole region in $d$ dimensions. However, their convex hull does not necessarily contain the black hole region. In this sense, the WGC (in its convex hull formulation for multiple $\Uone$ factors) is {\em not} automatically preserved under dimensional reduction~\cite{Heidenreich:2015nta}. However, if we start with an infinite tower of particles of different charge in $D$ dimensions, each obeying the WGC, their infinite set of towers of $d$-dimensional Kaluza-Klein excitations will satisfy the convex hull condition. This was an early motivation for postulating a strong form of the conjecture, requiring infinite towers of WGC-obeying particles.

\medskip

\noindent 
{\bf String theory examples; modular invariance.} String theory provides us with a wealth of examples of quantum gravity theories (or vacua of a single theory, depending on your perspective). The masses of various charged particles are sometimes calculable within string vacua, generally in two cases: when the theory is weakly coupled, or when the masses of states saturate a BPS bound due to supersymmetry. Whenever we are able to carry out checks in such examples, we find that there are infinite towers of charged particles that obey the WGC. In the perturbative regime, this can be proven quite generally with a string worldsheet argument based on the principle of modular invariance~\cite{Heidenreich:2016aqi, Montero:2016tif}. More recently, similar arguments have been extended to a much larger class of string compactifications known as F-theory vacua, which are essentially the strongly coupled limit of Type IIB string theory. Although F-theory is generically strongly coupled, an individual gauge group within an F-theory vacuum may be weakly coupled, and in this case spectra are again calculable, and again examples have towers of particles obeying the WGC~\cite{Lee:2018urn, Lee:2019tst, Lee:2019xtm}. Other interesting checks have been carried out for BPS states, providing examples that satisfy the tower form of the WGC away from the weakly-coupled limit~\cite{Alim:2021vhs, Gendler:2022ztv}.

\medskip

\noindent
{\bf Emergence of weak coupling.} Above, I mentioned the quantum gravity cutoff $\Lambda_\textsc{QG}$ associated with the species bound, where large numbers of particles drive gravity to strong coupling. We expect that, at energies near $\Lambda_\textsc{QG}$, there are no approximate global symmetries whatsoever. In particular, we expect that there are no weakly coupled gauge theories at such a scale. Instead, weak coupling should always ``emerge'' only in the infrared~\cite{Harlow:2015lma}. Weak coupling is associated with a large coefficient $1/e^2$ for a gauge field kinetic term. One way to achieve such a coefficient is to integrate out a tower of charged particles, each of which contributes to this coefficient. Heuristically, then (ignoring constant factors and logarithms), we expect a relationship like (in 4d, using the standard QED beta function~\eqref{eq:runningU1coupling})
\begin{equation}
\frac{1}{e^2} \sim \sum_{i | m_i < \Lambda_\textrm{gauge}} q_i^2 = N(\Lambda_\textrm{gauge}) \langle q^2\rangle_{\Lambda_\textrm{gauge}},
\end{equation}
where the sum is over all particles $i$ with mass below the gauge theory's UV cutoff $\Lambda_\textrm{gauge}$, with associated charge $q_i$. In the last step, $N(\Lambda_\textrm{gauge})$ refers to the number of particles below the energy $\Lambda_\textrm{gauge}$ and $\langle q^2 \rangle_{\Lambda_\textrm{gauge}}$ refers to the average charge-squared of these particles.  The assumption that the gauge theory is weakly coupled at the quantum gravity scale tells us that $\Lambda_\text{gauge} \lesssim \Lambda_\textsc{QG}$. From this together with the species bound, we derive that
\begin{equation}
\Lambda_\text{gauge}^2 \lesssim e^2 \langle q^2\rangle_{\Lambda_\textrm{gauge}} M_\mathrm{Pl}^2.
\end{equation}
This is an interesting WGC-like statement that shows that the basic emergence principle implies that the {\em average} particle below the gauge theory's UV cutoff obeys the WGC bound~\cite{Heidenreich:2017sim}.

A variety of similar statements can be derived in arbitrary numbers of dimensions, and for non-abelian gauge groups. These emergence arguments also connect closely to another idea known as the Swampland Distance Conjecture (SDC)~\cite{Ooguri:2006in}, which posits that in asymptotic limits of scalar field moduli spaces in quantum gravity, there is always an infinite tower of states whose mass $m(\phi)$ goes to zero exponentially, at least as fast as $\exp(- \lambda d(\phi)/M_\mathrm{Pl})$ with the geodesic distance $d(\phi)$ in field space (from some fixed reference point) measured in Planck units (with $\lambda$ expected to be $O(1)$; recent arguments have sharpened this to $\lambda \geq 1/\sqrt{d-2}$~\cite{Etheredge:2022opl}). One version of the emergence argument assumes that the kinetic term of $\phi$ is itself generated (or at least dominated) by loops of the particles in the tower, which can be shown to imply that the tower masses are exponential in the canonically normalized scalar field~\cite{Heidenreich:2018kpg, Grimm:2018ohb}; see also~\cite{Stout:2021ubb, Stout:2022phm}. Limits where a gauge coupling $g \to 0$ in quantum gravity are expected to be infinite-distance limits in moduli spaces, in which case the SDC tower and the WGC tower may be one and the same, although sometimes there are multiple towers becoming light and the WGC tower is not the lightest.

\medskip

\noindent
{\bf Strong breaking of global symmetries at the cutoff.} As discussed in \S\ref{subsec:explicitbreakingpform}, the existence of electrically charged particles explicitly breaks a 1-form global symmetry that acts on Wilson line operators in free $\Uone$ gauge theory. We expect that all global symmetries are broken in quantum gravity, which implies that some electrically charged particles should exist. But in fact, we have even stronger expectations: all global symmetries should be {\em badly} broken at the cutoff $\Lambda_\textsc{QG}$. If a gauge theory is weakly coupled, a single charged particle (like the electron) does not {\em badly} break the 1-form symmetry. To quantify this, one can consider how ``close to topological'' the symmetry operator is~\cite{Cordova:2022rer}. An unbroken symmetry has an associated topological symmetry operator. For a mildly broken symmetry, enlarging the surface on which the operator is inserted will {\em slowly} change a correlation function. For a badly broken symmetry, it will {\em rapidly} change a correlation function. It turns out that the deviation of the 1-form symmetry operator from being topological is precisely measured by the beta function of the gauge theory induced by charged particles running in loops. A tower of charged particles can badly break the 1-form symmetry, by driving the theory to strong coupling by the cutoff $\Lambda_\textsc{QG}$. This calculation then becomes essentially the same as the emergence calculation discussed above. Thus, towers of WGC-obeying charged particles are expected to appear in order to ensure that the 1-form electric symmetry is badly broken in the UV.

\medskip

Before closing the discussion on towers of particles, I want to highlight one more recent development that proposes a much sharper picture of what these towers are. The Emergent String Conjecture~\cite{Lee:2019wij}, formulated by Lee, Lerche, and Weigand based on evidence in F-theory~\cite{Lee:2018urn, Lee:2019tst, Lee:2019xtm, Alvarez-Garcia:2021pxo}, holds that weak-coupling limits in quantum gravity come in only two forms: decompactification limits and emergent string limits. In the former case, a tower of light states arises from Kaluza-Klein modes. In the latter, it arises from excitations of a string that becomes asymptotically tensionless at weak coupling. One reason why this is a very promising idea for phenomenology is that particles carrying Kaluza-Klein charge are never chiral. For many particle physics applications, we are interested in weakly coupled gauge groups coupled to chiral matter. The Emergent String Conjecture suggests that these are always described by low-tension strings. In the case of a Kaluza-Klein tower, a 4d gauge theory with coupling $g$ can have a quantum gravity cutoff as high as $\Lambda_\textsc{QG} \lesssim g^{1/3} M_\mathrm{Pl}$. However, we expect that in the case of a stringy tower, we have a quantum gravity cutoff near the WGC scale itself, $\Lambda_\textsc{QG} \lesssim g M_\mathrm{Pl}$. By providing a reason to focus on the latter case in many applications, the Emergent String Conjecture greatly strengthens the power of the WGC (at the cost of relying on stronger assumptions). We will have more to say about implications for phenomenology in the following part.

\subsection{The WGC for $p$-form gauge fields}
\label{subsec:WGCpform}

In the discussion above I have focused on the WGC for ordinary (1-form) gauge fields, but AMNV also formulated it for general $p$-form gauge fields. In this case, charged objects (called ``branes,'' in general, especially when $p \geq 3$) have $p$-dimensional worldvolume, and instead of being characterized by a mass these have a tension ${\cal T}_p$ with mass dimension $p$. A $p$-form gauge field has a coupling $e_p$ of mass dimension $p + 1 - D/2$, in $D$-dimensional spacetime. 
The WGC then says that there should exist a charged object whose tension obeys the inequality
\begin{equation} \label{eq:pformWGC}
{\cal T}_p^2 \leq \gamma e_p^2 Q^2 M_\mathrm{Pl}^{D-2},
\end{equation}
where $\gamma$ is a constant factor derived from the extremality bound for charged black branes (often $O(1)$, but not necessarily when there are scalar forces that are much stronger than gravity).

In some cases, the charged objects (strings or branes) may arise as solitons within an effective field theory. An example arises in the abelian Higgs model: in the limit where the Higgs mode is heavy, we can integrate it out and write the theory in terms of a BF mass (as in \S\ref{sec:CSStueckelberg}). The $B$ field dual to the Higgs phase is a 2-form gauge field, and the WGC tells us that there should exist a charged string with tension below a scale of order $v M_\mathrm{Pl}$. This string is the familiar ANO (Abrikosov-Nielsen-Olesen) vortex. In its core, the Higgs VEV goes to zero. Such a solitonic charged object has a tension that can be derived within effective field theory (in this case, ${\cal T} \sim v^2$), in which case the WGC bound is usually the simple constraint that all VEVs in the theory are below the Planck scale. By contrast, a fundamental string or brane that is not describable as a soliton within some EFT has a core that probes UV physics. In such a case, the tension of the object is generally a strong cutoff on the theory, in the sense that it is of order $\Lambda_\textsc{QG}$ or higher.

\newpage

\section*{\Large Part Five: Phenomenological Insights from Quantum Gravity}
\label{sec:lecturefive}
\addcontentsline{toc}{section}{\nameref{sec:lecturefive}}

Now that we've seen some examples of how quantum gravity might constrain effective quantum field theories, let's try to take a further step and connect these ideas to real-world particle phenomenology and potential experiments. I will allow myself somewhat more speculation in this part of the notes than in earlier parts. On the other hand, I will try to stick close to ideas that are relatively well-established, such as conjectures that are known to hold in a large collection of quantum gravity theories. It is possible that some of these conjectures can be falsified theoretically, with explicit counterexamples in string theory vacua. This would be very interesting. (A conjecture about quantum gravity can only be shown to be false with an explicit, consistent quantum gravity theory, a basic fact about logic that is sometimes overlooked by authors who present EFT counterexamples.) 

\section{Charge quantization}
\label{sec:chargequantization}

A simple black hole argument tells us that gauge groups in quantum gravity should be compact, as discussed in~\S\ref{sec:BHnoglobal}. This means that abelian, continuous gauge groups should be made up of products of $\Uone$ factors, not $\mathbb{R}$, and hence electric charge should be quantized. In the Standard Model, the $\Uone_\textsc{Y}$ charge assignments are fixed by anomaly cancelation arguments once the $\SU(3)_\textsc{C} \times \SU(2)_\textsc{L}$ representations are chosen,\footnote{There is some fine print: this argument assumes that all three generations have the same hypercharge assignments, and even then there is one other viable solution in which the right-handed up and down quarks have equal and opposite hypercharge and the other fermions are hypercharge-neutral. See \S22.4 of~\cite{Weinberg:1996kr}, or early discussion in~\cite{Geng:1989tcu, Minahan:1989vd}.} so we don't need to invoke quantum gravity to argue for why the charges are all multiples of a base unit. On the other hand, beyond the Standard Model, we could ask whether we might expect to see irrational charges, and quantum gravity gives us a clear answer. As explained in~\S\ref{sec:kineticmixing}, small, non-quantized effective ``millicharges'' can arise by kinetic mixing of the hypercharge gauge boson with a massless dark photon. This is a well-studied scenario in particle phenomenology and cosmology. Importantly, the dark photon will mediate self-interactions of the millicharged particles that are much stronger than their interaction through ordinary photons. In some cases, millicharged particles have been studied {\em without} including a dark photon, and hence omitting such strong self-interactions. For irrational values of millicharge, this is inconsistent with quantum gravity.

Of course, charge quantization in quantum gravity does leave open the alternative that the basic unit of $\Uone_\textsc{Y}$ charge is much smaller than the smallest value we see in the Standard Model (conventionally normalized to be $1/6$). Perhaps it is, for example, $1/6000$, and the Standard Model fields all have charge that is a multiple of 1000 in terms of the base unit, {\em without} violating charge quantization. This is a logical possibility, albeit one that seems implausible. However, the search for quantum gravity theories in which light fields have large integer $\Uone$ charges has recently turned up interesting examples~\cite{Li:2022vfj}, so this remains an active area of investigation that may yet have surprises in store for us.

\section{Weak Gravity Conjecture and phenomenology}

\subsection{The photon mass}
\label{subsec:photonmass}

In effective field theory, it is perfectly consistent to add a mass term for an abelian spin-1 boson. For example, we could add a mass term of the form $A_\mu A^\mu$ for the photon (which could arise by matching to a mass term for the hypercharge spin-1 boson in the full Standard Model). Doing this for {\em non-abelian} spin-1 fields is problematic: scattering amplitudes for the longitudinal modes of gluons of mass $m$ interacting through the non-abelian self-interaction with coupling $g$ give rise to amplitudes that grow with energy, signaling that the theory breaks down at or before a cutoff energy $E_\mathrm{max} \sim m/g$. This argument for the $W$ and $Z$ bosons of the Standard Model led to the LHC ``no-lose theorem''~\cite{Lee:1977eg}. The Higgs mechanism provides extra terms in the scattering amplitude that remedy the problem.  In the abelian case, however, no Higgs mechanism is needed because the dangerous self-interaction is absent. Nothing goes wrong whatsoever! This is counterintuitive, because adding a mass produces a discontinuous change in the number of degrees of freedom: the photon would have three propagating polarization states, instead of two. However, all interactions of the longitudinal mode turn off in the limit as the mass goes to zero.

Why, then, should we expect the Standard Model photon to be massless? A common answer is ``gauge invariance,'' but this is poor logic. A massless spin-1 boson must have an associated gauge redundancy to remove the longitudinal mode. A {\em massive} spin-1 boson is perfectly healthy without any such redundancy, so there is no need to include a gauge symmetry. One can do so if one wishes, writing the mass term in the Stueckelberg form $(A_\mu - \partial_\mu \theta)^2$ where $\theta$ shifts under a gauge transformation of $A$, but this is essentially putting an extra degree of freedom in and then promptly taking it right back out.

We know, in the real world, that if the photon has a mass it is a very tiny one. One of the easiest bounds to understand comes from Fast Radio Bursts: they emit radio waves over a range of frequencies, which arrive on Earth at about the same time. If the photon has a mass, the lower-frequency modes will travel more slowly than the higher-frequency modes, and a burst traveling over a very long distance would arrive on Earth with a noticeable delay between modes of different frequencies. Radio waves are useful for this test because they have relatively low frequency, meaning that a small photon mass has a larger effect on their speed than it would for, say, a gamma ray. Examining such a signal, one can conclude that~\cite{Wu:2016brq, Bonetti:2016cpo}
\begin{equation} \label{eq:photonmassbound}
m_\gamma \lesssim 10^{-14}\,\mathrm{eV}.
\end{equation}
This is not the strongest photon mass bound that you can find in the literature, it's just one of the easiest to understand and the least sensitive to any assumptions about modeling magnetic fields in the Solar System or the galaxy. And it is, compared to the mass of any known massive particle, a very stringent bound. Given our discussion above, one might think that an effective field theorist should think that, despite this stringent bound, if we keep doing more and more precise measurements we will find someday that the photon mass is nonzero. Just like the cosmological constant or the neutrino mass, it is not forbidden by any deep principle, so it will turn out to be not exactly zero.

\bigskip

Really?

\bigskip

Surely not!

\medskip

I think that any reasonable physicist {\em should} expect that the photon mass really is zero---{\em exactly} zero!---but it is surprisingly hard to give an airtight argument for this. My take on this question is that effective field theory is not sufficiently powerful to prove that the photon mass is zero, but that quantum gravity is~\cite{Reece:2018zvv}. Here is my argument, which is not completely free of loopholes, but which I find compelling.

The first case to discuss is the Higgs mechanism. We could imagine a new scalar field $\phi$, which is a singlet of $\SU(3)_\textsc{C}$ and $\SU(2)_\textsc{L}$ with nonzero hypercharge. In order to give a very tiny mass to the photon, consistent with~\eqref{eq:photonmassbound}, it must have either a very small charge or a very small VEV or both. If it has a large charge, we would see its effects; for instance, the particle could be directly produced in scattering experiments, and it would affect the running of the fine structure constant. So the only way to make this work is the kind of scenario that we called ``implausible'' in \S\ref{sec:chargequantization}, in which the fundamental unit of hypercharge is much, much smaller than the smallest hypercharge of any Standard Model fermion. This means that the QED coupling constant that we call $e$ should really be thought of as $e_0 N$, where $N$ is a very large integer and $e_0 \ll 1$. At energies well above $m_\phi$, where the photon mass is negligible, $\phi$ will essentially behave like a millicharged particle, and experimental constraints tell us that $N \gtrsim 10^{14}$ for $m_\phi \lesssim \mathrm{keV}$ (see, e.g.,~\cite{Davidson:2000hf}). We could have, for example, $m_\phi \sim \langle \phi \rangle \sim 1\,\mathrm{eV}$ and $N \sim 10^{15}$, which would give $m_\gamma \sim e_0 \langle \phi \rangle \sim 10^{-15}\,\mathrm{eV}$. There are at least two things that should make us uncomfortable here. One is that the scalar $\phi$ is extremely light, posing a severe fine-tuning or naturalness problem. Even {\em graviton} loops would produce a large correction to $m_\phi$ unless the cutoff is below about $\sqrt{m_\phi M_\mathrm{Pl}} \sim \mathrm{MeV}$. Of course, naturalness problems are not sharp no-go statements. The second source of discomfort is that we are postulating that the electron charge, in units of the fundamental electric charge, is on the order of $10^{15}$, which seems absurd (but which I don't currently know a completely convincing quantum gravity argument against). Furthermore, the magnetic WGC tells us that there would be a UV cutoff on the validity of the $\Uone$ gauge theory description, at a scale $\Lambda_\Uone \lesssim e_0 M_\mathrm{Pl} \sim 1\,\mathrm{TeV}$. This raises a concern that there could be trouble with LHC constraints (and we could use stronger bounds on the photon mass, like that arising from Jupiter's electric field, to find a sharper problem, albeit with more assumptions to examine). However, we really have to know more about what happens at the scale $\Lambda_\Uone$ to see if this is a problem. If there is simply a tower of extremely weakly interacting particles coupling to the Standard Model with strength $e_0$, these could easily have escaped experimental notice. The Tower or Sublattice WGCs tell us that the fundamental quantum gravity cutoff in such a scenario is below $e_0^{1/3} M_\mathrm{Pl} \sim 10^{13}\,\mathrm{GeV}$, which is not an obvious problem. However, the Emergent String Conjecture would suggest that, because there are chiral fermions charged under the photon, the $e_0 \to 0$ limit is not a decompactification limit but a tensionless string limit, and so $\Lambda_\textsc{QG} \lesssim e_0 M_\mathrm{Pl} \sim 1\,\mathrm{TeV}$. 

Suppose, then, that we give up on the Higgs mechanism and simply write down a bare (Proca or Stueckelberg) mass for the hypercharge boson. What could go wrong? I claim that in the quantum gravity context, we really should think of such a massive spin-1 boson as a genuine $\Uone$ gauge field that has eaten a (necessarily) compact scalar boson $\theta$ of charge $q$. This is because, if we consider black holes with radius much less than the Compton radius of the photon, the arguments of~\S\ref{sec:BHnoglobal} should still apply to good approximation, and we recover our conclusions about compactness. Thus, the longitudinal mode of the photon should be a compact boson, not a noncompact one. As discussed in \S\ref{sec:CSStueckelberg}, such a theory can be reinterpreted as a theory of a 2-form gauge field $B$ that is dual to the compact boson $\theta$, in the sense that $\frac{1}{2\cpi} \dif B = f^2 \star \dif\theta$. In this interpretation, the photon mass is due to a $\frac{1}{2\cpi} B \wedge F$ term. Now, we apply the Weak Gravity Conjecture to both the photon (with fundamental coupling $e_0$) and to the $B$ field (with coupling $2\cpi f$). The photon mass is $e_0 q f/(2\cpi)$. Thus, a small photon mass requires either $e_0 \ll 1$ (similar to the discussion in the Higgs case) or a small $f$ (or both). However, now a new problem arises for small $f$: the WGC for $p$-form gauge fields~\eqref{eq:pformWGC} tells us that there are strings charged under the $B$ field with tension ${\cal T} \lesssim 2\cpi f M_\mathrm{Pl}$. In the Higgs case, these strings are ANO vortices, solitonic solutions where the Higgs VEV goes to zero in the string core. However, in the Stueckelberg case, by definition there is no Higgs field and the string core is not describable within effective field theory. As a defining feature of a scenario where the UV completion is {\em fundamentally} of Stueckelberg type, rather than just a limit of the Higgs scenario, the string should be a fundamental object whose core probes deep UV physics, not resolvable within any EFT description. In this case, the tension of the string is at or above the quantum gravity cutoff. In other words, for a given UV cutoff $\Lambda_\textsc{QG}$, we obtain a {\em lower} bound on $f$ from the WGC:
\begin{equation}
\sqrt{2\cpi {\cal T}} \gtrsim \Lambda_\textsc{QG} \quad \Rightarrow \quad f \gtrsim \frac{\Lambda_\textsc{QG}^2}{4\cpi^2 M_\mathrm{Pl}}.
\end{equation} 
Now, very conservatively, we know that $\Lambda_\textsc{QG}$ is at least above the TeV scale, which tells us that $f \gtrsim 10^{-5}\,\mathrm{eV}$. This is not nearly small enough to explain~\eqref{eq:photonmassbound} without also invoking a huge integer $N$, as in the Higgs case. Thus, quantum gravity strongly disfavors the possibility that the photon has a nonzero mass. I expect that, in fact, it completely rules it out, though the argument is not rigorous (and in particular, we would have to close the large-integer loophole~\cite{Craig:2018yld}).

Similar conclusions apply to dark photons. In fact, there are interesting scenarios where string theory compactifications contain dark photons with Stueckelberg masses~\cite{Goodsell:2009xc, Cicoli:2011yh}, and the relationship between their mass and the UV cutoff is consistent with this discussion. These conclusions also apply to the magnetic photon mass discussed in \S\ref{subsubsec:magneticmass}.

\subsection{Can a light $\BminusL$ gauge boson exist?}
\label{subsec:BminusL}

The Standard Model with right-handed neutrinos admits a $\Uone_{\BminusL}$ global symmetry. By right-handed neutrinos, I mean fermions that do not carry any charges under the Standard Model gauge group. They can, potentially, have Dirac masses that pair them up with the ordinary neutrinos inside the lepton doublets, via the Higgs. However, in this context they should {\em not} have a Majorana mass term (which is allowed by all the gauge symmetries), because that would break the $\Uone_{\BminusL}$ symmetry. Beyond that, I assume that all renormalizable terms of the Standard Model are included, but only those non-renormalizable terms consistent with the $\BminusL$ symmetry are allowed. This symmetry is non-anomalous, both in the sense that it has no ABJ anomaly with the Standard Model (which is to say, it is a good global symmetry of the quantum theory) and it has no 't Hooft anomaly (so that it can be gauged). 

It is thus natural to ask: is $\Uone_\BminusL$ a gauge symmetry of our universe?

It is certainly possible that it is a {\em higgsed} gauge symmetry, perhaps with an $O(1)$ coupling, provided that it is broken above the TeV scale. But if it is a massless gauge symmetry, it mediates a new long-range force that acts on ordinary matter. Experimental constraints are very strong: $e_\textsc{B-L} \lesssim 10^{-24}$~\cite{Wagner:2012ui, Heeck:2014zfa}. The associated WGC scale is $e_\textsc{B-L} M_\mathrm{Pl} \lesssim\,\mathrm{keV}$. The minimal (electric) WGC can be satisfied by neutrinos. Even the form of the WGC calling for a {\em tower} of $\BminusL$ charged particles at the keV scale, as discussed in \S\ref{sec:cutofftowers}, is not obviously ruled out, because these particles would interact extremely weakly with ordinary matter. The scale at which the tower drives gravity to strong coupling could be as high as $e_\textsc{B-L}^{1/3} M_\mathrm{Pl} \lesssim 10^{10}\,\mathrm{GeV}$, safely out of reach of current experiments.

However, the Emergent String Conjecture is more powerful. It tells us that we should expect that an extraordinarily weak $e_\textsc{B-L}$ would arise only in one of two limits: a decompactification limit, with a tower of Kaluza-Klein modes, or an emergent string limit. The former would have a higher quantum gravity cutoff, but is incompatible with $\BminusL$, because the Standard Model contains chiral fermions carrying $\BminusL$ charge. Kaluza-Klein modes are never chiral. This leaves the emergent string case, but in that case, we expect the fundamental cutoff to be $\Lambda_\textsc{QG} \sim e_\textsc{B-L} M_\mathrm{Pl}$. This is in clear contradiction with the validity of local EFT well above the keV scale.

This example highlights the utility of sharpened conjectures that have emerged in recent years. It is important to put them on a more solid theoretical footing, so that we can argue that phenomenologically viable scenarios, like an ultralight $\BminusL$ gauge boson, are in fact imcompatible with the principles of quantum gravity as we currently understand them. Conversely, experiments that pursue such scenarios are useful, because they could falsify proposed principles of quantum gravity.

\section{Neutrino masses}

The origin of neutrino masses is not yet known. One possibility is that, in the absence of new degrees of freedom, they arise from a dimension five operator:
\begin{equation}
\frac{1}{M} c_{ij}  \left(h \cdot \ell_i\right) \left(h \cdot \ell_j\right) + \mathrm{h.c.}
\end{equation}
Here $M$ is a mass scale, $c_{ij}$ a dimensionless matrix, and the $i$ and $j$ indices label generations. We have (in unitary gauge)
\begin{equation}
h = \begin{pmatrix} 0 \\ \frac{1}{\sqrt{2}} (v + h^0)\end{pmatrix}, \qquad \ell_i = \begin{pmatrix} \nu_i \\ \ell^-_i \end{pmatrix},
\end{equation}
and $h \cdot \ell$ denotes the antisymmetric contraction of $\SU(2)_\textsc{L}$ indices, so this term becomes
\begin{equation}
\frac{1}{2M} c_{ij} \left(v + h^0\right)^2 \nu_i \nu_j + \mathrm{h.c.},
\end{equation} 
which is a Majorana mass term for the neutrinos. We don't know the absolute neutrino mass scale, but atmospheric neutrino oscillations tell us about a difference in squared masses, $\Delta m^2 \approx 2.4 \times 10^{-3}\,\mathrm{eV}^2$, which implies that at least one neutrino mass is larger than $0.05\,\mathrm{eV}$. If we take $c \lesssim 1$ this corresponds to a mass scale $M \lesssim v^2/(0.05\,\mathrm{eV}) \sim 10^{16}\,\mathrm{GeV}$. This is the well-known {\em seesaw scale} for Majorana neutrino masses. The physics at the scale $M$ could consist of heavy singlet fermions which themselves have Majorana masses, which is the standard seesaw mechanism. However, this is just one possibility, not a requirement. 

The Majorana mass scenario requires no special symmetries, and the associated scale $M$ is below the Planck scale. If the $c_{ij}$ are small (perhaps for whatever---currently unknown---reason many of the Standard Model Yukawa couplings are small), the scale could be even lower. None of this runs into any tension with anything that we know about quantum gravity. In many ways, Majorana neutrinos are the simplest option. This scenario is also testable through neutrinoless double beta decay experiments, though the favored parameter range will not be in reach of near-future experiments.

The other scenario is that neutrinos have {\em Dirac} masses: in this case, there are additional light degrees of freedom in the form of fermions ${\widetilde \nu}_i$ that are neutral under the Standard Model gauge group, with mass terms arising from Yukawa couplings of the form
\begin{equation}
\left(y_\nu\right)_{ij} \left(h \cdot \ell_i\right) {\widetilde \nu}_j + \mathrm{h.c.},
\end{equation}
directly analogous to the up-type quark masses in the Standard Model. These terms preserve a $\Uone_\textsc{L}$ ``lepton number'' symmetry, $\ell_i \mapsto \E^{-\iu \alpha} \ell_i$, ${\widetilde \nu} \mapsto \E^{+\iu \alpha} {\widetilde \nu}$. This symmetry has an ABJ anomaly, but it is can be extended to the non-anomalous $\BminusL$ symmetry. For this mechanism to explain the data, the parameters $y_\nu$ must be much smaller than any other Yukawa couplings in the Standard Model. However, the known Yukawas already span several orders of magnitude for unknown reasons (maybe flavor symmetries, maybe locality in extra dimensions, or some combination thereof). Perhaps a full understanding of this mechanism would also explain why the $y_\nu$ are small. (Another option is to consider ${\widetilde \nu}_j$ to be fundamentally a higher dimension operator, e.g., to realize these neutrinos as composite states.)

This scenario requires more care to embed in quantum gravity. In particular, it requires a symmetry to forbid large Majorana masses for the singlet fermions $\widetilde \nu$, of the form
\begin{equation} \label{eq:Majorana}
M_{ij} {\widetilde \nu}_i {\widetilde \nu}_j + \mathrm{h.c.}
\end{equation}
In quantum gravity, we can't invoke a global symmetry, so the absence of this term should somehow be explained as either a direct consequence or a side effect of gauge symmetry. The obvious candidate is to gauge $\BminusL$. As discussed in \S\ref{subsec:BminusL}, a massless $\BminusL$ gauge boson is highly constrained by data, and incompatible with plausible forms of the WGC. One possibility is that it is gauged but higgsed at some high energy scale. If it is higgsed to nothing, or to a $\mathbb{Z}_2$ subgroup, then the terms~\eqref{eq:Majorana} can be generated below the scale of higgsing. However, if these terms are sufficiently small, the neutrinos can be ``pseudo-Dirac,'' i.e., they can have mostly Dirac masses with small Majorana terms that slightly split each Dirac fermion into two nearly-degenerate Majorana mass eigenstates. However, in this case there are experimental constraints that require that the pseudo-Dirac mass terms be Dirac to {\em extremely} high approximation~\cite{deGouvea:2009fp}, which strongly restricts the possible means of generating Majorana terms.

A simpler scenario is to gauge a $\mathbb{Z}_k$ subgroup of $\BminusL$, with $k > 2$. (Gauging the non-anomalous discrete subgroup of $\BplusL$ is also an option.) This forbids the Majorana mass terms~\eqref{eq:Majorana}, without producing any obvious pathology. Thus, if experiments indicate that neutrino masses are Dirac, a very plausible explanation would be that there is a new discrete gauge symmetry in nature. This would be a profound discovery, as it differs from the known ingredients in the Standard Model. However, it is difficult to decisively test experimentally. Discrete gauge groups in quantum gravity imply the existence of cosmic twist strings, which have Aharonov-Bohm interactions with charged particles. Such strings must exist as dynamical objects in the theory to avoid a generalized 2-form global symmetry~\cite{Heidenreich:2021xpr}. If the $\mathbb{Z}_k$ gauge symmetry is fundamental, rather than a remnant of a higgsed $\Uone$, there is no obvious cosmological mechanism to produce the strings, so there is no straightforward observational constraint. 

Thus, I leave you with a challenge: if neutrino masses are Dirac (or pseudo-Dirac to very good approximation), quantum gravity leads us to expect that there is a new gauge symmetry of nature beyond the Standard Model. Is there a way to confront this expectation with experiment? In the meantime, my bet is on the Majorana scenario (though not with great confidence).

\section{The Strong CP Problem in quantum gravity}

In my opinion, the Strong CP problem that we introduced in \S\ref{sec:StrongCP} is currently the best place to look for an interface between ideas from quantum gravity and experimental particle physics. Solutions to this problem are all closely related to symmetries. One approach assumes that CP or some other generalized parity symmetry is a symmetry of nature~\cite{Nelson:1983zb,Barr:1984qx}. Because these are {\em spacetime} symmetries, the only way that they can be gauged is via quantum gravity. The symmetry breaking phase transition produces exactly stable domain walls~\cite{McNamara:2022lrw}, which must be inflated away for a viable cosmology. Such CP- or parity-based models are interesting, but in my opinion, less plausible than the axion solution.

Both the massless up quark solution and the axion solution to the Strong CP problem can be thought of as gauging the instanton number symmetry of QCD, in the sense discussed in \S\ref{sec:axionasgauge}: the equations of motion lead to a Gauss law constraint that sets the net instanton number in spacetime to zero. In the language of \S\ref{sec:generalizedsymmetry}, we can think of instanton number as a kind of $(-1)$-form symmetry charge: it must be integrated over {\em all} dimensions of spacetime, rather than over a slice of positive codimension. Such a putative symmetry is not as well-understood as higher $p$-form symmetries, but there are reasons to think that $(-1)$-form global symmetries are also forbidden in quantum gravity~\cite{McNamara:2019rup, Heidenreich:2020pkc}, and must be either broken (meaning that there must be some field configurations interpolating between those with different instanton number, which obviously requires a UV completion beyond the original gauge theory) or gauged. Gauging with a chiral fermion current, as in the massless up quark solution, seems unlikely to be realized in quantum gravity. It requires a chiral symmetry that is broken {\em only} by the ABJ anomaly. Such a broken symmetry cannot be gauged, so it is difficult to see why there would not be generic symmetry-violating terms in the action.\footnote{An anomalous chiral symmetry can be gauged by a massive gauge field, as in the Green-Schwarz mechanism. In this case, the gauge fields eats an axion-like field to obtain its mass. Examples of this type in string theory often nonetheless have a {\em different} axion in the low-energy EFT~\cite{Kim:1988dd, Honecker:2013mya, Choi:2014uaa, Buchbinder:2014qca}.} On the other hand, the axion can literally {\em be} a gauge field. We have many examples in string theory in which 4d axions arise from higher-dimensional gauge fields. It has long been appreciated that axions are ubiquitous in string theory~\cite{Witten:1984dg}. What we can add to that, from a more modern perspective, is that they appear to be there not just because we are looking under a lamppost of (for example) highly supersymmetric vacua, but because they have an important job to do: the axions eliminate the would-be $(-1)$-form global instanton number symmetry by gauging it. This gives us a reason to be optimistic that they will still exist even in less familiar corners of the quantum gravity landscape. When we think of axions as a type of gauge field, we are naturally led to ask what the Weak Gravity Conjecture has to say about axion physics.

\subsection{The WGC for axions}

In \S\ref{subsec:WGCpform}, we discussed the generalization of the WGC to $p$-form gauge fields. In \S\ref{sec:axionasgauge}, we argued that axions should be thought of as a type of $0$-form gauge field. This suggests that the WGC can be applied to axions. In fact, this application was already discussed in the original AMNV paper proposing the WGC~\cite{Arkani-Hamed:2006emk}. The analogy is that the gauge coupling $e_p$ becomes the inverse decay constant $1/f$, the objects charged under the axion are instantons, and the analogue of the mass or tension is just the {\em action} $S_\mathrm{inst}$ of the instanton. Thus, the axion WGC is the statement that an instanton should exist with
\begin{equation} \label{eq:axionWGC}
S_\mathrm{inst} \lesssim \frac{M_\mathrm{Pl}}{f}.
\end{equation}
The prefactor that turns $\lesssim$ into a sharp inequality depends, as usual, on the extremality bound, and is expected to be $O(1)$ if all scalar forces have strength comparable to gravity. In the absence of a saxion mode, there are no non-singular gravitational instanton solutions analogous to extremal black holes, so it is unclear if we can assign a precise constant prefactor at all. However, in cases with a saxion mode there is an analogue of the extremality bound for instantons~\cite{Gutperle:2002km, Bergshoeff:2004pg, Bergshoeff:2004fq}, so the analogy can be made precise. In many cases, the axion WGC is just the dimensional reduction of a more well-behaved $p$-form WGC in a higher dimensional theory. (See discussions in~\cite{Heidenreich:2015nta, Hebecker:2016dsw}.) 

The axion WGC can be directly applied to QCD, where we know that there are BPST instantons with $S_\mathrm{inst} = 8\cpi^2/g^2$. The axion WGC says that {\em some} instanton should obey the bound, but we expect that this should apply to the conventional gauge theory instantons. In particular, if the axion is to solve the Strong CP problem, there should not be other contributions to the axion potential that will dominate over the QCD contribution. Thus, a reasonable expectation for a QCD axion is that the axion WGC implies that
\begin{equation} \label{eq:WGCfbound}
f \lesssim \frac{g^2}{8\cpi^2} M_\mathrm{Pl} \sim 10^{16}\,\mathrm{GeV}.
\end{equation}
This is a nontrivial prediction about axion physics from the WGC! It disfavors axions with very high decay constant. Importantly, the range $f \lesssim 10^{16}\,\mathrm{GeV}$ of QCD axion decay constants can realistically be targeted by experiments in the foreseeable future~\cite{Jaeckel:2022kwg}.

Note that there is some ambiguity in the interpretation of~\eqref{eq:WGCfbound} because the coupling $g$ runs and becomes large in the infrared. We do not currently have a sharp enough understanding of the axion WGC to be very precise about this, but we expect that the right interpretation is that $g$ should be evaluated at a UV scale, such as $f$. In the numerical estimate, we have used the GUT value $\alpha_\textsc{GUT} \approx 1/25$. Because the running is slow in the UV, the precise details are not very important.

\subsection{The WGC and axion strings}

The natural next step after considering the WGC for axions is to consider the {\em magnetic} WGC for axions, or equivalently, the WGC for the 2-form gauge field $B$ that is Hodge dual to the axion, with $f^2 \star \dif \theta \equiv \frac{1}{2\cpi} \dif B$. (Recall the discussion in \S\ref{sec:CSStueckelberg}; there is also a subtlety related to the modified Bianchi identity, to which we will return in a moment.) The objects charged under $B$ are strings: these are the {\em axion strings} around which $\theta$ winds. The general $p$-form WGC becomes, in this case,
\begin{equation} \label{eq:axionstringWGC}
{\cal T} \lesssim \gamma 2\cpi f M_\mathrm{Pl},
\end{equation}
for some prefactor $\gamma$ which, as usual, we would read off from the extremality bound for black string solutions, and which we expect to be $O(1)$ unless there are scalar forces that are much stronger than gravity. There are several subtleties for the case of axion strings. The tension of an axion string in 4d has a logarithmic IR divergence associated with the energy cost of the winding of $\theta$ at large distances. We interpret the scale ${\cal T}$ in~\eqref{eq:axionstringWGC} as the tension of the core of the string, not this long-distance contribution. Gravitational backreaction is also important: cosmic strings in 4d (or codimension-2 objects in any number of dimensions) cause a {\em deficit angle} in space. If their tension becomes larger than the Planck scale, they would effectively eat up the entire space, so that no static string solution exists. In this case, there are interesting time-dependent string solutions, explored for example in~\cite{Dolan:2017vmn}. For our current purposes, in light of~\eqref{eq:WGCfbound}, we are interested in values of $f$ well below the Planck scale and we will assume that gravitational backreaction is small.

If we combine~\eqref{eq:WGCfbound} and~\eqref{eq:axionstringWGC}, we obtain a bound on the axion string tension that depends on the gauge coupling and the Planck scale:
\begin{equation} \label{eq:axionstringWGC2}
{\cal T} \lesssim \gamma \frac{g^2}{4\cpi} M_\mathrm{Pl}^2.
\end{equation}
We expect massive string modes (closed loops of string, interpreted as particles) to have mass at the scale $\sqrt{2\cpi {\cal T}} \lesssim g M_\mathrm{Pl}$. But this bounds their mass in terms of the WGC scale for the gauge theory, $g M_\mathrm{Pl}$! 

We believe that this is not a coincidence~\cite{Heidenreich:2021yda}. The reason lies in the subtlety we encountered when constructing the $B$ field dual to the axion when the axion couples to gauge fields~\eqref{eq:dualizeaxion}. We saw that to consistently reproduce the modified Bianchi identity $\dif{\star{\dif \theta}} \propto \mathrm{tr}(G \wedge G)$, it is necessary that the gauge field $B$ shifts under a gauge transformation of the ordinary gauge field $A$. This implies that we cannot simply write a coupling $\int_\Sigma B$ on the axion string worldsheet $\Sigma$: it would not be gauge invariant under an $A$ gauge transformation. This requires charged modes to exist on the axion string worldsheet, an example of the general phenomenon of anomaly inflow. In fact, this was one of the original examples~\cite{Callan:1984sa}.

\begin{figure}[!h]
\centering
\includegraphics [width = 0.45\textwidth]{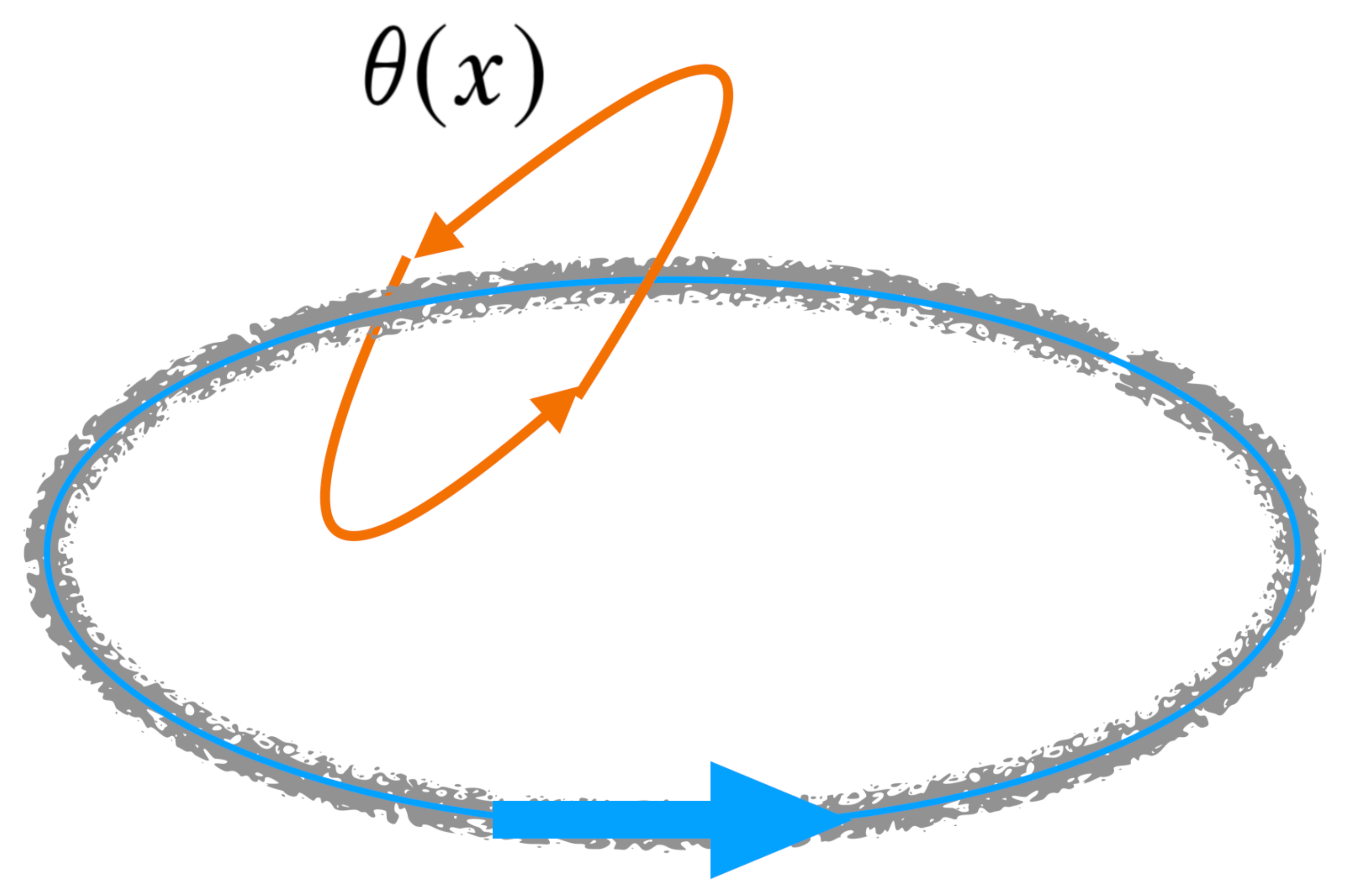}
\caption{Sketch of a closed axion string (fuzzy gray loop) with circulating electric charge under an ordinary gauge group. The orange loop indicates that $\theta$ winds from $0$ to $2\cpi$ when circling the string. The blue arrow indicates the direction of propagation of electric charge along the string.
} \label{fig:axionstringwithcharge}
\end{figure}

Following~\cite{Callan:1984sa}, let's explain this physics again, from a slightly different perspective. Around the axion string, we have $\oint \dif \theta = 2\cpi$. In particular, $\theta$ cannot be well-defined in the core of the string. This motivates rewriting the axion coupling to gauge fields by integrating by parts:
\begin{equation} \label{eq:thetaCS}
\frac{N}{8\cpi^2} \int \theta \,\mathrm{tr}(G \wedge G) = -\frac{N}{2\cpi} \int \dif \theta \wedge J_\textsc{CS},
\end{equation}
with the Chern-Simons current (recall \S\ref{subsec:CSterm3d})
\begin{equation}
J_\textsc{CS} = \frac{1}{4\cpi} \mathrm{tr}\left(A \wedge \dif A - \iu \frac{2}{3} A \wedge A \wedge A\right).
\end{equation}
The form~\eqref{eq:thetaCS} of the term that is most convenient in the presence of an axion string is not gauge invariant under $A$ gauge transformations! In fact, if $A \mapsto A + \dif \alpha$, we have
\begin{equation}
J_\textsc{CS} \mapsto J_\textsc{CS} + \frac{1}{4\cpi} \mathrm{tr}\left(\alpha G\right).
\end{equation}
This makes the action~\eqref{eq:thetaCS} ill-defined on its own, but the string worldsheet action can {\em also} not be gauge invariant, in such a way that the sum of~\eqref{eq:thetaCS} and the string worldsheet action {\em is} invariant. The way that this can happen is that the $(1+1)$d string worldsheet hosts an {\em anomalous} gauge theory with chiral, charge-carrying excitations.  Figure~\ref{fig:axionstringwithcharge} depicts a closed axion string loop with such a charge-carrying mode excited.

To summarize, if an axion couples to gauge fields, axion strings can always admit charge under the gauge field. The combination of the axion WGC bound on the decay constant~\eqref{eq:WGCfbound} and the magnetic axion WGC bound on the string tension~\eqref{eq:axionstringWGC} then tells us, as in~\eqref{eq:axionstringWGC2}, that the axion string modes obey the WGC for the {\em ordinary} gauge field (at least up to $O(1)$ prefactors; see~\cite{Cota:2022yjw}). Such a ``mixing'' of WGC bounds among different gauge fields that couple via Chern-Simons terms is quite general~\cite{Heidenreich:2021yda, Kaya:2022edp}.

This leads us to conclude:

\begin{framed}
\noindent
{\em Axion strings are WGC towers.} When a gauge field couples to an axion, the tower of charged particles satisfying the WGC for the gauge field can arise from closed loops of axion string with charge circulating around the string. 
\end{framed}

As discussed in \S\ref{subsec:WGCpform}, there is an important distinction between {\em solitonic} axion strings, which arise in conventional 4d axion theories like the KSVZ model and have a core described within effective field theory, and {\em fundamental} axion strings with cores that probe UV physics. In string theory, the latter often arise as either the fundamental F-string of string theory or as D-branes wrapped on internal cycles. It is the latter type of string that we expect to give rise to a tower of states corresponding to a fundamental UV cutoff on the theory.

A theory with a very light axion has a very good approximate global symmetry in the infrared: the shift symmetry of the axion. We expect that in quantum gravity all global symmetries are badly broken in the ultraviolet; for an axion arising as a mode of a higher-dimensional gauge field, the symmetry breaking is expected to be via towers of charged objects, as discussed in \S\ref{sec:cutofftowers}. One can ask: how good can an approximate global symmetry be in the infrared? It has been proposed that any symmetry will be broken by effects of order $\exp(-S_\textsc{BH})$ where $S_\textsc{BH}$ is the entropy of the smallest semiclassical black hole (see~\cite{Fichet:2019ugl, Daus:2020vtf} and an interesting recent semiclassical calculation~\cite{Bah:2022uyz}). If this black hole has radius $\Lambda_\textsc{QG}^{-1}$, this estimate corresponds to $\exp(-8\cpi^2 M_\mathrm{Pl}^2/\Lambda_\textsc{QG}^2)$. Comparing this to the suppression of an axion potential generated by gauge theory instantons, $\exp(-8\cpi^2/g^2)$, we see that the instanton effects are dominant provided $\Lambda_\textsc{QG} \leq g M_\mathrm{Pl}$, consistent with a cutoff at the scale where the WGC tower appears. 

\subsection{The expected axion}

I will now briefly summarize some expectations that I have about axions, based on work in progress that will be published elsewhere. In this subsection, I will allow myself to venture further out on a limb than I have elsewhere in these lecture notes. 

We have argued that {\em if} an axion exists in a theory of quantum gravity, the WGC implies an upper bound on its decay constant and a corresponding upper bound on the tension of axion strings, and that the axion strings may supply the WGC tower of charged states. We have also pointed out that the axion can be thought of as gauging the $(-1)$-form instanton number global symmetry, and that there are reasons to expect that quantum gravity requires that this symmetry be either gauged or broken. However, this leaves open the possibility that the symmetry is simply {\em broken}, without any axion field in the effective theory.

There are a number of examples of quantum gravity theories in which we find gauge fields that do not couple to an associated axion. These gauge fields include the graviphoton field in rigid Calabi-Yau compactifications~\cite{Cecotti:2018ufg} and Kaluza-Klein gauge fields (which may couple to axions in various ways, but not the standard $\theta F \wedge F$ coupling~\cite{Heidenreich:2020pkc,Heidenreich:2021yda,Grimm:2022xmj}). There are cases where an axion coupled to a gauge field exists, but has a large tree-level mass of monodromy type~\cite{Silverstein:2008sg}, i.e., a $\theta \dif C_3$ coupling of the general BF form discussed in \S\ref{sec:CSStueckelberg}. This allows an instanton to dissolve into $\dif C_3$ flux, which is tantamount to breaking the $(-1)$-form instanton number symmetry. Thus, one might conclude that some gauge fields in quantum gravity theories couple to light axions, and others do not, so a general argument will not shed light on whether or not we expect a light axion coupled to gluons to exist in our universe.

However, the examples of which I am aware in which one finds gauge fields without a light axion are {\em also} examples lacking light charged matter. For example, in Kaluza-Klein compactifications, fields with Kaluza-Klein charge necessarily have a mass at the cutoff scale at which the 4d EFT breaks down. In the rigid Calabi-Yau case, the charged objects are D3 branes wrapped on the holomorphic 3-cycle of the Calabi-Yau, with mass around the 4d Planck scale. Such examples leave open the conjecture that quantum gravity theories with gauge fields coupled to light charged matter must have a gauged, rather than broken, $(-1)$-form instanton number symmetry, with a corresponding light axion field. One heuristic reason for thinking this might be true is that if the instanton number symmetry is broken, we expect that it should be badly broken at the cutoff scale. This means that an instanton configuration of size $\Lambda_\textsc{QG}^{-1}$ should be easily deformed into configurations with no instanton number at all. However, if there are light charged fermions in the theory, such instanton field configurations have long-range fermionic zero modes attached, as discussed in \S\ref{sec:oneloopexactness}. These act to stabilize the instanton configuration against short-range deformations that could destroy it.

This perspective also resonates with the Emergent String Conjecture. For asymptotically weak-coupling limits, it states that the tower of light modes is either a set of Kaluza-Klein modes or a set of string modes. The Kaluza-Klein case is incompatible with light charged matter; the string case resembles an axion string (see other related comments on axion strings in~\cite{Lanza:2020qmt, Lanza:2021udy}). What I am suggesting here is stronger, in that my remarks are not restricted to asymptotic limits. This is likely necessary to make contact with phenomenology: the Standard Model gauge couplings are $O(1)$ numbers, suggesting that at least the moduli that control these couplings are far from any asymptotic regime.

The suggestion that gauge theories with light charged matter fields require a light axion is an example of a compelling phenomenological claim that may be derivable from quantum gravity. Any UV completion of the Standard Model that respects some plausible general principles, such as the absence of generalized global symmetries, may be required to contain a light axion field, with the crucial interaction with gluons that makes it at least a strong candidate for solving the Strong CP problem. This suggestion is consistent (in a nontrivial way) with a large body of evidence, and amenable to further study. A counterexample of a string compactification with gauge fields and light charged matter but no light axion would immediately imply that I'm on the wrong track, unless it has other unusual features that could motivate a refined conjecture.

My expectation, based on my current understanding of quantum gravity, is that a light axion field with a $\theta\,\mathrm{tr}(G \wedge G)$ coupling exists in our universe. The axion is likely to arise from a higher-dimensional gauge field. Furthermore, the axion decay constant should be near the fundamental UV cutoff of the theory, with the axion string as a fundamental object rather than a solitonic one. More specifically, I expect that
\begin{equation}
\Lambda_\textsc{QG} \lesssim \frac{4\pi^2}{g} f,
\end{equation} 
with $f$ the axion decay constant (which is experimentally measurable) and $g$ the QCD coupling evaluated at high energies. This is a stronger statement than~\eqref{eq:axionstringWGC2}, based on the expectation that if $f$ is far below the upper bound in~\eqref{eq:axionWGC} with $O(1)$ coefficient, the axion string tension will be correspondingly far below the upper bound~\eqref{eq:axionstringWGC} with $O(1)$ coefficient, because the prefactors in both cases become small in the same limit of strong scalar forces. There are several assumptions underlying these claims, which I will present in more detail elsewhere. What I hope that readers will take away is that there is at least the potential to extract nontrivial expectations about phenomenology from our current knowledge of quantum gravity. These are not rigorous theorems, but they rest on plausible general assumptions, and if they are falsified we will learn that quantum gravity in our universe behaves rather differently from the examples of consistent quantum gravity theories that we know so far.

\section{Closing remarks}

This section of these lecture notes might age more rapidly than the rest, but I would like to give some big-picture thoughts about the current state of particle physics, and how I would situate the topics discussed in these notes within that broader context.

Before 2012, a large fraction of the particle physics community agreed on one urgent goal: understand the mechanism of electroweak symmetry breaking. Since the LHC discovered a particle with all the expected properties of the Higgs boson, completing the Standard Model, the priorities have become less clear. We lack a theory of dark matter and of the matter-antimatter asymmetry in the universe. The Standard Model itself has many unexplained small parameters. These remain important problems, but the community's confidence that the answers will be discovered by current of near-future experiments has decreased. In this environment, what should we do?

Of course, opinions vary widely, but here are some of mine. On the experimental front, from the bottom up, there are several experiments that have the potential to push constraints on physics beyond the Standard Model into the many-TeV range. These include searches for electric dipole moments~\cite{Alarcon:2022ero} and for charged lepton flavor violation (e.g., $\mu \to e$ conversion when scattering on a nucleus)~\cite{Baldini:2018uhj}. These have in common that the Standard Model prediction is many orders of magnitude below current sensitivity, so there is a large territory in which any signal at all would constitute a definitive discovery of new physics. The reason they have such high sensitivity is that they search for new violation of symmetries that are broken in only very mild ways by known physics. 

In my opinion, it should be an urgent priority for the field to build a new high-energy collider, with reach well beyond that of the LHC, whether this is a hadron collider, a muon collider, of (if technology develops fast enough) a high-energy linear $e^+ e^-$ collider. If we put this off until several decades in the future, both the expertise and the enthusiasm for high-energy colliders that currently exists in the community will dwindle, and will be difficult to recreate. But it may be that the community will not rally behind such a high-energy machine until we get a clear signal from somewhere else, like an EDM, that there is new physics not too far away.

Efforts to study particle physics using astrophysics and cosmology have greatly expanded in recent years. It is possible that we will learn dramatically different information from such probes of new physics than we could learn from any terrestrial experiment. For example, a measurement of a nonzero tensor to scalar ratio $r$ in the CMB would tell us that the Hubble scale during inflation was large (say, $10^{13}\,\mathrm{GeV}$), providing one anchor at a very high energy scale that could be the entry point for learning more (e.g., through non-Gaussianities) about heavy particles that could have existed during inflation. We should enthusiastically pursue such well-motivated opportunities to learn about high energy physics from clues left behind from the early universe.
 
While there are a number of such well-motivated experimental and observational directions to pursue, recent years have also seen an explosion of phenomenological and experimental fishing expeditions, searching for models that have no particular theory motivation simply because we can. There is nothing wrong with this, to the extent that it can be done affordably and we might stumble across something unanticipated and exciting. But I think that we must also continue, from the theory side, to seek a better understanding of where we should {\em expect} to find new physics. The hubristic claims of some theorists in advance of the LHC have cast such pursuits in a bad light, but we should not forget that often in the history of physics, reasoning about theory from general principles has led to conclusions that were confirmed by experiments decades later.

Developments in quantum gravity provide an often overlooked tool for thinking about particle physics. Quantum gravity is extraordinarily difficult to connect with experiment, because it seems to offer no guarantee of new physics below the Planck scale, and there seem to be an enormous number of consistent quantum gravity theories (or vacua of one theory). But recent years have shown us that some of the oldest posited principles of quantum gravity, such as the absence of global symmetries, have much richer implications than anyone had previously anticipated. We should push such principles as far as we can, to try to find sharp confrontations between their predictions and experiment or observation. My current belief is that the Strong CP problem offers the most promising arena for bridging the gap between such principles and the real world. I think that it should be a high priority of the theory community to assess to what extent the existence of a light axion follows from general principles of quantum gravity, together with our existing knowledge about particle physics in our universe. More generally, axions and neutrinos with Majorana masses are two of the obvious places in particle physics where a very high energy scale appears in the denominator of a mass, potentially linking light particles with physics at the quantum gravity cutoff scale. This is a way that, even in standard quantum field theory, ultraviolet energies can leave behind clues in the far infrared. We should make the most of such opportunities.

I am optimistic that the next two decades will see a convergence between pure theory and dramatic experimental results. I hope that these lecture notes will provide some young theorists with useful tools to contribute to this effort.

\newpage

{\small
\bibliography{ref}
\bibliographystyle{utphys}
}

\end{document}